\documentclass[12pt,twoside,openany]{book}

\usepackage{color}
\definecolor{indigo}{RGB}{0,0,120}
 
\usepackage[top = 2.5cm, bottom = 2.5cm, left = 2.5cm, right = 2.5cm]{geometry}

\usepackage[nocompress]{cite}
\usepackage[pdftex]{graphicx}
\usepackage{calligra} 
\usepackage{subcaption}
\usepackage{wrapfig}
\usepackage{comment}
\usepackage{amsmath}
\usepackage[colorlinks=true, linkcolor=indigo, citecolor=blue, urlcolor=blue]{hyperref}
\usepackage{amssymb}
\usepackage{cancel}

\newcommand\abstractname{Abstract}

\makeatletter
\if@titlepage
  \newenvironment{abstract}{%
      \titlepage
      \null\vfil
      \@beginparpenalty\@lowpenalty
      \begin{center}%
        \bfseries \abstractname
        \@endparpenalty\@M
      \end{center}}%
     {\par\vfil\null\endtitlepage}
\else
  
\fi
\makeatother

\newcount\colveccount  
\newcommand*\colvec[1]{\global\colveccount#1  \begin{pmatrix} \colvecnext} \def\colvecnext#1{#1 \global\advance\colveccount-1
        \ifnum\colveccount>0 \\ \expandafter\colvecnext
        \else \end{pmatrix} \fi}
        
\DeclareMathAlphabet{\mathcalligra}{T1}{calligra}{m}{n}
\DeclareFontShape{T1}{calligra}{m}{n}{<->s*[2.2]callig15}{}
\newcommand{\scripty}[1]{\ensuremath{\mathcalligra{#1}}}

\def\tr{\;{\rm tr}\;}
\def\Tr{\;{\rm Tr}\;}
\def\sgn{\;{\rm sgn}\;}

\def\imply{\Rightarrow}

\def\fl{\noindent}

\newcommand{\bra}{\langle}
\newcommand{\ket}{\rangle}

\newcommand{\tl}[1]{\tilde{#1}}
\newcommand{\dd}[2]{\frac {\partial #1}{\partial #2}}
\newcommand{\deldel}[2]{\frac {\delta #1}{\delta #2}}
\newcommand{\pdr}{\partial}
\newcommand{\DD}[2]{\frac {d #1}{d #2}}

\newcommand{\beq}{\begin{equation}}
\newcommand{\eeq}{\end{equation}}
\newcommand{\beqs}{\begin{eqnarray}}
\newcommand{\eeqs}{\end{eqnarray}}

\newcommand{\half}{\frac{1}{2}}
\newcommand{\ov}[1]{\frac{1}{#1}}

\newcommand{\sign}[1]{{\rm sgn}({#1})}

\def\al{\alpha}		
\def\g{\gamma} 		
 
\def\del{\delta}	
\def\D{\Delta}		
\def\eps{\epsilon} 
\def\veps{\varepsilon}

\def\la{\lambda}	
\def\La{\Lambda}	
\def\sig{\sigma}		
		
\def\tht{\theta}	
		
\def\om{\omega}		
\def\Om{\Omega}

\newcommand{\bfv}{{\bf v}}

\newcommand{\bfr}{{\bf r}}

\newcommand{\bfE}{{\bf E}}
\newcommand{\bfB}{{\bf B}}

\newcommand{\Blue}{\color{blue}}

\begin{document}

\frontmatter
\thispagestyle{empty}
\begin{center}

\vspace*{5mm}

{\LARGE \bf Integrability and dynamics}

\vspace{.3cm}

{\LARGE \bf of the Rajeev-Ranken model}
\vspace{1cm}
\\{\Large by}\\ 
\vspace{1cm}
{\Large \textbf{T R Vishnu}} 
\large 

\vspace{2cm}

{\it A thesis submitted in partial fulfillment of the requirements for \\ the degree of Doctor of Philosophy in Physics} \\
\vspace*{0.8cm}
to  
\vspace*{0.5cm} 

Chennai Mathematical Institute

\vspace{0.5cm}

Submitted: May, 2021\\
Defended: September 15, 2021

\vspace{1cm}

\begin{figure}[h]
\begin{center}
\includegraphics[width = 7cm]{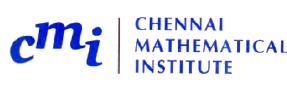}
\end{center}
\end{figure}

\vspace{0.1cm}

Plot H1, SIPCOT IT Park, Siruseri,\\
Kelambakkam, Tamil Nadu 603103, \\
India

\end{center}

\newpage
\thispagestyle{empty}

\begin{flushleft}

Advisor:

Prof.~Govind~S~Krishnaswami, \emph{Chennai Mathematical Institute (CMI).}\\ 

\vspace*{1cm}

Doctoral Committee Members: 

\begin{enumerate}

\item Prof.~V~V~Sreedhar, \emph{Chennai Mathematical Institute (CMI).}\\
\item Prof.~Ghanasyam~Date, \emph{Chennai Mathematical Institute (CMI).}

\end{enumerate}
\end{flushleft}

\newpage

\chapter*{\center Declaration}

This thesis is a presentation of my original research work, carried out under the guidance of Prof. Govind S Krishnaswami at the Chennai Mathematical Institute. This work has not formed the basis for the award of any degree, diploma, associateship, fellowship or other titles in Chennai Mathematical Institute or any other university or institution of higher education.

\vspace*{5mm}
\begin{flushright}
T R Vishnu\\
May 25, 2021
\end{flushright}

\vspace*{30mm}

In my capacity as the supervisor of the candidate's thesis, I certify that the above statements are true to the best of my knowledge.
\vspace*{5mm}
\begin{flushright}
Govind S Krishnaswami\\
May 25, 2021
\end{flushright}

\newpage

\chapter*{Acknowledgments}
\addcontentsline{toc}{chapter}{Acknowledgments}

First and foremost, I would like to thank my supervisor Govind S Krishnaswami. His constant guidance, wisdom and patience are the foundations of this thesis. The lengthy discussion sessions with him have helped me in approaching physics from a research perspective. It is worth learning how he breaks a problem into small solvable pieces and adds the results from each of them to answer a bigger question. I have been trying to incorporate this idea in my research. Moreover, he has helped me acquire skills for oral presentation of results and for writing manuscripts in a logical and precise manner by demanding more from me and criticising when needed.

I would like to extend my thanks to my doctoral committee members V V Sreedhar and Ghanasyam Date. I had fruitful interactions with them while discussing the progress of my research work. I would also like to thank K G Arun, Alok Laddha, H S Mani, N D Hari Dass, T R Govindarajan and Sujay Ashok, who taught me courses and gave me valuable insights. I thank K Narayan and Amitabh Virmani from the Physics department for their time and help during my research at the  Chennai Mathematical Institute (CMI). I would also like to thank S G Rajeev and Dileep Jatkar for carefully reading this thesis and for their questions and suggestions.

I thank the organisers of the conferences and schools: Integrable systems in Mathematics, Condensed Matter and Statistical Physics (ICTS, 2018), Conference on Nonlinear Systems and Dynamics (JNU, 2018), Young Researchers Integrability School and Workshop: A modern primer for 2D CFT (ESI, 2019), XXXIII SERB Main school-Theoretical High Energy Physics (SGTB Khalsa College, 2019) and Lecture series on Basics of nonlinear integrable systems and their applications (SASTRA University, 2021), for selecting me to participate in them. These programs have helped me to widen my knowledge in my area of research.

I thank my colleagues and friends from CMI for their love and support. In particular, I would like to acknowledge the support of my friends Krishnendu N V, Sonakshi Sachdev, Himalaya Senapati, Kedar S Kolekar, Sachin Phatak, Debangshu Mukherjee, Kishor Shalukhe, Navaneeth Mohan, Aswin P, A Manu, Ramadas N, Aneesh P B, Athira P V, Pratik Roy, Debodirna Ghosh, Sourav Roy Choudhary, Shanmugapriya and Saleem Muhammed from the Physics department. I also thank my friends from the Computer Science and Mathematics departments, especially Govind R, Sayan Mukherjee, Keerthan Ravi, Aashish Satyajith, Muthuvelmurugan, Anbu Arjunan, S P Murugan, Praveen Kumar Roy, Naveen Kumar, Abhishek Bharadwaj, Rajib Sarkar, Deepak K D and Sarjick Bakshi. The warm friendship and support from these people made my time in CMI a memorable one. I also thank all the other academic, administrative, housekeeping, mess and security staff of the Institute for sparing their time to help me during my time at CMI.

I would like to acknowledge the research scholarship support from the Science and Engineering Research Board, Govt. of India, the  Infosys Foundation, the J N Tata Trust and CMI. Moreover, I  am also grateful to CMI for supporting my travel to Vienna to attend the YRISW-2019 conference. 

Several people have given me support and encouragement that led me to pursue research in Physics. First of all, I would like to thank all my teachers who influenced and moulded me to fit in academia. These teachers include Swapna, Smitha, R Sreekumar, S I Issac, S Nagesh, Rukmani Mohanta, K P N Murthy, A K Kapoor, S Dutta Gupta, S N Kaul, Bindu Bambah, S Chaturvedi and E Harikumar. I would also like to thank my cousins Archana, Keerthana and Akshay for their support. I also thank all my friends, especially Vishnuduth, Jayaram, Vaisakh, Tom, Dins, Nimisha, Sanjay, Arya, Anjana, Zuhair, Anjali and Subhish for their love and encouragement. I extend my gratitude to all my family members who have encouraged me to pursue my dream.

Finally, I would like to thank my brother T R Jishnu and my sister-in-law R Gopika for their constant encouragement. Above all, I thank my parents, A Jayasree and K P Ravi, for their support and unconditional love. Without them, I would not be able to chase my dreams. I would like to dedicate this thesis to my parents and teachers. 

\newpage

\chapter*{\center Abstract}
\addcontentsline{toc}{chapter}{Abstract}

This thesis concerns the dynamics and integrability of the Rajeev-Ranken (RR) model, a mechanical system with 3 degrees of freedom describing screw-type nonlinear wave solutions of a scalar field theory dual to the 1+1D SU(2) Principal Chiral Model. This field theory is strongly coupled in the UV and could serve as a toy model to study nonperturbative features of theories with a perturbative Landau pole. 

We begin with a Lagrangian and a pair of Hamiltonian formulations based on compatible degenerate nilpotent and Euclidean Poisson brackets. Darboux coordinates, Lax pairs and classical $r$-matrices are found. Casimirs are used to identify the symplectic leaves on which a complete set of independent conserved quantities in involution are found, establishing Liouville integrability. Solutions are expressible in terms of elliptic functions and their stability is analyzed. The model is compared and contrasted with those of Neumann and Kirchhoff.

Common level sets of conserved quantities are generically 2-tori, though horn tori, circles and points also arise. On the latter, conserved quantities develop relations and solutions degenerate from elliptic to hyperbolic, circular and constant functions. The common level sets are classified using the nature of roots of a cubic polynomial. We also discover a family of action-angle variables that are valid away from horn tori. On the latter, the dynamics is expressed as a gradient flow.

In Darboux coordinates, the model is reinterpreted as an axisymmetric quartic oscillator. It is quantized and variables are separated in the Hamilton-Jacobi and Schr\"odinger equations. Analytic properties and weak and strong coupling limits of the radial equation are studied. It is shown to reduce to a generalization of the Lam\'e equation. Finally, we use this quantization to find an infinite dimensional reducible unitary representation of the above nilpotent Lie algebra.


\newpage

\chapter*{List of publications}
\addcontentsline{toc}{chapter}{List of publications}

This thesis is based on the following papers co-authored with my supervisor Prof. Govind S. Krishnaswami.

\begin{enumerate}

\item On the Hamiltonian formulation and integrability of the Rajeev-Ranken model, \href{https://doi.org/10.1088/2399-6528/ab02a9}{J. Phys. Commun. {\bf 3}, 025005 (2019)}, \href{https://arxiv.org/abs/1804.02859}{arXiv:1804.02859[hep-th]}.

\item Invariant tori, action-angle variables and phase space structure of the Rajeev-Ranken model, \href{https://aip.scitation.org/doi/10.1063/1.5114668}{J. Math. Phys. {\bf 60}, 082902 (2019)}, \href{https://arxiv.org/abs/1906.03141}{arXiv:1906.03141[nlin.SI]}. 

\item An introduction to Lax pairs and the zero curvature representation, \href{https://arxiv.org/abs/2004.05791}{arXiv:2004.05791[nlin.SI]}.

\item The idea of a Lax pair - Part I: Conserved quantities for a dynamical system, \href{https://www.ias.ac.in/describe/article/reso/025/12/1705-1720}{Resonance {\bf 25}, 1705 (2020).}

\item The idea of a Lax pair - Part II: Continuum wave equations, \href{https://doi.org/10.1007/s12045-021-1124-1}{Resonance {\bf 26}, 257 (2021).}

\item The quantum Ranjeev-Ranken model as an anharmonic oscillator, Preprint in preparation.

\end{enumerate}

\small

\tableofcontents

\newpage
\thispagestyle{empty}

\normalsize

\mainmatter

\chapter[Introduction]{Introduction}
\chaptermark{Introduction}
\label{chapter:Introduction}

\section[Motivation and background]{Motivation and background \sectionmark{Motivation and background}}
\sectionmark{Motivation and background}
\label{s:motivation-and-background}

In this thesis we investigate the dynamics and integrability of a mechanical system describing a class of nonlinear wave solutions of a 1+1-dimensional (1+1D) scalar field theory.  This scalar field theory was introduced in the work of Zakharov and Mikhailov \cite{Z-M} and Nappi \cite{Nappi}. It is `pseudodual'\footnote{This scalar field is obtained from a noncanonical transformation of the principal chiral field. Moreover, while the models are classically equivalent, their quantum theories are qualitatively different. This motivates the term `pseudodual'.} to the 1+1D SU(2) principal chiral model (PCM), which is equivalent to the 1+1D SO(4) nonlinear sigma model. The latter is an effective theory for pions, displays asymptotic freedom and possesses a mass gap \cite{Polyakov}. It serves as a good toy model for 3+1D Yang-Mills theory, which describes the physics of strong interactions. The PCM and nonlinear sigma model are prime examples of integrable field theories and nonperturbative results concerning their $S$-matrix and spectrum have been obtained using the methods of integrable systems by Zamolodchikov and Zamolodchikov \cite{Z-Z} (factorized $S$-matrices), by Polyakov and Wiegmann \cite{P-W} (fermionization) and by Faddeev and Reshetikhin \cite{F-R} (quantum inverse scattering method).

Unlike the PCM, its pseudodual scalar field theory is strongly coupled in the ultraviolet and displays particle production. Thus, as pointed out by Rajeev and Ranken {\cite{R-R}}, this scalar field theory could serve as a lower-dimensional toy model for studying certain nonperturbative aspects of theories with a perturbative Landau pole (such as 3+1D $\la \phi^4$ theory, which appears in the scalar sector of the Standard Model of particle physics). In particular, one wishes to identify degrees of freedom appropriate to the description of the dynamics of such models at high energies (if indeed, a UV completion can be defined). Though the pseudodual scalar field theory has been shown by  Curtright and Zachos \cite{C-Z} to possess infinitely many {\it nonlocal} conservation laws, it has not yet been possible to solve it in anywhere near the way that the PCM has been solved.  The pseudodual scalar field theory is also interesting for other reasons. Unlike the PCM, which is based on the semi-direct product of an $\mathfrak{su}(2)$ current algebra and an abelian current algebra, its pseudodual is based on a nilpotent current algebra and a quadratic Hamiltonian. Theories that admit a formulation in terms of quadratic Hamiltonians and nilpotent Lie algebras are particularly interesting: they include the harmonic and anharmonic oscillators as well as field theories such as Maxwell, $\la \phi^4$ and Yang-Mills. Based on this structural similarity, it is plausible that some common techniques of analysis may apply to several of these models.

There are yet other reasons to be interested in the PCM, its pseudodual scalar field theory and more generally the pseudoduality transformation. For instance, a generalization of the PCM to a centrally-extended Poincar\'e group leads to a model for gravitational plane waves \cite{N-W}. On the other hand, a generalization to other compact Lie groups shows that the pseudodual models have 1-loop beta functions with opposite signs \cite{Alvarez}. Interestingly, the sigma model for the noncompact Heisenberg group \cite{B-Y} is also closely connected to the above pseudodual scalar field theory that we study. Similar duality transformations have also been employed in the $AdS_5 \times S^5$ superstring sigma model in connection with the Pohlmeyer reduction \cite{G-T} and in integrable $\la$-deformed sigma models \cite{G-S-S}. The above dual scalar field theory also arises in a large-level and weak-coupling limit of the 1+1D SU(2) Wess-Zumino-Witten model. This field theory is also of interest in connection with the theory of hypoelliptic operators \cite{R-R}. In another direction of some relevance, attempts have been made to understand the connection (or lack thereof) between the absence of particle production, integrability and factorization of the tree-level S-matrix in massless 2D sigma models \cite{H-L-T}.

As a step towards understanding the 1+1D scalar field theory dual to the SU(2) PCM, Rajeev and Ranken \cite{R-R} obtained a consistent mechanical reduction to a class of nonlinear constant energy-density classical waves. These novel `screw-type' continuous waves could play a role similar to solitary waves in other field theories. The restriction of the scalar field theory to these nonlinear waves is governed by a Hamiltonian system with 3 degrees of freedom, which we refer to as the Rajeev-Ranken (RR) model.

In this thesis, we will explore the integrability and dynamics of the RR model and obtain results on both its classical and quantum versions. Aside from its intrinsic interest, we hope that understanding the mechanical model in detail will shed light on its parent scalar field theory.  Moreover, comparing the RR model and its  integrable features with other dynamical systems has been very helpful in discovering common features and transplanting ideas between these models. We next outline the major results of this thesis.

\section[Outline and summary of results]{Outline and summary of results\sectionmark{Outline and summary of results}}
\sectionmark{Outline and summary of results}
\label{s:outline-and-summary-of-results}

In Chapter \ref{chapter:Nilpotent-field-theory-to-the-Rajeev-Ranken-model}, we introduce the 1+1D scalar field theory pseudodual to the SU(2) principal chiral model. The SU(2)-group valued principal chiral field $g(x,t)$ is related to the $\mathfrak{su}(2)$-Lie algebra valued scalar field $\phi(x,t)$ via the noncanonical transformations
	\beq
	g^{-1} g' = \la \dot \phi \quad \text{and} \quad g^{-1} \dot g = \la \phi'.
	\label{e:relation-principal-chiral-and-scalar-fields}
	\eeq
Here, primes and dots denote space and time derivatives respectively and $\la > 0$ is a dimensionless coupling constant. We then discuss the Hamiltonian-Poisson bracket formulations of the PCM and its dual scalar field theory. We briefly mention salietnt features of the models and point out that unlike the `Euclidean' current algebra of the PCM, the scalar field theory is based on a step-3 nilpotent current algebra. Next, we sketch the way Rajeev and Ranken obtained a mechanical system by reducing the scalar field theory to screw-type waves of the form:
	\beq
	\phi(x,t) = e^{Kx}R(t)e^{-Kx} + mKx\quad \text{with} \quad 
	K = \frac{i}{2} k \sigma_3.
	\label{e:screw-type-nonlinear-wave}
	\eeq
Here, $R(t)$ is a dynamical traceless $2 \times 2$ anti-hermitian $\mathfrak{su}$(2) matrix, while $K$ is a constant matrix. In (\ref{e:screw-type-nonlinear-wave}), $m$ is a dimensionless parameter, $k$ a constant wavenumber and $\sigma_3$ the third Pauli matrix. The dynamics of these screw-type waves is described by a Hamiltonian system with three degrees of freedom and its equations of motion (EOM) are
	\beq
	\dot{L} = \left[K, S\right] 
	\quad \text{and} \quad 
	\dot{S} = \la \left[S, L\right].
	\eeq
Here $S(t)$ and $L(t)$ are dynamical $\mathfrak{su}$(2) matrices related to $R(t)$ via
	\beq
	L = \left[ K, R \right] + mK
	\quad \text{and}
	\quad S = \dot{R} + \ov{\la} K.
	\eeq
The matrices $L$ and $S$ may also be regarded as a pair of dynamical  vectors in 3D Euclidean space ($\vec L = \Tr (L \vec \sigma/2i), \vec S = \Tr (S \vec \sigma/2i)$) equipped with the cross-product Lie bracket. Thus the phase space of the RR model is six-dimensional.

In Chapter \ref{chapter:Hamiltonian-formulation-and-Liouville-integrability}, we discuss the Hamiltonian formulation and Liouville integrability of the RR model. In Section \ref{s:Hamiltonian-PB-Lagrangian-RR-model}, we find a Lagrangian as well as a pair of distinct Hamiltonian-Poisson bracket formulations for the RR model. The corresponding nilpotent and Euclidean Poisson brackets are shown to be compatible and to generate a (degenerate) Poisson pencil. In Section \ref{s:Lax-pairs-r-matrices-and-conserved-quantities-RR-model}, Lax pairs (see Refs.~\cite{G-V-3, G-V-4, G-V-5} for an exposition on Lax pairs) and $r$-matrices associated with both Poisson structures are obtained and used to find four generically independent conserved quantities ${\mathfrak{c}}, m, s$ and $h$. They are related to the $S$ and $L$ variables via 
	\beqs
	{\mathfrak{c}} k^2 &=& \Tr \left(\frac{L^2}{2} - \frac{1}{\la}KS \right) = \half L_a L_a + \frac{k}{\la} S_3, \cr
	m k^2 &=& \Tr KL = - k L_3, \quad 
	s^2 k^2 = \Tr S^2 \quad \text{and} \quad
	h k^2 = \Tr SL.
	\eeqs
Here, ${\mathfrak{c}}$ and $m$ may be shown to be Casimirs of the nilpotent Poisson algebra. The value of the Casimir $L_3$ is written as $-m$ in units of $k$ by analogy with the eigenvalue of the angular momentum component $L_z$ in units of $\hbar$. The conserved quantity $\Tr SL$ is called $h$ for helicity by analogy with other such projections. The quantity $s^2 k^2$ is the square of the radius of the $S$-sphere in the 3D Euclidean $S$-space. These conserved quantities are in involution with respect to both Poisson structures on the 6D phase space. The symmetries and canonical transformations generated by these conserved quantities are identified and three of their combinations are related to Noether charges of the nilpotent scalar field theory. Two of these conserved quantities ${\mathfrak{c}}$ and $m$ (or $s$ and $h$) are shown to lie in the center of the nilpotent (or Euclidean) Poisson algebra. Thus, by assigning numerical values to the Casimirs, we may go from the 6D phase space of the model to its 4D symplectic leaves $M^4_{{\mathfrak{c}} m}$ (or $M^4_{s h}$). On the latter, we have two generically independent conserved quantities in involution, thereby rendering the system Liouville integrable. This explains how we can have four independent conserved quantities in involution for a system with a 6D phase space. Though all four conserved quantities are shown to be generically independent, there are singular submanifolds of the phase space where this independence fails. In fact, we find the submanifolds where pairs, triples or all four conserved quantities are dependent and identify the relations among conserved quantities on these singular submanifolds. Pleasantly, these submanifolds are shown to coincide with the ‘static’ and ‘circular/trigonometric’ submanifolds\footnote{Static submanifolds consist of static solutions while the trigonometric submanifolds are the ones on which the solutions are expressible in terms of trigonometric functions of time.} of the phase space and to certain nongeneric common level sets of conserved quantities. In Section \ref{s:stability-of-static-solutions}, we analyze the stability of classical static solutions of the RR model and of the corresponding nonlinear wave solutions of the scalar field theory. Finally, the weak coupling limit ($\la \to 0$) of the classical continuous screw-type waves is examined. They are shaped like a screw with axis along the third internal direction suggesting the name `screwons'.

One may wonder whether the Rajeev-Ranken model is related to any other integrable systems. In Appendix \ref{A:RR-model-and-Neumann-Model}, we compare and contrast the RR model with the ($N = 3$) Neumann model \cite{B-B-T, B-T}, which is an integrable system describing the dynamics of a particle moving on an $N$-sphere subject to harmonic forces. Though the models are not quite the same (as the corresponding dynamical variables live in different spaces), this comparison allows us to discover a new Hamiltonian formulation for the Neumann model \cite{G-V-1}. In Appendix \ref{B:Kirchhoff-equations}, we give the EOM of the RR model a new interpretation as Euler equations for a centrally extended Euclidean algebra with a quadratic Hamiltonian. Thus, they bear a kinship to Kirchhoff's equations for a rigid body moving in a perfect fluid \cite{MT}. The latter is an integrable system whose equations are Euler equations for a Euclidean algebra \cite{D-K-N, Sokolov, B-M-S}. Roughly, $\vec L$ and $\vec P = \vec S - \vec K/\la$ play the roles of total angular momentum and linear momentum of the body-fluid system in a body-fixed frame. However, while the Poisson brackets of the Kirchhoff system are given by the Euclidean $L$-$P$ Lie algebra, the RR model involves its central extension. Solutions of the RR model are also interpreted as a special family of flat $\mathfrak{su}(2)$ connections on 1+1D Minkowski space. Indeed, the currents  $r_0 = g^{-1}\dot g$ and $r_1 = g^{-1} g'$ of the PCM (for the SU(2) group-valued principal chiral field $g(x,t)$) are components of a flat $\mathfrak{su}$(2) connection in 1+1-dimensions, satisfying the additional condition $\dot{r_0}  = r_1'$. Solutions of the dual scalar field theory thus furnish a special class of flat connections $r_{\mu} = \la \epsilon_{\mu \nu} \pdr^{\nu} \phi$. This is to be contrasted with certain other integrable systems (investigated for instance in \cite{A-M, Audin, F-Ro}), which describe Hamiltonian dynamics on the space of flat connections on a Riemann surface. Evidently, while solutions to the RR model are very special classes of flat connections, the latter models deal with evolution on the space of all flat connections.

Though analytic solutions in terms of elliptic functions had been found in \cite{R-R}, questions about the structure of the phase space of the RR model and its dynamics were open. In Chapter \ref{chapter:Phase-space-structure-and-action-angle-variables}, we use the Casimirs of the (nilpotent) Poisson algebra to find all symplectic leaves on the $S$-$L$ phase space and a convenient set of Darboux coordinates on them. The system is Liouville integrable on each symplectic leaf and the generic common level sets of conserved quantities are shown to be 2-tori. Going beyond the generic cases, we find three more types of common level sets: horn tori (tori with equal major and minor radii - see Fig.~\ref{f:theta-phi-dynamics-3D-horn-torus}), circles and points. These three arise when the conserved quantities develop relations and are associated to the degeneration of solutions from elliptic to hyperbolic and circular functions. An elegant geometric construction allows us to realize each common level set as a fibre bundle with base determined by the roots of a cubic polynomial. We show that the union of common level sets of a given type may be treated as the phase space of a self-contained dynamical system. By contrast with the dynamics on tori and circles, which is Hamiltonian, that on horn tori is shown to be a gradient flow. In fact, horn tori behave like separatrices and are also associated to a transition in the topology of energy level sets. By a careful use of the Poisson structure and elliptic function solutions, we also discover a family of action-angle variables for the model away from horn tori. A more detailed sectionwise summary of this chapter is given in the beginning of Chapter \ref{chapter:Phase-space-structure-and-action-angle-variables}.

In Chapter \ref{chapter:Quantum-Rajeev-Ranken-model-and-anharmonic-oscillator}, we discuss some aspects of the quantum version of the Rajeev-Ranken model. In Section \ref{s:Electromagnetic-Hamiltonian}, we begin with Rajeev and Raken's mechanical interpretation of the model in terms of a charged particle moving in a static electromagnetic field \cite{R-R}. They used this viewpoint to quantize the model in the Schr\"odinger picture and obtained dispersion relations for the quantized nonlinear waves in the weak and strong coupling limits. However, their radial equation and its associated strong coupling dispersion relation appear to have some errors. In Section \ref{s:Mechanical-interpretation-of-the-Rajeev-Ranken-Model}, we take a complementary approach by interpreting the Rajeev-Ranken model as a 3D cylindrically symmetric anharmonic oscillator. This interpretation follows from rewriting the Hamiltonian in terms of the Darboux coordinates introduced in Section \ref{s:Darboux-coordinates-M6} and identifying the coordinates and momenta as those of a massive nonrelativistic particle. In Section \ref{s:Quantum-RR-model}, we exploit this mechanical interpretation to canonically quantize the model and separate variables in the Schr\"odinger equation. Though the radial equation is in general not exactly solvable, its analytic properties are studied and it is shown to be reducible to a generalization of the Lam\'e equation. As with the classical model, the quantum RR model resembles the quantum Neumann model, as we observe by examining properties of the corresponding radial equations \cite{B-T}. We obtain the energy spectrum at weak coupling and its dependence on the wavenumber in a suitably defined strong coupling limit. In Section \ref{s:Separatio-of-variables-and-the-WKB-approximation}, we separate variables in the Hamilton-Jacobi equation and use this to find the WKB quantization condition, though in an implicit form. In another direction, we notice that the EOM of the RR model can also be interpreted as Euler equations for a step-3 nilpotent Lie algebra (see Appendix \ref{C:RR-equations-as-Euler-equations-for-a-nilpotent-Lie-algebra}). In Section \ref{s:Unitary-representation-of-nilpotent-Lie-algebra}, we exploit our canonical  quantization to uncover an infinite dimensional reducible unitary representation of this nilpotent algebra, which is then decomposed using its Casimir operators. 

Finally, in Chapter \ref{chapter:Discussion}, we discuss some of the results of this thesis and mention possible directions for  further research. 

It is satisfying that a detailed  and explicit analysis of the dynamics and phase space structure of this model has been possible using fairly elementary methods. Our results should be helpful in understanding other aspects of the model's integrability (bi-Hamiltonian formulation on symplectic leaves, spectral curve etc.), the stability of its solutions, effects of perturbations and its quantization (for instance via our action-angle variables, through the representation theory of nilpotent Lie algebras or via path integrals using our Lagrangian obtained from Darboux coordinates, to supplement the Schr\"odinger picture results in \cite{R-R} and in Chapter \ref{chapter:Quantum-Rajeev-Ranken-model-and-anharmonic-oscillator}). Quite apart from its physical origins and possible applications, we believe that the elegance of the Rajeev-Ranken model justifies a detailed study. It is hoped that the insights gained can then also be usefully applied to understanding the parent scalar field theory.

\chapter[Principal chiral model to the Rajeev-Ranken model]{Principal chiral model to the Rajeev-Ranken model}
\chaptermark{Principal chiral model to the Rajeev-Ranken model}
\label{chapter:Nilpotent-field-theory-to-the-Rajeev-Ranken-model}

In this chapter, we introduce the nilpotent scalar field theory dual to the principal chiral model. Then we show how Rajeev and Ranken obtained a consistent reduction of this field theory to a mechanical system with three degrees of freedom which describes certain screw-type nonlinear wave solutions of the field theory. This chapter is based on \cite{R-R} and \cite{G-V-1}.

\section[Nilpotent scalar field theory dual to the PCM]{Nilpotent scalar field theory dual to the PCM \sectionmark{Nilpotent scalar field theory dual to the PCM}}
\sectionmark{Nilpotent scalar field theory dual to the PCM}
\label{s:Nilpotent-scalar-field-theory-dual-to-the-PCM}

As mentioned in the Introduction (Chapter \ref{chapter:Introduction}), a  scalar field theory pseudodual to the 1+1-dimensional SU(2) principal chiral model was introduced in the work of Zakharov and Mikhailov \cite{Z-M} and Nappi \cite{Nappi}.  The 1+1D principal chiral model is defined by the action
	\beq
	S_{\rm PCM} = \frac{1}{2\la^2} \int \Tr \left(\pdr_{\mu}g \pdr^{\mu}g^{-1}\right) dx dt
	= \frac{1}{2\la^2} \int \Tr \left[ (g^{-1}\dot g )^2 - (g^{-1}g')^2 \right] dx dt,
	\label{e:action-PCM}
	\eeq
with primes and dots denoting $x$ and $t$ derivatives. Here, $\la > 0$ is a dimensionless coupling constant and $\Tr = -2 \tr$. The corresponding equations of motion (EOM) are nonlinear wave equations for the components of the SU(2)-valued field $g(x,t)$ and may be written in terms of the $\mathfrak{su}(2)$ Lie algebra-valued time and space components of the right current, $r_0 = g^{-1}\dot{g}$ and $r_1 = g^{-1}g'$:
	\beq
	\ddot g - g'' = \dot g g^{-1} \dot g - g' g^{-1}g' \qquad \text{or} \qquad	
	\dot r_0 - r_1' = 0.
	\eeq
An equivalent formulation is possible in terms of left currents $l_{\mu} = (\partial_{\mu} g) g^{-1}$. Note that $r_0$ and $r_1$ are components of a flat connection; they satisfy the zero curvature `consistency' condition
	\beq
	\dot{r_1} - r_0' + \left[r_0, r_1\right] = 0.
	\eeq
Following Rajeev and Ranken \cite{R-R}, we define right current components rescaled by $\la$, which are especially useful in discussions of the strong coupling limit:
	\beq
	I = \frac{1}{\la^2}r_1 \quad \text{and} \quad J = \frac{1}{\la}r_0.
	\eeq
In terms of these currents, the EOM and zero-curvature condition become 
	\beq
	\dot{J} = \la I' \quad \text{and} \quad
	\dot{I} = \la\left[I,J\right] +\frac{1}{\la} J'.
	\label{e:eom-ZC-currents}
	\eeq
These EOM may be derived from the Hamiltonian following from $S_{\rm PCM}$ (upon dividing by $\la$),
	\beq
	H_{\rm PCM}  = \half \Tr \int dx \left(\la I^2 + \ov{\la} J^2 \right) 
	\label{e:H-PCM} 
	\eeq
and the PBs:
	\beqs
	\{ I_a(x), I_b(y) \} &=& 0, \quad \{ J_a(x), J_b(y) \} = - \la^2 \eps_{abc} J_c(x) \del(x-y) \cr
	\text{and} \quad \{ J_a(x), I_b(y) \} &=& -\la^2 \eps_{abc} I_c(x) \del(x-y) + \del_{ab} \pdr_x \del(x-y) 
	\label{e:PB-PCM} 
	\eeqs
for $a, b = 1, 2, 3$. Since both $I$ and $J$ are anti-hermitian, their squares are negative operators, but the minus sign in $\Tr$ ensures that $H_{\rm PCM} \geq 0$. The Poisson algebra (\ref{e:PB-PCM}) is a central extension of a semi-direct product of the abelian algebra generated by the $I_a$ and the $\mathfrak{su}(2)$ current algebra generated by the $J_a$. It may be regarded as a (centrally extended) `Euclidean' current algebra. These PBs follow from the canonical PBs between $I$ and its conjugate momentum in  the  action (\ref{e:action-PCM}) \cite{F-T}. The multiplicative constant in $\{ J_a, J_b \}$ is not fixed by the EOM. It has been chosen for convenience in identifying Casimirs of the reduced mechanical model (see Section \ref{s:semi-direct-product-PB}). 

The EOM $\dot J = \la I'$ is identically satisfied if we express the currents in terms of a Lie algebra-valued potential $\phi$: 
	\beqs
	I &=& \frac{\dot \phi}{\la} 
	\quad \text{and} \quad J = \phi' \quad \text{or} \quad 
	r_{\mu} = \la \eps_{\mu \nu} \partial^{\nu} \phi \cr 
	\text{with} \quad 
	g_{\mu \nu} &=& \colvec{2}{1 & 0 }{0 & -1} \quad \text{and} \quad \eps^{0 1} = 1.	
	\label{e:currents-NFT}
	\eeqs
The zero curvature condition ($\dot{I} -  J'/ \la = \la\left[I,J\right]$) now becomes a $2^{\rm nd}$-order nonlinear wave equation for the scalar $\phi$ (with the speed of light re-instated):
	\beq
	\ddot\phi = c^2 \phi'' + c \la [ \dot{\phi}, \phi' ].
	\label{e:Nonlinear-wave-equation}
	\eeq
The field $\phi$ is an anti-hermitian traceless $2 \times 2$ matrix in the $\mathfrak{su}(2)$ Lie algebra, which may be written as a linear combination of the generators $t_a = \sigma_a / 2i$ where $\sig_a$ are the Pauli matrices:
	\beq
	\phi = \phi_a t_a = \frac{1}{2i} \phi \cdot \sigma 
	\quad \text{with} \quad \phi_a = i \tr(\phi \sigma_a) = \Tr(\phi t_a)
	\eeq 
for $a = 1, 2, 3$. The generators are normalized according to $\Tr(t_a t_b) = \del_{ab}$ and satisfy $\left[ t_a , t_b \right] = \eps_{abc} t_c$. As noted in \cite{R-R}, a strong-coupling limit of (\ref{e:Nonlinear-wave-equation}) where the $\la [\dot \phi, \phi']$ term dominates over $\phi''$, may be obtained by introducing the rescaled field $\tl \phi (\xi, \tau) = \la^{2/3} \phi(x,t)$, where $\xi = x$ and $\tau = \la^{1/3} t$. Taking $\la \to \infty$ holding $c$ fixed gives the Lorentz noninvariant equation $\tl \phi_{\tau \tau} = c[\tl \phi_{\tau} , \tl \phi_{\xi}]$. Contrary to the expectations in \cite{R-R}, the `slow-light' limit $c \to 0$ holding $\la$ fixed is not quite the same as this strong-coupling limit.

The wave equation (\ref{e:Nonlinear-wave-equation}) follows from the Lagrangian density (with $c=1$)
	\beq
	\mathcal{L} = \Tr \left( \frac{1}{2\la}(\dot{\phi}^2 - \phi'^2) + \frac{1}{3}\phi [\dot{\phi}, \phi'] \right)
	 = \ov{2\la} \pdr_{\mu} \phi_a\pdr^{\mu} \phi_a + \frac{1}{6} \eps_{abc}\epsilon^{\mu \nu}\phi_a \pdr_{\mu} \phi_b \pdr_{\nu}\phi_c.
	\label{e:Lagrangian-scalar-field}
	\eeq
The momentum conjugate to $\phi$ is $\pi =  \dot \phi/ \la  - (1/3) \left[\phi , \phi'\right]$ and satisfies 
	\beq
	\dot \pi = \frac{\phi''}{\la} + \frac{2}{3} [ \dot \phi, \phi' ] + \frac{1}{3} [\dot \phi' , \phi ] 
	= \frac{\phi''}{\la} + \frac{2 \la }{3} [ \pi , \phi' ] + \frac{\la}{3}[ \pi' , \phi ] + \frac{2 \la}{9} [ [ \phi , \phi'], \phi'] + \frac{\la}{9} [[\phi , \phi''], \phi].
	\label{e:pi-field}
	\eeq
The conserved energy and Hamiltonian coincide with $H_{\rm PCM}$ of (\ref{e:H-PCM}):
	\beqs
	E &=& \frac{1}{2\la} \Tr \int dx \left[ \dot \phi^2 + \phi'^2 \right] \cr 
	\quad \text{and} \quad
	H  &=& \half \Tr \int  dx\: \left[ \la \left(\pi + \frac{1}{3}[ \phi, \phi']\right)^2 + \ov \la \phi'^2\right]. 
	\label{e:H-Nilpotent-field-theory}
	\eeqs
If we postulate the canonical PBs
	\beq
	\{ \phi_a(x), \phi_b(y) \} = 0, \quad \{ \pi_a(x) , \pi_b(y) \} = 0 \quad
	\text{and} \quad  \{ \phi_a(x), \pi_b(y) \} = \del_{ba} \del(x-y),
	\label{e: Canonical-PB-field}
	\eeq
then Hamilton's equations $\dot \phi = \{ \phi , H \}$ and $\dot \pi = \{ \pi , H \}$ reproduce (\ref{e:pi-field}).
The canonical PBs between $\phi$ and $\pi$ imply the following PBs among the currents $I, J$ and $\phi$:
	\beqs
	\{ J_a(x) , J_b(y) \} &=& 0, \quad \{ I_a(x), J_b(y) \} =  \delta_{a b} \pdr_x \delta(x-y), \cr
	\{ \phi_a(x) , I_b(y) \} &=&  \del_{ab} \del (x-y), \quad
	\{ \phi_a(x), J_b(y) \} = 0 \quad \text{and} \cr 
	\{ I_a(x), I_b(y) \} &=&  \frac{\eps_{abc}}{3} \left( 2 J_c(x) + (\phi_c(x) - \phi_c(y)) \pdr_y \right) \del(x-y).
	\label{e:PB-IJ-phi}
	\eeqs
These PBs define a step-3 nilpotent Lie algebra in the sense that all triple PBs such as 
	\beq
	\{ \{ \{I_a(x), I_b(y)\} , I_c(z)\}, I_d(w) \}
	\label{e:triple-PBs}
	\eeq
vanish. Note however that the currents $I$ and $J$ {\it do not} form a closed subalgebra of (\ref{e:PB-IJ-phi}). Interestingly, the EOM (\ref{e:eom-ZC-currents}) also follow from the same Hamiltonian (\ref{e:H-PCM}) if we postulate the following closed Lie algebra among the currents
	\beqs
	\{ J_a(x) , J_b(y) \} &=& 0, \quad
	\{ I_a(x), J_b(y) \} =  \delta_{a b} \pdr_x \delta(x-y) \quad
	\text{and} \cr
	\{ I_a(x) , I_b(y) \} &=& \epsilon_{abc}J_c\delta(x-y).
	\label{e:PB-currents}
	\eeqs
Crudely, these PBs are related to (\ref{e:PB-IJ-phi}) by `integration by parts'. As with (\ref{e:PB-IJ-phi}), this Poisson algebra of currents is a nilpotent Lie algebra of step-3 unlike the Euclidean algebra of Eq.~(\ref{e:PB-PCM}).

The scalar field with EOM (\ref{e:Nonlinear-wave-equation}) and Hamiltonian (\ref{e:H-Nilpotent-field-theory}) is classically related to the PCM through the change of variables $r_{\mu} = \la \eps_{\mu \nu} \partial^{\nu} \phi$. However, as noted in \cite{C-Z}, this transformation is not canonical, leading to the moniker `pseudodual'. Though this scalar field theory has not been shown to be integrable, it does possess infinitely many (nonlocal) conservation laws \cite{C-Z}. Moreover, the corresponding quantum theories are different. While the PCM is asymptotically free, integrable  and serves as a toy-model for 3+1D Yang-Mills theory, the quantized scalar field theory displays particle production (a nonzero amplitude for $2 \to 3$ particle scattering), has a positive $\beta$ function \cite{Nappi} and could serve as a toy-model for 3+1D $\la \phi^4$ theory \cite{R-R}.

\section[Reduction of the nilpotent field theory and the RR model]{Reduction of the nilpotent field theory and the RR model \sectionmark{Reduction of the nilpotent field theory and the RR model}}
\sectionmark{Reduction of the nilpotent field theory and the RR model}
\label{s:NFT-to-RR}

Before attempting a challenging nonperturbative study of the nilpotent field theory, it is interesting to study its reduction to finite dimensional mechanical systems obtained by considering special classes of solutions to the nonlinear wave equation (\ref{e:Nonlinear-wave-equation}). The simplest such solutions are traveling waves $\phi(x,t) = f(x-vt)$ for constant $v$. However, for such $\phi$, the commutator term $ - \la [v f', f'] = 0$ so that traveling wave solutions of (\ref{e:Nonlinear-wave-equation}) are the same as those of the linear wave equation. Nonlinearities play no role in similarity solutions either. Indeed, if we consider the scaling ansatz $\tl \phi\left(\xi, \tau \right) = \La^{-\gamma} \phi(x,t)$ where $\xi = \La^{-\alpha} x$ and $\tau = \La^{-\beta}t$, then (\ref{e:Nonlinear-wave-equation}) takes the form:
	\beq
	\La^{\gamma - 2\beta} \tl \phi_{\tau \tau} - \La^{\gamma - 2 \alpha} \tl \phi_{\xi \xi} - \La^{2 \gamma - (\beta + \alpha)} \la [ \tl \phi_{\tau} , \tl \phi_{\xi} ] = 0. 
	\eeq
This equation is scale invariant when $\alpha = \beta$ and $\gamma = 0$. Hence similarity solutions must be of the form $\phi(x,t) =  \psi(\eta)$ where $\eta = x/t$ and $\psi$ satisfies the {\it linear} ODE
	\beq
	\eta^2 \psi'' - \psi'' + 2 \eta \psi' = - \la \eta [ \psi', \psi'] = 0.	
	\eeq

Recently, Rajeev and Ranken \cite{R-R} found a mechanical reduction of the nilpotent scalar field theory for which the nonlinearities play a crucial role. They considered the wave ansatz:
	\beq
	\phi(x,t) = e^{Kx}R(t)e^{-Kx} + mKx\quad \text{with} \quad 
	K = \frac{i}{2} k \sigma_3
	\label{e:ansatz}
	\eeq
which leads to `continuous wave' solutions of (\ref{e:Nonlinear-wave-equation}) with constant energy-density.  These screw-type configurations are obtained from a Lie algebra-valued matrix $R(t)$ by combining an internal rotation (by angle $\propto x$) and a translation. The constant traceless anti-hermitian matrix $K$ has been chosen in the $3^{\rm rd}$ direction. The ansatz (\ref{e:ansatz}) depends on two parameters: a dimensionless real constant $m$ and the constant $K_3 = -k$ with dimensions of a wave number which could have either sign. When restricted to the submanifold of such propagating waves, the field equations (\ref{e:Nonlinear-wave-equation}) reduce to those of a mechanical system with 3 degrees of freedom which we refer to as the Rajeev-Ranken model. The currents (\ref{e:currents-NFT}) can be expressed in terms of $R$ and $\dot R$:
	\beq
	I = \frac{1}{\la}e^{Kx}\dot{R}e^{-Kx} \quad \text{and} \quad 
	J = e^{Kx} \left( {\left[K,R\right] +mK} \right) e^{-Kx}.
	\label{e:currents-I-J}
	\eeq
These currents are periodic in $x$ with period $2\pi/|k|$. We work in units where $c=1$ so that $I$ and $J$ have dimensions of a wave number. If we define the traceless anti-hermitian matrices
	\beq
	L = \left[ K, R \right] + mK
	\quad \text{and}
	\quad S = \dot{R} + \ov{\la} K,
	\label{e: L-and-S}
	\eeq
then it is possible to express the EOM and consistency condition (\ref{e:eom-ZC-currents}) as the pair
	\beq
	\dot{L} = \left[K, S\right] 
	\quad \text{and} \quad 
	\dot{S} = \la \left[S, L\right].
	\label{e:EOM-LS}
	\eeq
In components $(L_a = \Tr( L t_a)$ etc.), the equations become 
	\beqs
	\dot L_1 &=& k S_2, \qquad \dot L_2 = - k S_1, \qquad \dot L_3 = 0, \cr 
	\dot S_1 &=& \la (S_2 L_3 - S_3 L_2),
	\quad \dot S_2 = \la (S_3 L_1 - S_1 L_3) \quad \text{and} \quad
	\;\; \dot S_3 = \la (S_1 L_2 - S_2 L_1).
	\label{e:EOM-LS-explicit}
	\eeqs
Here, $L_3 = - m k$ is a constant, but it will be convenient to treat it as a coordinate. Its constancy will be encoded in the Poisson structure so that it is either a conserved quantity or a Casimir. Sometimes it is convenient to express $L_{1,2}$ and $S_{1,2}$ in terms of polar coordinates:
	\beq
	L_1 = kr \cos \tht, \quad L_2 = kr \sin \tht, \quad S_1 = k \rho \cos \phi \quad \text{and} \quad S_2 = k \rho \sin \phi.
	\label{e:L-S-polar}
	\eeq 
Here, $r$ and $\rho$ are dimensionless and positive. We may also express $L$ and $S$ in terms of coordinates and velocities (here $u = \dot{R_3}/k - 1/\la$):
	\beqs
	L &=& \frac{k}{2i} \colvec{2}{-m & R_2 + i R_1}{R_2 - iR_1 & m}
\quad \text{and} \quad 
	S = \frac{1}{2i} \colvec{2}{ uk & \dot{R_1} - i \dot{R_2}}{\dot{R_1} + i\dot{R_2} & -uk} \quad \text{or} \cr
	L_1 &=& k R_2, \;\;\; L_2 = -k R_1,
	\;\;\; L_3 = -mk, \;\;\; S_1 = \dot R_1, \;\;\; S_2 = \dot R_2 \;\;\; \text{and} \;\;\; S_3 = uk.
	\label{e:EOM-R_3}
	\eeqs
It is clear from (\ref{e: L-and-S}) that $L$ and $S$ do not depend on the coordinate $R_3$. The EOM (\ref{e:EOM-LS}, \ref{e:EOM-R_3}) may be expressed as a system of three second order ODEs for the components of $R(t)$: 
	\beqs
	\ddot R_1 &=& \la k (R_1 \dot R_3 -  m \dot R_2) - k^2 R_1, \quad 
	\ddot R_2 = \la k (R_2 \dot R_3 +  m  \dot R_1) - k^2 R_2 \quad 
	\text{and} \cr
	\ddot R_3 &=& \frac{-\la k}{2} ( R_1^2 + R_2^2)_{t}.
	\label{e:EOM-R}
	\eeqs
Rajeev and Ranken used conserved quantities to simplify these equations of motion and express the solutions to (\ref{e:EOM-R}) in terms of elliptic functions.

\chapter[Rajeev-Ranken model: Hamiltonian formulation and Liouville integrability]{Rajeev-Ranken model: Hamiltonian formulation and Liouville integrability}
\chaptermark{Liouville integrability}
\label{chapter:Hamiltonian-formulation-and-Liouville-integrability}

We begin this chapter by introducing a pair of Hamiltonian-Poisson bracket formulations for the RR model. Then we find a Poisson pencil, Lax pairs, $r$-matrices and a complete set of conserved quantities in involution, thereby establishing its Liouville integrability. These conserved quantities are then related to the Noether charges of the parent scalar field theory. Static and trigonometric submanifolds of the phase space are introduced, where the generally elliptic function solutions simplify. Then, we investigate the functional independence of the conserved quantities by examining the linear independence of the associated one-forms. Finally, we discuss the stability of static solutions of the RR model and the corresponding solutions of the field theory. This chapter is based on  \cite{G-V-1}.

\section[Hamiltonian, Poisson brackets and Lagrangian]{Hamiltonian, Poisson brackets and Lagrangian \sectionmark{Hamiltonian, Poisson brackets and Lagrangian}}
\sectionmark{Hamiltonian, Poisson brackets and Lagrangian}
\label{s:Hamiltonian-PB-Lagrangian-RR-model}
\subsection{Hamiltonian and PBs for the RR model}
\label{s:Hamiltonian-mechanical}

The Rajeev-Ranken model, which is a mechanical system with 3 degrees of freedom and phase space $M^6_{S \text{-} L}$ ($\mathbb{R}^6$ with coordinates $L_a, S_a$) can be given a Hamiltonian-Poisson bracket formulation. A Hamiltonian is obtained by a reduction of that of the nilpotent field theory (\ref{e:H-Nilpotent-field-theory}). For the nonlinear screw wave (\ref{e:ansatz}), we have $\Tr \dot \phi^2 = \Tr \dot R^2 $ and $\Tr \phi'^2 = \Tr ([K,R] + m K)^2$. Thus the ansatz (\ref{e:ansatz}) has a constant energy density and we define the reduced Hamiltonian to be the energy (\ref{e:H-Nilpotent-field-theory}) per unit length (with dimensions of 1/area):
	\beq
	H = \half \Tr \left[ \left(S-\frac{1}{\la}K\right)^2 +  L^2 \right]
	=  \frac{S_a^2 + L_a^2}{2} + \frac{k}{\la} S_3 + \frac{k^2}{2\la^2} 
	= \half \left[ \dot R_a^2 + k^2 \left(R_1^2 + R_2^2 + m^2 \right) \right].
	\label{e: H-mechanical}
	\eeq
We have multiplied by $\la$ for convenience. PBs among $S$ and $L$ which lead to the EOM (\ref{e:EOM-LS}) are given by 
	\beq
	\left\{ L_a, L_b \right\}_{\nu} = 0, 
	\quad \left\lbrace S_a, S_b \right\rbrace_{\nu} = \la \epsilon_{abc} L_c
	\quad \text{and} \quad 
	\left\lbrace S_a, L_b \right\rbrace_{\nu} = -\epsilon_{abc} K_c.
	\label{e: PB-SL}
	\eeq
We may view this Poisson algebra as a finite-dimensional version of the nilpotent Lie algebra of currents $I$ and $J$ in (\ref{e:PB-currents}) with $K$ playing the role of the central $\del'$ term. In fact, both are step-3 nilpotent Lie algebras (indicated by $\{ \cdot , \cdot \}_\nu$ in the mechanical model) and we may go from (\ref{e:PB-currents}) to (\ref{e: PB-SL}) via the rough identifications (up to conjugation by $e^{Kx}$):
	\beq
	J_a \to L_a, \quad I_a \to \ov{\la} \left( S_a - \frac{K_a}{\la} \right), \quad  
	\del_{ab} \pdr_x\del(x-y) \to -\eps_{abc} K_c \quad \text{and} \quad \{ \cdot , \cdot \} \to \la \{ \cdot , \cdot \}_{\nu}.
	\eeq
Note that the PBs (\ref{e: PB-SL}) have dimensions of a wave number. They may be expressed as $\{ f, g \}_{\nu} = \scripty{r}^{a b}_0 \pdr_a f \pdr_b g$ where the anti-symmetric Poisson tensor field $\scripty{r}_0 = (0 \: A | A \: B)$ with the $3 \times 3$ blocks $A_{a b} = -\eps_{abc} K_c$ and $B_{a b} = \la \eps_{abc} L_c$.
This Poisson algebra is degenerate: $\scripty{r}_0$ has rank four and its kernel is spanned by the exact 1-forms $d L_3$ and $d\left( S_3 + (\la/k)(L_1^2 + L_2^2)/2 \right)$. The corresponding center of the algebra can be taken to be generated by the Casimirs $m k^2 \equiv \Tr KL $ and ${\mathfrak{c}} k^2 \equiv \Tr \left( (L^2/2) - (KS/\la) \right)$. 
\vspace{.25cm}

{\bf \fl Euclidean PBs:} The $L$-$S$ EOM (\ref{e:EOM-LS}) admit a second Hamiltonian formulation with a nonnilpotent Poisson algebra arising from the reduction of the Euclidean current algebra of the PCM (\ref{e:PB-PCM}). It is straightforward to verify that the PBs
	\beq
	\{ S_a, S_b \}_{\varepsilon} = 0, \quad \{ L_a, L_b \}_{\varepsilon} = - \la \eps_{abc} L_c \quad \text{and} \quad \{ L_a , S_b \}_{\varepsilon} = - \la \eps_{abc} S_c
	\label{e:PB-SL-dual}
	\eeq
along with the Hamiltonian (\ref{e: H-mechanical}) lead to the EOM (\ref{e:EOM-LS}). This Poisson algebra is isomorphic to the Euclidean algebra in 3D (${\mathfrak e}(3)$ or ${\mathfrak{iso}}(3)$) a semi-direct product of the simple $\mathfrak{su}(2)$ Lie algebra generated by the $L_a$ and the abelian algebra of the $S_a$. Furthermore, it is easily verified that $s^2 k^2 \equiv \Tr S^2$ and $h k^2 \equiv \Tr SL$ are Casimirs of this Poisson algebra whose Poisson tensor we denote $\scripty{r}_1$. It follows that the EOM (\ref{e:EOM-LS}) obtained from these PBs are unaltered if we remove the $\Tr S^2$ term from the Hamiltonian (\ref{e: H-mechanical}). The factor $\la$ in the $\{ L_a, S_b\}_{\varepsilon}$ PB is fixed by the EOM while that in the $\{ L_a, L_b \}_{\varepsilon}$ PB is necessary for $h$ to be a Casimir.  
\vspace{.25 cm}

{\fl \bf Formulation in terms of real antisymmetric matrices:} It is sometimes convenient to re-express the $2 \times 2$ anti-hermitian $\mathfrak{su}(2)$ Lie algebra elements $L, S$ and $K$ as $3 \times 3$ real anti-symmetric matrices (more generally we would contract with the structure constants):
	\beq
	\tl L_{k l} = \half \eps_{k l m} L_m \quad \text{with} \quad 
	L_j = \eps_{j k l} \tl L_{k l} \quad \text{and similarly for $\tl S$ and $\tl K$}.
	\eeq 
The EOM (\ref{e:EOM-LS})  and the Hamiltonian (\ref{e: H-mechanical}) become:
	\beq
	{\dot {\tl L}} = -2 [ \tl K , \tl S ], \quad 
	{\dot {\tl S}} = -2 \la [ \tl S, \tl L ] \quad \text{and} \quad 
	H = -\tr \left( \left(\tl S - \tl K/\la \right)^2 + \tl L^2 \right).
	\label{e:EOM-tilde-LS-and-H-RR-real-anti-symm}
	\eeq 
Moreover, the nilpotent $(\nu)$ (\ref{e: PB-SL}) and Euclidean $(\varepsilon)$ (\ref{e:PB-SL-dual}) PBs become
	\beqs
	\{ \tl S_{kl}, \tl S_{pq} \}_{\nu} &=& \frac{\la}{2} \left( \del_{kq} \tl L_{pl} - \del_{pl} \tl L_{kq}  + \del_{ql} \tl L_{kp} - \del_{kp} \tl L_{ql} \right), \cr
	\{ \tl S_{kl}, \tl L_{pq} \}_{\nu} &=& -\half \left( \del_{kq} \tl K_{pl} - \del_{pl} \tl K_{kq} + \del_{ql} \tl K_{kp} - \del_{kp} \tl K_{ql} \right) \quad \text{and} \quad \{ \tl L_{kl} , \tl L_{pq} \}_{\nu} = 0 \qquad 
	\label{e:PB-nilpotent-tilde-LS}\\
	\text{and} \quad \{ \tl L_{kl} , \tl L_{pq} \}_{\varepsilon} &=& -\frac{\la}{2} \left( \del_{kq} \tl L_{pl} - \del_{pl} \tl L_{kq}  + \del_{ql} \tl L_{kp} - \del_{kp} \tl L_{ql} \right), \cr
	\{ \tl S_{kl} , \tl L_{pq} \}_{\varepsilon} &=& -\frac{\la}{2} \left( \del_{kq} \tl S_{pl} - \del_{pl} \tl S_{kq} + \del_{ql} \tl S_{kp} - \del_{kp} \tl S_{ql} \right) \quad \text{and} \quad \{ \tl S_{kl}, \tl S_{pq} \}_{\varepsilon} = 0. 
	\label{e:PB-semi-diect-tilde-LS}
	\eeqs
Interestingly, we notice that both (\ref{e:PB-nilpotent-tilde-LS}) and (\ref{e:PB-semi-diect-tilde-LS}) display the symmetry $\{ \tl S_{kl}, \tl L_{pq} \} = \{ \tl L_{kl}, \tl S_{pq} \}$. The Hamiltonian (\ref{e:EOM-tilde-LS-and-H-RR-real-anti-symm}) along with either of the PBs (\ref{e:PB-nilpotent-tilde-LS}) or (\ref{e:PB-semi-diect-tilde-LS}) gives the EOM in (\ref{e:EOM-tilde-LS-and-H-RR-real-anti-symm}).

\subsection{Poisson pencil from nilpotent and Euclidean PBs}
\label{s:semi-direct-product-PB}

The Euclidean $\{ \cdot, \cdot \}_{\varepsilon}$ (\ref{e:PB-SL-dual}) and nilpotent $\{ \cdot, \cdot \}_{\nu}$ (\ref{e: PB-SL}) Poisson structures among $L$ and $S$ are compatible and together form a Poisson pencil. In other words, the linear combination
	\beq
	\{ f,g \}_\al = (1- \al) \{ f, g \}_\nu + \al \{ f, g\}_\varepsilon
	\label{e:Poisson-pencil}
	\eeq
defines a Poisson bracket for any real $\al$. The linearity, skew-symmetry and derivation properties of the $\al$-bracket follow from those of the individual PBs. As for the Jacobi identity, we first prove it for the coordinate functions $L_a$ and $S_a$. There are only four independent cases:
	\beqs
	\{ \{ S_a , S_b\}_\al, S_c \}_\al + \text{cyclic} &=& -(1 - \al) \la \eps_{abd} \left((1 - \al) \eps_{dce} K_e + \al \la \eps_{dce} S_e \right) + \text{cyclic} = 0, \cr
	\{ \{ L_a, L_b\}_\al , L_c \}_\al + \text{cyclic} &=& \al^2 \la^2 \eps_{abd} \eps_{dce} L_e + \text{cyclic} = 0, \cr
	\{ \{ S_a, S_b \}_\al, L_c \}_\al + \text{cyclic} &=& -(1- \al) \al \la^2 \eps_{abd} \eps_{dce} L_e + \text{cyclic} = 0 \quad  \text{and} \cr
	\{ \{ L_a, L_b\}_\al , S_c \}_\al + \text{cyclic} &=& \al \la \eps_{abd} \left( (1- \al)\eps_{dce} K_e + \al \la \eps_{dce} S_e \right) + \text{cyclic} = 0.
	\label{e:Jacobi-identity}
	\eeqs 
The Jacobi identity for the $\al$-bracket for linear functions of $L$ and $S$ follows from (\ref{e:Jacobi-identity}). For more general functions of $L$ and $S$, it follows by applying the Leibniz rule ($\xi_i = (L_{1,2,3},S_{1,2,3})$):
	\beq
	\{ \{ f, g\}_\al, h \}_\al + \text{cyclic} = \dd{f}{\xi_i} \dd{g}{\xi_j} \dd{h}{\xi_k} \left( \{ \{ \xi_i, \xi_j \}_\al , \xi_k \}_\al + \text{cyclic} \right) = 0.
	\eeq 

As noted, both the nilpotent and Euclidean PBs are degenerate: ${\mathfrak{c}}$ and $m$ are Casimirs of $\{ \cdot , \cdot \}_\nu$ while those of $\{ \cdot, \cdot \}_\veps$ are $s^2$ and $h$. In fact, the Poisson tensor $\scripty{r}_\al = (1- \al) \scripty{r}_{0} + \al \scripty{r}_{1}$ is degenerate for any $\alpha$ and has rank 4. Its independent Casimirs may be chosen as $(1- \al)( m/ \la ) + \al h$ and $(1 - \al) {\mathfrak{c}} - \al  s^2/2$, whose exterior derivatives span the kernel of $\scripty{r}_\al$. The $\nu$ and $\varepsilon$ PBs become nondegenerate upon reducing the 6D phase space to the 4D level sets of the corresponding Casimirs. Since the Casimirs are different, the resulting symplectic leaves are different, as are the corresponding EOM. Thus these two PBs do not directly lead to a bi-Hamiltonian formulation.

\subsection{Darboux coordinates and Lagrangian from Hamiltonian}
\label{s:Darboux-coordinates-M6}
	
Though they are convenient, the $S$ and $L$ variables are noncanonical generators of the nilpotent degenerate Poisson algebra (\ref{e: PB-SL}). Moreover, they lack information  about the coordinate $R_3$. It is natural to seek canonical coordinates that contain information on all six generalized coordinates and velocities $(R_a, \dot R_a)$ (see (\ref{e:currents-I-J})). Such Darboux coordinates will also facilitate a passage from Hamiltonian to Lagrangian. Unfortunately, as discussed below, the naive reduction of (\ref{e:Lagrangian-scalar-field}) does not yield a Lagrangian for the EOM (\ref{e:EOM-R}). 

It turns out that momenta conjugate to the coordinates $R_a$ may be chosen as (see (\ref{e:EOM-R_3}))
	\beqs
	kP_1 &=& S_1 + \frac{\la}{2} m L_1 = \dot R_1 + \frac{\la}{2}m k R_2, 
	\quad kP_2 = S_2 + \frac{\la}{2} m L_2 = \dot R_2 - \frac{\la}{2}m k R_1  \quad \text{and} \cr
	kP_3 &=& \frac{k\la}{2} ( 2 {\mathfrak{c}} - m^2 ) + \frac{k}{\la} = S_3 + \frac{k}{\la} + \frac{\la}{2k} \left( L_1^2 + L_2^2 \right) = \dot R_3 + \frac{\la k}{2}(R_1^2 + R_2^2).
	\label{e:canonical-momenta-P_a}
	\eeqs
We obtained them from the nilpotent algebra (\ref{e: PB-SL}) by requiring the canonical PB relations
	\beq
	\{ R_a, R_b \} = 0, 
	\quad \{ P_a, P_b \} = 0 
	\quad \text{and} \quad \{ R_a, k P_b \} = \delta_{a b} 
	\quad \text{for} \quad a,b = 1,2,3.
	\label{e:Canonical-PB-RP}
	\eeq
Note that $R_a$ cannot be treated as coordinates for the Euclidean PBs (\ref{e:PB-SL-dual}), since $\{ R_1, R_2 \} = (1/k^2) \{ L_1, L_2 \}_{\varepsilon} \neq 0$. Darboux coordinates associated to the Euclidean PBs, may be analogously obtained from the coordinates $Q$ in the wave ansatz for the mechanical reduction of the principal chiral field $g = e^{\la s K x} Q(t) e^{-Kx}$ given in Table I of \cite{R-R}. 

Since $R_3$ does not appear in the Hamiltonian (\ref{e: H-mechanical}) (regarded as a function of $(S, L)$ or $(R, \dot R)$), we have taken the momenta in (\ref{e:canonical-momenta-P_a}) to be independent of $R_3$ so that it will be cyclic in the Lagrangian as well. However, the above formulae for $P_a$ are not uniquely determined. For instance, the PBs (\ref{e:Canonical-PB-RP}) are unaffected if we add to $P_a$ any function of the Casimirs $({\mathfrak{c}}, m)$ as also certain functions of the coordinates (see below for an example). In fact, we have used this freedom to pick $P_3$ to be a convenient function of the Casimirs. Moreover, $\{ R_3, k P_3 \} = 1$ is a new postulate, it is not a consequence of the $S$-$L$ Poisson algebra.
  
The Hamiltonian (\ref{e: H-mechanical}) can be expressed in terms of the $R$'s and $P$'s:
	\beq
	 \frac{H}{k^2} = \sum_{a=1}^3 \frac{P_a^2}{2} + \frac{\la m}{2} \left( R_1 P_2 - R_2 P_1 \right) + \frac{\la^2}{8} \left( R_1^2 + R_2^2 \right) \left[ R_1^2 + R_2^2 + m^2 - \frac{4}{\la} \left( P_3 - \frac{1}{\la} \right) \right] + \frac{m^2}{2}.
	\label{e:Hamiltonian-mech-R-P}
	\eeq
The EOM (\ref{e:EOM-LS}), (\ref{e:EOM-R_3}) follow from (\ref{e:Hamiltonian-mech-R-P}) and the PBs (\ref{e:Canonical-PB-RP}). Thus $R_a$ and $kP_b$ are Darboux coordinates on the 6D phase space $M^6_{R \text{-} P} \cong \mathbb{R}^6$. Note that the previously introduced phase space $M^6_{S \text{-} L}$ is different from $M^6_{R \text{-} P}$, though they share a 5D submanifold in common parameterized by $(L_{1,2}, S_{1,2,3})$ or $(R_{1,2}, P_{1,2,3})$. $M^6_{S \text{-} L}$ includes the constant parameter $L_3 = - m k$ as its sixth coordinate but lacks information on $R_3$ which is the `extra' coordinate in $M^6_{R \text{-} P}$.

\vspace{.25cm}

{\fl \bf Lagrangian for the RR model:} A Lagrangian $L_{\rm mech}(R,\dot R)$ for our system may now be obtained via a Legendre transform by extremizing $kP_a \dot R_a - H$ with respect to all the components of $kP$:
	\beq
	L_{\rm mech} = \half \left[ \sum_{a=1}^3 \dot R_a^2 - \la m k \left( R_1 \dot{R_2} - R_2 \dot{R_1} \right) +  k \left( R_1^2 + R_2^2 \right) ( \la \dot{R_3} -  k) -  m^2 k^2 \right].
	\label{e:Lagrangian-Mech}
	\eeq 
$R_3$ is a cyclic coordinate leading to the conservation of $kP_3$. However $L_{\rm mech}$ does not admit an invariant form as the trace of  a polynomial in $R$ and $\dot R$. Such a form may be obtained by subtracting the time derivative of $(\la k/6) \left( R_3 (R_1^2 + R_2^2) \right)$ from $L_{\rm mech}$ to get: 
	\beqs
	L_{\rm mech}' &=& \Tr \left( \frac{\dot R^2}{2} - \half  ([K, R] + mK)^2 + \frac{\la}{2} R [\dot R, mK] + \frac{\la}{3} R \left[\dot R , [ K, R ] \right] \right) \cr
	 &=& \half \Tr \left( \left(S - \frac{K}{\la} \right)^2 - L^2 + \la R \left[ S - \frac{K}{\la} , L \right] - \frac{\la}{3} R \left[ S - \frac{K}{\la} , [K, R] \right] \right).
	\label{e:Lagrangian-mech}
	\eeqs 
The price to pay for this invariant form is that $R_3$ is no longer cyclic, so that the conservation of $P_3$ is not manifest. The Lagrangian $L_{\rm mech}'$ may also be obtained directly from the Hamiltonian (\ref{e:Hamiltonian-mech-R-P}) if we choose as conjugate momenta $k\Pi_a$ instead of the $kP_a$ of (\ref{e:canonical-momenta-P_a}):
	\beq
	\Pi_1 = P_1 - \frac{\la}{3} R_1 R_3,\quad  
	\Pi_2 = P_2 - \frac{\la}{3} R_2 R_3 \quad \text{and} \quad
	\Pi_3 = P_3 - \frac{\la}{6}(R_1^2 + R_2^2). 
	\eeq
Interestingly, while both $L_{\rm mech}$ and $L_{\rm mech}'$ give the correct EOM (\ref{e:EOM-R}), unlike with the Hamiltonian, the naive reduction $L_{\rm naive}$ of the field theoretic Lagrangian (\ref{e:Lagrangian-scalar-field}) does not. This discrepancy was unfortunately overlooked in Eq.~(3.7) of \cite{R-R}. Indeed $L_{\rm naive}$ differs from $L_{\rm mech}'$ by a term which is {\it not} a time derivative: 
	\beq
	L_{\rm naive} = L_{\rm mech}' + \frac{\la m}{6} \Tr  K \:[ \dot R, R].
	\eeq
To see this, we put the ansatz (\ref{e:ansatz}) for $\phi$ in the nilpotent field theory Lagrangian (\ref{e:Lagrangian-scalar-field}) and use
	\beqs
	\Tr \dot \phi^2 &=& \Tr \dot R^2 , \quad \Tr \phi'^2 = \Tr ([K,R] + m K)^2 \quad \text{and} \cr
	\Tr \phi [\dot \phi, \phi'] &=& \Tr R \left[\dot R ,[K,R]+ mK \right] + \frac{m x k^2}{2} \DD{}{t} (R_1^2 + R_2^2) 
	\eeqs
to get the naively reduced Lagrangian  
	\beq
	L_{\rm naive} = \Tr \left( \half \dot R^2 + \frac{\la}{3} R \left[ \dot R , \left[ K , R \right] + mK \right] - \half (\left[K, R\right] + mK)^2\right).
	\eeq
In obtaining $L_{\rm naive}$ we have ignored an $x$-dependent term as it is a total time derivative, a factor of the length of space and multiplied through by $\la$. As mentioned earlier, $L_{\rm naive}$ does {\it not} give the correct EOM for $R_1$ and $R_2$ nor does it lead to the PBs among $L$ and $S$ (\ref{e: PB-SL}) if we postulate canonical PBs  among $R_a$ and their conjugate momenta. However the Legendre transforms of $L_{\rm mech}, L_{\rm mech}'$ and $L_{\rm naive}$ all give the same Hamiltonian (\ref{e: H-mechanical}). 

One may wonder how it could happen that  the naive reduction of the scalar field gives a suitable Hamiltonian  but not a suitable Lagrangian for the mechanical system. The point is that while a Lagrangian encodes the EOM, a
Hamiltonian by itself does not. It needs to be supplemented with PBs. In the present case, while we used a naive reduction of the scalar field Hamiltonian as the Hamiltonian for the RR model, the relevant PBs ((\ref{e: PB-SL}) and (\ref{e:Canonical-PB-RP})) are not a simple reduction of those of the field theory ((\ref{e:PB-currents}) and (\ref{e: Canonical-PB-field})). Thus, it is not surprising that the naive reduction of the scalar field Lagrangian does not furnish a suitable Lagrangian for the mechanical system. This possibility was overlooked in \cite{R-R} where the former was proposed as a Lagrangian for the RR model.

\section[Lax pairs, $r$-matrices and conserved quantities]{Lax pairs, $r$-matrices and conserved quantities \sectionmark{Lax pairs, $r$-matrices and conserved quantities}}
\sectionmark{Lax pairs, $r$-matrices and conserved quantities}
\label{s:Lax-pairs-r-matrices-and-conserved-quantities-RR-model}
\subsection{Lax Pairs and $r$-matrices}
\label{s:Lax-pair-and-r-matrix}

The EOM (\ref{e:EOM-LS}) admit a Lax pair $(A, B)$ with complex spectral parameter $\zeta$ \cite{Lax}. In other words, if we choose
 	\beq
	A(\zeta) = -K\zeta^2 + L\zeta + \frac{S}{\la} 
	\quad \text{and} \quad	
	B(\zeta) = \frac{S}{\zeta},
	\label{e:Lax-pair}
	\eeq
then the Lax equation $\dot A = [B, A]$ at orders $\zeta^1$ and $\zeta^0$ are equivalent to (\ref{e:EOM-LS}). The Lax equation implies that $\Tr A^n(\zeta)$ is a conserved quantity for all $\zeta$ and every $n = 1,2,3 \ldots$. To arrive at this Lax pair we notice that $\dot A = [B,A]$ can lead to (\ref{e:EOM-LS}) if $L$ and $S$ appear linearly in $A$ as coefficients of different powers of $\zeta$. The coefficients have been chosen to ensure that the fundamental PBs (FPBs) between matrix elements of $A$ can be expressed as the commutator with a nondynamical $r$-matrix proportional to the permutation operator. In fact, the FPBs with respect to the nilpotent PBs (\ref{e: PB-SL}) are given by
	\beqs
	\{ A(\zeta) \stackrel{\otimes}{,} A(\zeta') \}_\nu
	&=& -\frac{1}{4 \la}\left(\epsilon_{abc}L_c -\epsilon_{abc} K_c \left(\zeta + \zeta'\right) \right) \sig_a \otimes \sig_b \cr
	&=& \frac{i}{2 \la}\left(L_3 - \left(\zeta + \zeta'\right) K_3  \right)  \left(\sigma_- \otimes \sigma_+ - \sigma_+ \otimes \sigma_- \right) \cr 
	&& + \frac{1}{4 \la} \sum_\pm \left( {L_2 \pm i L_1} \right)\left(\sigma_\pm \otimes \sigma_3 - \sigma_3 \otimes \sigma_\pm \right).
	\label{e:FPB-nilpotent}
	\eeqs
Here, $\sig_\pm = (\sig_1 \pm i \sig_2)/2$. These FPBs can be expressed as a commutator
	\beqs
	\left\lbrace A(\zeta) \stackrel{\otimes}{,} A(\zeta') \right\rbrace_\nu &=& \left[ r(\zeta, \zeta') , A(\zeta)\otimes I + I \otimes A(\zeta')\right]  \quad \text{where}\cr
	r(\zeta, \zeta') &=& - \frac{P}{2\la (\zeta - \zeta')} 
	\quad \text{with} \quad 
	P = \half \left(I + \sum_{a=1}^3 \sigma_a \otimes \sigma_a \right).
	\label{e:r-matrix-nilpotent}
	\eeqs
To obtain this $r$-matrix we used the following identities among Pauli matrices:
	\beqs
	\sigma_- \otimes \sigma_+ - \sigma_+ \otimes \sigma_- &=& \half [ P, \sigma_3 \otimes I ] = -\half [ P , I \otimes \sigma_3 ]
	\quad \text{and} \cr
	\sigma_\pm \otimes \sigma_3 - \sigma_3 \otimes \sigma_\pm &=& \pm\left[ P , \sigma_\pm \otimes I \right] = \mp \left[ P, I \otimes \sigma_\pm \right].
	\eeqs
We may now motivate the particular choice of Lax matrix $A$ (\ref{e:Lax-pair}). The  nilpotent $S$-$L$ PBs (\ref{e: PB-SL}) do not involve $S$, so the PBs between matrix elements of $A$ are also independent of $S$. Since $P(A \otimes B) =  (B \otimes A) P$, the commutator $\left[P, A \otimes I + I \otimes A \right] = 0$ if $A$ is independent of $\zeta$. Thus for $r \propto P$, $S$ can only appear as the coefficient of $\zeta^0$ in $A$.

The same commutator form of the FPBs (\ref{e:r-matrix-nilpotent}) hold for the Euclidean PBs (\ref{e:PB-SL-dual})  if we use
	\beq
	r_{\varepsilon}(\zeta, \zeta') = \la^2 r(\zeta, \zeta') =  -\frac{\la P}{2 (\zeta - \zeta')},
	\label{e:r-matrix-semi-direct}
	\eeq
provided we define a new Lax matrix $A_\varepsilon = A/\zeta^2$. The EOM for $S$ and $L$ are then equivalent to the Lax equation $\dot A_{\varepsilon} = [B, A_{\varepsilon}]$ at order $\zeta^{-2}$ and $\zeta^{-1}$. In this case, the FPBs are
	\beq	
	\{ A_{\varepsilon}(\zeta) \stackrel{\otimes}{,} A_{\varepsilon}(\zeta') \}_\varepsilon
	= \frac{1}{4 \zeta \zeta'} \left(\la \epsilon_{abc}L_c + \left(\ov{\zeta} + \ov{\zeta'} \right) \epsilon_{abc} S_c \right) \sig_a \otimes \sig_b. 
	\label{e:FPB-semi-direct}
	\eeq

\subsection{Conserved quantities in involution for the RR model}
\label{s:conserved-quantities}

The existence of a classical $r$-matrix implies that the conserved quantities are in involution. In other words, Eq.~(\ref{e:r-matrix-nilpotent}) for the FPBs implies that the conserved quantities $\Tr A^n(\zeta)$ are in involution: 
	\beq
	\left\{\Tr A^m(\zeta) \stackrel{\otimes}{,} \Tr A^n(\zeta') \right\} = mn \Tr \left[ r(\zeta,\zeta') , A^m(\zeta)\otimes A^{n-1}(\zeta') + A^{m-1}(\zeta)\otimes A^{n}(\zeta') \right] = 0
	\eeq
for $m,n = 1,2,3 \ldots$. Each coefficient of the $2n^{\rm th}$ degree polynomial $\Tr A^n(\zeta)$ furnishes a conserved quantity in involution with the others. However, they cannot all be independent as the model has only 3 degrees of freedom. For instance, $\Tr A(\zeta) \equiv 0$ but 
	\beq
	\Tr A^2(\zeta) = \zeta^4 \: K_a K_a - 2 \zeta^3 \: L_a K_a + 2 \zeta^2 \left( \frac{L_a L_a}{2} - \frac{S_a K_a}{\la} \right) + \frac{2\zeta}{\la} \: S_a L_a + \frac{1}{\la^2} S_a S_a.
	\eeq
In this case, the coefficients give four conserved quantities in involution:
	\beqs
	s^2 k^2 &=& \Tr S^2, \quad
	h k^2 = \Tr SL, \quad
	m k^2 = \Tr KL = - k L_3 \cr
	\text{and} \quad
	{\mathfrak{c}} k^2 &=& \Tr \left(\frac{L^2}{2} - \frac{1}{\la}KS \right) = \half L_a L_a + \frac{k}{\la} S_3.
	\label{e:conserved-quantities}
	\eeqs
Factors of $k^2$ have been introduced so that ${\mathfrak{c}}$, $m$, $h$ and $s^2$ (whose positive square-root we denote by $s$) are dimensionless. In \cite{R-R}, $h$ and ${\mathfrak{c}}$ were named $C_1$ and $C_2$. ${\mathfrak{c}}$ and $m$ may be shown to be Casimirs of the nilpotent Poisson algebra (\ref{e: PB-SL}). The value of the Casimir $L_3$ is written as $-m$ in units of $k$ by analogy with the eigenvalue of the angular momentum component $L_z$ in units of $\hbar$. The conserved quantity $\Tr SL$ is called $h$ for helicity by analogy with other such projections. The Hamiltonian (\ref{e: H-mechanical}) can be expressed in terms of $s^2$ and $\mathfrak{c}$:
	\beq
	H = k^2 \left(\half s^2+ {\mathfrak{c}} + \ov{2\la^2} \right).
	\label{e:Hamiltonian-s}
	\eeq
It will be useful to introduce the 4D space of conserved quantities $\cal Q$ with coordinates $\mathfrak{c}$, $s$, $m$ and $h$ which together define a many-to-one map from $M^6_{S \text{-} L}$ to $\cal Q$. The inverse images of points in $\cal Q$ under this map define common level sets of conserved quantities in $M^6_{S \text{-} L}$. By assigning arbitrary real values to the Casimirs $\mathfrak{c}$ and $m$ we may go from the 6D $S$-$L$ phase space to its nondegenerate $4$D symplectic leaves $M^4_{{\mathfrak{c}} m}$ given by their common level sets. For the reduced dynamics on $M^4_{{\mathfrak{c}} m}$, $s^2$ (or $H$) and $h$ define two conserved quantities in involution. 

\begin{figure}[h]
	\centering
		\begin{subfigure}[b]{5cm}
		\includegraphics[width=5cm]{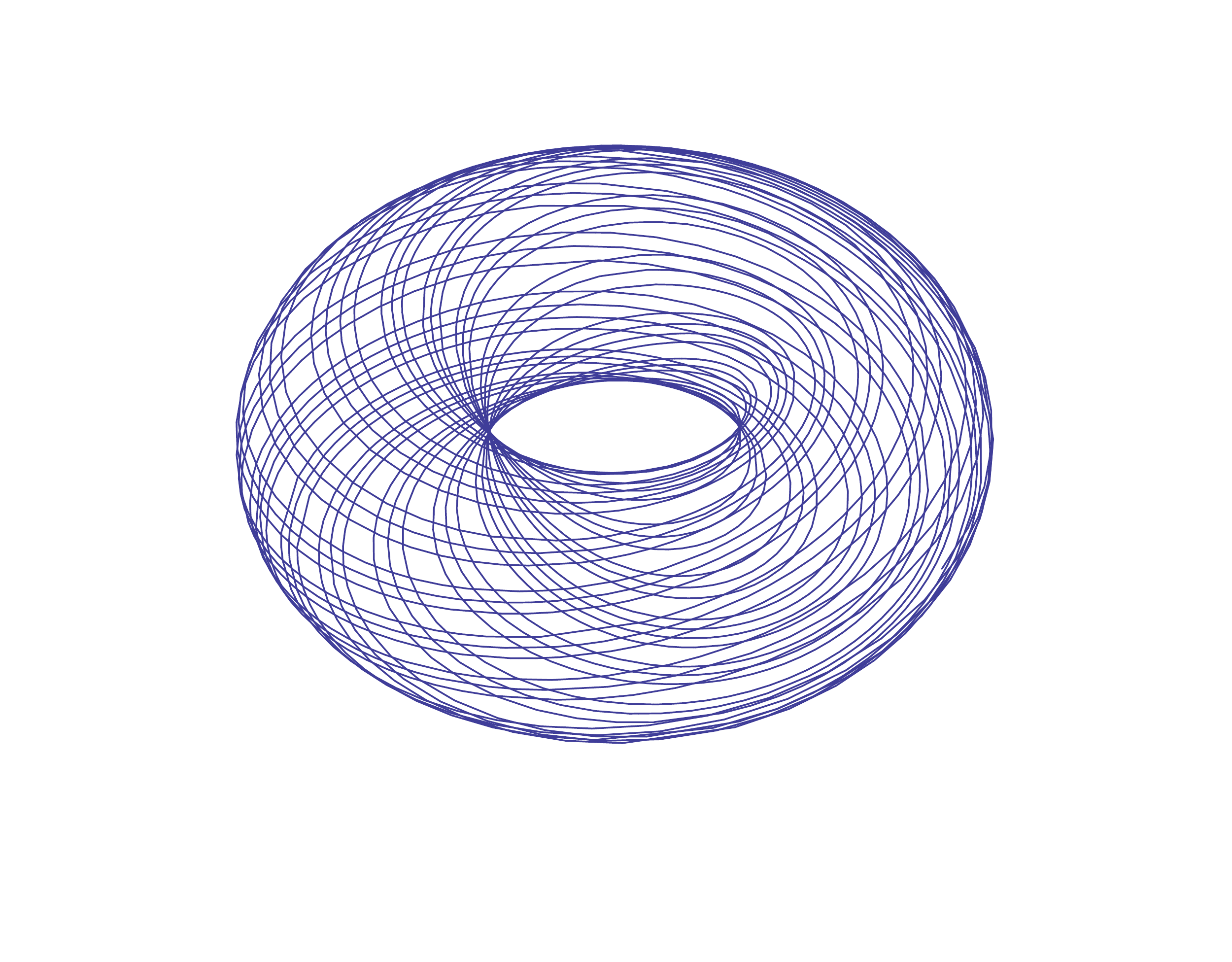}
		\end{subfigure}
	\caption{\footnotesize A trajectory  with initial conditions $\tht(0)= 0.1$ and $\phi(0) = 0.2$ plotted for $0 \leq t \leq 200/k$ on a generic common level set of the conserved quantities ${\mathfrak{c}}, m, s$ and $h$. The common level set is a 2-torus parameterized by the polar and azimuthal angles $\tht$ and $\phi$ and has been plotted for the values ${\mathfrak{c}} = 1/2, h = 0, m = s = 1$ with $k = \la = 1$. It is plausible that the trajectory is quasi-periodic and dense on the torus so that any additional conserved quantity would have to be a constant.}
	\label{f:theta-phi-dynamics-3D}
	\end{figure}

The independence of ${\mathfrak{c}}, m , h$ and $s$ is discussed in Section \ref{s:Independence-of-conserved-quantities}. However, higher powers of $A$ do not lead to new conserved quantities. $\Tr A^3 \equiv 0$ since $\Tr (t_a t_b t_c) = \half \epsilon_{abc}$ for $t_a = \sig_a/2i$. The same applies to other odd powers. On the other hand, the expression for $A^4(\zeta)$ given in Appendix \ref{D:A^4}, along with the identity $\Tr (t_a t_b t_c t_d) = -\frac{1}{4} (\del_{ab} \del_{cd} - \del_{ac} \del_{bd} + \del_{ad} \del_{bc})$ gives
	\beqs
	\ov{k^4} \Tr A^4(\zeta) &=& -\ov 4 s^4 - h s^2 \zeta - \left( \frac{{\mathfrak{c}}s^2 + h^2 }{\la^2} \right) \zeta^2 -  \left(\frac{2 h {\mathfrak{c}}}{\la} - \frac{ms^2}{\la^2} \right) \zeta^3 -  \left( {\mathfrak{c}}^2  + \frac{s^2 }{\la^2} - \frac{2}{\la} m h  \right) \zeta^4 \cr && +  \left( m{\mathfrak{c}} - \ov \la h \right) \zeta^5 - \left( {\mathfrak{c}} + \half m + 2 m^2 \right)\zeta^6 + \ov 4 m \zeta^7 - \ov 4  \zeta^8.
	\label{e:trace-A4}
	\eeqs
Evidently, the coefficients of various powers of $\zeta$ are functions of the known conserved quantities (\ref{e:conserved-quantities}). It is possible to show that the higher powers $\Tr A^6, \Tr A^8 , \ldots$ also cannot yield new conserved quantities by examining the dynamics on the common level sets of the known conserved quantities. In fact, we find that a generic trajectory (obtained by solving (\ref{e:theta-phi-dynamics})) on a generic common level set of all four conserved quantities is dense (see Fig. \ref{f:theta-phi-dynamics-3D} for an example). Thus, any additional conserved quantity would have to be constant almost everywhere and cannot be independent of the known ones.

\vspace{.25cm}

{\fl \bf Canonical vector fields on $M^6_{S \text{-} L}$:} On the phase space, the canonical vector fields ($V_f^a = \scripty{r}_0^{a b} \pdr_b f$) associated to conserved quantities, follow from the nilpotent Poisson tensor $\scripty{r}_0$ of Section \ref{s:Hamiltonian-mechanical}. They vanish for the Casimirs ($V_{\mathfrak{c}} = V_{m} = 0$) while for helicity and the Hamiltonian $(H = E k^2)$,
	\beqs
	k V_h &=&  L_2 \pdr_{L_1} -  L_1 \pdr_{L_2} + S_2 \pdr_{S_1} -  S_1 \pdr_{S_2}    \quad \text{and}  \cr
	k V_E &=& S_2 \pdr_{L_1} - S_1 \pdr_{L_2} + \frac{\la}{k} \left[(S_2 L_3 - L_2 S_3) \pdr_{S_1} + (S_3 L_1 - S_1 L_3) \pdr_{S_2} + (S_1 L_2 - S_2 L_1) \pdr_{S_3} \right]. \qquad
	\label{e:Hamiltonian-V-F}
	\eeqs
The coefficient of each of the coordinate vector fields in $V_E$ gives the time derivative of the corresponding coordinate (upto a factor of $k^2$) and leads to the EOM (\ref{e:EOM-LS-explicit}). These vector fields commute, since $[ V_E, V_h ] = - V_{\{E, h \}}$.

\vspace{.25cm}

{\fl \bf Conserved quantities for the Euclidean Poisson algebra:} As noted, the same Hamiltonian (\ref{e: H-mechanical}) with the $\{ \cdot, \cdot \}_\varepsilon$ PBs leads to the $S$-$L$ EOM (\ref{e:EOM-LS}). Moreover, it can be shown that ${\mathfrak{c}}, m, s$ and $h$ (\ref{e:conserved-quantities}) continue to be in involution with respect to $\{ \cdot, \cdot \}_\varepsilon$ and to commute with $H$. Interestingly, the Casimirs (${\mathfrak{c}}, m$) and non-Casimir conserved quantities $(s^2, h)$ exchange roles in going from the nilpotent to the Euclidean Poisson algebras. 

\vspace{.25cm}

{\fl \bf Simplification of EOM  using conserved quantities:} Using the conserved quantities we may show that $\dot{u}, \dot{\tht}$ and $\dot{\phi}$ are functions of $u = S_3/k$ alone. Indeed, using (\ref{e: PB-SL}) and (\ref{e:L-S-polar}) we get
	\beqs
	\dot{u}^2 &=& \frac{\dot S_3^2}{k^2} = \la^2  k^2 \rho^2 r^2 \sin^2(\theta - \phi), \qquad
	\dot \tht = \frac{L_1 \dot L_2 - \dot L_1 L_2}{L_1^2 + L_2^2}  = - \frac{k \rho}{r} \cos(\tht - \phi) \cr
\text{and} \quad 
	\dot \phi &=& \frac{S_1 \dot S_2 - \dot S_1 S_2}{S_1^2 + S_2^2} = k m \la + k \la   \frac{r u}{\rho} \cos(\tht - \phi).
	\label{e:theta-phi-u-EOM} 
	\eeqs
Now $r,\rho$ and $\tht - \phi$ may be expressed as functions of $u$ and the conserved quantities. In fact,
	\beq
	\rho^2 = s^2 -u^2, \quad  r^2 = 2{\mathfrak{c}} - m^2 - \frac{2u}{\la} \quad  \text{and} \quad h = \frac{\Tr S L}{k^2} = -m u + r \rho  \cos(\tht - \phi).
	\label{e:relation-theta-phi-u}
	\eeq
Thus we arrive at
	\beq
	\dot{u}^2 = \la^2  k^2 \left[ (s^2 - u^2)\left(2{\mathfrak{c}} -m^2 - \frac{2u}{\la}\right) - (h + mu)^2\right] = 2 \la k^2 \chi(u),
	\label{e:EOM-u}
	\eeq
	\beq
	\dot{\theta} = -k\left(\frac{h + m u}{2{\mathfrak{c}} - m^2 - \frac{2u}{\la}}\right) \quad \text{and} \quad
	\dot{\phi} = k m \la + k \la u \left(\frac{h + mu}{s^2 - u^2}\right).
	\label{e:theta-phi-dynamics}
	\eeq
Moreover, the formula for $h$ in (\ref{e:relation-theta-phi-u}) gives a relation among $u, \tht$ and $\phi$ for given values of conserved quantities. Thus, starting from the 6D $S$-$L$ phase space and using the four conservation laws, we have reduced the EOM to a pair of ODEs on the common level set of conserved quantities. For generic values of the conserved quantities, the latter is an invariant torus parameterized, say, by $\tht$ and $\phi$. Furthermore, $\dot u^2$ is proportional to the cubic $\chi(u)$ and may be solved in terms of the $\wp$ function while $\tht$ is expressible in terms of the Weierstrass $\zeta$ and $\sigma$ functions as shown in Ref. \cite{R-R}.

\subsection{Symmetries and associated  canonical transformations}
\label{s:Noether-symmetries-CT}

Here, we identify the Noether symmetries and canonical transformations (CT) generated by the conserved quantities. The constant $m = -L_3/k$ commutes (relative to $\{ \cdot, \cdot \}_{\nu}$) with all observables and acts trivially on the coordinates $R_a$ and momenta $P_b$ of the mechanical system.

The infinitesimal CT $R_3 \to R_3 + \varepsilon$ corresponding to the cyclic coordinate in $L_{\rm mech}$ (\ref{e:Lagrangian-Mech}) is generated by $ (\varepsilon \la k/2) (2 {\mathfrak{c}} - m^2) = \varepsilon k (P_3 - 1/ \la)$ (\ref{e:canonical-momenta-P_a}). $L_{\rm mech}$ is also invariant under infinitesimal rotations in the $R_1$-$R_2$ plane. This corresponds to the infinitesimal CT
	\beq
	\del R_a = \varepsilon \eps_{ab} R_b, \quad
	\del P_a = \varepsilon \eps_{ab} P_b \quad \text{for} \quad a,b = 1,2 \quad \text{and} \quad
	\del R_3 = \del P_3 = 0,
	\eeq
with generator (Noether charge) $\varepsilon k \left[ h + (\la m/ 2) (2 {\mathfrak{c}} - m^2 ) \right]$. The additive constants involving $m$ may of course be dropped from these generators. Thus, while $P_3$ (or equivalently $\mathfrak{c}$) generates translations in $R_3$, $h$ (up to addition of a multiple of $P_3$) generates rotations in the $R_1$-$R_2$ plane. In addition to these two point-symmetries, the Hamiltonian (\ref{e:Hamiltonian-mech-R-P}) is also invariant under an infinitesimal CT that mixes coordinates and momenta:
	\beqs
	\del R_{a} &=& 2 \varepsilon P_a, 
	\quad \del P_a = \varepsilon \la^2 \left[\frac{2}{\la}\left( P_3 - \frac{1}{\la} \right) - (R_1^2 + R_2^2) - \frac{m^2}{2} \right] R_a \quad
	\text{for} \quad a = 1,2
	\cr \text{while} \quad \del R_3 &=& \varepsilon \left[2 P_3 - \la (R_1^2 + R_2^2)\right] \quad \text{and} \quad \del P_3 = 0.
	\eeqs 
This CT is generated by the conserved quantity 
	\beq
	\varepsilon k \left[ s^2 + 2 {\mathfrak{c}} + \la m \left( h + \left(\frac{\la m}{2} \right)(2 {\mathfrak{c}} - m^2 )\right)  \right]
	\label{e:generator-momenta}
	\eeq
which differs from $s^2$ by terms involving $h$ and $\mathfrak{c}$ which serve to simplify the CT by removing an infinitesimal rotation in the $R_1$-$R_2$ plane as well as a constant shift in $R_3$. Here, upto Casimirs, (\ref{e:generator-momenta}) is related to the Hamiltonian via $s^2 + 2 {\mathfrak{c}} = (1/k^2)(2 H - k^2/\la^2)$. 

The above assertions follow from using the canonical PBs, $\{ R_a , k P_b \} = \del_{ab}$ to compute the changes $\del R_a = \{ R_a , Q \}$ etc., generated by the three conserved quantities $Q$ expressed as:
	\beqs
	h &=& P_1 R_2 - P_2 R_1 - m P_3, \qquad
	{\mathfrak{c}} = \frac{1}{\la} \left( P_3 - \frac{1}{\la} \right) + \frac{m^2}{2} \quad \text{and} 
	\cr
	s^2 &=& \sum_{a=1}^3 P_a^2  
	+ \la m \eps_{ab} R_a P_b - \frac{2}{\la} P_3 + \frac{\la^2}{4} \left(R_1^2 + R_2^2 \right) \left[ R_1^2 + R_2^2 - \frac{4}{\la} \left( P_3 - \frac{1}{\la} \right) + m^2 \right] + \frac{1}{\la^2}. \qquad
	\eeqs

\subsection{Relation of conserved quantities to Noether charges of the field theory}
\label{s:conserved-qtys-Noether-charges}

Here we show that three out of four combinations of conserved quantities ($P_3, h - m/\la$ and  $H$) are reductions of scalar field Noether charges, corresponding to symmetries under translations of $\phi$, $x$ and $t$. The fourth conserved quantity $L_3 = -mk$ arose as a parameter in (\ref{e:ansatz}) and is not the reduction of any Noether charge. By contrast, the charge corresponding to internal rotations of $\phi$ does not reduce to a conserved quantity of the RR model. 



Under the shift symmetry $\phi \to \phi + \eta $ of (\ref{e:Nonlinear-wave-equation}), the PBs (\ref{e: Canonical-PB-field}) preserve their canonical form as $\del \pi = (1/3) [\eta , \phi']$ commutes with $\phi$. This leads to the conserved Noether density and current
	\beq
	j_t =  \Tr  \eta \left( \frac{\dot \phi}{\la}  - \frac{ [ \phi , \phi']}{2}\right) 
	\quad \text{and} \quad
	j_x = \Tr \eta \left( -\frac{\phi'}{\la} + \frac{ [ \phi,  \dot \phi ]}{2}\right).
	\eeq
The conservation law $\pdr_t j_t + \pdr_x j_x = 0$ is equivalent to (\ref{e:Nonlinear-wave-equation}) \cite{C-Z}. Taking $\eta \propto \la$, all matrix elements of $Q^{\rm s} = \int \left( \dot \phi - (\la/2) [ \phi , \phi']\right) \: dx$ are conserved. To obtain $P_3$ (\ref{e:canonical-momenta-P_a}) as a reduction of $Q^{\rm s}$ we insert the ansatz (\ref{e:ansatz}) to get 
	\beq
	Q^{\rm s} = \int e^{Kx} \tl Q^{\rm s} e^{-Kx} \: dx \quad \text{where} \quad  \tl Q^{\rm s} = \dot R - \frac{\la}{2} [R, [ K, R] + mK].
	\eeq	
Expanding $\tl Q^{\rm s} = \tilde{Q}^{\rm s}_a t_a$ and using the Baker-Campbell-Hausdorff formula we may express
	\beq
	Q^{\rm s} = \int \left( \cos kx \sig_2 - \sin kx \sig_1 \right) \frac{\tilde{Q}^{\rm s}_1}{2i} \: dx + \int \left( \cos kx \sig_1 + \sin kx \sig_2 \right) \frac{\tilde{Q}^{\rm s}_2}{2i} \: dx  + \int \tilde{Q}^{\rm s}_3 \frac{\sig_3}{2i} \:dx.
	\eeq
The first two terms vanish while $\tl{Q}^{\rm s}_3 = P_3$ so that $Q^{\rm s} = l P_3 t_3$, where $l$ is the spatial length.



The density $({\cal P} = \Tr \dot{\phi} \phi'/\la)$ and current $(-{\cal E} = -  (1/2\la) \Tr ( \dot \phi^2 + \phi'^2))$ (\ref{e:H-Nilpotent-field-theory}) corresponding to the symmetry $x \to x + \eps$ of (\ref{e:Nonlinear-wave-equation}) satisfy $\pdr_t {\cal P} - \pdr_x {\cal E} = 0$ or $\Tr \left( \ddot \phi - \phi'' \right) \phi' = 0$. The conserved momentum $P = \Tr \int I J \: dx$ per unit length upon use of (\ref{e: L-and-S}) reduces to
	\beq
	P = \Tr \int \ov \la e^{Kx} \dot R \left( [K, R] + mK \right) e^{-Kx} dx = \frac{l}{\la} \Tr \left(S - \ov \la K \right)L = \frac{l k^2}{\la} \left( h - \frac{m}{\la} \right).
	\eeq 
As shown in Section \ref{s:Hamiltonian-mechanical}, the field energy per unit length reduces to the RR model Hamiltonian (\ref{e: H-mechanical}).



Infinitesimal internal rotations $\phi \to \phi + \tht [n, \phi]$ (for $n \in \mathfrak{su}(2)$ and small angle $\tht$) are symmetries of (\ref{e:Lagrangian-scalar-field}) leading to the Noether density and current:
	\beq
	j_t = \Tr \left( \frac{n}{\la} [ \phi, \dot \phi ] -  \frac{n}{3} [\phi, [\phi, \phi']] \right) \quad \text{and} \quad 
	j_x = \Tr \left( -\frac{n}{\la}[ \phi , \phi'] +  \frac{n}{3} [\phi, [\phi, \dot \phi]] \right)
	\label{e:Q-internal-rotation}
	\eeq
and the conservation law $\Tr \left( n \left[ \phi, \frac{\ddot\phi - \phi''}{\la} - [\dot \phi, \phi']\right]\right) = 0$. However, the charges $ Q^{\rm rot}_n = \int j_t \: dx$ do not reduce to conserved quantities of the RR model. This is because the space of mechanical states is {\it not} invariant under the above rotations as $K = i k \sig_3/2$ picks out the third direction.

\subsection{Static and circular submanifolds}
\label{s:Static-and-trigonometric-solutions}

In general, solutions of the EOM of the RR model (\ref{e:EOM-LS}) are expressible in terms of elliptic functions \cite{R-R}. Here, we discuss the `static' and `circular' (or `trigonometric') submanifolds of the phase space where solutions to (\ref{e:EOM-LS}) reduce to either constant or circular functions of time. Interestingly, these are precisely the places where the conserved quantities fail to be  independent as will be shown in Section \ref{s:Independence-of-conserved-quantities}.

\subsubsection*{Static submanifolds}

By a static solution on the $L$-$S$ phase space we mean that the six variables $L_a$ and $S_b$ are time-independent. We infer from (\ref{e:EOM-LS-explicit}) that static solutions occur precisely when $S_1 = S_2 = 0$ and $S_3 L_2 = S_3 L_1 = 0$. These conditions lead to two families of static solutions $\Sigma_3$ and $\Sigma_2$. The former is a 3-parameter family defined by $S_{1,2,3}= 0$ with the $L_a$ being arbitrary constants. The latter is a 2-parameter family where $L_3$ and $S_3$ are arbitrary constants while $L_{1,2} = S_{1,2} = 0$. We will refer to $\Sigma_{2,3}$ as `static' submanifolds of $M^6_{S \text{-} L}$. Their intersection is the $L_3$ axis. Note however, that the `extra coordinate' $R_3(t)$ corresponding to such solutions evolves linearly in time, $R_3(t) = R_3(0) + (S_3 + k/\la)t$.

The conserved quantities satisfy interesting relations on $\Sigma_2$ and $\Sigma_3$. On $\Sigma_2$ we must have $h = \mp \sign{k} \: m s$ and ${\mathfrak{c}} = m^2/ 2 \pm \sign{k} \: s/ \la$ with $s \geq 0$ where the signs correspond to the two possibilities $S_3 = \pm s |k|$. Similarly, on $\Sigma_3$ we must have $s = h = 0$ with $2 {\mathfrak{c}} - m^2 \geq 0$. While $\Sigma_3$ may be regarded as the pre-image (under the map introduced in Section \ref{s:conserved-quantities}) of the submanifold $s = 0$ of the space of conserved quantities $\cal Q$, $\Sigma_2$ is {\it not} the inverse image of any submanifold of $\cal Q$. In fact, the pre-image of the submanifold of $\cal Q$ defined by the relations that hold on $\Sigma_2$ also includes many interesting nonstatic solutions that we shall discuss elsewhere.

\subsubsection*{Circular or Trigonometric submanifold}

As mentioned in Section \ref{s:conserved-quantities} the EOM may be solved in terms of elliptic functions \cite{R-R}. In particular, since from (\ref{e:EOM-u}) $\dot u^2 = 2 \la k^2 \chi(u)$, $u$ oscillates between a pair of adjacent zeros of the cubic $\chi$, between which $\chi > 0$. When the two zeros coalesce $u = S_3/k$ becomes constant in time. From (\ref{e:EOM-LS-explicit}) this implies $S_1 L_2 = S_2 L_1$, which in turn implies that $\tan \tht = \tan \phi$ or $\tht - \phi = n \pi$ for an integer $n$. Moreover, $\rho, r$ and $\dot{\tht} = \dot{\phi}$ become constants as from (\ref{e:theta-phi-dynamics}), they are functions of $u$. Thus the EOM for $S_1 = k \rho \cos \phi$ and $S_2 = k \rho \sin \phi$ simplify to $\dot{S_1} = -\dot{\phi} S_2$ and $\dot{S_2} = \dot{\phi} S_1$ with solutions given by circular functions of time. The same holds for $L_1 = k r \cos \tht$ and $L_2 = k r \sin \tht$ as $\dot{L_1} = k S_2$ and $\dot{L_2} = -k S_1$ (\ref{e:EOM-LS-explicit}). Thus, we are led to introduce the circular submanifold of the phase space as the set on which solutions degenerate from elliptic to circular functions. In what follows, we will express it as an algebraic subvariety of the phase space. Note first, using (\ref{e:L-S-polar}), that on the circular submanifold
	\beq
	\dot{\tht} = \dot{\phi} = (-1)^{n+1}\frac{k \rho}{r} = -\frac{k S_1}{L_1} = -\frac{k S_2}{L_2}.
	\eeq
Thus EOM on the circular submanifold take the form
	\beq
	\dot L_3 = \dot S_3 = 0, \quad
	\dot L_1 = k S_2, \quad
	\dot L_2 = -k S_1, \quad \dot S_1 = \frac{k S_2^2}{L_2}
	\quad  \text{and} \quad
	\dot S_2 = -\frac{k S_1^2}{L_1}.
	\label{e:EOM-trigonometric-submanifold}
	\eeq
The nonsingular nature of the Hamiltonian vector field $V_E$ ensures that the above quotients make sense. Interestingly, the EOM (\ref{e:EOM-LS-explicit}) reduce to (\ref{e:EOM-trigonometric-submanifold}) when $S$ and $L$ satisfy the following three relations
	\beq
	\Xi_1: \; (S \times L)_3 = 0, \quad
	\Xi_2: -\la L_1(S \times L)_2 = k S_1^2 
	\quad \text{and} \quad
	\Xi_3: \: \la L_2 (S \times L)_1 = k S_2^2.
	\label{e:trigonometric-submanifold-conditions}
	\eeq
Here $(S \times L)_3 = S_1 L_2 - S_2 L_1$ etc. The conditions (\ref{e:trigonometric-submanifold-conditions}) define a singular subset $\bar {\cal C}$ of the phase space. $\bar {\cal C}$ may be regarded as a disjoint union of the static submanifolds $\Sigma_2$ and $\Sigma_3$ as well as the three submanifolds ${\cal C}$, ${\cal C}_1$ and ${\cal C}_2$ of dimensions four, three and three, defined by:
	\beqs
	{\cal C}: && S_1 \neq 0, \quad S_2 \neq 0, \quad \Xi_1 \;\; \text{and either} \;\; \Xi_2 \;\; \text{or} \;\; \Xi_3 , \cr
	{\cal C}_1: && S_1 =0,\quad S_2 \neq 0, \quad  L_1 = 0  \quad \text{and} \quad \Xi_3 \cr
	\text{and} \quad {\cal C}_2: && S_1 \neq 0,\quad S_2 = 0, \quad  L_2 = 0 \quad \text{and}\quad \Xi_2. \qquad
	\label{e:singular-submanifolds-for-4wedge-vanish}
	\eeqs
${\cal C}_1$, ${\cal C}_2$, $\Sigma_2$ and $\Sigma_3$ lie along boundaries of ${\cal C}$. The dynamics on ${\cal C}$ (where $L_{1,2}$ and $S_{1,2}$ are necessarily nonzero) is particularly simple. We call ${\cal C}$ the circular submanifold, it is an invariant submanifold on which $S$ and $L$ are circular functions of time. Indeed, to solve (\ref{e:EOM-trigonometric-submanifold}) note that the last pair of equations may be replaced with $\dot L_1/L_1 = \dot S_1/S_1$ and $\dot L_2/L_2 = \dot S_2/S_2$ which along with $S_1 L_2 = S_2 L_1$ implies that $S_{1,2} = \al L_{1,2}$ for a constant $\al > 0$. Thus we must have $\dot S_1 = k \al S_2$ and $\dot S_2 = - k \al S_1$ with the solutions
	\beq
	S_1/k = A \sin(k \al t) + B \cos(k \al t) \quad \text{and} 
	\quad S_2/k = - B \sin( k \al t) + A\cos(k \al t).
	\eeq
$A$ and $B$ are dimensionless constants of integration. As a consequence of $\Xi_2$ or $\Xi_3$ (\ref{e:trigonometric-submanifold-conditions}), the constant values of $L_3 = -k m$ and $S_3 = u k$ must satisfy the relation $u = -\alpha (\alpha + \la m)/\la$. The other conserved quantities are given by 
	\beqs
	{\mathfrak{c}} &=& \frac{1}{2} \left(m^2 + 
	\frac{A^2 + B^2}{\al^2} - \frac{2 \al (\al + \la m)}{\la^2} \right),
	\quad
	 h = \frac{A^2+B^2}{\al} + \frac{\al m (\al +\la m)}{\la} \quad \text{and} \cr
	 s^2 &=& A^2+B^2+\frac{\al^2 (\al + \la m)^2}{\la^2}.
	 \eeqs
Though we do not discuss it here, it is possible to show that these trigonometric solutions occur precisely when the common level set of the four conserved quantities is a circle as opposed to a 2-torus. Unlike $\Sigma_2$ and $\Sigma_3$, the boundaries ${\cal C}_1$ and ${\cal C}_2$ are {\it not} invariant under the dynamics. The above trajectories on ${\cal C}$ can reach points of ${\cal C}_1$ or ${\cal C}_2$, say when $S_1$ or $S_2$ vanishes. On the other hand, in the limit $A = B = 0$ and $\al \neq 0$, the above trigonometric solutions reduce to the $\Sigma_2$ family of static solutions. What is more, $\Sigma_2$ lies along the common boundary of ${\cal C}_1$ and ${\cal C}_2$. Finally, when $A$, $B$ and $\al$ are all zero, $S_1, S_2$ and $S_3$ must each vanish while $L_1, L_2$ and $L_3$ are arbitrary constants. In this case, the trigonometric solutions reduce to the $\Sigma_3$ family of static solutions.

\subsection{Independence of conserved quantities and singular submanifolds} 
\label{s:Independence-of-conserved-quantities}

We wish to understand the extent to which the above four conserved quantities are independent. We say that a pair of conserved quantities, say $f$ and $g$, are independent if $df$ and $dg$ are linearly independent or equivalently  if $df \wedge dg$ is not identically zero. Similarly, three conserved quantities are independent if $df \wedge dg \wedge dh \not \equiv 0$ and so on. In the present case, we find that the pairwise, triple and quadruple wedge products of $d{\mathfrak{c}}, dh, dm$ and $ds^2$ do not vanish identically on the whole $L$-$S$ phase space. Thus the four conserved quantities are generically independent. However, there are some `singular' submanifolds of the phase space where these wedge products vanish and relations among the conserved quantities emerge. This happens precisely on the static submanifolds $\Sigma_{2,3}$ and $\bar {\cal C}$ which includes the circular submanifold and its boundaries discussed in Section \ref{s:Static-and-trigonometric-solutions}.

A related question is the independence of the canonical vector fields obtained through contraction of the 1-forms with the (say, nilpotent) Poisson tensor $\scripty{r}_0$. The Casimir vector fields $V_{\mathfrak{c}}$ and $V_m$ are identically zero as $d{\mathfrak{c}}$ and $dm$ lie in the kernel of $\scripty{r}_0$. Passing to the symplectic leaves $M^4_{{\mathfrak{c}} m}$, we find that the vector fields corresponding to the non-Casimir conserved quantities $V_E$ and $V_h$ are generically linearly independent.  Remarkably, this independence fails precisely where $M^4_{{\mathfrak{c}} m}$ intersects $\bar {\cal C}$.

\subsubsection*{Conditions for pairwise independence of conserved quantities}
 
The 1-forms corresponding to our four conserved quantities are
	\beq
	k^2 ds^2 = 2 S_a dS_a, \quad k^2 d{\mathfrak{c}} = L_a dL_a + \frac{k}{\la} dS_3, \quad
	-k \: dm = dL_3 \quad \text{and} \quad  k^2 dh = S_a dL_a + L_a dS_a.
	\eeq
None of the six pairwise wedge products is identically zero:
	\beqs
	\frac{k^4}{2} ds^2 \wedge dh &=& S_a S_b dS_a \wedge dL_b + \half (S_a L_b - S_b L_a) dS_a \wedge dS_b, 
	\quad
	\frac{k^3}{2} dm \wedge ds^2  = S_a dS_a \wedge dL_3 \cr
	k^3 dm \wedge dh &=& S_a dL_a \wedge dL_3 + L_a dS_a \wedge dL_3, 
	\quad  
	k^3 d{\mathfrak{c}} \wedge dm = L_a dL_3 \wedge dL_a + \frac{k}{\la} dL_3 \wedge dS_3 \cr 
	\frac{k^4}{2} ds^2 \wedge d{\mathfrak{c}} &=&  S_a L_b dS_a \wedge dL_b + \frac{kS_a}{\la} dS_a \wedge dS_3 \cr
	k^4 dh \wedge d{\mathfrak{c}} &=& \half(S_a L_b - S_b L_a) dL_a \wedge dL_b - \sum _{b \neq 3} L_a L_b dL_a \wedge dS_b + \frac{k L_a}{\la} dS_a \wedge dS_3 \cr
	&& + \left( \frac{k S_a}{\la} - L_a L_3 \right) dL_a \wedge dS_3.  
	\eeqs
Though no pair of conserved quantities is dependent on $M^6_{S \text{-} L}$, there are some relations between them on certain submanifolds. For instance, $ds^2 \wedge dh = ds^2 \wedge dm = 0$ on the $3$D submanifold $\Sigma_3$ (where $s = 0$) while $dh \wedge dm = 0$ on the curve defined by $S_{1,2} = L_{1,2,3} = 0$ where $h = m = 0$. Similarly, $ds^2 \wedge d{\mathfrak{c}} = 0$ on both these submanifolds where $s = 0$ and $\la^2 {\mathfrak{c}}^2 = k^2 s^2$ respectively. Moreover, $dh \wedge d{\mathfrak{c}} = 0$ on the curve defined by $S_{1,2} = L_{1,2} = L_3^2 - k S_3 / \la = 0$ where $k^2 h^2 = \la^2 {\mathfrak{c}}^3$. However, the dynamics on each of these submanifolds is trivial as each of their points represents a static solution. On the other hand, the Casimirs $m$ and ${\mathfrak{c}}$ are independent on all of $M^6_{S \text{-} L}$ provided $1/\la k^2 \ne 0$.

\subsubsection*{Conditions for relations among triples of conserved quantities:}
 
The four possible wedge products of three conserved quantities are given below.
	\beqs
	\frac{k^5}{2} dh \wedge ds^2 \wedge dm &=& S_a S_b dS_a \wedge dL_b \wedge dL_3 + \half (S_a L_b - S_b L_a) dS_a \wedge dS_b \wedge dL_3 \cr 
	\frac{k^6}{2} ds^2 \wedge dh \wedge d{\mathfrak{c}} &=& \half S_a (S_b L_c - S_c L_b) dS_a \wedge dL_b \wedge dL_c + (S_1 L_2 - S_2 L_1) \frac{k}{\la} dS_1 \wedge dS_2 \wedge dS_3 \cr
	&& + \left[(S_a L_3 - S_3 L_a)L_c - \frac{S_a S_c k}{\la} \right] dS_a \wedge dS_3 \wedge dL_3  \cr
	&& + \sum_{a, b \neq 3} \half( S_a L_b - S_b L_a) L_c dS_a \wedge dS_b \wedge dL_c \cr 
	\frac{k^5}{2} dm \wedge ds^2 \wedge d{\mathfrak{c}} &=& S_a L_b dS_a \wedge dL_3 \wedge dL_b + \frac{k S_a}{\la} dS_a \wedge dL_3 \wedge dS_3 \cr
	k^5 dm \wedge dh \wedge d{\mathfrak{c}} &=& (S_2 L_1 - S_1 L_2) dL_1 \wedge dL_2 \wedge dL_3 + \left( \frac{k S_a}{\la} - L_a L_3 \right) dL_a \wedge dL_3 \wedge dS_3 \cr
	&& - \sum_{b \neq 3} L_a L_b dL_a \wedge dL_3 \wedge dS_b + \frac{k L_a}{\la} dS_a \wedge dL_3 \wedge dS_3.
	\eeqs
It is clear that none of the triple wedge products is identically zero, so that there is no relation among any three of the conserved quantities on all of $M^6_{S \text{-} L}$. However, as before, there are relations on certain submanifolds. For instance, $ds^2 \wedge dm \wedge d{\mathfrak{c}} = ds^2 \wedge dh \wedge d{\mathfrak{c}} = ds^2 \wedge dh \wedge dm = 0$ on both the static submanifolds $\Sigma_3$ and $\Sigma_2$ of Section \ref{s:Static-and-trigonometric-solutions}. On $\Sigma_2$ we have the three relations $s^2 = (\la^2/4) (2{\mathfrak{c}} - m^2)^2$, $\la^2 (2{\mathfrak{c}} s^2 - h^2)^2 = 4s^6$ and $h^2 = m^2 s^2$. On the other hand, $dh \wedge dm \wedge d{\mathfrak{c}} = 0$ only on the static submanifold $\Sigma_2$ on which the relation $4h^2 = \la^2 m^2 (2{\mathfrak{c}} - m^2)^2$ holds. 

\subsubsection*{Vanishing of four-fold wedge product and the circular submanifold}

Finally, the wedge product of all four conserved quantities is
	\beqs
	\frac{k^7}{2}dh \wedge ds^2 \wedge dm \wedge d{\mathfrak{c}} &=& (S_1 L_2 - S_2 L_1)\bigg[S_b dL_1 \wedge dL_2 \wedge dL_3 \wedge dS_b \cr 
	&& - \frac{k}{\la} dS_1 \wedge dS_2 \wedge dS_3 \wedge dL_3 - L_b dS_1 \wedge dS_2 \wedge dL_b \wedge dL_3 \bigg] \cr 
	&& + \left[\frac{S_a S_b k}{\la} + (L_a S_3 - S_a L_3) L_b\right] dS_a \wedge dS_3 \wedge dL_b \wedge dL_3.
	\label{e:four-fold-wedge-product}
	\eeqs
This wedge product is {\it not} identically zero on the $L$-$S$ phase space so that the four conserved quantities are independent in general. It does vanish, however, on the union of the two static submanifolds $\Sigma_2$ and $\Sigma_3$. This is a consequence, say, of $ds^2 \wedge dm \wedge d{\mathfrak{c}}$ vanishing on both these submanifolds. Alternatively, if $S_1 = S_2 = 0$, then requiring $ dh \wedge ds^2 \wedge dm \wedge d{\mathfrak{c}} = 0$ implies either $S_3 = 0$ or $L_1 = L_2 = 0$. Interestingly, the four-fold wedge product also vanishes elsewhere. In fact, the necessary and sufficient conditions for it to vanish are $\Xi_1, \Xi_2$ and $\Xi_3$ introduced in (\ref{e:trigonometric-submanifold-conditions}) which define the submanifold $\bar {\cal C}$ of the phase space that includes the circular submanifold ${\cal C}$ and its boundaries ${\cal C}_{1,2}$ and $\Sigma_{2,3}$. 
 
Consequent to the vanishing of the four-fold wedge product $dh \wedge ds^2 \wedge dm \wedge d{\mathfrak{c}}$, the conserved quantities must satisfy a new relation on ${\cal C}$ which may be shown to be the vanishing of the discriminant $\D ({\mathfrak{c}}, m , s^2, h)$ of the cubic polynomial 
	\beq
	\chi(u) = u^3 - \la {\mathfrak{c}} u^2 - \left( s^2 + \la h m \right) u + \frac{\la}{2}\left( 2 {\mathfrak{c}}s^2 - h^2 - m^2 s^2 \right). 
	\label{e:cubic-equation-S3}
	\eeq
The properties of $\chi$ help to characterize the common level sets of the four conserved quantities. In fact, $\chi$ has a double zero when the common level set of the four conserved quantities is a circle (as opposed to a 2-torus) so that it is possible to view ${\cal C}$ as a union of circular level sets. Note that $\D$ in fact vanishes on a submanifold of phase space that properly contains $\bar {\cal C}$. However, though the conserved quantities satisfy a relation on this larger submanifold, their wedge product only vanishes on $\bar {\cal C}$. The nature of the common level sets of conserved quantities will be examined in the next Chapter.

\subsubsection*{Independence of Hamiltonian and helicity on symplectic leaves $M^4_{{\mathfrak{c}}m}$}

So far, we examined the independence of conserved quantities on $M^6_{S \text{-} L}$ which, however, is a degenerate Poisson manifold. By assigning arbitrary real values to the Casimirs $\mathfrak{c}$ and $m$ (of $\{ \cdot , \cdot \}_\nu$) we go to its symplectic leaves $M^4_{{\mathfrak{c}} m}$. $L_{1,2}$ and $S_{1,2}$ furnish coordinates on $M^4_{{\mathfrak{c}} m}$ with
	\beq
	S_3 (L_1, L_2) =  \frac{\la k}{2}\left( (2 {\mathfrak{c}} - m^2) - \frac{1}{k^2} (L_1^2 + L_2^2) \right)  \quad \text{and} \quad L_3 = -m k.
	\label{e:S3-and-L3-on-M4-cm}
	\eeq
The Hamiltonian $H = E k^2$ (or $k^2 s^2 = 2 (H - {\mathfrak{c}} k^2 - k^2/ 2 \la^2) $) and helicity $h$ are conserved quantities for the dynamics on $M^4_{{\mathfrak{c}} m}$. Here we show that the corresponding vector fields $V_E$ and $V_h$ are generically independent on each of the symplectic leaves and also identify where the independence fails. On $M^4_{{\mathfrak{c}} m}$, the Poisson tensor $\scripty{r}_0$ is nondegenerate so that $V_E$ and $V_h$ are linearly independent iff $dE \wedge dh \neq 0$. We find 
	\beqs
	k^5 dE \wedge dh &=& (S_1 L_2 - S_2 L_1)\left( k dS_1 \wedge dS_2 + \la S_3 dL_1 \wedge dL_2 \right) \cr
	&& + \sum_{a,b =1,2}\left(\la (S_b L_3 - S_3 L_b)L_a - k S_a S_b \right) dL_a \wedge dS_b
	\eeqs 
Here $S_3$ and $L_3$ are as in (\ref{e:S3-and-L3-on-M4-cm}). Interestingly, the conditions for $dE \wedge dh$ to vanish are the same as the restriction to $M^4_{{\mathfrak{c}}m}$ of the conditions for the vanishing of the four-fold wedge product $dh \wedge ds^2 \wedge dm \wedge d{\mathfrak{c}}$ (\ref{e:four-fold-wedge-product}).  It is possible to check that this wedge product vanishes on $M^4_{{\mathfrak{c}} m}$ precisely when $S_{1,2}$ and $L_{1,2}$ satisfy the relations $\Xi_1, \Xi_2$ and $\Xi_3$ of (\ref{e:trigonometric-submanifold-conditions}), where $S_3$ (\ref{e:S3-and-L3-on-M4-cm}) and $L_3 = -m k$ are expressed in terms of the coordinates on $M^4_{{\mathfrak{c}}m}$. Recall from Section \ref{s:Static-and-trigonometric-solutions} that (\ref{e:trigonometric-submanifold-conditions}) is satisfied on the singular set $\bar {\cal C} \subset M^6_{S \text{-} L}$ consisting of the union of the circular submanifold ${\cal C}$ and its boundaries ${\cal C}_{1,2}$ and $\Sigma_{2,3}$. We note in passing that the $E$ and $h$ when regarded as functions on $M^6_{S \text{-} L}$ (rather than $M^4_{{\mathfrak{c}} m}$) are independent everywhere except on a curve that lies on the static submanifold $\Sigma_2$. In fact, we find that $dE \wedge dh$ vanishes iff $S_{1,2} = L_{1,2} = 0$ and $S_3^2 + kS_3/ \la = L_3^2$. Thus, on $M^4_{{\mathfrak{c}} m}$, $V_E$ and $V_h$ are linearly independent away from the set (of measure zero) given by the intersection of $\bar {\cal C}$ with $M^4_{{\mathfrak{c}} m}$. For example, the intersections of ${\cal C}$ with $M^4_{{\mathfrak{c}} m}$ are in general $2$D manifolds defined by four conditions among $S$ and $L$: $\Xi_1$ and $\Xi_2$ (with $S_{1,2} \neq 0$) as well as the condition (\ref{e:S3-and-L3-on-M4-cm}) on $S_3$ and finally $L_3 = -m k$. {\it This independence along with the involutive property of $E$ and $h$ allows us to conclude that the system is Liouville integrable on each of the symplectic leaves.}

\section[Stability of classical static solutions]{Stability of classical static solutions\sectionmark{Stability of classical static solutions}}
\sectionmark{Stability of classical static solutions}
\label{s:stability-of-static-solutions}

In this section, we discuss the stability of classical static solutions of the RR model. In Section \ref{s:Static-and-trigonometric-solutions}, we found the static submanifolds $\Sigma_2$ ($S_{1,2} = L_{1,2} = 0$) and $\Sigma_3$ ($S_{1,2,3} = 0$) on the $L$-$S$ phase space of the RR model. Viewed on the $R$-$P$ phase space, these solutions are static except for a possible linear time-dependence of $R_3$ ($\dot R_3 = S_3 + k/ \la$). Here we examine the stability of these solutions on the $L$-$S$ and $R$-$P$ phase spaces as well as in the parent scalar field theory. These solutions are in general neutrally stable centers with some additional flat directions as well as a possible direction of linear growth in time.

\subsection{Static solutions in the $L$-$S$ phase space and their stability}

Recall that the Hamiltonian of the RR model in the $L$-$S$ variables is
	\beq
	H = \half \left[ \left(S - \frac{K}{\la} \right)^2 + L^2 \right] = \sum_{a =1}^3 \frac{S_a^2 + L_a^2}{2} + \frac{k S_3}{\la} + \frac{k^2 }{2 \la^2} \geq \frac{L_3^2}{2}.
	\label{e:Hamiltonian-RR-LS-variables}
	\eeq 
Here $L_3$ is a Casimir of the nilpotent Poisson algebra. For each value of $L_3 = - mk$, $H$ attains its global minimum $H  = m^2 k^2/2$ at a unique ground state which lies on $\Sigma_2$:
	\beq
	L_{1,2} = S_{1,2} = ,0 \quad L_3 = -mk \quad \text{and} \quad S_3 = \frac{K_3}{\la} = -\frac{k}{\la}.
	\eeq
When elevated to the canonical $R$-$P$ phase space each of these ground states corresponds to a one parameter family of static ground states parametrized by the arbitrary constant value of $R_3$, which is a cyclic coordinate in the Hamiltonian (see Eq.~(\ref{e:Hamiltonian-mech-R-P})).

We now examine the stability of all the static solutions on $\Sigma_2$ by considering the small perturbations:
	\beq
	L_{1,2} = 0 + l_{1,2}, \quad L_3 = -m k+ l_3 \quad S_{1,2} = 0 + s_{1,2} \quad \text{and} \quad S_3 = a k + s_3.
	\eeq
Notice that, $a = -1/\la$ for this to be a ground state. The linearization of the $L$-$S$ equations of motion $\dot L = [K, S]$ and $\dot S = \la [ S, L]$ are
	\beq
	\dot l_1 = k s_2, \quad \dot l_2 = -k s_1, \quad \dot l_3 = 0, \quad \dot s_1 = \la k ( - m s_2 - a l_2), \quad \dot s_2 = \la k (a  l_1 + m s_1) \quad \text{and} \quad \dot s_3 = 0.
	\label{e:linearized-equations-L-S-m-nonzero}
	\eeq
The directions $l_3$ and $s_3$ are flat and $l_3$-$s_3$ plane is a plane of fixed points of this linear system. The remaining variables $l_{1,2}$ and $s_{1,2}$ satisfy a homogenous linear system ($\dot x = A x$) with a nonsingular matrix $A$. The eigenvalues of $A$ are 
	\beq
	\pm \la_{\pm} = \pm \frac{k\sqrt{2 a \lambda  - m^2 \la^2 \pm m \la \sqrt{ - 4 a \lambda + m^2 \la^2}}}{\sqrt{2}}.
	\label{e:imaginary-eigenvalues-Sigma-2}
	\eeq
When $m = 0$ and $a = -1/ \lambda$, $A$ may be diagonalized with eigenvalues $\pm i k$, each with multiplicity two. It is possible to see that $\pm \la_{\pm}$ are imaginary for all values of $m$ and $a$. {\it Thus every point of $\Sigma_2$ is a neutrally stable static solution (a 4D center with two flat directions).}

A similar stability analysis can be done for the static submanifold $\Sigma_3$ defined by $S_{1,2,3} = 0$. We consider small perturbations around any point of $\Sigma_3$:
	\beq
	S_{1,2,3} = 0 + s_{1,2,3}, \quad L_1 = a k + l_1, \quad L_2 = b k + l_2 \quad \text{and} \quad L_3 = -m k + l_3,
	\eeq 
which lead to the linearized equations
	\beqs
	\dot l_1 &=& k s_2, \quad \dot l_2 = -k s_1, \quad \dot l_3 = 0, \quad \dot s_1 = \la k ( -m s_2 - b s_3), \cr
	\dot s_2 &=& \la k ( a s_3 + m s_1 ) \quad \text{and} \quad \dot s_3 = \la( b s_1 - a s_2).
	\eeqs
This system has a three parameter family of fixed points corresponding to $s_{1,2,3} = 0$ and $l_i$ arbitrary. The dynamics along the flat $l_3$ direction decouples. The coefficient matrix $A$ for the remaining five equations has a pair of imaginary eigenvalues ($\pm i k \la \sqrt{a^2 + b^2 + m^2}$) with corresponding imaginary eigenvectors. However, $A$ is a deficient. Its other eigenvalue zero has algebraic multiplicity 3 but only two linearly independent eigenvectors which are in the $l_1$ and $l_2$ directions. Linearized equations become simple in the Jordan basis where $S^{-1} A S = J$. The Jordan block corresponding to the zero eigenspace can be taken as $((0,0,0),(0,0,1),(0,0,0))$. {\it Each of the above fixed points behaves as a center in two directions with oscillatory time dependence. In addition, there are three flat directions and one direction with linear growth in time as in the case of the free particle.} 

\subsection{Static solutions in the $R$-$P$ phase space and their stability}

Now, we examine the stability of static solutions in the $R$-$P$ variables. The equations of motion of the RR model in terms of $R$-$P$ variables are
	\beqs
	\dot R_1 &=& k P_1 - \frac{\la m k}{2} R_2, \quad \dot R_2 = k P_2 + \frac{\la m k}{2} R_1, \quad \dot R_3 = k P_3 - \frac{\la k}{2}(R_1^2 + R_2^2), \quad k \dot P_3 = 0, \cr
	k \dot P_1 &=& -\frac{\la m k^2}{2} P_2 - \left( \frac{\la^2 m^2 k^2}{4} - \la k^2 P_3 + k^2 \right) R_1 - \frac{\la^2 k^2}{2} (R_1^2 + R_2^2) R_1, \quad \text{and}\cr
	k \dot P_2 &=& \frac{\la m k^2}{2} P_1 - \left( \frac{\la^2 m^2 k^2}{4} - \la k^2 P_3 + k^2 \right) R_2 - \frac{\la^2 k^2}{2} (R_1^2 + R_2^2) R_2.
	\eeqs 
The static solutions of these equations are a one parameter family in the $R$-$P$ phase space with values $R_{1,2} = 0$, $k P_{1,2,3} = 0$ and $R_3$ an arbitrary constant parameter (same as the unique ground state in $\Sigma_2$). We consider small perturbations around these static solutions
	\beq
	R_{1,2} = 0 + r_{1,2}, \quad R_3 = R_3(0) + r_3, \quad k P_{1,2,3} = 0 + k p_{1,2,3}.
	\label{e:static-solutions-R-P}
	\eeq
This leads to the linearized equations 
	\beqs
	\dot r_1 &=& k p_1 - \frac{\la m k}{2} r_2, \quad \dot r_2 = k p_2 + \frac{\la m k}{2} r_1, \quad \dot r_3 = k p_3, \quad k \dot p_3 = 0, \cr
	k \dot p_1 &=& -\frac{\la m k^2}{2} p_2 - \frac{\la^2 m^2 k^2}{4} r_1 - k^2 r_1 \quad \text{and} \cr
	k \dot p_2 &=&  \frac{\la m k^2}{2} p_1 - \frac{\la^2 m^2 k^2}{4} r_2 - k^2 r_2.
	\label{e:linearized-equations-R-P}
	\eeqs
Using the map between $R$-$P$ and $L$-$S$ variables (see Eqs.~(\ref{e:EOM-R_3}) and (\ref{e:canonical-momenta-P_a})), it is easy to show that these equations reduce to Eq.~(\ref{e:linearized-equations-L-S-m-nonzero}) when $S_3 = -k/\la$ or $a = -1/\la$. The dynamics in $r_3$-$kp_3$ subspace decouples 
	\beq
	\colvec{2}{\dot r_3}{k \dot p_3} = \colvec{2}{0 & 1}{ 0 & 0} \colvec{2}{r_3}{k p_3}.
	\eeq
$r_3$ and $kp_3$ are like the position and momentum of a free particle: $p_3$ is constant and $r_3$ is linear in time. The dynamics in the $r_{1,2}$-$kp_{1,2}$ space is oscillatory corresponding to the four imaginary eigenvalues $\pm \la_{\pm}$ (with $a = -1/\la$ in Eq.(\ref{e:imaginary-eigenvalues-Sigma-2})). {\it Thus the above fixed points behave as four dimensional centers with an additional flat direction and a direction of linear growth in time.}

\subsection{Stability of static continuous waves in the scalar field theory}

Here we examine the stability of static `continuous waves' regarded as solutions of the scalar field theory. These static solutions of the RR model form a one parameter family and are given by $R_{1,2} = 0, R_3 = R_3(0)$ and $k P_{1,2,3} = 0$ (see Eq.~(\ref{e:static-solutions-R-P})). The corresponding scalar field configurations 
	\beq
	\phi_0(x, t) = e^{K x} R(t) e^{-K x} + m K x= R_{3} (0) \frac{\sig_3}{2 i} + m K x,
	\eeq
are static solutions of the equations of motion $\ddot{\phi} = \phi'' + \la [ \dot \phi, \phi']$. For small perturbations $\phi = \phi_0 + \varphi$, the linearized equations of motion reduce to the wave equation $\ddot \varphi = \varphi''$. The latter can be written as a first-order system 
	\beq
	\colvec{2}{\dot \varphi}{\dot \psi} = \colvec{2}{0 & 1}{ \frac{\pdr^2}{\pdr x^2} & 0} \colvec{2}{\varphi}{\psi},
	\eeq
which may be regarded as an infinite collection of equations for the Fourier mode $\tl \psi(l) = \int e^{-i l x} \psi(x) dx$. Each Fourier mode evolves independently via the coefficient matrix $A_l = ((0,1),(-l^2,0))$. For nonzero real $l$, $A_l$ has eigenvalues $\pm i l$ and $\tl \varphi(l)$,$\tl \psi(l)$ are oscillatory. When $l = 0$, $A_0$ is not diagonalizable and $\tl \varphi (0), \tl \psi(0)$ are like the position and momentum of a free particle. {\it Thus perturbations to static solutions of the RR model are oscillatory in time in all but two directions: $\tl \varphi(0)$ is a flat direction while $\tl \psi(0)$ displays linear growth in time.}

\subsection{Weak coupling limit of classical continuous waves}
\label{d:Weak-coupling-limit-of-classical-continous-waves}

In the weak coupling limit $\la \to 0$, the classical equations of motion of the RR model (\ref{e:EOM-R}) become
	\beq
	\ddot{R}_1 = -k^2 R_1,  \quad \ddot{R}_2 = -k^2 R_2 \quad \text{and} \quad \ddot{R_3} = 0,
	\eeq
with the general solution 
	\beq
	R_{1} = A \cos{kt} + B \sin{kt}, \quad R_2 = C \cos{kt} + D \sin{kt}  \quad \text{and} \quad R_3 = E t + F,
	\eeq
for constants $A, \cdots, F$. The corresponding continuous wave solutions of the weakly coupled field equations $\phi_{tt} = \phi_{xx}$ for the $\mathfrak{su}(2)$ valued field $\phi(x,t)$ are:
\footnotesize
	\beqs
	\phi(x,t) &=& e^{Kx} R(t) e^{-Kx} + m K x = \phi_a \frac{\sig_a}{2i} \cr
	 &=& \frac{1}{2i} \colvec{2}{Et + F - m k x & e^{i kx}( (C + iA) \cos{kt} + (D + i B) \sin{kt})}{ e^{-i kx} ((C - i A) \cos{kt} + (D - i B) \sin{kt})) & -Et - F + m k x }. \qquad
	\eeqs 
\normalsize
From this, we have the components of the classical field
\small
	\beqs
	\phi_1 &=& \cos k x (C \cos kt + D \sin kt) - \sin k x (A \cos kt + B \sin kt), \cr
	\phi_2 &=& \sin k x (C \cos kt + D \sin kt) + \cos k x (A \cos kt + B \sin kt) \quad \text{and} \quad
	\phi_3 =  Et + F - m k x. \qquad
	\eeqs
\normalsize
Though these are not travelling waves, $\phi_{1,2}$ are periodic in $x$ and $t$ while $\phi_3$ is linear corresponding to free particle behaviour in the $z$-direction, which will be discussed while comparing the RR model to an anharmonic oscillator in Section \ref{s:Mechanical-interpretation-of-the-Rajeev-Ranken-Model}. These continuous waves are not localized like solitons but shaped like a screw with axis along the third internal direction. In fact, they have a constant energy density 
	\beq
	{\cal E} = \frac{1}{2 \la} \Tr  (\dot{\phi}^2 + \phi'^2) =\frac{1}{2\la}( E^2 + k^2 (A^2 + B^2 + C^2 + D^2 + m^2)).
	\eeq
Thus we propose the name `screwons' for these weak coupling continuous waves and their nonlinear counterparts.

\chapter[Phase space structure and action-angle variables]{Phase space structure and action-angle variables}
\chaptermark{Phase space structure and action-angle variables}
\label{chapter:Phase-space-structure-and-action-angle-variables}

In this chapter, we discuss the phase-space structure, dynamics and a set of action-angle variables for the Rajeev-Ranken model. This chapter is based on \cite{G-V-2}.  A brief outline of the results obtained in this chapter was  given in Section \ref{s:outline-and-summary-of-results}. Here we begin with a more detailed summary of the results in each section and briefly mention the methods adopted.

In Section \ref{s:redution-of-dynamics}, we use the conserved quantities $\mathfrak{c}, m, s$ and $h$ of the model to reduce the dynamics to their common level sets. To begin with, in Section \ref{s:M4-cm-vectorfields-and-coordinates}, assigning numerical values to the Casimirs $\mathfrak{c}$ and $m$ of the nilpotent Poisson algebra (see Section \ref{s:Hamiltonian-mechanical}), enables us to reduce the 6D degenerate Poisson manifold of the $S$-$L$ variables ($M^6_{S \text{-} L}$) to its nondegenerate 4D symplectic leaves $M^4_{\mathfrak{c} m}$. We also find Darboux coordinates on $M^4_{\mathfrak{c} m}$ and use them to obtain a Lagrangian. Next, assigning numerical values to energy $E$, we find the generically 3D energy level sets $M^E_{\mathfrak{c} m}$ and use Morse theory to discuss the changes in their topology as the energy is varied (see Section \ref{s:Hill-region-Morse-theory}). In Section \ref{s:Reduction-2D} we consider the common level sets $M^{s h}_{\mathfrak{c} m}$ of all four conserved quantities and argue that they are generically diffeomorphic to 2-tori. This is established by showing that they admit a pair of commuting tangent vector fields (the canonical vector fields $V_E$ and $V_h$ associated to the conserved energy and helicity $h$) that are linearly independent away from certain singular submanifolds. Section \ref{s:common-level-set-conserved-qtys} is devoted to a systematic identification of all common level sets of the conserved quantities $\mathfrak{c}, m, s$ and $h$. We find that the condition for a common level set to be nonempty is the positivity of a cubic polynomial $\chi(u)$, which also appears in the nonlinear evolution equation $\dot u^2 =  2 \la k^2 \chi(u)$ for $u = S_3/k$. Each common level set of conserved quantities may be viewed as a  bundle over a band of latitudes of the $S$-sphere $(\vec S \cdot \vec S = s^2 k^2)$, with fibres given by a pair of points that coalesce along the extremal latitudes (which must be zeros of $\chi$) (see Fig.~\ref{f:Common-level-set-L-space}). By analyzing the graph of the cubic $\chi$ (see Fig. \ref{f:Examples-of-common-level-set}) we show that the common level sets are compact and connected and can only be of four types: 2-tori (generic), horn tori, circles and single points (nongeneric). The nongeneric common level sets arise as limiting cases of 2-tori when the major and minor radii coincide, minor radius shrinks to zero or when both shrink to zero.

In Section \ref{s:foliation-of-phase-space}, we study the dynamics on each type of common level set. The union of single point common level sets comprises the static subset: it is the union of a 2D and a 3D submanifold ($\Sigma_2$ and $\Sigma_3$) of phase space. In Section \ref{s:circular-level-sets}, we discuss the 4D union $\cal C$ of all circular level sets. Circular level sets arise when $\chi$ has a double zero at a non polar latitude of the $S$-sphere. On $\cal C$, solutions reduce to trigonometric functions, the wedge product $dh \wedge ds^2 \wedge dm \wedge d\mathfrak{c}$ vanishes and the conserved quantities satisfy the relation $\D = 0$, where $\D$ is the discriminant of $\chi$. Geometrically, $\cal C$ may be realized as a circle bundle over a 3D submanifold $\cal Q_{\cal C}$ of the space of conserved quantities. Finally, we find a set of canonical variables on $\cal C$ comprising the two Casimirs $\mathfrak{c}$ and $m$ and the action-angle pair $-kh$ and $\tht = \arctan(L_2/L_1)$.

In Section \ref{s:horn-toroidal-level-sets}, we examine the 4D union $\cal{\bar{H}}$ of horn toroidal level sets. It may be viewed as a horn torus bundle over a 2D space of conserved quantities. Horn tori arise when the cubic $\chi(u)$ is positive between a simple zero and a double zero at a pole of the $S$-sphere. Solutions to the EOM degenerate to hyperbolic functions on $\cal{\bar{H}}$ and every trajectory is a homoclinic orbit which starts and ends at the center of a horn torus (see Fig.~\ref{f:theta-phi-dynamics-3D-horn-torus}). As a consequence, the dynamics on $\cal{\bar H}$ is not Hamiltonian, though we are able to express it as a gradient flow, thus providing an example of a lower-dimensional gradient flow inside a Hamiltonian system. Interestingly, though the conserved quantities are functionally related on horn tori, the wedge product $dh \wedge ds^2 \wedge dm \wedge d\mathfrak{c}$ is nonzero away from their  centers.

In Section \ref{s:toroidal-level-sets}, we discuss the 6D union $\cal T$ of 2-toroidal level sets, which may be realized as a torus bundle over the subset $\D \ne 0$ of the space of conserved quantities. We use two patches of  the local coordinates $\mathfrak{c}, m, h, s, \tht$ and $u$ to cover $\cal T$. The solutions of the EOM are expressed in terms of elliptic functions and the trajectories are generically quasi-periodic on the tori (see Fig~\ref{f:torus-plot-theta-phi}). By inverting the Weierstrass-$\wp$ function solution for $u$, we discover one angle variable. Next, by imposing canonical Poisson brackets, we arrive at a system of PDEs for the remaining action-angle variables, which remarkably reduce to ODEs. The latter are reduced to quadrature allowing us to arrive at a fairly explicit formula for a family of action-angle variables. In an appropriate limit, these action-angle variables are shown to degenerate to those on the circular submanifold $\cal C$.

\section[Using conserved quantities to reduce the dynamics]{Using conserved quantities to reduce the dynamics\sectionmark{Using conserved quantities to reduce the dynamics}}
\sectionmark{Using conserved quantities to reduce the dynamics}
\label{s:redution-of-dynamics}

In this section, we discuss the reduction of the six-dimensional $S$-$L$ phase space ($M^6_{S \text{-} L}$) by successively assigning numerical values to the conserved quantities $\mathfrak{c}, m, s$ and $h$. For each value of the Casimirs $\mathfrak{c}$ and $m$ we obtain a four-dimensional manifold $M^4_{\mathfrak{c} m}$ with nondegenerate Poisson structure, which is expressed in local coordinates along with the equations of motion. Next, we identify the (generically three-dimensional) constant energy submanifolds $M^E_{\mathfrak{c} m} \subset M^4_{\mathfrak{c} m}$, where $E$ is a function of $s$ and $\mathfrak{c}$ (see Eq.~(\ref{e:Hamiltonian-s})). Moreover, we use Morse theory to study the changes in topology of $M_E^{\mathfrak{c} m}$ with  changing energy. Finally, the conservation of helicity $h$ allows us to reduce the dynamics to generically two-dimensional manifolds $M^{s h}_{\mathfrak{c} m} $, which are the common level sets of all four conserved quantities. By analysing the nature of the canonical vector fields $V_{E}$ and $V_h$, the latter are shown to be 2-tori in general. We also argue that there cannot be any additional independent integrals of motion. Though the common level sets of all four conserved quantities $M_{\mathfrak{c} m}^{s h}$ are generically 2-tori, there are other possibilities. We show that $M_{\mathfrak{c} m}^{s h}$ has the structure of a bundle over a portion of the sphere $\Tr S^2 = s^2 k^2$, determined by the zeros of a cubic polynomial $\chi(u)$. By analyzing the possible graphs of $\chi$ we show that $M_{\mathfrak{c} m}^{s h}$ is compact, connected and of four possible types: tori, horn tori, circles and points. In another words, we found all possible types of common level sets of conserved quantities of the RR model.

\subsection{Using Casimirs $\mathfrak{c}$ and $m$ to reduce to 4D phase space $M^4_{\mathfrak{c} m}$}
\label{s:M4-cm-vectorfields-and-coordinates}

\subsubsection{Symplectic leaves $M^4_{\mathfrak{c} m}$ and energy and helicity vector fields}
\label{s:canonical-vector-fields}

The common level sets of the Casimirs $\mathfrak{c}$ and $m$ are the four-dimensional symplectic leaves $M^4_{\mathfrak{c} m} \cong \mathbb{R}^4$   of the phase space $M^6_{S \text{-} L}$. On $M^4_{\mathfrak{c} m}$, the Poisson tensor $\scripty{r}^{ab}$ corresponding to the nilpotent Poisson algebra (\ref{e: PB-SL}) is nondegenerate and may be inverted to obtain the symplectic form $\omega_{a b}$. In Cartesian coordinates $\xi^a = (L_1, L_2, S_1, S_2)$,
	\beq
	\scripty{r}^{a b} = i k \colvec{2}{0 & \sig_2}{\sig_2 & -\la m \sig_2}
	\quad \text{and} \quad
	\omega_{a b} = (\scripty{r}^{-1})_{a b} = -\frac{i}{k}\colvec{2}{m \la \sig_2 &  \sig_2}{\sig_2 & 0}.
	\eeq
This symplectic form $\omega = (1/2) \omega_{ab} d\xi^a \wedge d\xi^b$ is the exterior derivative of the canonical 1-form $\al = -(1/2) \omega_{ab} \xi^b d\xi^a$. Expressing helicity $h$ (\ref{e:conserved-quantities}) and $E$ (\ref{e:Hamiltonian-s}) as functions on $M^4_{\mathfrak{c} m}$ by eliminating 
	\beq
	S_3 (L_1, L_2) = \frac{\la k}{2} \left( \left(2\mathfrak{c} -m^2\right) -\ov{k^2}(L_1^2 + L_2^2) \right) 
	\quad \text{and} \quad 
	L_3 = -m k
	\label{e:S3-L12}
	\eeq
we obtain the helicity and Hamiltonian vector fields on $M^4_{\mathfrak{c} m}$:
	\beqs
	k V_h &=&  L_2 \pdr_{L_1} - L_1 \pdr_{L_2} + S_2 \pdr_{S_1} - S_1 \pdr_{S_2} \quad \text{and} \cr
	k V_E &=&  S_2 \pdr_{L_1} - S_1 \pdr_{L_2} - \left[\la \frac{S_3 L_2}{k} + \la m S_2 \right] \pdr_{S_1} + \left[\la \frac{S_3 L_1}{k} + \la m S_1 \right] \pdr_{S_2}.
	\label{e:Hamiltonian-vector-fields}
	\eeqs
Since $E$ and $h$ commute, $\omega(V_E, V_h) = \{ E, h \} = 0$. It is notable that $V_h$ is nonzero except at the origin ($S_{1,2} = L_{1,2} = 0$), while $V_E$ vanishes at the origin and on the circle ($L_1^2 + L_2^2 = k^2 (2 \mathfrak{c} - m^2), S_{1,2} = 0$). The points where $V_E$ and $V_h$ vanish turn out be the intersection of $M^4_{\mathfrak{c} m}$ with the static submanifolds  
	\beq
	\Sigma_{2} = \{ \vec S, \vec L \:|\: S_{1,2} = L_{1,2} = 0 \} \quad \text{and} \quad 
	\Sigma_3 = \{ \vec S, \vec L \:|\: \vec S = 0 \}
	\label{e:static-submanifolds}
	\eeq
introduced in Section \ref{s:Static-and-trigonometric-solutions}, where  $S$ and $L$ are time-independent. The points where $V_E$ vanish will be seen in Section \ref{s:Hill-region-Morse-theory} to be critical points of the energy function. 

\subsubsection{ Darboux coordinates on symplectic leaves $M^4_{\mathfrak{c} m}$}

Since $M^4_{\mathfrak{c} m} \cong \mathbb{R}^4$ it is natural to look for global canonical coordinates. In fact, the canonical coordinates $(R_a, k P_b)$ on the six-dimensional phase space $M^6_{R \text{-} P}$ (see Section \ref{s:Darboux-coordinates-M6}) restrict to Darboux coordinates on $M^4_{\mathfrak{c} m}$:
	\beq
	k R_a = - \eps_{ab}L_b \quad \text{and} \quad
	k P_a = S_a + \frac{\la m}{2} L_a \quad \text{for} \quad a,b = 1,2
	\label{e:R-P-relation-to-L-S}
	\eeq
with $\{ R_a, k P_b \} = \del_{ab}$ and $\{ R_a, R_b \} = \{ P_a, P_b \} = 0$. The Hamiltonian is a quartic function in these coordinates:
	\beq
	\frac{H}{k^2} = \frac{ P_1^2 + P_2^2}{2} + \frac{\la m}{2}(R_1 P_2 -R_2 P_1) + \frac{\la^2}{8} (R_1^2 + R_2^2) \left(R_1^2 + R_2^2 + 3 m^2 - 4 \mathfrak{c} \right) + \frac{\la^2}{8} (2 \mathfrak{c} - m^2)^2 + \mathfrak{c} + \frac{1}{2 \la^2}.
	\eeq
The equations of motion resulting from these canonical Poisson brackets and Hamiltonian are cubically nonlinear ODEs. In fact, for $a = 1, 2$:
	\beq
	k^{-1} \dot{R_a} = P_a - \frac{\la m}{2} \eps_{a b} R_b \quad \text{and} \quad
	k^{-1} \dot{P_a} = -\frac{\la m}{2} \eps_{a b} P_b - \frac{\la^2}{4} \left(3 m^2  - 4 \mathfrak{c} + 2 R_b R_b \right)R_a.
	\eeq
A Lagrangian $L_{\mathfrak{c} m}(R, \dot R)$,  leading to these equations of motion can be obtained by extremizing $k P_a \dot{R_a} - H$ with respect to $P_1$ and $P_2$:
	\beqs
	L_{\mathfrak{c} m} &=& \half \left( \dot{R_1}^2 + \dot{R_2}^2 - \la m k (R_1 \dot{R_2} - R_2 \dot{R_1}) \right) - \frac{\la^2 k^2}{8} (R_1^2 + R_2^2) \left(R_1^2 + R_2^2 + 2 m^2 - 4 \mathfrak{c} \right) \cr
	&&- k^2 \left(\frac{\la^2}{8} (2 \mathfrak{c} - m^2)^2 + \mathfrak{c} + \frac{1}{2 \la^2} \right).
	\eeqs

\subsection{Reduction to tori using conservation of energy and helicity}
\label{s:Reduction-2D}

So far, we have chosen (arbitrary) real values for the Casimirs $\mathfrak{c}$ and $m$ to arrive at the reduced phase space $M^4_{\mathfrak{c} m}$. Now assigning numerical values to the Hamiltonian $H = E k^2$ we arrive at the generically three-dimensional constant energy submanifolds $M^E_{\mathfrak{c} m}$ which foliate $M^4_{\mathfrak{c} m}$. It follows from the formula for the Hamiltonian (\ref{e:Hamiltonian-s}) that each of the $S_a$ is bounded above in magnitude by $|k|s = \sqrt{2 k^2 (E - \mathfrak{c} - 1/2 \la^2)}$. Moreover, $M^E_{\mathfrak{c} m}$ is closed as it is the inverse image of a point. Thus, constant energy manifolds are compact. Interestingly, the topology of $M^E_{\mathfrak{c} m}$ can change with energy: this will be discussed in Section \ref{s:Hill-region-Morse-theory}. In addition to the Hamiltonian and Casimirs $\mathfrak{c}$ and $m$, the helicity $h k^2 = \Tr SL$ is a fourth (generically independent) conserved quantity (see Section \ref{s:conserved-quantities}). Thus each trajectory must lie on one of the level surfaces $M^{E h}_{\mathfrak{c} m}$ of $h$ that foliate $M^E_{\mathfrak{c} m}$. Note that since $s \geq 0$ is uniquely determined by $E$ (and vice versa), the level sets of the conserved quantities $M^{E h}_{\mathfrak{c} m}$ and $M^{s h}_{\mathfrak{c} m}$ are in 1-1 correspondence and we will use the two designations interchangeably. 

We will see in Section \ref{s:canonical-vector-fields-topology} that these common level sets  of conserved quantities $M^{E h}_{\mathfrak{c} m}$ are generically 2-tori, parameterized by the angles $\tht$ and $\phi$ which (as shown in Section \ref{s:conserved-quantities}) evolve according to
	\beq
	\dot{\theta} =  -k\left(\frac{h + m u}{2\mathfrak{c} - m^2 - 2u/\la}\right) \quad \text{and} \quad
	\dot{\phi} = k m \la + k \la u \left(\frac{h + mu}{s^2 - u^2}\right).
	\label{e:theta-phi-dynamics-torus}
	\eeq
Here, $u = S_3/k$ is related to $\tht$ and $\phi$  via helicity $h k^2 = \Tr S L$ and other conserved quantities (\ref{e:conserved-quantities})
	\beq
	\sqrt{\left(s(E,\mathfrak{c})^2-u^2 \right) \left(2\mathfrak{c} - m^2 - 2u/\la \right)} \: \cos(\theta - \phi) = h + mu.
	\label{e:relation-theta-phi-u-torus}
	\eeq
In other words, the components $V_E^\tht = \dot{\theta}/k^2$ and $V_E^\phi = \dot{\phi}/k^2$ of the Hamiltonian vector field $V_E = V_E^\tht \pdr_\tht + V_E^\phi \pdr_\phi$ are functions of $u$ alone. Though the denominators in (\ref{e:theta-phi-dynamics-torus}) could vanish, the quotients exist as limits, so that $V_E$ is nonsingular on $M^{s h}_{\mathfrak{c} m}$. Interestingly, as pointed out in \cite{R-R}, $u$ evolves by itself as we deduce from (\ref{e:EOM-LS}):
	\beq
	\dot{u}^2 = \la^2  k^2 \rho^2 r^2 \sin^2(\theta - \phi) = \la^2  k^2 \left[ (s^2 - u^2)\left(2 \mathfrak{c} - m^2 - \frac{2u}{\la}\right) - (h + mu)^2\right] = 2 \la k^2 \chi(u).
	\label{e:EOM-u}
	\eeq
This cubic $\chi(u)$ will be seen to play a central role in classifying the invariant tori in Section \ref{s:common-level-set-conserved-qtys}. The substitution $u = a v + b$, reduces this ODE to Weierstrass normal form
	\beq
	\dot v^2 = 4 v^3 - g_2 v - g_3, \quad \text{where} \quad a = 2/k^2 \la \quad \text{and} \quad  b = \mathfrak{c} \la/ 3, 
	\label{e:Weierstrass-normal-form}
	\eeq
with solution $v(t) = \wp(t + \alpha; g_2, g_3)$. Here, the Weierstrass invariants are: 
	\beq
	g_2 = \frac{ k^4 \la^2}{3} (3 \la h m + \la^2 \mathfrak{c}^2 + 3 s^2), \quad
	g_3 = \frac{k^6 \la^4}{108}  (27 h^2 + 18 \la \mathfrak{c} m h + 4 \la^2 \mathfrak{c}^3 - 36 \mathfrak{c} s^2 + 27 m^2 s^2).
	\label{e:Weierstrass-invariants}
	\eeq
Thus we obtain
	\beq
	u(t) = \frac{2}{k^2 \la} \wp(t + \alpha) + \frac{\mathfrak{c} \la}{3},
	\label{e:u-wp-function}
	\eeq
which oscillates periodically in time between $u_{\rm min}$ and $u_{\rm max}$, which are neighbouring zeros of $\chi$ between which $\chi$ is positive. Choosing $\al$ fixes the initial condition, with its real part fixing the origin of time. In particular, if $\al =\omega_I$ (the imaginary half-period of $\wp$), then $u(0) = u_{\rm min}$. On the other hand, $u(0) = u_{\rm max}$ if $\al = \om_R + \om_I$, where $\om_R$ is the real half-period. The formula (\ref{e:u-wp-function}) will be used in Section \ref{s:toroidal-level-sets} to find a set of action-angle variables for the system.

\subsubsection{Reduction of canonical vector fields to $M_{\mathfrak{c} m}^{s h}$ and its topology}
\label{s:canonical-vector-fields-topology}

In this section, we use the coordinates $(s^2, h , \tht, \phi)$ to show that the canonical vector fields $V_E$ and $V_h$ are tangent to the level sets $M^{s h}_{\mathfrak{c} m}$, which are shown to be compact connected Lagrangian submanifolds of the symplectic leaves $M^4_{\mathfrak{c} m}$. Moreover, $V_E$ and $V_h$ are shown to be generically linearly independent and to commute, so that $M^{s h}_{\mathfrak{c} m}$ are generically 2-tori.

On $M^4_{\mathfrak{c} m}$, the coordinates $(s^2, h , \tht, \phi)$ (as opposed to $(L_1,L_2,S_1,S_2)$) are convenient since the common level sets $M^{s h}_{\mathfrak{c} m} \subset M^4_{\mathfrak{c} m}$ arise as intersections of the $s^2$ and $h$ coordinate hyperplanes. The remaining variables $\tht$ and $\phi$ furnish coordinates on $M^{s h}_{\mathfrak{c} m}$. The Poisson tensor on $M^4_{\mathfrak{c} m}$ in these coordinates has a block structure, as does the symplectic form:
	\beq
	\scripty{r}^{a b} = \ov k \colvec{2}{0 & \al}{- \al^t & \beta} \quad \text{and} \quad \omega_{a b} = k \colvec{2}{-\g & -\delta^t}{ \delta & 0},
	\label{e:r-omega-s-h-tht-phi-coordinate}
	\eeq
where $\al, \beta, \g$ and $\delta$ are the dimensionless $2 \times 2$  matrices:
	\beqs
	\al &=& \colvec{2}{ \frac{-2 \dot \tht}{k} & \frac{-2 \dot \phi}{k}}{1 & 1}, \quad \beta =-i\frac{s_{\tht \phi}}{r \rho} \sigma_2, \quad \g = \left(- \al^t\right)^{-1} \beta \alpha^{-1} = -\frac{\beta}{\det \alpha} \cr
	 \text{and} \quad \delta &=& \al^{-1} = \frac{1}{\det \al}\colvec{2}{1 & \frac{2 \dot \phi}{k} }{-1 & \frac{-2 \dot \tht}{k}} \quad \text{with} \quad 
	 \det \al = k^2\sqrt{\det \scripty{r}} = \frac{-2}{k} \left(\dot \tht - \dot \phi \right).
	\eeqs
Here $s_{\tht \phi } = \sin(\tht - \phi)$ and $\dot \tht$ and $\dot \phi$ are as in (\ref{e:theta-phi-dynamics-torus}), subject to the relation (\ref{e:relation-theta-phi-u-torus}). From (\ref{e:conserved-quantities}), it follows that $\rho$ and $r$ may be expressed in terms of $s^2, h, \tht$ and $\phi$, by solving the pair of equations
	\beq
	h = r \rho \, c_{\tht \phi} - \frac{\la m}{2} \left( 2\mathfrak{c} - (r^2 + m^2) \right) \quad \text{and} \quad
	s^2 = \rho^2 + \frac{\la^2}{4} \left(2 \mathfrak{c} - (r^2 + m^2)\right)^2.
	\eeq 
Here $c_{\tht \phi } = \cos(\tht - \phi)$. In these coordinates, $V_h$ and $V_E$ (\ref{e:Hamiltonian-vector-fields}) have no components along $\pdr_s$ or $\pdr_h$: 
	\beq
	k V_h = -(\pdr_\tht + \pdr_\phi) \quad \text{and} \quad
	k V_{E} = -\frac{\rho}{r} c_{\tht \phi} \pdr_{\tht}  + \left(\la m + \frac{\la^2 r}{ 2 \rho} c_{\tht \phi} \left(2 \mathfrak{c} - (r^2 + m^2)\right) \right) \pdr_{\phi}.
	\label{e:hamiltonian-vector-field-s-h-tht-phi}
	\eeq
Thus, $V_h$ and $V_E$ are tangent to $M^{s h}_{\mathfrak{c} m}$. Moreover, the restriction of $\omega$ to $M^{s h}_{\mathfrak{c} m}$ is seen to be identically zero as it is given by the $\tht$-$\phi$ block in (\ref{e:r-omega-s-h-tht-phi-coordinate}) so that $M^{s h}_{\mathfrak{c} m}$ is a Lagrangian submanifold. Trajectories on $M^{s h}_{\mathfrak{c} m}$ are the integral curves of $V_E$.

To identify the topology of the common level set $M^{s h}_{\mathfrak{c} m}$, it is useful to investigate the linear independence (over the space of functions) of the vector fields $V_E$ and $V_h$. On $M^4_{\mathfrak{c} m}$, $\omega$ is nondegenerate so that $V_E$ and $V_h$ are linearly independent iff $dE \wedge dh \neq 0$. We know that this wedge product vanishes on $M^4_{\mathfrak{c} m}$ precisely when $S_{1,2}$ and $L_{1,2}$ satisfy the relations (see Section \ref{s:Independence-of-conserved-quantities}):
	\beq
	\Xi_1: \; (S \times L)_3 = 0, \quad
	\Xi_2: -\la L_1(S \times L)_2 = k S_1^2 
	\quad \text{and} \quad
	\Xi_3: \: \la L_2 (S \times L)_1 = k S_2^2.
	\label{e:trigonometric-submanifold-conditions-Xi}
	\eeq
Here $(S \times L)_3 = S_1 L_2 - S_2 L_1$ etc., and $S_3$ and $L_3$ are expressed using (\ref{e:S3-L12}). It was shown in Section \ref{s:Independence-of-conserved-quantities} that (\ref{e:trigonometric-submanifold-conditions-Xi}) are the necessary and sufficient conditions for the four-fold wedge product $dh \wedge ds^2 \wedge dm \wedge d\mathfrak{c}$ to vanish on  $M^6_{S\text{-}L}$. Moreover, it was shown that this happens precisely on the singular set $\bar {\cal C} \subset M^6_{S\text{-}L}$ which consists of the circular/trigonometric submanifold ${\cal C}$ and its boundaries ${\cal C}_{1,2}$  and $\Sigma_{2,3}$. Thus, $V_E$ and $V_h$ are linearly independent away from the set (of measure zero) given by the intersection of $\bar{\cal C}$ with $M^4_{\mathfrak{c} m}$. [For given $\mathfrak{c}$ and $m$, the intersection of $\cal C$ with $M^4_{\mathfrak{c} m}$ is in general a two-dimensional manifold defined by four conditions among the six variables $\vec S$ and $\vec L$: $\Xi_1$ and $\Xi_2$ (with $S_{1,2} \neq 0$) as well as the conditions in Eq. (\ref{e:S3-L12}).] Furthermore, since $E$ and $h$ Poisson commute, $[V_E, V_h] = - V_{\{E,h \}} = 0$. {\it So, as long as we stay away from these singular submanifolds, $V_E$ and $V_h$ are a pair of commuting linearly independent vector fields tangent to $M^{s h}_{\mathfrak{c} m}$ }(see Lemma 1 in Chapter 10 of {\cite{Arnold}}). Additionally, we showed at the beginning of Section \ref{s:Reduction-2D}  that the energy level sets $M^E_{\mathfrak{c} m} \subset M^4_{\mathfrak{c} m}$ are compact manifolds. Now, $M^{s h}_{\mathfrak{c} m}$ must also be compact as it is a closed subset of $M^E_{\mathfrak{c} m}$ (the inverse image of a point). Finally, we will show in Section \ref{s:Examples-of-CL-sets-of-conserved-quantities} that $M^{s h}_{\mathfrak{c} m}$ is connected. Thus, for generic values of the conserved quantities, $M_{\mathfrak{c} m}^{s h}$ is a compact, connected surface with a pair of linearly independent tangent vector fields. By Lemma 2 in Chapter 10 of {\cite{Arnold}}, it follows that the common level sets of conserved quantities $M^{s h}_{\mathfrak{c} m}$ are generically diffeomorphic to 2-tori. 

We observed in Section \ref{s:conserved-quantities} that a generic trajectory on a 2-torus common level set $M_{\mathfrak{c} m}^{sh}$ is dense (see Figs.~\ref{f:theta-phi-dynamics-3D} and \ref{f:torus-plot-theta-phi}). This implies that any additional continuous conserved quantity would have to be constant everywhere on the torus and cannot be independent of the known ones. Thus, we may rule out additional independent conserved quantities.

\subsection{Classifying all common level sets of conserved quantities}
\label{s:common-level-set-conserved-qtys}

In Section \ref{s:Reduction-2D} we showed that the common level sets of the conserved quantities $\mathfrak{c}, m, s$ and $h$ are generically 2-tori. However, this leaves out some singular level sets. These nongeneric common level sets occur when the conserved quantities fail to be independent and also correspond to the degeneration of the  elliptic function solutions (\ref{e:u-wp-function}) to hyperbolic and circular functions. Here,  we use a geometro-algebraic approach to classify all common level sets and show that there are only four possibilities: 2-tori, horn tori, circles and single points. Interestingly, the analysis relies on the properties of the cubic $\chi(u)$ that arose in the equation of motion for $u$ (\ref{e:EOM-u}).

\subsubsection{Common level sets as bundles and the cubic $\chi$}

We wish to identify the submanifolds of phase space $M^6_{S\text{-}L}$ obtained by successively assigning numerical values to the four conserved quantities $s, h, \mathfrak{c}$ and $m$. Not all real values of these conserved quantities lead to nonempty common level sets. From (\ref{e: H-mechanical}), we certainly need the Hamiltonian  $H \geq 0$ and $s^2 \geq 0$. It follows that $- s^2/2 - 1/2\la^2 \leq \mathfrak{c} \leq H/k^2 - 1/2\la^2$. However, these conditions are not always sufficient; additional conditions will be identified below. The situation is analogous to requiring the energy ($L_1^2 / 2 I_1 + L_2^2/ 2 I_2 + L_3^2 / 2 I_3$ in the principle axis frame) and square of angular momentum $(L_1^2 + L_2^2 + L_3^2)$ to be non negative for force-free motion of a rigid body. These two conditions are necessary but not sufficient to ensure that the angular momentum sphere and inertia ellipsoid intersect.
\begin{figure}[h]
	\centering
		\begin{subfigure}[t]{5cm}
		\centering
		\includegraphics[width=5cm]{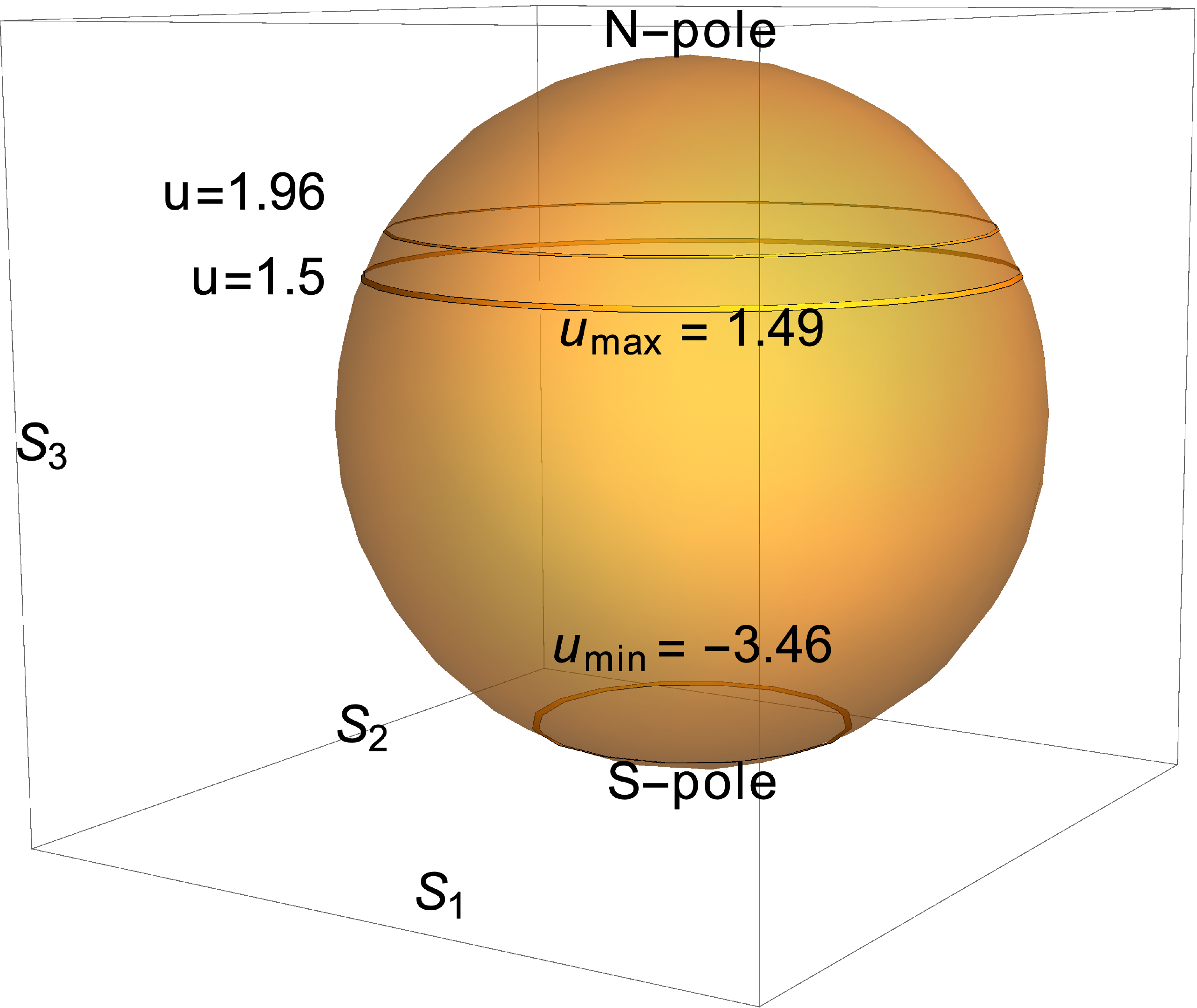}
		\caption{}
		\label{f:base-space}
		\end{subfigure}
		\qquad \quad
		\begin{subfigure}[t]{6cm}
		\centering
		\includegraphics[width= 6cm]{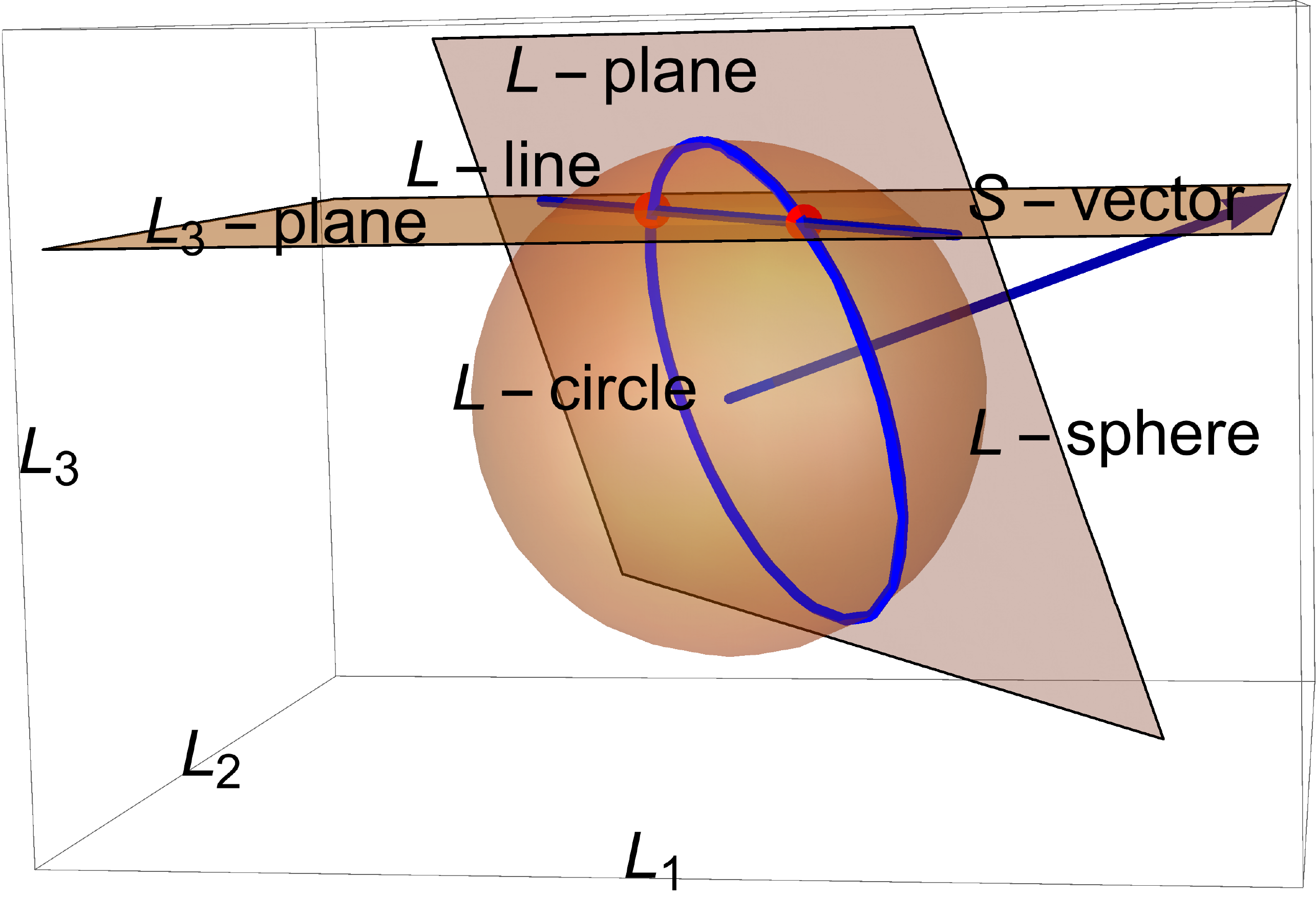}
		\caption{}
		\end{subfigure}
	\caption{ \footnotesize (a) The $S$-sphere $S_1^2 + S_2^2 + S_3^2 = 14 = s^2$ for $k = 1$. For $h = 1, \mathfrak{c} =2$ and $\la =1$, only latitudes below $u = S_3/k = 1.96$ (\ref{e:Intersection-L-plane-and-sphere}) are allowed if the $L$-sphere and $L$-plane are to intersect. However, if we take $m = -1$, the upper bound $u \leq (\la/2)(2 \mathfrak{c} - m^2)$ following from $L_1^2 + L_2^2 \geq 0$ and (\ref{e:conserved-quantities}) further restricts $u$ to lie below $1.5$. Finally, the condition $\chi \geq 0$ for nonempty fibres restricts $u$ to lie between the simple zeros $u_{\rm min} = -3.46$ and  $u_{\rm max} = 1.49$. (b) The $L$-space above the base point $\vec S = (3,2,1)$ for the same values of constants. The $L$-plane normal to $\vec S$ at a distance of $1/\sqrt{14}$ from $(0,0,0)$ is the level set $h = 1$. The $L$-sphere of radius $\sqrt{2}$ (the level set $\mathfrak{c}= 2$) intersects the $L$-plane along the $L$-circle. The horizontal $L_3$-plane $(L_3 = -m  = 1)$ intersects the $L$-plane along the $L$-line. The fibre over $\vec S$ is the pair of points where the $L$-line intersects the $L$-circle. The corresponding common level set is a 2-torus as in (C1) of Section \ref{s:Examples-of-CL-sets-of-conserved-quantities}.}
	\label{f:Common-level-set-L-space}
\end{figure}

First, putting $S_a S_a = s^2 k^2$ defines a $2$-sphere (the $`S$-sphere') in the $S$-space as in Fig.~\ref{f:base-space}. We may regard $u$ (or $S_3 = k u$) for $|u| \leq s$ as the latitude on the $S$-sphere with $u = \pm \sign{k}s$ representing the North $({\cal N})$ and South $({\cal S})$ poles. At each point on the $S$-sphere, the conservation of helicity $S_a L_a = k^2 h$ forces $\vec{L}$ to lie on a plane (the $`L$-plane') perpendicular to the numerical vector $\vec{S}$ at a distance $|h k|/s$ from the origin of the $L$-space. At this point, we have assigned numerical values to $s$ and $h$, which happen to be Casimirs of the Euclidean Poisson algebra (\ref{e:PB-SL-dual}). It remains to impose the conservation of $\mathfrak{c}$ and $m$.

For each point on the $S$-sphere, the condition $L_a^2/2 + k S_3/\la = \mathfrak{c} k^2$ (\ref{e:conserved-quantities}) defines an $L$-sphere of radius $\sqrt{2}|k|\left(\mathfrak{c} - u/\la\right)^\half  $in the $L$-space provided $\mathfrak{c} \geq u/\la$. Since $u \geq -s$, the conserved quantities must be chosen to satisfy $\mathfrak{c} \geq - s/\la$. In fact, this ensures that $H \geq 0$ and thus subsumes the latter. The $L$-sphere and the $L$-plane intersect along an $L$-circle provided the radius of the $L$-sphere exceeds the distance of the $L$-plane from the origin, i.e.,
	\beq
	|k|^{-1} \: {\rm rad}(L\text{-sphere}) = \sqrt{2\left(\mathfrak{c} - \frac{u}{\la} \right)} \geq \frac{ |h|}{s} = |k|^{-1} \: {\rm dist}(L\text{-plane}, {\bf 0})
	\quad \text{or} \quad  u \leq \la \left(\mathfrak{c} - \frac{h^2}{2s^2}\right).
	\label{e:Intersection-L-plane-and-sphere}
	\eeq
Thus, for the intersection to be nonempty, depending on the sign of $k$, $\vec S$ must lie below or above a particular latitude determined by (\ref{e:Intersection-L-plane-and-sphere}). Furthermore, since $u \geq -s$, we must choose 
	\beq
	\mathfrak{c} \geq \mathfrak{c}_{\rm min} = -  s/ \la + h^2 / 2 s^2.
	\label{e:C-lower-bound-strong}
	\eeq
When the inequality (\ref{e:Intersection-L-plane-and-sphere}) is saturated, the $L$-plane is tangent to the $L$-sphere and the $L$-circle shrinks to a point. In summary, the common level set of the three conserved quantities $s, h$ and $\mathfrak{c}$ can be viewed as a sort of fibre bundle with base given by the portion of the $S$-sphere lying above or below a given latitude. The fibres are given by $L$-circles of varying radii which shrink to a point along the extremal latitude.

The final conserved quantity $\Tr KL = mk^2$ restricts $\vec L$ to the horizontal plane $L_3 = -m k$. For each nonpolar point on the $S$-sphere, this $L_3$-plane intersects the above $L$-plane along the $L$-line $S_1 L_1 + S_2 L_2 = hk^2 + mk S_3$ (assuming $S_1, S_2$ are not both zero). This line intersects the $L$-sphere at a pair of points, provided the radius of the $L$-sphere is greater than the distance of the $L$-line from the origin of the $L$-space, i.e.	
	\beq
	|k|^{-1} \: {\rm rad}(L\text{-sphere}) = \sqrt{2 \left(\mathfrak{c} - \frac{u}{\la}\right)} \geq \left( m^2 + \frac{(h + mu)^2 }{s^2 - u^2} \right)^{\half} 
	= |k|^{-1} \: {\rm dist}(L\text{-line}, {\bf 0}).
	\label{e:Intersection-L-line-and-L-sphere}
	\eeq
The two points of intersection coincide if the inequality is saturated so that the $L$-line is tangent to the $L$-sphere. Note that inequality (\ref{e:Intersection-L-line-and-L-sphere}) implies (\ref{e:Intersection-L-plane-and-sphere}), provided the $L$-sphere is nonempty ($\mathfrak{c} \geq u/\la $). This is geometrically evident since the distance of the $L$-line from the origin is bounded below by the distance $|k h|/s$ of the $L$-plane (which contains the $L$-line) from the origin.

\vspace{.25cm}

\footnotesize

{\fl \bf Remark:} Another way to see that (\ref{e:Intersection-L-line-and-L-sphere}) implies (\ref{e:Intersection-L-plane-and-sphere}) is to note that if $g =  m^2  + ((h + m u)^2/(s^2 - u^2)) - (h^2/s^2)$, then
	\beq
	\ov{k^2} \text{dist}(L\text{-line}, {\bf 0})^2 =  m^2  + \frac{(h + m u)^2}{s^2 - u^2} = \frac{h^2}{s^2} + g(u) = \ov{k^2} \text{dist}(L\text{-plane}, {\bf 0})^2 + g(u).
	\eeq	
Eq. (\ref{e:Intersection-L-line-and-L-sphere}) would then imply (\ref{e:Intersection-L-plane-and-sphere}), if we can show that $g(u) \geq 0$ on the sphere $|u| \leq s$. To see this, we first note that $g(u) \to + \infty$ at the poles $u = \pm s$ so that it suffices to show that the quadratic polynomial $\tilde{g}(u) =  g(u)(s^2 - u^2)$ is nonnegative for $|u| < s$. This is indeed the case since the global minimum of $\tl g(u)$ attained at $u^* = -m s^2/h$ is simply zero.

\vspace{.25cm}
\normalsize
Assuming (\ref{e:Intersection-L-line-and-L-sphere}) holds, the common level set of all four conserved quantities may be viewed as a sort of fibre bundle with base given by the part of the $S$-sphere satisfying (\ref{e:Intersection-L-line-and-L-sphere}) and fibres given by either one or a pair of points (this is the case for nonpolar latitudes, see below for the special circumstance that occurs above the poles). In other words, provided $\mathfrak{c} \geq \mathfrak{c}_{\rm min}$, the `base' space is the part of the $S$-sphere consisting of all latitudes $u$ lying in the interval $-s \leq u \leq \min(s, \la (\mathfrak{c} - h^2/2s^2))$ and satisfying the cubic inequality following from (\ref{e:Intersection-L-line-and-L-sphere})
	\beq
	\chi(u) = u^3 - \la \mathfrak{c} u^2 - \left( s^2 + \la h m \right) u + \frac{\la}{2}\left( 2 \mathfrak{c}s^2 - h^2 - m^2 s^2 \right) \geq 0.
	\label{e:cubic-equation-S3}
	\eeq
The roots of the cubic equation $\chi(u) = 0$ resulting from the saturation of this inequality determine the extremal latitudes where the two-point fibres degenerate to a single point (provided the extremal latitude does not correspond to a pole of the $S$-sphere). If an extremal latitude is at one of the poles then $\chi(\pm s) = -(\la/2)(h \pm ms)^2$ must vanish there and the determination of the fibre over the pole is treated below. 

Recall that the discriminant $\D = b^2 c^2 - 4 c^3 - 4 b^3 d - 27 d^2 + 18 bcd$ of the cubic $x^3 + b x^2 + c x + d$ is the product of squares of differences between its roots. It vanishes iff a pair of roots coincide. The discriminant of the cubic $\chi(u)$ will be useful in the analysis that follows. It is a function of the four conserved quantities: \small
	\beqs
	\D &=& \la^4 \mathfrak{c}^2 \left(\frac{s^2}{\la} +hm\right)^2 + 4 \la^3 \left(\frac{s^2}{\la} + hm\right)^3 + 2 \la^4 \mathfrak{c}^3(2\mathfrak{c} s^2 - h^2 - m^2 s^2)  -\frac{27}{4}\la^2 (2\mathfrak{c} s^2 - h^2 - m^2 s^2)^2 \cr
	&& + 9\la^3 \mathfrak{c}\left(\frac{s^2}{\la} + hm\right)(2 \mathfrak{c} s^2 - h^2 - m^2 s^2).
	\label{e:discriminant}
	\eeqs
	\normalsize	

\subsubsection{Fibres over the poles of the $S$-sphere}
\label{s:fibres-over-poles-of-S-sphere}

At the ${\cal N}$ and ${\cal S}$ poles $(u = \pm \sign{k}\, s)$ of the $S$-sphere, the $L$-plane $(S_3 L_3 = h k^2)$ and $L_3$-plane $(L_3 = -m k)$ are both horizontal: their intersection does not define an $L$-line. For the common level sets of $h$ and $L_3$ to be nonempty, the planes must coincide:
	\beq
	h = \mp m\, \sign{k} \: s
	\label{e:L-L3-plane}
	\eeq
with upper/lower signs corresponding to the ${\cal N}/{\cal S}$ poles. This condition ensures that $\chi$ vanishes at the corresponding pole, implying that it cannot be positive at a physically allowed pole of the $S$-sphere.

Now, for the $L$-sphere to intersect the $L_3$-plane, its radius must be bounded below by $|m k|$:
	\beq
	|k|^{-1} \: {\rm rad}(L\text{-sphere}) = \sqrt{2}\left( \mathfrak{c} \mp \frac{\sign{k}\, s}{\la}\right)^{\half}\geq |m| = |k|^{-1}\: {\rm dist}(L_3\text{-plane}, {\bf 0}).
	\label{e:condition-for-fibre-at-poles-of-S-sphere}
	\eeq
When this inequality is strict, the fibre over the pole is a circle ($L$-circle) while it is a single point when the inequality is saturated. Interestingly, in the latter case, the discriminant $\Delta$ (\ref{e:discriminant}) vanishes, so that the pole must either be a double or triple zero of $\chi$. On the other hand, when the inequality is strict, $\chi$ must have a simple zero at the pole. This structure of fibres over the poles is in contrast to the two point fibres over the non polar latitudes of the $S$-sphere when $\chi > 0$. For example, suppose $k = \la = s = 1$ and take $h = -m = 1$ so that the $L_3$ and $L$-planes over the ${\cal N}$ pole $(S_3 = 1)$ coincide. These planes intersect the $L$-sphere provided $\mathfrak{c} \geq 3/2$ (see (\ref{e:condition-for-fibre-at-poles-of-S-sphere})). Moreover, the fibre over the ${\cal N}$ pole is a single point if $\mathfrak{c} = 3/2$ and a circle if $\mathfrak{c} > 3/2$.   

\subsubsection{Properties of $\chi$ and the closed, connectedness of common level sets}

We observed in Section \ref{s:fibres-over-poles-of-S-sphere} that $\chi$ must vanish at a physically allowed pole of the $S$-sphere and that we must have $h = \pm m \: \sign{k} \: s$ for this to happen. Here, we investigate the possible behaviour of $\chi$ near a pole, which helps in restricting the allowed graphs of $\chi$. We find that the sign of $\chi'$ at  an allowed pole is fixed and also that the allowed latitudes must form a closed  and connected set. As a consequence, we deduce that some graphs of $\chi$ are disallowed. For example, $\chi$ cannot have a triple zero at a nonpolar latitude. We also deduce that the common level sets must be both closed and connected. 


\vspace{.25cm}

{\fl \bf Result 1: Sign of $\chi'$ at a pole which is a {\it simple} zero:} Suppose $\chi$ has a simple zero  at the pole $u =  \pm s$ with nonempty fibre over it, then $\chi'(\pm s) \lessgtr 0$.
\vspace{.25cm}

\begin{figure}[h]
	\centering
		\begin{subfigure}[b]{13cm}
		\includegraphics[width=13cm]{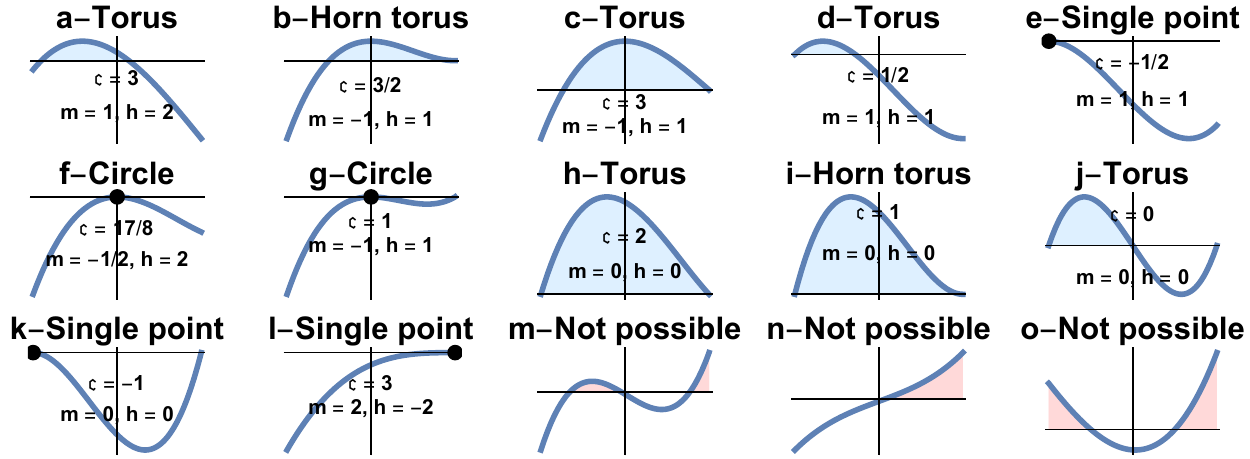}
		\end{subfigure}
	\caption{\footnotesize (a) - (l) Plots of the cubic $\chi(u)$ for latitudes between the south and north poles $- s \leq u \leq s$ for $k = \la = s = 1$ and $\mathfrak{c}, m$ and $h$ as indicated. The physically allowed latitudes with $\chi \geq 0$ are shaded in blue. The black dots indicate a single allowed latitude with $\chi$ necessarily having zeros of order more than one. The corresponding common level sets of conserved quantities (see Section \ref{s:Examples-of-CL-sets-of-conserved-quantities}) are a $2$-torus [(a), (c), (d), (h), (j)], a horn torus [(b), (i)], a circle [(f),(g)], and  a single point [(e), (k), (l)]. In (c), (d), (h) and (j) the fibre over the physically allowed poles (where $\chi$ has a simple zero) are circles while they are single points in (b), (e), (k) (double zero) and (l) (triple zero). In (i) the fibre over the ${\cal S}$ pole (simple zero) is a circle and is a point over the ${\cal N}$ pole (double zero). Similar figures with ${\cal N}$ and ${\cal S}$ exchanged arise when $k < 0$. Figures (m)-(o) show cases that {\it cannot} occur for any set of physically allowed conserved quantities as a consequence of {\bf Result} 2.}
	\label{f:Examples-of-common-level-set}
\end{figure}

{\fl \bf Proof of $\chi'(s) < 0$:} Suppose $h = -m s$, so that $\chi(u)$ has a simple zero at the pole $u = s$ with circular fibre over it (see Eq.(\ref{e:L-L3-plane})). Then (\ref{e:cubic-equation-S3}) implies
	\beq
	\chi'(s) = 2 s^2 - \la s (2 \mathfrak{c} - m^2).
	\eeq
Suppose $\chi'(s) > 0$, then $\mathfrak{c} < s/ \la + m^2/2$. But in this case, the upper bound on the latitude $u \leq \min[s, \la \mathfrak{c} - \la h^2/ (2 s^2)] < s$ so that $u = s$ could not have been an allowed latitude. On the other hand, if $\chi'(s) < 0$, then $u=s$ is an allowed latitude. Thus, when the ${\cal N}/{\cal S}$ pole for $k \gtrless 0$ is a simple zero of $\chi$ with nonempty fibre, it is always surrounded by other allowed latitudes. In particular, the north poles in Fig.~\ref{f:Examples-of-common-level-set}g, j and k are not allowed latitudes, while they {\it are} in Fig.~\ref{f:Examples-of-common-level-set}c and h.
\vspace{.25cm}

{\fl \bf Proof of $\chi'(-s) > 0$:} On the other hand, suppose $h = m s$ so that $\chi$ has a simple zero at $u = -s$ with nonempty fibre. Suppose $\chi'(-s) < 0$, then as before (\ref{e:cubic-equation-S3}) implies $\mathfrak{c} < -s/ \la + m^2/2 \leq \mathfrak{c}_{\rm min}$ which violates (\ref{e:C-lower-bound-strong}). Thus $\chi'(-s)$ must be positive. In other words, when the pole $u= -s$ is a simple zero of $\chi$ with nonempty fibre, it must be surrounded by other allowed latitudes. So the poles cannot be simple zeros unless the neighbouring latitudes are allowed. In particular, the south poles in Fig.~\ref{f:Examples-of-common-level-set}d, h, i and j are allowed latitudes.

\vspace{.25cm}

{\bf \fl Result 2: Set of allowed latitudes and common level set must be closed:} The conserved quantities $\mathfrak{c}, m, s$ and $h$ define continuous functions (quadratic in $S$ and $L$) from the phase space $M^6_{S\text{-}L}$ to the four-dimensional space $\cal Q$ of conserved quantities (which is a subset of $\mathbb{R}^4$ consisting of the 4-tuples $(\mathfrak{c}, m ,h, s)$ subject to the conditions $s \geq 0$ and $\mathfrak{c} \geq \mathfrak{c}_{\rm min}$ (\ref{e:C-lower-bound-strong})). Each of their common level sets must be a closed subset of $M^6_{S\text{-}L}$ as it is the inverse image of a point in $\cal Q$. We may use this to deduce that $\chi$ cannot approach a positive value at a pole. We have already observed that if a pole is an allowed latitude then $\chi$ must vanish there. On the other hand, suppose a pole $P$ is not an allowed latitude but $\chi$ is positive in a neighbourhood of $P$. Then the set of allowed latitudes would be an open set and so would the common level set. In particular, $\chi$ cannot have (i) only one simple zero on the $S$-sphere and be nonvanishing elsewhere (as in Fig.~\ref{f:Examples-of-common-level-set}n) (ii) three simple zeros between the poles (see Fig.~\ref{f:Examples-of-common-level-set}m) (iii) a double zero and a simple zero between the poles (iv) a triple zero at a nonpolar latitude (v) two simple zeros between the poles with $\chi > 0$ at the poles (as in Fig.~\ref{f:Examples-of-common-level-set}o) or (vi) a double zero between the poles with $\chi > 0$ at the poles.

\vspace{.25cm}

{\fl \bf Common level set of conserved quantities must be connected:} For the common level set to be disconnected, the set of allowed latitudes on the $S$-sphere must be disconnected. The only remaining way that this could happen is for $\chi$ to have three distinct simple zeros on latitudes $u \in [- s, s]$ of the $S$-sphere. Let us show that this is disallowed. Now {\bf Result} 2 prevents $\chi$ from having three simple zeros at nonpolar latitudes. It only remains to consider the cases where either of the poles is a simple zero of $\chi$. If $\chi$ has a simple zero at $s$, then by {\bf Result} 1, $\chi'(s) < 0$. Since $\chi(\infty) = \infty$, $\chi$ can have at most one more zero on the $S$-sphere so that the set of allowed latitudes is connected. On the other hand, suppose $\chi$ has a simple zero at $-s$, then $\chi'(-s) > 0$ by {\bf Result} 1. Suppose further that $\chi$ has two more simple zeros $-s < u^* < u^{**} \leq s$ on the $S$-sphere, then by {\bf Result} 2, $u^{**}$ must equal $s$ as otherwise $\chi$ would be positive at the pole $u = s$ as in the disallowed Figs.~\ref{f:Examples-of-common-level-set}m, n and o. So $u^{**} = s$ with $\chi'(s) > 0$ as in Fig.~\ref{f:Examples-of-common-level-set}j. But in this case, {\bf Result} 1 forbids $u^{**}$ from being an allowed latitude, so that the set of allowed latitudes is again a single interval $[-s, u^*]$.

\vspace{.25cm}

{\fl \bf Triple zeros of $\chi$:} For $\chi(u)$ (\ref{e:cubic-equation-S3}) to have a triple zero, i.e., to be of the form $(u - z)^3$, we must have $z = \la \mathfrak{c}/3$ and the conserved quantities must satisfy two conditions:
	\beq
	\mathfrak{c}^2 = -\frac{3}{\la} \left(hm +\frac{s^2}{\la}\right) \quad \text{and} \quad 2 \la^2 \mathfrak{c}^3= -27(2\mathfrak{c}s^2 - h^2 - m^2s^2).
	\label{e:chi-triple-0-condition-on-cons-qty}
	\eeq
These conditions define a two-dimensional surface in the space $\cal Q$ of conserved quantities. {\bf Result} 2 implies that $\chi$ cannot have a triple zero at a nonpolar latitude. On the other hand, $\chi$ {\it can} have a triple zero at ${\cal N}$ or ${\cal S}$ provided both (\ref{e:L-L3-plane}) and (\ref{e:chi-triple-0-condition-on-cons-qty}) are satisfied. Putting $h = \mp \sign{k} \, ms$ in (\ref{e:chi-triple-0-condition-on-cons-qty}), the conditions for $\cal N$ or $\cal S$ to be a triple zero become 
	\beq
	\pm 3 \la \, \sign{k} \, s m^2 = \la^2 \mathfrak{c}^2 + 3 s^2 \quad \text{and} \quad  
	\la \mathfrak{c} = 3 s.
	\label{e:triple-zero-condition-chi}
	\eeq
The first condition implies that $\chi$ cannot have a triple zero at ${\cal S}$ for $k > 0$ or at ${\cal N}$ for $k < 0$. On the other hand, $\chi$ {\it can} have a triple zero at $\cal N$ for $k > 0$ as in Fig.~\ref{f:Examples-of-common-level-set}l.

\subsubsection{Possible types of common level sets of all four conserved quantities}
\label{s:Examples-of-CL-sets-of-conserved-quantities}

Here we combine the above results on the connectedness of common level sets, slope of $\chi$ at the poles and on the structure of the fibres over polar and nonpolar latitudes of the $S$-sphere to identify all possible common level sets of conserved quantities. There are only four possibilities: the degenerate or singular level sets (horn tori, circles and single points) and the generic common level sets (2-tori). These possibilities are distinguished by the location of roots of $\chi$. They are discussed below and illustrated in Fig \ref{f:Examples-of-common-level-set}. In (C1)-(C5) below we take $k > 0$ so that $u = \pm s$ correspond to the $\cal N$ and $\cal S$ poles. Similar results hold for $k < 0$ with $\cal N$ and $\cal S$ interchanged.



(C1) For generic values of conserved quantities, $\chi(u)$ is positive between two neighbouring nonpolar simple zeros $u_{\rm min} < u_{\rm max}$ lying in $(-s, s)$ (E.g. $k = \la = s = m = 1$, $h = 2$ and $\mathfrak{c} = 3$ as in Fig.~\ref{f:Examples-of-common-level-set}a). The base space of Section \ref{s:common-level-set-conserved-qtys} is the portion of the $S$-sphere lying between the latitudes $u_{\rm min}$ and $u_{\rm max}$, with the two-point fibres shrinking to single point fibres along the extremal latitudes $u_{\rm min}$ and $u_{\rm max}$. The resulting common level set is homeomorphic to a pair of finite coaxial cylinders with top as well as bottom edges identified, i.e., a $2$-torus.


To visualize the above toroidal common level sets and some of its limiting cases which follow, it helps to qualitatively relate the separation between zeros of $\chi$ to the geometric parameters of the torus embedded in three dimensions. For instance, the minor diameter of the torus grows with the distance between $u_{\rm min}$ and $u_{\rm max}$. Thus, when the simple zeros coalesce at a double zero, the minor diameter vanishes and the torus shrinks to a circle. Similarly (for $k > 0$) the major diameter of the torus grows with the distance between $u_{\rm min}$ and $\cal N$. Thus, when $u_{\rm max} \to {\cal N}$, the major and minor diameters become equal and we expect the torus to become a horn torus. However, this requires the fibre over ${\cal N}$ to be a single point, which is true only when $\cal N$ is a double zero of $\chi$. 


(C2) A limit of (C1) where either $u_{\rm min} \to {\cal S}$ or $u_{\rm max} \to {\cal N}$ and $\chi$ is positive between them. For instance, if $u_{\rm max} \to {\cal N}$ and the fibre over $\cal N$ is a single point, then the common level set is homeomorphic to a horn torus (E.g. $\la = k = s= h = 1$, $m =-1$ and $\mathfrak{c} = 3/2$ as in Fig.~\ref{f:Examples-of-common-level-set}b). On the other hand, for $\mathfrak{c} > 3/2$ the fibre over ${\cal N}$ is a circle and we expect the common level set to be a $2$-torus (see Fig.~\ref{f:Examples-of-common-level-set}c). It is as if the circular fibre over the single-point latitude $\cal N$ plays the role of an extremal circular latitude with single point fibre in (C1), thus the roles of base and fibre are reversed. Similarly, when $u_{\rm min} \to {\cal S}$ with circular fibre over ${\cal S}$, the common level set is homeomorphic to a 2-torus (E.g. $k = \la = s = m = h = 1$ and $\mathfrak{c} > -1/2$ as in Fig.~\ref{f:Examples-of-common-level-set}d). In the limiting case where $\mathfrak{c} = \mathfrak{c}_{\rm min} = -1/2$, the two simple zeros $u_{\rm min}$ and $u_{\rm max}$ merge at $\cal S$. The fibre over $\cal S$ becomes a single point and the common level shrinks to a point (see Fig.~\ref{f:Examples-of-common-level-set}e).
 

(C3) Another limit of (C1) where the roots $u_{\rm min}$ and $u_{\rm max}$ coalesce at a double root $u_d \in (-s, s)$ of $\chi$. $\chi$ is negative on the $S$-sphere except along the latitude $u_d$ and the fibre over it is a single point. The discriminant $\D$ (\ref{e:discriminant}) must vanish for this to happen. The common level set becomes a circle corresponding to the latitude $u_d$. For example, if $k = \la =1$ and $s = 1, m = -1/2, h = 2$ and $\mathfrak{c} = 17/8$, then the equator $u_d = 0$ is the allowed latitude as shown in Fig.~\ref{f:Examples-of-common-level-set}f. Another example of a circular common level set appears in Fig.~\ref{f:Examples-of-common-level-set}g. In this case {\bf Results} 1 and 2  exclude the north pole ensuring the connectedness of the common level set.


(C4) A limit of (C1) where the simple zeros $u_{\rm min}$ and $u_{\rm max}$ move to ${\cal S}$ and ${\cal N}$ respectively, with $\chi > 0$ in between. In this case, both poles have circular fibres and the common level set is a 2-torus. This happens, for instance, when $\mathfrak{c} \to \infty$, irrespective of the values of $m, h$ and $s > 0$. Another way for this to happen is for $m$ and $h$ to vanish so that the poles are automatically zeros of
	\beq
 	\chi(u) = u^3 - \la \mathfrak{c} u^2 - s^2u + \la \mathfrak{c} s^2 = (u - s)(u + s)(u - \la \mathfrak{c}) \quad [\text{for} \quad m = h = 0]
	\eeq
and to choose $\mathfrak{c} > s/\la$ to ensure there is no zero in between. Holding $s, h$ and $m$ fixed, three more possibilities arise as we decrease $\mathfrak{c}$. When $\mathfrak{c} = s/\la$, $\chi$ has a double zero at ${\cal N}$ (Fig.~\ref{f:Examples-of-common-level-set}i) with a single point fibre over it and the common level set becomes a horn torus.  For $-s/\la < \mathfrak{c} < s/\la$, the third zero of $\chi$ moves from $\cal N$ to the latitude  $u = \la \mathfrak{c}$. By {\bf Result} 1, the allowed latitudes go from $u= -s$ to $u = \la \mathfrak{c}$ (see Fig.~\ref{f:Examples-of-common-level-set}j), and the common level set returns to being a  $2$-torus. Finally, when $\mathfrak{c} = \mathfrak{c}_{\rm min} = -s/\la$, the only allowed latitude $(\cal S)$ is a double zero and the common level set shrinks to a point (see Fig.~\ref{f:Examples-of-common-level-set}k).   


(C5) $\chi$ has a zero at just one of the  poles and is negative elsewhere on the $S$-sphere. The common level set is then a single point. We encountered this as a limiting case of (C2) where $\chi$ has a double zero at $\cal S$ as in Fig.~\ref{f:Examples-of-common-level-set}e. This can also happen when $\chi$ is negative on the $S$-sphere except for a triple zero at either ${\cal S}$ ($k < 0$) or ${\cal N}$ ($k > 0$) (see Eq. (\ref{e:triple-zero-condition-chi})). For example, when $k= \la = s = 1$, $\mathfrak{c} = 3, m = 2$ and $h=-2$, $\chi$ has a triple zero at ${\cal N}$ as in Fig.~\ref{f:Examples-of-common-level-set}l.

\subsection{Nature of the `Hill' region and energy level sets using Morse theory}
\label{s:Hill-region-Morse-theory}

In this section, we study the `Hill' region $W^{E}_{\mathfrak{c} m}$, which we define as the set of points on the symplectic leaf $M^4_{\mathfrak{c} m}$ with energy less than or equal to $E$:	
	\beq
	W^{E}_{\mathfrak{c} m} = \{ p \in M^4_{\mathfrak{c} m} | H(p) \leq E \}.
	\eeq
The $H = E k^2$ energy level set $M^E_{\mathfrak{c} m}$ is then the boundary of $W^{E}_{\mathfrak{c} m}$. Taking $R_{1,2}$ and $P_{1,2}$ (\ref{e:R-P-relation-to-L-S}) as coordinates on $M^4_{\mathfrak{c} m}$, we treat the Hamiltonian
	\beq
	\frac{H}{k^2} = \frac{P_1^2 + P_2^2}{2} + \frac{\la m}{2}(R_1 P_2 -R_2 P_1) + \frac{\la^2}{8} (R_1^2 + R_2^2) \left(R_1^2 + R_2^2 + 3 m^2 - 4 \mathfrak{c} \right) + \frac{\la^2}{8} (2 \mathfrak{c} - m^2)^2 + \mathfrak{c} + \frac{1}{2 \la^2}
	\eeq
as a Morse function \cite{Milnor}. The nature of critical points of $H$ depends on the value of $2 \mathfrak{c} - m^2$. There are two types of critical points: (a) an isolated critical point at $R_{1,2} = P_{1,2} = 0$ which exists for all values of $2 \mathfrak{c} - m^2$ and (b) a ring of critical points 
	\beq
	R_1^2 + R_2^2 = 2 \mathfrak{c} - m^2 \quad \text{with} \quad ( P_1, P_2 ) = \frac{\la m}{2} \left( R_2, -R_1 \right),
	\eeq
which exists only for $2 \mathfrak{c} - m^2 > 0$ and shrinks to the isolated critical point when $2 \mathfrak{c} - m^2  = 0$. The energy at these critical points is
	\beq
	E_{\rm iso} = \frac{\la^2}{8} (2 \mathfrak{c} - m^2)^2 + \mathfrak{c} + \frac{1}{2 \la^2} \quad \text{and} \quad
	E_{\rm ring} =  \mathfrak{c} + \ov{2 \la^2}.
	\eeq
Upon varying $\mathfrak{c}$ and $m$, the isolated critical points cover all of the static submanifold $\Sigma_2$ while the rings of critical points cover the static submanifold $\Sigma_3$. By finding the eigenvalues of the Hessian of the Hamiltonian at these critical points, we find that for $2 \mathfrak{c} - m^2 < 0$ the isolated critical point G is a local minimum of energy (four +ve eigenvalues). In fact, for $2 \mathfrak{c} - m^2 < 0$, the isolated critical point has to be the global minimum of energy as the energy is bounded below and there are no other extrema of energy. For $2 \mathfrak{c} - m^2 > 0$, the isolated critical point becomes a saddle point (two +ve and two -ve eigenvalues) with energy $E_{\rm sad} = E_{\rm iso}$. On the other hand, the ring of critical points are degenerate global minima (three +ve and one zero eigenvalue). To apply Morse theory, we need the indices of the critical points of $H$ (number of negative eigenvalues of the Hessian). From the foregoing, we see that the ground state G has index zero, the saddle point has index two and the degenerate critical points on the ring may be nominally assigned a vanishing index.

\vspace{.25cm}

{\fl \bf Change in topology of the Hill region:} According to Morse theory \cite{Milnor}, the topology of the Hill region can change only at critical points of the Hamiltonian. (a) For $2 \mathfrak{c} - m^2 < 0$, there is only one critical point, the global minimum G with index zero and energy $E_{\rm G} = E_{\rm iso}$. Thus, as $E$ increases beyond $E_{\rm G}$, the Hill region $W^E_{\mathfrak{c} m}$ goes from being empty to being homeomorphic to a 4-ball $(B^4 = \{ {\bf x} \in \mathbb{R}^5 \quad \text{with} \quad  \Vert {\bf x} \Vert \leq 1 \})$ arising from the addition of a 0-cell. (b) For $2 \mathfrak{c} - m^2 > 0$, there are two critical values of energy $E_{\rm ring} < E_{\rm sad}$ corresponding to the ring of critical points and the saddle point. The index vanishes along the ring of critical points, so when $E$ crosses $E_{\rm ring}$, the Hill region acquires a 3-ball (0-cell) for each point on the ring corresponding to the 3 positive eigenvalues of the Hessian. Thus $W^{E}_{\mathfrak{c} m}  \cong  B^3 \times S^1$ for $E_{\rm ring} < E < E_{\rm sad}$. The saddle point with $E = E_{\rm sad}$ has index two, so the topology of $W^{E}_{\mathfrak{c} m}$ changes to $B^4$ upon adding a 2-cell to $B^3 \times S^1$ (the analogous statement in one lower dimension is that adding a 2-cell to the hole of the solid torus $(B^2 \times S^1)$ gives a $B^3$).

\vspace{.25cm}

{\fl \bf Nature of energy level sets:} The energy level set $M^E_{\mathfrak{c} m}$ is the boundary of the Hill region, i.e. $M^E_{\mathfrak{c} m} = \pdr W^E_{\mathfrak{c} m}$. It is a 3-manifold except possibly at the critical energies. Thus for $2\mathfrak{c} - m^2 < 0$, $M^E_{\mathfrak{c} m} \cong \pdr B^4 \cong S^3$ for all energies $E > E_{\rm G}$. On the other hand, when $2\mathfrak{c} - m^2 > 0$ the energy level set undergoes a change in topology from $S^2 \times S^1$ to $S^3$ as $E$ crosses $E_{\rm sad}$. 

The energy level sets at the critical values $E_{\rm G}, E_{\rm sad}$ and $E_{\rm ring}$ are exceptional. For given $\mathfrak{c}$ and $m$ with $2 \mathfrak{c} - m^2 < 0$ and $E = E_{\rm G}$, $M^E_{\mathfrak{c} m}$ is a single point on $\Sigma_2$ (the critical point), since G is the nondegenerate global minimum of energy. When $2 \mathfrak{c} - m^2 > 0$, $E = E_{\rm sad}$ fixes $s = (\la/2)(2 \mathfrak{c} - m^2)$ leaving a range of possible values of $h \in (h_{\rm min} , h_{\rm max})$, whose values are determined by eliminating $u$ from the conditions $\chi(u) = \chi'(u) = 0$. This leads to a three-dimensional energy level set. $M^{E_{\rm sad}}_{\mathfrak{c} m}$ includes one horn torus with its center as the saddle point for $h = h_{\rm sad}$ as well as a one parameter family of toroidal level sets for $h_{\rm min} < h \neq h_{\rm sad}  < h_{\rm max}$ and a pair of  circular level sets occurring at $h_{\rm min}$ and $h_{\rm max}$. Interestingly, horn tori arise only  when $E= E_{\rm sad}$, since $s = (\la/2)(2 \mathfrak{c} - m^2)$ is a necessary condition for horn tori (see Section \ref{s:horn-toroidal-level-sets}). Thus, the horn torus is a bit like the figure-8 shaped separatrix one encounters in particle motion in a double well potential. Finally, the $E = E_{\rm ring}$ level manifold consists of a ring of single point common level sets, each lying on the static submanifold $\Sigma_3$. Unlike static solutions and horn tori, circular and 2-toroidal level sets also arise at noncritical energies.

\section[Foliation of phase space by tori, horn tori, circles and points]{Foliation of phase space by tori, horn tori, circles and points \sectionmark{Foliation of phase space}}
\sectionmark{Foliation of phase space}
\label{s:foliation-of-phase-space}

For generic allowed values of the conserved quantities $\mathfrak{c}, m, s$ and $h$, their common level set in the $M^6_{S \text{-} L}$ phase space is a $2$-torus. As noted, this happens when $\chi$ has simple zeros along a pair of latitudes of the $S$-sphere and is positive between them. However, this $4$-parameter family of invariant tori does not completely foliate the phase space: there are some other `singular' level sets as well: horn tori, circles and points. The union of single-point level sets is $\Sigma_2 \cup \Sigma_3$ (\ref{e:static-submanifolds}), consisting of static solutions. They occur when $\chi(u)$ has a triple zero at $u = s$ or is a local maximum at a double zero at $u = \pm s$. We will now discuss the other cases in increasing order of complexity. In each case, we view the union of common level sets of a given type as the state space of a self-contained dynamical system which has the structure of a fibre bundle over an appropriate submanifold of the space $\cal Q$ of conserved quantities. The fibres in each case are circles, horn tori and tori. The dynamics on the union of circles and tori is Hamiltonian and we identify action-angle variables on them. On the other hand, we show that the dynamics on the union of horn tori is a gradient flow.

\subsection{Union $\cal C$ of circular level sets: Poisson structure \& action-angle variables}
\label{s:circular-level-sets}

In this section, we show that the union of circular level sets is the same as the trigonometric/circular submanifold $\cal C$ (introduced in Section \ref{s:Static-and-trigonometric-solutions}) where the solutions are sinusoidal functions of time. Local coordinates on $\cal C$ are furnished by $\mathfrak{c}, m, u$ and $\tht$ (or equivalently $\phi$) and we express the Hamiltonian in terms of them. The Poisson structure on $\cal C$ is degenerate with $\mathfrak{c}$ and $m$ generating the center and their common level sets being the symplectic leaves. While $u$ is a constant of motion, $\tht$ evolves linearly in time. We exploit these features to obtain a set of action-angle variables for the dynamics on $\cal C$.

\subsubsection{$\cal C$ as a circle bundle and dynamics on it}

As pointed out in example (C3) of Section \ref{s:Examples-of-CL-sets-of-conserved-quantities}, the common level set of conserved quantities is a circle when the cubic $\chi(u)$ (\ref{e:cubic-equation-S3}) has a double zero at a nonpolar latitude of the $S$-sphere and is negative on either side of it. In this case, the latitude $u$ is restricted to the location of the double zero. To identify the three-dimensional hypersurface ${\cal Q}_{\cal C}$ in the four-dimensional space $\cal Q$ of conserved quantities, where $\chi$ has a double zero at a nonpolar latitude, we will proceed in two steps.  First, we compare the equation $\chi = 0$ with $(u - u_2)^2(u - u_1) = 0$ to arrive at the three conditions:
	\beq
	2 u_2 + u_1 = \la  \mathfrak{c}, \quad
	u_2^2 + 2 u_2 u_1 = -\left( s^2 + h m \la \right) \quad \text{and} \quad
	-u_2^2 u_1 = \frac{\la}{2} \left((2\mathfrak{c} - m^2) s^2 - h^2 \right).
	\label{e:relations-for-chi-double-zero}
	\eeq
The first two may be used to express the roots $u_2$ and $u_1$ in terms of conserved quantities:
	\beq
	u_2^\pm = (1/3) \left( \la \mathfrak{c}  \pm \sqrt{\la^2 \mathfrak{c}^2 + 3( s^2 + \la h m)}\right) \quad \text{and} \quad 
	u_1^\pm = \la \mathfrak{c} - 2u_2.
	\label{e:roots-u1-u2}
	\eeq
The third equation in (\ref{e:relations-for-chi-double-zero}) then leads to the following conditions among conserved quantities
	\beq
	27 \la h^2 - 36 \la \mathfrak{c} s^2  + 27 \la m^2 s^2  + 18 \la^2 \mathfrak{c} h m  + 4  \la^3 \mathfrak{c}^3  = \mp 4(3 s^2 + \la (3 h m + \la \mathfrak{c}^2 ))^{3/2}.
	\label{e:relations-circle-level-set}
	\eeq
Squaring, these conditions are equivalent to $\Delta = 0$, where $\D$ is the discriminant (\ref{e:discriminant}) of $\chi$. The three-dimensional submanifold of $\cal Q$ defined by $\D = 0$, however, includes 4-tuples ($\mathfrak{c}$, $m$, $s$, $h$) corresponding to horn toroidal (double zero at the pole $u=s$) or single-point (triple zero at $u = s$ or double zero at $u= s$ or $-s$) common level sets, in addition to circular level sets. To eliminate the former, we must impose the further conditions $u_2 \neq u_1$, $|u_2| < s$ and $\chi''(u_2) < 0$. This last condition, which says $u_2 <\la \mathfrak{c}/3$, selects the roots $u_{1,2} = u_{1,2}^-$ in (\ref{e:roots-u1-u2}). These conditions define the three-dimensional hypersurface ${\cal Q}_{\cal C} \subset \cal Q$ corresponding to circular level sets. Now, $\mathfrak{c}, m$ and $s$ may be chosen as coordinates on ${\cal Q}_{\cal C}$, with (\ref{e:relations-circle-level-set}) allowing us to express $h$ in terms of them. Interestingly, we find by studying examples, that  for values of $\mathfrak{c}, m$ and $s$ corresponding to a circular level set, there are generically two distinct values of $h$; so we would need two such coordinate patches to cover ${\cal Q}_{\cal C}$. The union of all these circular level sets may be viewed as a sort of circle bundle over ${\cal Q}_{\cal C}$ and forms a four-dimensional `circular' submanifold $\cal C$ of $M^6_{S\text{-}L}$. As shown in Section \ref{s:Static-and-trigonometric-solutions} and Section \ref{s:Independence-of-conserved-quantities}, this circular submanifold along with its boundary coincides with the set where the four-fold wedge product $dh \wedge ds^2 \wedge dm \wedge d\mathfrak{c}$ vanishes. 

The equations of motion (\ref{e:EOM-LS}) simplify on the circular submanifold $\cal C$. Indeed, since $S_3 = ku$ is a constant, $\dot S_3 = 0$ so that $S_1/S_2 = L_1/L_2$ implying that $\tht - \phi = n \pi$ where $n \in {\mathbb Z}$. As shown in Section \ref{s:Static-and-trigonometric-solutions}, the equations of motion then simplify to 
	\beq
	\dot{S_1} = -\dot \phi S_2, \qquad
	\dot{S_2} = \dot\phi S_1, \qquad
	\dot{L_1} = k S_2 \qquad \text{and} \qquad
	\dot{L_2} = -k S_1
	\eeq 
with sinusoidal solutions:
	\beq
	S_1/k = A \sin k \om t + B \cos k \om t \quad \text{and} \quad 
	S_2/k = A \cos k \om t - B \sin k \om t.
	\label{e:circular-level-set-solutions}
	\eeq
Here, using (\ref{e:L-S-polar}), $\omega = S_{1,2}/L_{1,2} = (-1)^{n} \rho/r = -\dot \phi/k = -\dot \tht/k$, which varies with location on the base ${\cal Q}_{\cal C}$. It is the nondimensional angular velocity for motion in the circular fibres. Since $\rho$ and $r$ are positive, $(-1)^n \omega = |\omega|$. Here both $\tht$ and $\phi$ evolve linearly in time and the equality of 
	\beq
	\dot \tht = (-1)^{n+1} \frac{k \rho}{r} \quad \text{and} \quad \dot \phi = k \la \left(m + (-1)^n \frac{u r}{\rho} \right)
	\label{e:circular-levelset-dynamics}
	\eeq 
implies that the constant of motion $u$ may be expressed in terms of $\omega$ and $m$: 
	\beq
	u = - \om (m + \om/\la).
	\label{e:quadratic-alpha}
	\eeq
\footnotesize
{\fl \bf Remark:} If the $S$-sphere shrinks to a point $(s = h = 0)$ then one still has circular level sets consisting of latitudes of the $L$-sphere determined by $m$, provided $2\mathfrak{c} \geq m^2$. However, each point on these exceptional circular level sets is a static solution lying on $\Sigma_3$ (\ref{e:static-submanifolds}).
\normalsize
	
\subsubsection{Canonical coordinates on $\cal C$}

{\fl \bf Local coordinates on $\cal C$:} For the analysis that follows, a convenient set of coordinates on the `circle bundle' $\cal C$ consists of $\mathfrak{c}, m$ and $\omega$ for the base ${\cal Q}_{\cal C}$ and $\tht$ for the fibres. The dynamics on $\cal C$ admits three independent conserved quantities as there is one relation among $\mathfrak{c}, m , s$ and $h$ following from (\ref{e:relations-circle-level-set}). Since the common level sets of the conserved quantities on $\cal C$ are circles, rather than tori, it is reasonable to expect there to be two Casimirs (say $\mathfrak{c}$ and $m$) for the Poisson structure on $\cal C$, as we show below. In fact, $\cal C$ is foliated by the common level surfaces of $\mathfrak{c}$ and $m$ (symplectic leaves) which serve as phase spaces (with coordinates $\omega$ and $\tht$) for a system with one degree of freedom. $\tht$  is then the coordinate along the circular level sets of the Hamiltonian on these two-dimensional symplectic leaves. 

To find the reduced Hamiltonian on $\cal C$ we express the remaining variables in terms of $\mathfrak{c}, m, \omega$ and $\tht$. The formula for $\mathfrak{c}$ (\ref{e:conserved-quantities}) along with (\ref{e:quadratic-alpha}) determines $r^2 \equiv 2\mathfrak{c} - m^2 + (2 \om/\la) (m + \om/\la)$ and consequently $\rho = |\omega| r$ as well. The remaining conserved quantities are given by
	\beqs
	h &=& (-1)^n \rho r - m u = \omega \left( 2\mathfrak{c} - m^2 + \frac{2 \om}{\la}\left( m + \frac{\om}{\la} \right)  \right) + m \om \left(m + \frac{\om}{\la} \right)  \quad \text{and} \cr
	s^2 &=& \rho^2 + u^2 = 2 \om^2 \left(\mathfrak{c}  + \frac{2 m \om}{\la} + \frac{3 \omega^2}{2 \la^2}\right). 
	\label{e:h-and-s-on-circular-LS}
	\eeqs
Thus, the reduction of the Hamiltonian (\ref{e: H-mechanical}) to the trigonometric submanifold is
	\beq
	H(\mathfrak{c}, m, \omega) = k^2  \left( \om^2 \left(\mathfrak{c} + \frac{2 m \omega}{\la} + \frac{3 \omega^2}{2 \la^2}  \right) + \mathfrak{c} + \frac{1}{2 \la^2} \right).
	\label{e:H-trigonometric}
	\eeq
As remarked, for given values of $\mathfrak{c}, m$ and $s$, there are generically two possible values of $h$ corresponding to two points on ${\cal Q}_{\cal C}$. By considering examples, we verified that for each of them, there is a unique $\om$ that satisfies (\ref{e:quadratic-alpha}) and both the equations in (\ref{e:h-and-s-on-circular-LS}).

\vspace{.25 cm}

{\fl \bf Poisson structure on $\cal C$:} We wish to identify Poisson brackets among the coordinates $\mathfrak{c}, m, \omega$ and $\tht$ that along with the reduced Hamiltonian (\ref{e:H-trigonometric}) gives the equation of motion $\dot{\tht}= -\om k$ on $\cal C$. As noted, it is natural to take $\mathfrak{c}$ and $m$ as Casimirs so that $\{ \mathfrak{c} , m\} = \{ \mathfrak{c} , \om \} = \{m, \om \} = \{ \mathfrak{c}, \tht \} = \{ m , \tht \} = 0$. The only nontrivial Poisson bracket $\{ \tht, \omega \}$ is then determined as follows from (\ref{e:H-trigonometric}):
	\beq
	\dot \tht = - k \omega = \{ \tht, H \} = \pdr_\om H \{ \tht, \omega \} \quad \imply \quad
	\{ \tht, \omega \} = - \frac{k \omega}{\pdr_\om H} = -\frac{1}{2 k} \left(\mathfrak{c} + \frac{3 \om}{\la}\left( m + \frac{\om}{\la} \right) \right)^{-1}.
	\label{e:PB-phi-omega}
	\eeq
Moreover, this implies $\{ \tht, u \} =  (2 \om + m \la)/(k(2 \la \mathfrak{c} - 6 u))$, which notably differs from the original nilpotent Poisson bracket $\{ \tht, u \}_{\nu} = 0$ (\ref{e: PB-SL}).

\vspace{.25cm}

{\bf \fl Canonical action-angle variables on $\cal C$:} Since $\tht$ evolves linearly in time, it is a natural candidate for an angle variable. The corresponding canonically conjugate action variable $I$ must be a  function of $\omega, m$ and $\mathfrak{c}$ and is determined from (\ref{e:PB-phi-omega}) by the condition $\{ \tht , I(\omega) \} = I'(\omega) \{ \tht, \omega \} = 1$. We thus obtain, up to an additive constant,  the action variable 
	\beq
	I(\omega) = -k \om \left( 2 \mathfrak{c}  + \frac{3 m \omega}{\la} + \frac{2 \omega^2}{\la^2}  \right) = -k h.
	\label{e:circular-action-variable}
	\eeq
Thus we arrive at the remarkably simple conclusion that (aside from the Casimirs $\mathfrak{c}$ and $m$) $-k h$ and $\tht$ are action-angle variables on $\cal C$. Moreover, the canonical Poisson bracket $\{ \tht, -k h \} = 1$ agrees with that on the full phase space (see (\ref{e:PB-on-union-of-tori})). Our reason to work with $\om$ rather than $h$ as a coordinate is that the solutions (\ref{e:circular-level-set-solutions}) and the Hamiltonian (\ref{e:H-trigonometric}) have simple expressions in terms of $\om$. By solving the cubic (\ref{e:circular-action-variable}), $\om$ can be expressed in terms of $h$, which would allow us to write the Hamiltonian in terms of the action variable $- k h$.

\subsection{Union $\bar{\cal H}$ of horn toroidal level sets: Dynamics as gradient flow}
\label{s:horn-toroidal-level-sets}

Just as with the union of circular level sets $\cal C$, the union of horn toroidal level sets $\bar{\cal H}$ serves as the phase space for a self-contained dynamical system. However, unlike the sinusoidal periodic trajectories on $\cal C$, all solutions on $\bar{\cal H}$ are hyperbolic functions of time and are in fact homoclinic orbits joining the center of a horn torus to itself (see Fig.~\ref{f:theta-phi-dynamics-3D-horn-torus}). The centers themselves are static solutions. Horn tori arise only when the energy is equal to the critical value $E = E_{\rm sad}$ given in Section \ref{s:Hill-region-Morse-theory}. Thus, the horn tori are like the figure-8 shaped separatrices in the problem of a particle in a double well potential, separating two families of 2-tori. Interestingly, though the conserved quantities satisfy a relation on each horn torus, the four-fold wedge product $dh \wedge ds^2 \wedge dm \wedge d\mathfrak{c}$ vanishes only at its center. Finally, unlike on the circular submanifold, the flow on the horn-toroidal submanifold is {\it not} Hamiltonian, though we are able express it as a gradient flow.

The family of horn toroidal level sets is a two-dimensional submanifold ${\cal Q}_{\bar H}$ of the four-dimensional space of conserved quantities $\cal Q$. To see this, note that a horn torus arises when the cubic $\chi(u)$ of (\ref{e:cubic-equation-S3}) is positive between a simple zero and a double zero at the pole $u = s$ of the $S$-sphere. Thus, $\chi(u)$ must be of the form $\chi(u) = (u-u_1) (u-s)^2$ where $u_1 = \la m^2/2-s $ with $-s \leq u_1 \leq s$. These requirements imply $\chi(s) = \chi'(s) = 0$ and $\chi''(s) \geq 0$. Note that each nontrivial horn torus is a smooth two-dimensional surface except at its center which lies at the pole $u = s$. Trivial horn tori are those that have shrunk to the points at their centers and arise when $\chi''(s) = 0$. The conditions $\chi(s) = 0$ and  $\chi'(s) = 0$ lead to two relations among conserved quantities
	\beq
	h = -m s \quad \text{and} \quad \mathfrak{c} = \frac{m^2}{2} + \frac{s}{\la}, 
	\label{e:cons-qtys-Horn-tori-relations}
	\eeq
which together imply that $\D = 0$. The inequality $\chi''(s) \geq 0$ along with (\ref{e:cons-qtys-Horn-tori-relations}) restricts us to points above a parabola in the $m$-$s$ plane:
	\beq
	4s \geq \la m^2.
	\label{e:inequality-Horn-tori}
	\eeq 
The space ${\cal Q}_{\bar H}$ is given by  the set of such $(m,s)$ pairs. For each $(m,s) \in {\cal Q}_{\bar H}$ we get a horn torus $\bar H_{ms}$. The union of all horn tori is then given by $\bar {\cal H} = \cup_{4s \geq \la m^2} \bar H_{ms}$.  

\subsubsection{$\bar{\cal H}$ as a four-dimensional submanifold of $M^6_{S \text{-} L}$} 

Equations (\ref{e:cons-qtys-Horn-tori-relations}) and ({\ref{e:inequality-Horn-tori}}) when expressed in terms of $\vec S$ and $\vec L$ allow us to view the union of all horn tori $\bar {\cal H}$ as a four-dimensional submanifold of $M^6_{S \text{-} L}$:
	\beq
	S_1 L_1 + S_2 L_2 + ( S_3 - k s ) L_3 = 0, \quad \half (L_1^2 + L_2^2) + \frac{k S_3}{\la} = \frac{k^2 s}{\la} \quad \text{and} \quad L_3^2 \leq \frac{4 k^2 s}{\la}.
	\label{e:S-L-space-Horn-tori}
	\eeq
For any choice of $\vec S$, the first two conditions define a plane through the origin (normal to $(S_1, S_2, S_3 - s k)$) and a cylinder (of radius $r = \sqrt{(2 k/\la) (s k - S_3)}$ with axis along $L_3$) in the $L$-space. In general, this plane and cylinder intersect along an ellipse so that $\bar{\cal H}$ may be viewed as a kind of ellipse bundle over the $S$-space (subject to the inequality). The centers of the horn tori are the points where $S_{1,2} = L_{1,2} = 0$, $u = S_3/k = s$ and $|L_3/k| = |m| \leq \sqrt{4s/ \la}$ (see Section \ref{s:centers-of-horn-tori-punctured-horn-tori} below). Interestingly, it turns out that the inequality in (\ref{e:S-L-space-Horn-tori}) restricting the range of $L_3$ is automatically satisfied at all points of the base space other than when $u = s$ (which correspond to centers of horn tori). Indeed, let us find the range of values of $L_3$ allowed by the first two relations in (\ref{e:S-L-space-Horn-tori}) by parameterizing the elliptical fibre by the cylindrical coordinate $\tht$. Then $L_1 = r k \cos \tht$, $L_2 = r k \sin \tht$ and $L_3 = (2/ \la r) (S_1 \cos \tht + S_2 \sin \tht)$. The extremal values of $L_3$ on the ellipse occur at $\tht_{\rm ext} = \arctan S_2/S_1$ which implies that
	\beq
	|L_3|^2 \leq 
	\frac{2k}{\la} (sk + k \la) = \frac{4 k^2 s}{\la} - r^2.
	\eeq 
Thus the inequality in (\ref{e:S-L-space-Horn-tori}) is automatically satisfied away from the axis $r = 0$ which corresponds to the centers of horn tori.


\subsubsection{Centers of horn tori and punctured horn tori}
\label{s:centers-of-horn-tori-punctured-horn-tori}

It turns out that the centers of horn tori are static solutions and may therefore be regarded as forming the boundary of $\bar {\cal H}$. In particular, a trajectory on a horn torus $\bar H_{ms}$ can reach its center only when $t \to \pm \infty$. To find the space of centers $\cal O$ we note that they lie at the pole $u = s$ corresponding to $S_1 = S_2 = 0$ and $S_3 / k \geq 0$. The conditions (\ref{e:S-L-space-Horn-tori}) then become
	\beq
	(S_3 - k s) L_3 = 0, \quad \frac{L_1^2 + L_2^2}{2} + \frac{k S_3 }{\la} = \frac{k^2 s }{\la} \quad \text{and} \quad 4s \geq \la m^2 \quad \text{where} \quad s = \frac{S_3}{k}.
	\eeq 
The first condition is automatic, the second implies $L_{1,2} = 0$ while the inequality becomes $S_3 \geq (\la/4k) L_3^2$. Thus $\cal O$ is the two-dimensional subset of the static submanifold $\Sigma_2$ consisting of points on the $L_3$-$S_3$ plane, on or within the parabola $S_3 = (\la/4k) L_3^2$. The points on the parabola correspond to trivial horn tori. By eliminating their centers we obtain (nontrivial) punctured horn tori $H_{ms}$ which are smooth noncompact surfaces with the topology of infinite cylinders on which the dynamics is everywhere non static. We let ${\cal H} = \bar{\cal H} \setminus {\cal O} = \cup_{4s > \la m^2} H_{ms}$ denote the four-dimensional space consisting of the union of punctured horn tori. Thus $\cal H$ may be regarded as a cylinder bundle over the base ${\cal Q}_H = \{ (m, s) | 4s > \la m^2 \}$. Some possible coordinates on $\cal H$ are (a) $s, m, \tht, \phi$ (b) $s, m, u, \tht$ and (c) $S_{1,2,3}$ and either $L_1$ or $L_2$.


\subsubsection{Nonvanishing four-fold wedge product on $\bar{\cal H}$} 

We have argued that the conserved quantities satisfy the relations  (\ref{e:cons-qtys-Horn-tori-relations}) on $\bar{\cal H}$. Despite this, {\it we show that the wedge product $\Om_4 = dh \wedge ds^2 \wedge dm \wedge d\mathfrak{c}$ does not vanish on $\bar {\cal H}$ except on its  boundary ${\cal O} = \bar{\cal H} \setminus {\cal H}$}. To see this, note that in addition to the condition $\D(\mathfrak{c}, m, h, s) = 0$ (due to the presence of the double zero at the pole $u = s$), all four partial derivatives of $\D$ may be shown to vanish on $\bar{\cal H}$ by virtue  (\ref{e:cons-qtys-Horn-tori-relations}). In other words, the relation $\D_\mathfrak{c} d\mathfrak{c} + \D_{m} dm + \D_{h} dh + \D_s ds = 0$ following from $\D = 0$ is vacuous on $\bar{\cal H}$ (if not, we could wedge it, say, with $ds^2 \wedge dm \wedge d\mathfrak{c}$ to show that $\Om_4 = 0$). On the other hand, we showed in Section \ref{s:Independence-of-conserved-quantities} that $\Om_4$ vanishes precisely on the closure of the circular submanifold $\bar{\cal C}= {\cal C} \sqcup {\cal C}_1 \sqcup {\cal C}_2 \sqcup (\Sigma_2 \cup \Sigma_3)$. Thus, to show that $\Om_4$ is nonvanishing on $\cal H$, it suffices to find the points common to $\bar{\cal H}$ and $\bar{\cal C}$. Now $\bar{\cal H} \cap {\cal C}$ is empty as $\chi$ has a double/triple zero at $u = s$ for points on $\bar{\cal H}$ and a double zero away from the poles for points on ${\cal C}$. In fact, we find that $\bar {\cal H} \cap \bar{\cal C}$ is contained in the static submanifold $\Sigma_2$ so that $\Om_4$ is nowhere zero on $\cal H$ and vanishes only on its boundary $\cal O$. To see that $\bar{\cal H}$ does not have any points in common with either ${\cal C}_1$ or  ${\cal C}_2$ we observe that the conditions $h = - ms$, $\mathfrak{c} = s/\la + m^2/2$ (\ref{e:cons-qtys-Horn-tori-relations}) and the relations ($S_1 = L_1 = 0$ and $\Xi_3$) or ($S_2 = L_2 = 0$ and $\Xi_2$) that go into the definitions of $\bar{\cal H}$ and ${\cal C}_1$ or ${\cal C}_2$ (see Section \ref{s:Independence-of-conserved-quantities}), together define a parabola in phase space
	\beq
	4 k S_3 = \la L_3^2 
	\quad \text{with} \quad k S_3 \geq 0 \quad 
	\text{and}  \quad L_{1,2} = S_{1,2} = 0.
	\eeq
This parabola is contained in $\Sigma_2$ but does not lie on ${\cal H}, {\cal C}_1$ or ${\cal C}_2$ as the inequalities $4s > \la m^2, |S_2| > 0$ and $|S_1| > 0$ appearing in the definitions of ${\cal H}, {\cal C}_1$ and ${\cal C}_2$ are saturated along it. Points on this parabola correspond to horn tori that have shrunk to the single point at their centers and correspond to cubics $\chi$ with a triple zero at $u=s$. Thus, this parabola lies along the common boundary of ${\cal H}, {\cal C}_1$ and ${\cal C}_2$. Combining these results we see that $\Om_4 \ne 0$ on $\cal H$, but vanishes identically on its boundary consisting of the space of centers $\cal O$.

\subsubsection{Equations of motion on the horn torus:} 

On the horn torus $H_{ms}$ the evolution equation for $u$ (\ref{e:EOM-u}) simplifies: 
	\beq
	\dot{u}^2 = 2 \la k^2 \chi(u) = \la^2 k^2 (s-u)^2 \left[\frac{2}{\la}(s + u)- m^2 \right].
	\eeq
We may interpret this equation as describing the zero energy trajectory of a nonrelativistic particle of mass 2 with position $u(t)$ moving in a one-dimensional potential $V(u) = - 2 \la k^2 \chi(u)$. Since $V(u)$ is negative between the simple and double zeros at $u_1$ and $s$, the former is a turning point while the particle takes infinitely long to reach/emerge from $u=s$. Thus, the trajectory is like a solitary wave of depression. Choosing $u(0)$ to be its minimal value $u_1 = -s + \la m^2/2$, the trajectory of the particle is given by
	\beq
	u(t) = u_1 + (s-u_1) \tanh^{2} \left(\frac{t}{2\tau}\right)
	\quad \text{where} \quad \tau = \ov{\sqrt{\la k^2 (4s - \la m^2)}}.
	\label{e:particle-trajectory-horn-torus}
	\eeq
Notice that as $t \to \pm \infty$, $u(t) \to s$ and the solution approaches the center of the horn torus. Interestingly, the vector field $\dot u = \sqrt{-V(u)}$ is not smooth at $u = u_1$, which is a square root branch point. Thus, there is {\it another} solution $u(t) \equiv u_1$ with the same initial condition (IC) $u(0) = u_1$, which however is consistent with the $L$-$S$ equations of motion (\ref{e:EOM-LS}) only when $s = 0$. Note that (\ref{e:particle-trajectory-horn-torus}) can be obtained  as a limit of the $\wp$-function solution given in Section \ref{s:Reduction-2D}. On a horn torus, one of the half periods of the $\wp$-function is imaginary while the other diverges leading to the aperiodic solution (\ref{e:particle-trajectory-horn-torus}). 

\begin{figure}[h]
	\centering
		\begin{subfigure}[t]{5cm}
		\centering
		\includegraphics[width=5cm]{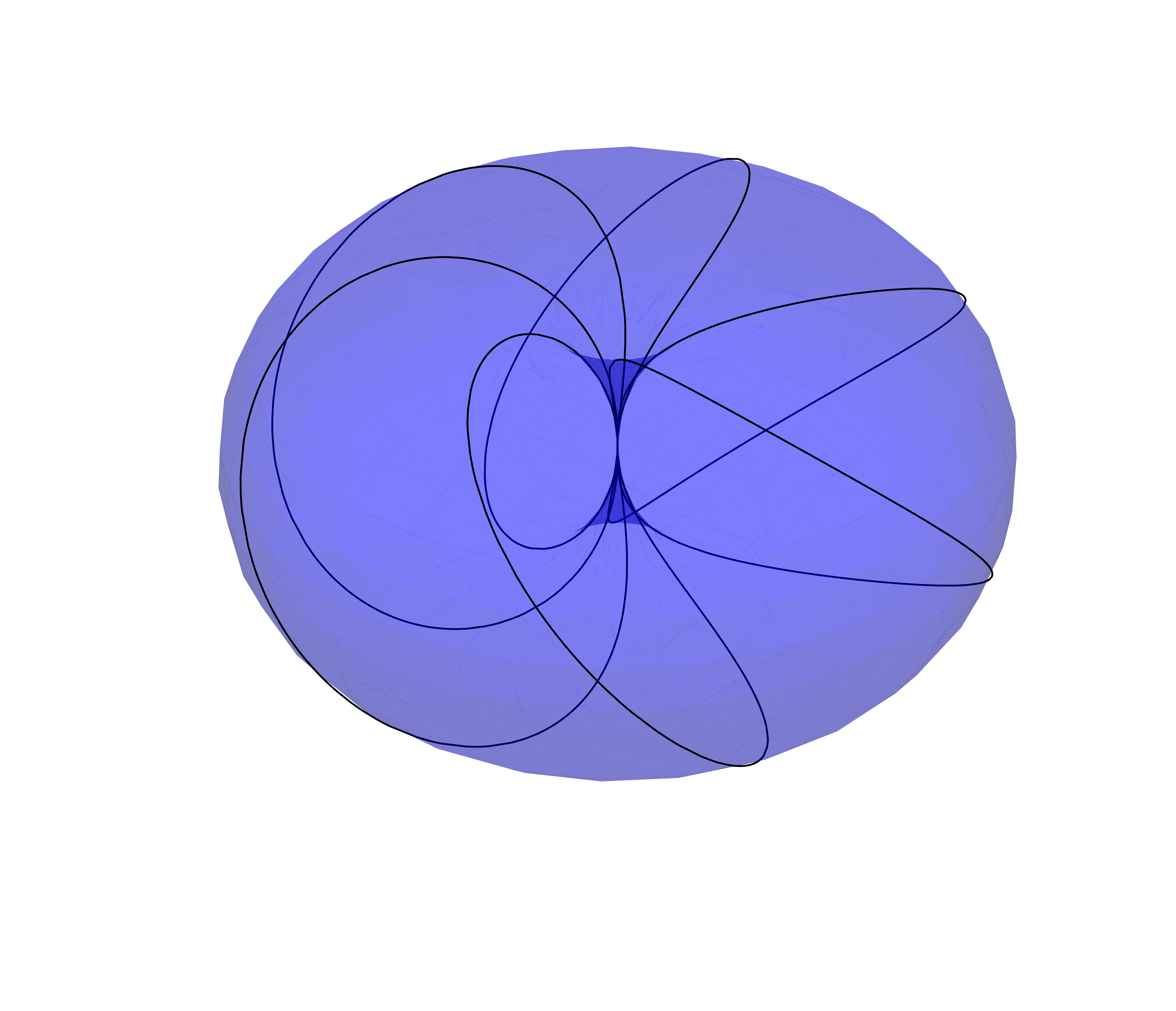}
		\caption{}
		\label{f:horn-torus-trajectories}
		\end{subfigure}
		\qquad \quad
		\begin{subfigure}[t]{3cm}
		\centering
		\includegraphics[width= 3cm]{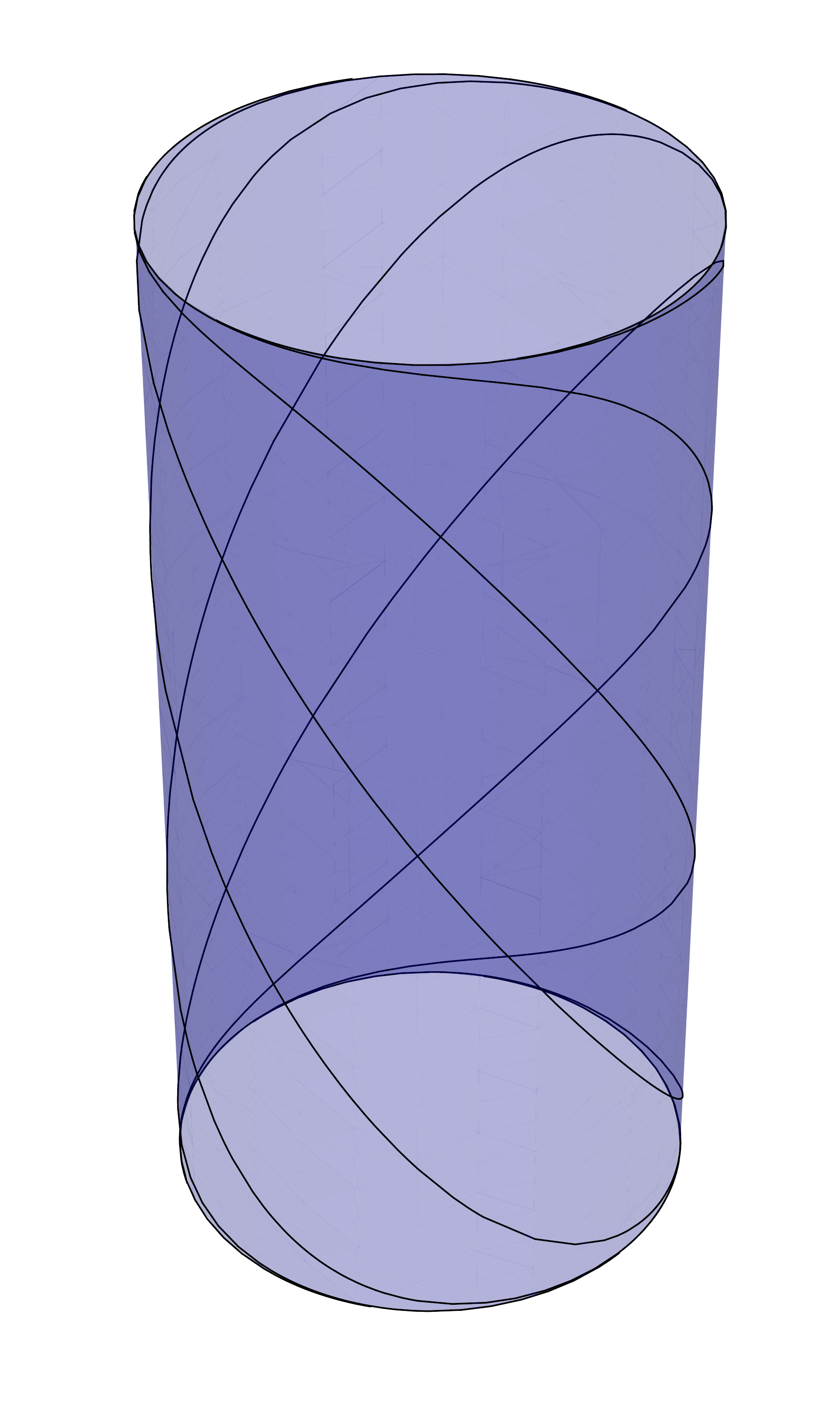}
		\caption{}
		\label{f:theta-phi-integration-cylinder-horn-torus}
		\end{subfigure}
	\caption{\footnotesize Six trajectories on a punctured horn torus (with $s = 1, m = -1$ and $\la = k = 1$) displayed in two embeddings [(a) Eq. (\ref{e:horn-torus-embedding}) and (b) Eq. (\ref{e:horntorus-cylinder-embedding}) with $R = 1.5$] passing through the points $\tht_e(0)= 0$ and $\phi_e(0) = 0, \pi/3, 2\pi/3, \pi, 4\pi/3, 5\pi/3$ extended indefinitely forward and backward in time. Trajectories emerge from the center (at $t = -\infty$) and approach the attractor at the center as $t \to \infty$ showing that the phase space volume cannot be preserved. In (b), the top and bottom rims of the cylinder correspond to the center of the horn torus.}
	\label{f:theta-phi-dynamics-3D-horn-torus}
\end{figure}

To describe the trajectories on a horn torus $H_{m s}$ we use the coordinates $\tht = \arctan(L_2/L_1)$ and $\phi = \arctan(S_2/S_1)$ in terms of which the equations of motion (\ref{e:theta-phi-dynamics}) simplify to
	\beq
	\dot \tht = \frac{k m \la}{2} \quad \text{and} \quad \dot \phi = \frac{k m \la s}{s+u} = \frac{2 k s \cos^2(\tht - \phi)}{m}. 
	\eeq
Notice that $\tht$ is monotonic in time: increasing/decreasing according as $\sign{km} = \pm 1$. It is convenient to pick ICs on the curve $u = u_1$ resulting in the solution
	\beq
	\tht(t) = \tht(0) + \frac{k m \la t}{2} \quad \text{and} \quad 
	\phi(t) = \phi(0) + \frac{k m \la t}{2} + 
	\arctan\left(\frac{ \tanh\left(\frac{t}{2\tau}\right)}{k \tau m \la}\right).
	\label{e:theta-phi-dynamics-horn-torus}
	\eeq
Though $\tht$ and $\phi$ are both ill-defined at the center of the horn torus ($L_{1,2} = S_{1,2} = 0$), we notice from (\ref{e:relation-theta-phi-u}) that the difference $\tht-\phi$ is well defined at the center: 
	\beq
	\lim_{t \to \pm \infty} (\tht(t) - \phi(t)) = \arccos \sqrt{\frac{\la m^2}{4s}} = \lim_{u \to s} (\tht - \phi).
	\eeq	
Since $\tht$ is ill-defined at the center $u=s$, it is convenient to switch to the `embedding' variables:
	\beq
	\tht_e = \frac{\pi (\phi - \tht)}{\sign{m k} \arctan \left(1/ m \la k \tau \right)}  \quad \text{and} \quad \phi_e = \phi.
	\eeq
The advantage of $\tht_e$ is that it approaches $\pm \pi/\sign{mk}$ as $t \to \pm \infty$ on any trajectory on $H_{ms}$. We may visualize the dynamics via the following embedding of the horn torus in Euclidean 3-space:
	\beq
	x = R(1+\cos \tht_e) \cos \phi_e, \quad y = R(1+\cos \tht_e) \sin \phi_e \quad \text{and} \quad z = R \sin \tht_e.
	\label{e:horn-torus-embedding}
	\eeq	
Here $R$ is the major (as well as the minor) radius of the horn torus (see Fig.~\ref{f:horn-torus-trajectories}). Alternatively, we may realize the punctured horn torus as a cylinder in three-dimensional space via the embedding
	\beq
	x = R \cos \phi_e, \quad y = R \sin \phi_e \quad \text{and} \quad z = \tht_e.
	\label{e:horntorus-cylinder-embedding}
	\eeq 
The center of the horn torus lies at $\tht_e = \pm \pi \: (\text{mod} \: 2\pi)$ with $\phi_e$ arbitrary (see Fig.~\ref{f:theta-phi-integration-cylinder-horn-torus}). As $t \to \pm \infty$ all trajectories spiral into the center of the horn torus as shown in Fig. \ref{f:theta-phi-dynamics-3D-horn-torus}. Thus, every trajectory is homoclinic, beginning and ending at the center of the horn torus.

As noted in Section \ref{s:Hill-region-Morse-theory}, horn tori arise only at the saddle points of the Hamiltonian $H = k^2 E_{\rm sad}$. Thus, they are analogs of the figure-8 shaped separatrix at energy $g a^4$ familiar from particle motion in the one-dimensional potential $V(x) = g(x^2 - a^2)^2$. For fixed $\mathfrak{c}, m$ with $2 \mathfrak{c} - m^2 > 0$, and $E = E_{\rm sad}$, $h$ can take a range of values from $h_{\rm min}$ to $h_{\rm max}$. There is a critical value $h_{\rm sad}$  in this range at which the common level set is a horn torus. It is flanked by 2-tori on either side. Thus, horn tori separate two families of toroidal level sets with the real half-period $\om_R$ of the $\wp$-function diverging as $h \to h_{\rm sad}^{\pm}$. 

\subsubsection{Flow on $\cal H$ is not Hamiltonian}

 The equations of motion on $\cal H$		
	\beq
	\dot s = \dot m = 0, \quad 
	\dot \tht = \half k m \la \quad \text{and} \quad 
	\dot \phi = \frac{k m \la s}{s + u} = \frac{2 k s \cos^2(\tht - \phi)}{m},
	\label{e:EOM-horn-tori}
	\eeq
 do not follow from any Hamiltonian and Poisson brackets on $\cal H$. This is because time-evolution does not satisfy the Liouville property of preserving phase volume: every initial condition is attracted to the center of a horn torus. Said differently, the flow can map a subset $I_0$ of $\cal H$ into a proper subset $I_t \subsetneq I_0$. To show this, it suffices to consider the dynamics on each $H_{ms}$ separately since the dynamics preserves individual punctured horn tori. Thus, consider the `upper cylinder' subset of $H_{ms}$: $I_0 = \{ (\phi_e, \tht_e)| \: \tht_e \geq \tht_0 \; \text{for some} \: -\pi < \tht_0 < \pi \}$. Then
 	\beq
	I_t = \left\{(\phi_e, \tht_e)| \tht_e > \tht_0 - \frac{\pi(\tht(t)-\phi(t))}{\sign{km} \arctan(1/k \tau m \la)} \right\}
	\eeq
is its image under evolution to time $t$. Since $\tht_e$ is monotonic in time, we observe that for $km > 0$, $I_t$ form a 1-parameter family of subsets with decreasing volume (relative to any reasonable volume measure on $H_{ms}$) while $\text{vol}(I_{t})$ grows if $km < 0$. Thus, the Liouville theorem would be violated if the dynamics on $H_{ms}$ or $\cal H$ were Hamiltonian. 

Interestingly, time evolution on $\cal H$ may be realized as a gradient flow. As before, we focus on the dynamics on each $H_{ms}$ separately. Since $W = -\sign{km} \tht$ is monotonically decreasing in time (\ref{e:EOM-horn-tori}), we choose it as the potential function for the gradient flow
	\beq
	\dot \xi^i = (\dot{\phi}, \dot \tht ) = V^i(\xi) = -g^{ij} \dd{W}{\xi^j} \quad
	\text{where} \quad  
	V^\phi = \frac{2 k s \cos^2(\tht - \phi)}{m} \quad \text{and} \quad
	V^\tht = \frac{k m \la}{2}.
	\eeq
	The inverse-metric on $H_{ms}$ that leads to this gradient flow must be of the form
	\beq
	g^{ij} = \sign{k m}\colvec{2}{\Upsilon & \dot{\phi}}{\dot{\phi} & \dot{\tht}}. 
	\eeq
Here $\Upsilon$ is an arbitrary function on $H_{ms}$ which we may choose so that the metric is, for simplicity, Riemannian (positive definite). This is ensured if
	\beq
	\det{g^{-1}}  > 0  \quad \Leftrightarrow \quad \Upsilon \dot{\tht} > \dot{\phi}^2 \quad \text{and} \quad \tr{g^{-1}} > 0 \quad \Leftrightarrow \quad \sign{k m} (\Upsilon + \dot{\tht} ) > 0.
	\eeq
The second condition is implied by the first, so a simple choice that ensures a Riemannian metric is $\Upsilon = (\dot{\phi}^2 / \dot{\tht}) + \sign{km}\: \epsilon$, for any $\epsilon > 0$. It might come as a surprise that this gradient flow admits homoclinic orbits beginning and ending at the center.
Such orbits are typically forbidden in gradient flows. Our horn tori evade this `no-go theorem' since the potential $W \propto \tht$ is not defined at the centers of horn tori.

\subsection{Dynamics on the union $\cal T$ of toroidal level sets}
\label{s:toroidal-level-sets}

For generic values of $\mathfrak{c}, m, s$ and $h$, i.e., for which the discriminant $\D \ne 0$ (\ref{e:discriminant}), the common level sets are 2-tori as shown in Section \ref{s:canonical-vector-fields-topology} and Section \ref{s:common-level-set-conserved-qtys}. The union  $\cal T$ of these 2-tori may be viewed as the state space of a self-contained dynamical system. Here, we express $\cal T$ as a torus bundle over a space ${\cal Q}_{\cal T}$ of conserved quantities, and find a convenient set of local coordinates on it along with their Poisson brackets implied by (\ref{e: PB-SL}). We use this Poisson structure and the time evolution of $u$ in terms of the $\wp$ function (\ref{e:u-wp-function}) to find a family of action-angle variables on $\cal T$. Finally, we  show that these action-angle variables degenerate to those on the union $\cal C$ of circular level sets when the tori degenerate to circles.

\subsubsection{Union of toroidal level sets}

Let us denote by ${\cal Q}_{\cal T}$, the subset $\D(\mathfrak{c}, m, s, h) \neq 0$ of the space $\cal Q$ of conserved quantities for which the common level sets are 2-tori. On ${\cal Q}_{\cal T}$ the cubic $\chi(u)$ (\ref{e:cubic-equation-S3}) is positive between two adjacent simple zeros $u_{\rm min}$ and $ u_{\rm max}$ and the common level set $M^{s h}_{\mathfrak{c} m}$ is a torus. Thus, on ${\cal Q}_{\cal T}$ the cubic takes the form $\chi(u) = (u - u_{\rm min})(u_{\rm max} -u)(u_3 - u)$ with $-s \leq u_{\rm min} < u_{\rm max} \leq s$ and $u_{\rm max} < u_3$. In this case, when $\chi(u)$ is written in Weierstrass normal form using $u = a v + b$, the   invariants $g_2$ and $g_3$ are real and the discriminant of the cubic is nonzero. It follows that the half periods $\om_R$ and $\om_I$ of Section \ref{s:Reduction-2D} are respectively real and purely imaginary. We designate the union of these tori ${\cal T} \subset M^6_{S \text{-} L}$ and the corresponding union for fixed $\mathfrak{c}$ and $m$, ${\cal T}^4_{\mathfrak{c} m}$. Here, ${\cal T}$ may be visualised as a torus bundle over ${\cal Q}_{\cal T}$. While $\tht$ and $\phi$ furnish global coordinates on the torus $M^{s h}_{\mathfrak{c} m}$, it is more convenient, when formulating the dynamics, to work with the local coordinates $(u, \tht)$ where $\cos (\tht - \phi) = (h + m u)/r \rho$. An advantage of $u$ is that unlike $\phi$, it commutes with $h$. However, since the cosine is a 2:1 function on $[0,2\pi]$, we need two patches $U_\pm$ with local coordinates $(u_\pm, \tht)$ to cover the torus with $u_{\rm min} \leq u_\pm \leq u_{\rm max}$ and $0 \leq \tht \leq 2 \pi$. In the $U_\pm$ patches, the formula for $\phi$ is
	\beq
	\phi = \tht \pm \arccos \left( \frac{h + m u}{r \rho}\right)_{[0, \pi]},
	\eeq
where the $\arccos$ function is defined to take values between $0$ and $\pi$. Whenever $u$ reaches either $u_{\rm min}$ or $u_{\rm max}$, the trajectory crosses over from one patch to the other.

\begin{figure}[h]
	\centering
		\begin{subfigure}[b]{5cm}
		\includegraphics[width=5cm]{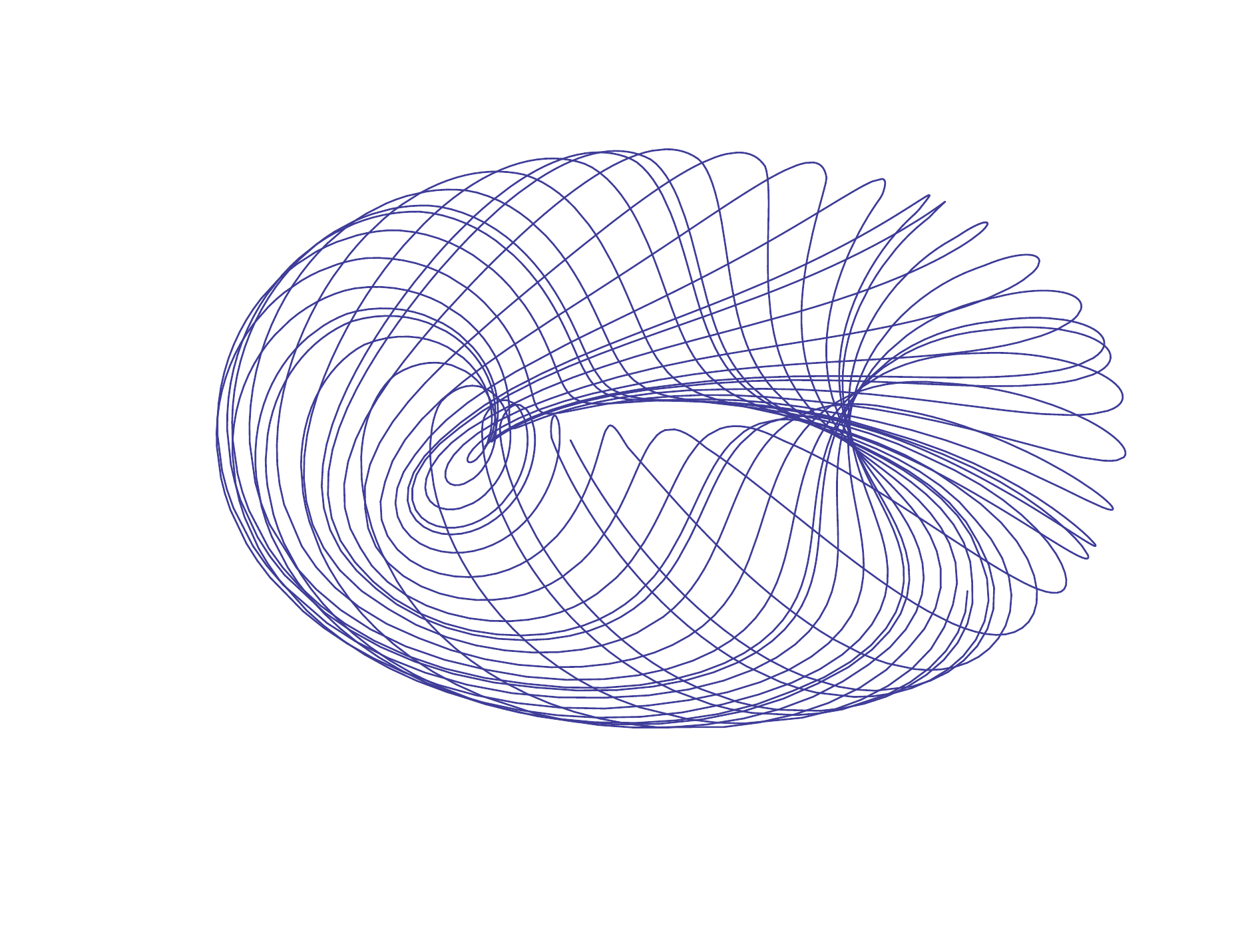}
		\end{subfigure}
	\caption{\footnotesize Trajectory on an invariant torus for the parameters $k = \la = 1, \mathfrak{c} = 3, h= 1, m = -1, s=1$ and $R = 2$ for $0 < t < 75 \om_R$ ($\om_R \approx 1.41$ is the real half-period of $u$ (\ref{e:u-wp-function})) displayed via the embedding $x = (R + \varrho \cos\tht_e)\cos \phi_e$, $y = (R + \varrho \cos \tht_e) \sin \phi_e$ and $z = \varrho \sin \tht_e$.  The poloidal and toroidal angles are $\tht_e =  \arcsin \left( (u - \bar{u})/\varrho  \right)$ and $\phi_e = \phi$ with $\bar{u} = (u_{\rm min} + u_{\rm max})/2$ and $\varrho = (u_{\rm max} - u_{\rm min})/2$. Unlike the angle variables $\tht^1$ and $\tht^2$ (\ref{e:action-angle-variables-torus}),  which are periodic on account of their linearity in time, neither $\tht_e$ nor $\phi_e$ is periodic.}
	\label{f:torus-plot-theta-phi}
\end{figure}


\subsubsection{Poisson structure on $\cal T$}

On ${\cal T}$, we use the local coordinates $\mathfrak{c}, m, s, h, \tht$ and $u$. The Poisson structure following from the nilpotent Poisson brackets (\ref{e: PB-SL}) is degenerate with the Casimirs $\mathfrak{c}$ and $m$ generating the center. The Poisson brackets among the remaining coordinates (on ${\cal T}^4_{\mathfrak{c} m}$) are:
	\beqs
	\{ s, h \} &=& \{ h, u \} = \{ \tht, u \} = 0 , \quad \{ h, \tht \} = \ov k, \quad  
	\{ s, \tht \} = \frac{h + m u}{k s r^2} = \frac{\rho}{k s r} \cos(\tht - \phi) = -\frac{\dot{\tht}}{k^2 s}, \cr
	\{ s, u_{\mp} \} &=& \mp \frac{\la}{k s} \sqrt{r^2 \rho^2 - (h + m u)^2} = \mp \frac{ \sqrt{2 \la k^2 \chi(u)}}{k^2 s} = -\frac{r \rho \la}{k s} \sin(\tht - \phi) = -\frac{\dot{u}}{k^2 s}.
	\label{e:PB-on-union-of-tori}
	\eeqs	
All the Poisson brackets other than $\{ s, u \}$ have a common expression on both patches $U_{\pm}$. Here $r^2 = 2 \mathfrak{c} - m^2 - 2u/\la$ and $\rho^2 = s^2 - u^2$. 

\subsubsection{Action-angle variables on $\cal T$}

We seek angle-action variables $(\tht^1,\tht^2, I_1,I_2)$ on ${\cal T}^4_{\mathfrak{c} m}$ satisfying canonical Poisson brackets
	\beq
	\{ \tht^i, \tht^j \} = \{ I_i , I_j \} = 0 \quad \text{and} \quad \{ \tht^i, I_j \} = \delta^i_j.
	\label{e:canonical-act-ang-PB}
	\eeq
The action variables $I_1$ and $I_2$ must be conserved and therefore functions of $s$ and $h$ alone, while the angles $\tht^1$ and $\tht^2$ must evolve linearly in time: $\dot \tht^j = \Om_j(s,h)$. Here we suppress the parametric dependence of $\tht^i$ and $I_j$ on the Casimirs $\mathfrak{c}$ and $m$ which specify the symplectic leaf. In what follows, we use the $\wp$-function solution (\ref{e:u-wp-function}) along with the requirement of canonical Poisson brackets to find a family of action-angle variables. Despite some long expressions in the intermediate steps, the final formulae  (\ref{e:action-angle-variables-torus}) for $(\tht^i, I_j)$ are relatively compact. Though we work here with the nilpotent Poisson structure (\ref{e:PB-on-union-of-tori}), it should be possible to generalize the resulting action-angle variables to the other members of the Poisson pencil (\ref{e:Poisson-pencil}).  

\vspace{.25cm}

{\fl \bf Determination of $\tht^1$ and $I_1$:} The evolution of $u$ (\ref{e:u-wp-function}) gives us one candidate for an angle variable evolving linearly in time 
	\beq
	\tht^1 =  k\left( \wp^{-1}\left(\frac{k^2 \la}{2} \left(u - \frac{\mathfrak{c} \la}{3} \right); g_2, g_3 \right) - \alpha(\mathfrak{c}, m, s, h) \right)  = k( t + t_0).
	\label{e:angle-variable-theta-1}
	\eeq
The factor of $k$ is chosen to make $\tht^1$ dimensionless. Here, $g_2$ and $g_3$ (\ref{e:Weierstrass-invariants}) are functions of the conserved quantities. From the definition of $\tht^1$, it follows that the frequency $\Omega_1 = k$. Choosing $\al$ to be the imaginary half-period $\om_I$ of the $\wp$-function in (\ref{e:u-wp-function}) ensures that $\tht^1$ is real. An action variable conjugate to $\tht^1$ is 
	\beq
	I_1(s, h) = \frac{k s^2}{2}  + f(h),
	\label{e:I1-action-avariable}
	\eeq
where $f'(h) \neq 0$ is an arbitrary function of $h$ (and possibly $\mathfrak{c}$ and $m$) to be fixed later. Upto the function $f$, $I_1$ is proportional to the Hamiltonian (\ref{e:Hamiltonian-s}). Eq. (\ref{e:I1-action-avariable}) is obtained by requiring
	\beq
	\{\tht^1, I_1 \} = \dd{\tht^1}{u} \dd{I_1}{s} \{ u, s \} = k\dd{(\wp^{-1}(v) - \al)}{v} \dd{v}{u}\dd{I_1}{s} \{ u, s \} = \frac{k}{\dot{u}} \dd{I_1}{s} \frac{\dot{u}}{k^2 s} = 1.
	\eeq
Here, $v = (u-b)/a$ (see Section \ref{s:Reduction-2D}) and we used the relation
	\beq
	\dd{\wp^{-1}(v; g2, g3)}{v} = \frac{1}{\dot{v}} = \frac{a}{\dot{u}}.
	\label{e:derivative-wp-inverse}
	\eeq 
For future reference we also note that as a consequence, $\pdr \tht^1/ \pdr u = k/ \dot{u}$. This derivative diverges at $u_{\rm min}$ and $u_{\rm max}$, which are the roots of $\chi$.

\vspace{.25cm}

{\fl \bf Determination of $\tht^2$ and $I_2$:} To identify the remaining action-angle variables $I_2(s,h)$ and $\tht^2(u,\tht,s,h)$ we first consider the constraints coming from the requirement that their Poisson brackets be canonical. While $\{ I_1, I_2 \} = 0$ is automatic, $\{ \tht^1 , I_2 \} = 0$ implies that $I_2(s,h)$ must be independent of $s$:
	\beq
	0 = \{ \tht^1 , I_2 \} = \dd{I_2}{s} \{ \tht^1, s \} + \dd{I_2}{h} \cancel{\{ \tht^1, h \}} \quad \imply \quad \dd{I_2}{s} = 0. 
	\eeq
The remaining Poisson brackets help to constrain $\tht^2$. For instance, $\{ \tht^2, I_2(h) \} = 1$ forces $\tht^2$ to be a linear function of $\tht$:
	\beq
	\{ \tht^2, I_2(h) \} = \dd{\tht^2}{\tht} I_2'(h) \{ \tht, h \} = -\dd{\tht^2}{\tht} \frac{I_2'(h)}{k} = 1  \quad \imply \quad \tht^2 =  -\frac{k}{I_2'(h)} \tht + g(u,s,h).
	\label{e:tht2-I2-PB}
	\eeq
Here $g$ is an arbitrary function which we will now try to determine. Next, $\{ \tht^2, I_1 \} = 0$ implies that $\tht^2$ evolves linearly in time: 
	\beqs	
	\{ \tht^2, I_1 \} = \dd{\tht^2}{u} \{ u, I_1 \} + \dd{\tht^2}{\tht} \{ \tht, I_1 \} = 0 \quad &\imply& \quad   \dot{\tht^2} =   \dd{\tht^2}{u} \dot{u} + \dd{\tht^2}{\tht} \dot{\tht}  =  f'(h) \dd{\tht^2}{\tht} \equiv \Omega_2 \cr
	&\imply& \quad \tht^2 = \frac{\Omega_2}{f'(h)} \tht + g(u,s,h).
	\label{e:linear-time-theta-2}
	\eeqs	
Comparing (\ref{e:tht2-I2-PB}) and (\ref{e:linear-time-theta-2}), it follows that $ \Omega_2 = -k f'(h)/I_2'(h)$ is independent of $s$. We may use (\ref{e:linear-time-theta-2}) to reduce the determination of the dependence of $\tht^2$ on $u$ to quadratures: 
	\beq
	\dot{\tht^2} = \Omega_2 = \frac{\Omega_2}{f'(h)} \dot{\tht} + \dd{g(u,s,h)}{u} \dot{u}.
	\label{e:time-evolution-tht-2}
	\eeq
Using (\ref{e:theta-phi-dynamics}) and (\ref{e:EOM-u}) we get
	\beq
	\dd{\tht^2}{u} = \dd{g}{u} 
	= \pm\Om_2  \frac{1+ \frac{k}{f'(h)}\left(\frac{h + m u}{2\mathfrak{c} - m^2 - {2 u}/{\la}}\right)}{\sqrt{2 \la k^2 \chi(u)}}.
	\label{e:u-dependence-tht-2}
	\eeq
Integrating,
\small
	\beq
	\frac{g(u, s, h)}{\Om_2} = \frac{ \pm 1}{\sqrt{2 \la k^2}} \left[ \left(1 - \frac{k m \la}{2 f'(h)} \right) \int_{u_{\rm min}}^{u} \frac{du'}{\sqrt{\chi(u')}} - \frac{k m \la}{2 f'(h)}\left( \frac{h}{m} + u_0 \right) \int_{u_{\rm min}}^u \frac{du'}{\left( u' - u_0 \right) \sqrt{\chi(u')}} \right]  + \tl{g}(s, h), \qquad
	\label{e:u-dependence-tht-2-integral}
	\eeq
\normalsize
where $u_0 = \mathfrak{c}/ \la - m^2 \la/ 2$. Recognizing these as incomplete elliptic integrals of the first and third kinds ($F$ and $\Pi$), we get (see Section 3.131, Eq. (3) and Section 3.137, Eq. (3) of \cite{G-R})
\small
	\beq
	\frac{g}{\Om_2} = \pm \sqrt{\frac{2}{\la k^2}}\frac{k m \la}{2 f'(h)} \left[ \left( \frac{2 f'(h)}{k m \la} - 1 \right) \frac{F(\g , q)}{\sqrt{u_3 - u_{\rm max}}} + \left( \frac{h}{m} + u_0 \right) \frac{\Pi \left( \gamma, \frac{u_{\rm max} - u_{\rm min}}{u_0 - u_{\rm min}}, q \right)}{(u_0 - u_{\rm min}) \sqrt{u _3 - u_{\rm min}} }  \right] + \tl{g}(s, h). 
	\label{e:angle-variable-elliptic-integral}
	\eeq
\normalsize
Here, $\tl{g}(s,h)$ is an integration constant, $u \in [u_{\rm min},u_{\rm max}]$ where $-s \leq u_{\rm min} < u_{\rm max}  < u_3$ (which are functions of $\mathfrak{c}, m, s$ and $h$) are the roots of the cubic $\chi(u)$. Moreover, the amplitude and elliptic modulus are
	\beq
	\gamma = \arcsin \sqrt{\frac{u - u_{\rm min}}{u_{\rm max} - u_{\rm min}}} \quad \text{and} \quad 
	q = \sqrt{\frac{u_{\rm max} - u_{\rm min}}{u_3 - u_{\rm min}}}.
	\eeq
To find the $s$ dependence of $\tht^2$, we notice that the last Poisson bracket $\{ \tht^1, \tht^2 \} = 0$ gives the following relation among derivatives of $\tht^2$:
	\beqs
	\{ \tht^1 , \tht^2 \} &=& \dd{\tht^2}{u} \{ \tht^1 , u \} + \dd{\tht^2}{\tht} \{ \tht^1 , \tht \} + \dd{\tht^2}{s} \{ \tht^1 , s \}  + \dd{\tht^2}{h} \cancel{\{ \tht^1 , h \}} = 0 \cr
	\imply  && \dd{\tht^2}{\tht} \dd{\tht^1}{s} \{ s, \tht \}  + \dd{\tht^2}{\tht} \dd{\tht^1}{h} \{ h , \tht \}  + \left( \dd{\tht^2}{u} \dd{\tht^1}{s} - \dd{\tht^2}{s} \dd{\tht^1}{u} \right) \{ s, u \} = 0.  
	\label{e:-tht1-tht2-PB-torus}
	\eeqs
Using the known formulae for the partial derivatives (\ref{e:angle-variable-theta-1}, \ref{e:derivative-wp-inverse}, \ref{e:linear-time-theta-2}, \ref{e:time-evolution-tht-2})
	\beq
	\frac{\partial \tht^2}{\partial \tht} = \frac{\Om_2}{f'(h)}, \quad 
	\dd{\tht^2}{u} = \frac{\Om_2}{\dot u}\left( 1 - \frac{\dot \tht}{f'(h)}\right), \quad
	\dd{\tht^1}{u} = \frac{k}{\dot u} \quad \text{and} \quad
	\pdr_{s,h} \tht^1 = k \pdr_{s, h} (\wp^{-1} - \om_I), 
	\eeq
we find the $s$ dependence of $\tht^2$ from (\ref{e:-tht1-tht2-PB-torus}):
	\beq
	\dd{\tht^2}{s} =\Om_2(h) \left[ \pdr_{s} (\wp^{-1} - \om_I) - \frac{k s}{f'(h)} \pdr_{h} (\wp^{-1} - \om_I) \right].
	\label{e:s-dependence-theta-2}
	\eeq
In effect, we have two expressions ((\ref{e:angle-variable-elliptic-integral}) and (\ref{e:s-dependence-theta-2})) for $\pdr_s\tht^2$. We exploit them to reduce the determination of the $s$ dependence of $\tht^2$ to quadrature. Comparing $\pdr_s$(\ref{e:angle-variable-elliptic-integral}) with (\ref{e:s-dependence-theta-2}) gives 
\footnotesize
	\beqs
	\pdr_s {\tl g} &=& \frac{\pdr}{\pdr s} \left[ \wp^{-1} - \om_I \mp \sqrt{\frac{2}{\la k^2}}\frac{k m \la}{2 f'(h)} \left\{ \left( \frac{2 f'(h)}{k m \la} - 1 \right) \frac{F(\g , q)}{\sqrt{u_3 - u_{\rm max}}} + \left( \frac{h}{m} + u_0 \right) \frac{\Pi \left( \gamma, \frac{u_{\rm max} - u_{\rm min}}{u_0 - u_{\rm min}}, q \right)}{(u_0 - u_{\rm min}) \sqrt{u _3 - u_{\rm min}} }  \right\} \right] \cr
	&& - \frac{k s}{f'(h)} \pdr_h(\wp^{-1} - \om_I).
	\eeqs
\normalsize
Thus
\footnotesize
	\beqs
	\tl{g}(s,h) &=& \wp^{-1} - \om_I \mp \sqrt{\frac{2}{\la k^2}}\frac{k m \la}{2 f'(h)} \left[ \left( \frac{2 f'(h)}{k m \la} - 1 \right) \frac{F(\g , q)}{\sqrt{u_3 - u_{\rm max}}} + \left( \frac{h}{m} + u_0 \right) \frac{\Pi \left( \gamma, \frac{u_{\rm max} - u_{\rm min}}{u_0 - u_{\rm min}}, q \right)}{(u_0 - u_{\rm min}) \sqrt{u _3 - u_{\rm min}} }  \right] \cr
	&& - \int^s_{\infty} \frac{k s'}{f'(h)} \pdr_h(\wp^{-1} - \om_I) \: ds' + \eta(h).
	\label{e:s-h-dependence-of -tht-2}
	\eeqs
\normalsize
Here $\eta(h)$ is an arbitrary `constant' of integration. Now, using (\ref{e:s-h-dependence-of -tht-2}) in (\ref{e:angle-variable-elliptic-integral}) results in some pleasant cancellations leading to a relatively simple formula for $g$:
	\beq
	\frac{g(u,s,h)}{\Om_2} = \wp^{-1} - \om_I - \frac{k}{f'(h)} \int_{\infty}^s  s' \:\pdr_h (\wp^{-1} - \om_I) \: ds' + \eta(h).
	\label{e:g-u-s-dependence-of-tht-2}
	\eeq
This determines the angle variable $\tht^2(\tht, u, s, h) = \Om_2 \tht/ f'(h) + g(u, s, h)$. It is noteworthy that $ \wp^{-1} - \om_I$ is simply $\tht^1/k$. The integral over $s'$ is from $\infty$ since, for sufficiently large $s$, $\D$ (\ref{e:discriminant}) is always positive so that $M_{\mathfrak{c} m}^{s h}$ is a torus. However, we must take $s > s_{\rm min}$, which is the value at which $\D$ vanishes and the torus $M_{\mathfrak{c} m}^{s h}$ shrinks to a circle.

\vspace{.25cm}
\footnotesize 

{\fl \bf Remark:} Consistency requires that the RHS of (\ref{e:s-h-dependence-of -tht-2}) be independent of $u$, which enters through $\wp^{-1}$ and $\gamma$. We verify this by showing that $\pdr_u$(\ref{e:g-u-s-dependence-of-tht-2}) agrees with (\ref{e:u-dependence-tht-2}). In fact, from (\ref{e:g-u-s-dependence-of-tht-2}) and using $\dot{u} = \pm \sqrt{2 \la k^2 \chi(u)}$ and (\ref{e:cubic-equation-S3}),
	\beq
	\ov{\Om_2} \dd{g}{u} =  \ov{\dot{u}} - \frac{k}{f'(h)} \int_{\infty}^s s' \: \pdr_h(1/\dot{u}) \: ds' =  \ov{\dot{u}} \mp \frac{k}{f'(h)} \int_{\infty}^s \frac{\la (h + m u)}{ 2 \sqrt{2 \la k^2} (\chi(u))^{3/2}} s' \: ds'
	= \pm \frac{1+ \frac{k}{f'(h)}\left(\frac{h + m u}{2\mathfrak{c} - m^2 - \frac{2 u}{\la}}\right)}{\sqrt{2 \la k^2 \chi(u)}},
	 \eeq
which agrees with (\ref{e:u-dependence-tht-2}). As $\chi$ (\ref{e:cubic-equation-S3}) is a quadratic function of $s'$, the integrand behaves as $1/s'^2$ for large $s'$, so that the lower limit does not contribute.

\vspace{.25cm}
\normalsize

{\fl\bf Summary:} Thus, aside from the Casimirs $\mathfrak{c}$ and $m$, the action-angle variables on the union of toroidal level sets ${\cal T}$ are given by the following functions of $s$, $h$, $u$ and $\tht$:
	\beqs
	I_1 &=& \frac{k s^2}{2} + f_{\mathfrak{c} m}(h), \quad 
	\tht^1= k \left( \wp^{-1}\left(\frac{k^2 \la}{2} \left(u - \frac{\mathfrak{c} \la}{3}\right); g_2, g_3  \right) - \om_I(\mathfrak{c}, m, s, h) \right), \cr
	I_2 &=& I_2(h; \mathfrak{c}, m)  \quad \text{and} \quad
	\tht^2 = \Omega_2 \left( \frac{\tht}{f_{\mathfrak{c} m}'(h)} +  \frac{\tht^1}{k} - \ov{f_{\mathfrak{c} m}'(h)} \int^s_{\infty} s' \: \pdr_h \tht^1 \: ds' + \eta(h) \right).\quad
	\label{e:action-angle-variables-torus}
	\eeqs
We have verified by explicit calculation that these variables are canonically conjugate. As a function of $u \in [u_{\rm min}, u_{\rm max}]$, $\tht^1$   increases from zero to $k \om_R$ (\ref{e:u-wp-function}). As noted, $\tht^2$ depends linearly on $\tht$, but finding its dependence on $u,s$ and $h$ requires the evaluation of the integral over $s'$ in (\ref{e:action-angle-variables-torus}). We have not been able to do this analytically but could evaluate it numerically for given $\mathfrak{c}$ and $m$. Here, $f, \eta$ and $I_2$ are arbitrary functions of $h$, with $f'$ and $I_2'$ nonzero and the frequency $\Om_2 = -k f'(h)/I_2'(h)$. A {\it simple choice} is to take 
	\beq
	f(h) = -I_2(h) = k h 
	\quad \text{and} \quad
	\eta(h)= 0. 
	\label{e:simple-choice-f-I2}
	\eeq
	For this choice, the Hamiltonian (\ref{e:Hamiltonian-s}) acquires a simple form in terms of the action variables
	\beq
	H = k( I_1 + I_2) + k^2 \left(\mathfrak{c} + \ov{2 \la^2} \right).
	\eeq
The corresponding frequencies $\Om_j = \pdr H/ \pdr I_j$ are then both equal to $k$. Though the frequencies are equal, the periodic coordinates $\tht^1$ and $\tht^2$ generally have different and incommensurate ranges, so that the trajectories are quasi-periodic (see Fig.~\ref{f:torus-plot-theta-phi}). While we do not have a simple formula for the range of $\tht^2$, that of $\tht^1$ is $2 k \om_R$ (twice its increment as $u$ goes from $u_{\rm min}$ to $u_{\rm max}$, see Eq. (\ref{e:u-wp-function})), which depends on the symplectic leaf and invariant torus via the four conserved quantities.

\vspace{.25cm}

{\fl \bf Relation to action-angle variables on the circular submanifold:} Finally, we show how the action-angle variables obtained above degenerate to those on the circular submanifold $\cal C$ of Section \ref{s:circular-level-sets}, where the elliptic function solutions reduce to trigonometric functions with the imaginary half-period $\om_I$ diverging. For given $\mathfrak{c}, m$ and $h$, we must let $s \to s_{\rm min}$ to reach the circular submanifold. On $\cal C$, the simple zeros of $\chi$, $u_{\rm min}$ and $u_{\rm max}$ coalesce at a double zero so that $u$ becomes a constant. Thus, the angle variable $\tht^1$ (\ref{e:action-angle-variables-torus}) ceases to be dynamical. In the same limit, from (\ref{e:action-angle-variables-torus}), the surviving angle variable $\tht^2$ becomes a linear function of $\tht$ with constant coefficients. Moreover, for the simple choices of Eq. (\ref{e:simple-choice-f-I2}), we get $I_2 = - k h$ and $\tht^2 = \tht$ upto an additive constant. Pleasantly, these action-angle variables are seen to agree with those obtained earlier on $\cal C$ (\ref{e:circular-action-variable}).

\chapter[Quantum Rajeev-Ranken model and anharmonic oscillator]{Quantum Rajeev-Ranken model as an anharmonic oscillator}
\chaptermark{Quantum RR model as an anharmonic oscillator}
\label{chapter:Quantum-Rajeev-Ranken-model-and-anharmonic-oscillator}

In this Chapter, which is based on \cite{G-V-6}, we discuss some aspects of the quantum version of the Rajeev-Ranken model. We begin with Rajeev and Ranken's mechanical interpretation of the model in terms of a charged particle moving in an electromagnetic field and its quantization. We find an error in their calculation of the effective potential seen by the particle which unfortunately affect their results on the spectrum and strong coupling dispersion relation. We derive the corrected effective potential in Section \ref{s:Electromagnetic-Hamiltonian} and to be doubly sure we also take a complementary approach by interpreting the RR model as a quartic oscillator. Using this new mechanical interpretation we canonically quantize the model and separate variables in the Schr\"odinger equation. Its radial equation is shown to be an ODE of type $[0,1,1_6]$ (in Ince's classification, see Appendix \ref{E:Singularities-of-second-order-ordinary-differential-equations}, Section \ref{s:Ince-classification}), which may be regarded as a generalization of the Lam\'e equation. We analyze a weak and a novel strong coupling limit of the radial equation to obtain dispersion relations for the corresponding quantized screw-type waves. In another direction, we interpret the EOM of the RR model as Euler equations for a step-3 nilpotent algebra and exploit our canonical quantization to find a unitary representation of this algebra.

\section[Electromagnetic Hamiltonian]{Electromagnetic interpretation of the RR model \sectionmark{Electromagnetic interpretation of the RR model}}
\sectionmark{Electromagnetic interpretation of the RR model}
\label{s:Electromagnetic-Hamiltonian}

Before interpreting the RR model as an anharmonic oscillator, we revisit the mechanical interpretation given by Rajeev and Ranken in terms of a charged particle moving in a static electromagnetic field. Here, we implement this general idea and derive the classical and quantum equations of motions. In the process, we notice certain errors in the analysis of Rajeev and Ranken, which affect their results on the spectrum and dispersion relations. This unfortunately cast doubts on their results.

The Hamiltonian of the RR model in Darboux coordinates $\{ R_a, k P_b \} = \delta_{a b}$ for $a,b = 1,2,3$ is:
	\beq
	 \frac{H}{k^2} = \sum_{a=1}^3 \frac{P_a^2}{2} + \frac{\la m}{2} \left( R_1 P_2 - R_2 P_1 \right) + \frac{\la^2}{8} \left( R_1^2 + R_2^2 \right) \left[ R_1^2 + R_2^2 + m^2 - \frac{4}{\la} \left( P_3 - \frac{1}{\la} \right) \right] + \frac{m^2}{2}.
	\label{e:Hamiltonian-mech-Darboux}
	\eeq
Suppose the Cartesian position and momentum coordinates of a charged particle are
	\beq
	x, y, z = R_{1,2,3} \quad \text{and} \quad
	p_{x,y,z} = k P_{1,2,3},
	\label{e:mapping-RR-EM}
	\eeq
then the Hamiltonian in (\ref{e:Hamiltonian-mech-Darboux}) can be rewritten as:
	\beq
	H = \frac{1}{2 \mu} \left[ \left( p_x - \frac{q \la m k y}{2c} \right)^2 + \left( p_y + \frac{q \la m k x}{2c} \right)^2 + \left( p_z - \frac{q \la k}{2c}(x^2 + y^2) \right)^2 \right] + \frac{q k^2}{2}(x^2 + y^2 + m^2),
	\label{e:Hamiltonian-EM-Cartesian}
	\eeq
in units where the charged particle has mass $\mu = 1$, charge $q =1$ and the speed of light $c =1$. This describes a charged particle moving in an EM field arising from the vector and scalar potentials: 
	\beqs
	A_x &=& \frac{\la m k y}{2}, \quad 
	A_y = -\frac{\la m k x}{2}, \quad 
	A_z = \frac{\la k}{2}(x^2 + y^2) \quad \text{and} \cr
	V(x,y,z) &=& \frac{k^2}{2}(x^2 + y^2 + m^2).
	\eeqs
The corresponding electromagnetic field is axisymmetric with ${\rm \bf E}$ pointing radially inward and ${\rm \bf B}$ having both azimuthal and axial components:
	\beq
	\bfE = -k^2 ( x \hat x + y \hat y) \quad \text{and} \quad 
	\bfB = \la k \left(  y \hat x - x \hat y - m \hat z \right).
	\eeq
The Hamiltonian in (\ref{e:Hamiltonian-EM-Cartesian}) along with the canonical PBs $ \{ x, p_x \} = \{ y, p_y \} = \{ z, p_z \} = 1$ gives the Newton-Lorentz equations  $\mu \ddot \bfr = q (\bfE + {\bfv}/c \times \bfB)$. In fact, using 
	\beq
	\dot x = \ov{\mu} \left( p_x - \frac{q \la m k y}{2 c} \right), \quad \dot y = \ov{\mu} \left( p_y + \frac{q \la m k x}{2 c} \right) \quad \text{and} \quad \dot z = \ov{\mu} \left( p_z - \frac{q \la k}{2 c}(x^2 + y^2) \right), 
	\eeq
we get the NL equations in component form
	\beqs
	\mu \ddot{x} &=& -q k^2 x + \frac{q}{c} \la k \left( -m \dot{y} + x \dot{z} \right), \quad
	\mu \ddot{y} = -q k^2 y + \frac{q}{c} \la k \left( m \dot{x} + y \dot{z} \right)  \quad \text{and} \cr
	\mu \ddot{z} &=& -\frac{q}{c} \la k \left( x \dot{x} + y \dot{y} \right).  \qquad
	\label{e:EOM-EM-component-form}
	\eeqs
These are seen to agree with (\ref{e:EOM-R}) in units where $\mu = q= c = 1$ upon use of (\ref{e:mapping-RR-EM}).

\subsection{Classical Hamiltonian in terms of cylindrical coordinates}
\label{s:classical-Hamiltonian-in-terms-of-cylindrical-coordinates}

The Hamiltonian (\ref{e:Hamiltonian-EM-Cartesian}) is invariant under rotation about and translation along the $z$-axis, so we make a canonical transformation to cylindrical coordinates $r = \sqrt{x^2 + y^2}, \tht = \arctan(y/x)$, and $z$ and their momenta $p_r = (x p_x + y p_y)/r $, $p_\tht = -y p_x + x p_y$ and $p_z$ satisfying the PBs
	\beq
	\{ r, p_r \} = \{ \tht, p_{\tht} \} = \{ z, p_z \} = 1.
	\label{e:canonical-PB-cylindrical}
	\eeq
The Hamiltonian corresponding to  (\ref{e:Hamiltonian-EM-Cartesian}) is \cite{R-R}
	\beq
	H = \frac{1}{2 \mu} \left[ p_r^2 +  \frac{\left( p_{\tht} - \frac{q A_\tht}{c} \right)^2}{r^2} + \left(p_z - \frac{q A_z}{c} \right)^2 \right] + q V(r),
	\label{e:Hamiltonian-EM-plane-polar}
	\eeq
where the scalar and the magnetic vector potentials are
	\beq
	V(r) = \frac{k^2 (r^2 + m^2)}{2} \quad \text{and} \quad {\bf A} = \frac{\la k}{2} (-m r \hat{\tht} + r^2 \hat{z}).
	\eeq 
The resulting electric and magnetic fields are 
	\beq
	{\bf E} = -k^2 r \hat{r} \quad \text{and} \quad {\bf B} = -\la k \left( r \hat{\tht} + m \hat{z} \right).
	\eeq
In terms of velocities the canonical conjugate momenta are 
	\beq
	p_{r} = \mu \dot r, \quad
	p_{\tht} = \mu r^2 \dot \tht + \frac{q A_{\tht}}{c} \quad \text{and} \quad
	p_z = \mu \dot z + \frac{q A_z}{c}.
	\eeq
Note that $p_{\tht} = r {\bf p} \cdot \hat \tht$ and $A_{\tht} = -(\la k m r^2)/2 = r {\bf A} \cdot \hat \tht$ are not simply the $\tht$ components. In terms of these coordinates (\ref{e:EOM-EM-component-form}) become
	\beq
	\mu \ddot{r} = \mu r \dot{\tht}^2 - q k^2 r + \frac{q}{c} ( \la k r \dot z - \la m k r \dot{\tht} ), \qquad
	\mu \ddot{z} = -\frac{q}{c} \la k r \dot{r}  \quad \text{and} \quad 
	\mu r \ddot{\tht} = \frac{q}{c} \la m k \dot{r} - 2 \mu \dot{r} \dot{\tht} .
	\eeq
	
\subsection{Quantization of the electromagnetic Hamiltonian}	

To quantize in Cartesian coordinates we represent the canonical momenta by the differential operators:
	\beq
	p_x = -i \hbar \pdr_x, \quad p_y = -i \hbar  \pdr_y \quad \text{and} \quad p_z = -i\hbar  \pdr_z,
	\eeq
satisfying the canonical commutation relations $[ x, p_x ] = [ y, p_y ] = [ z, p_z ] = i \hbar$. The Hamiltonian (\ref{e:Hamiltonian-EM-Cartesian}) becomes the operator
	\beqs
	\hat{H} &=& \frac{1}{2 \mu} \left[ \left(  -i \hbar \pdr_x - \frac{q\la mky}{2c} \right)^2 + \left(  -i \hbar \pdr_y - \frac{q\la mkx}{2c} \right)^2 + \left(  -i \hbar \pdr_z - \frac{q\la k (x^2 + y^2)}{2c} \right)^2 \right] \cr
	&&+ \frac{q k^2}{2}(x^2 + y^2 + m^2).
	\eeqs
To facilitate separation of variables, we transform to cylindrical coordinates using
	\beq
	\pdr_x = \cos \tht \pdr_r - \frac{\sin \tht}{r} \pdr_\tht \quad \text{and} \quad \pdr_y = \sin \tht \pdr_r - \frac{\cos \tht}{r} \pdr_\tht.
	\eeq
Thus we have 
\footnotesize
	\beqs
	 \left(  -i \hbar \pdr_x - \frac{q\la mky}{2c} \right)^2 = &&-\hbar^2 \cos^2 \tht \pdr_{r}^2 - \hbar^2 \sin^2 \tht \ov{r^2} \pdr_{\tht}^2 + 2 \hbar^2 \cos \tht \sin \tht \ov{r} \pdr_r \pdr_\tht \cr
	 && - \left( \frac{2 \hbar^2 \cos \tht \sin \tht}{r^2} + \frac{i \hbar q \la m k \sin^2 \tht}{c} \right) \pdr_\tht + \left( \frac{i \hbar q \la m k r \cos \tht \sin \tht}{c} - \frac{\hbar^2 \sin^2 \tht}{r} \right) \pdr_r \cr
	 && + \frac{q^2 \la^2 m^2 k^2 r^2 \sin^2 \tht}{4 c^2}.
	 \eeqs
\normalsize
Similarly,
\footnotesize
	\beqs
	 \left(  -i \hbar \pdr_y - \frac{q\la mkx}{2c} \right)^2 = &&-\hbar^2 \sin^2 \tht \pdr_{r}^2 - \hbar^2 \cos^2 \tht \ov{r^2} \pdr_{\tht}^2 - 2 \hbar^2 \cos \tht \sin \tht \ov{r} \pdr_r \pdr_\tht \cr
	 && + \left( \frac{2 \hbar^2 \cos \tht \sin \tht}{r^2} - \frac{i \hbar q \la m k \cos^2 \tht}{c} \right) \pdr_{\tht} - \left( \frac{\hbar^2 \cos^2 \tht}{r} + \frac{i \hbar q \la m k r \cos \tht \sin \tht}{c} \right) \pdr_r \cr
	 && + \frac{q^2 \la^2 m^2 k^2 r^2 \cos^2 \tht}{4 c^2}.
	 \eeqs
\normalsize
Adding these terms, we get
\footnotesize
	\beqs
	 \left(  -i \hbar \pdr_x - \frac{q\la mky}{2c} \right)^2 + \left(  -i \hbar \pdr_y - \frac{q\la mkx}{2c} \right)^2 &=& - \hbar^2 \left( \pdr_{r}^2 + \ov{r}\pdr_r \right) + \ov{r^2} \left( - \hbar^2 \pdr_{\tht}^2 \right. \cr \
	&&  \left. - \frac{i \hbar q \la mk r^2}{c} \pdr_{\tht} + \frac{q^2 \la^2 m^2 k^2 r^4}{4 c^2} \right) \cr 
	 &=& -\hbar^2 \left( \ov{r} \pdr_r \left( r \pdr_r \right) \right) + \ov{r^2} \left( - i \hbar \pdr_\tht + \frac{q \la m k r^2}{2 c} \right)^2. \quad
	\eeqs
\normalsize
Thus, the Hamiltonian in cylindrical coordinates is:
	\beq
	\hat{H} = -\frac{\hbar^2}{2 \mu } \ov{r} \pdr_r [ r \pdr_r] + \ov{2 \mu r^2} \left[-i \hbar \pdr_{\tht} - \frac{q A_{\tht}(r)}{c} \right]^2 + \frac{1}{2 \mu} \left[-i \hbar\pdr_z - \frac{q A_z(r)}{c} \right]^2  + q V(r),
	\label{e:Quantum-Hamiltonian}
	\eeq
where
	\beq
	A_\tht = -\frac{\la m k r^2}{2}, \quad A_z = \frac{\la k r^2}{2} \quad \text{and} \quad V(r) = \frac{k^2(r^2 + m^2)}{2}.
	\label{e:canonical-components-of-electro-magnetic-potentials}
	\eeq
If we introduce the momentum operators \cite{Liboff}
	\beq
	\hat p_r = -i \hbar \frac{1}{\sqrt{r}} \pdr_r \sqrt{r} = -i \hbar \left( \pdr_r +  \ov{2 r}  \right), \quad \hat p_{\tht} = - i \hbar \pdr_{\tht} \quad \text{and} \quad  \hat p_z = -i \hbar \pdr_z
	\eeq
which furnish a representation of the canonical commutation relations $[r, p_r] = [\tht, p_\tht] = [z, p_z] = i \hbar$ and are hermitian with respect to the inner product $\bra \phi |  \psi  \ket = \int r dr d\tht dz \:  \phi^* \psi$, then the Hamiltonian (\ref{e:Quantum-Hamiltonian}) may be written as\footnote{ Here, 
	\beq
	\hat p_r^2 = -\hbar^2 \left( \frac{\pdr^2}{\pdr r^2} + \ov{r} \dd{}{r} - \ov{4r^2} \right).
	\eeq}:
	\beq
	\hat{H} = \frac{1}{2 \mu} \left[ \hat p_r^2 +  \frac{\left( \hat p_{\tht} - \frac{q A_\tht}{c} \right)^2 - \frac{\hbar^2}{4}}{r^2} + \left(\hat p_z - \frac{q A_z}{c} \right)^2 \right] + q V(r).
	\label{e:quantum-Hamiltonian-cylindrical}
	\eeq
We note that this Hamiltonian differs from the direct quantization of the classical cylindrical Hamiltonian (\ref{e:Hamiltonian-EM-plane-polar}) by a centripetal potential $-\hbar^2/ 8 \mu r^2$. Thus, we choose to define the quantum theory via the canonical quantization in Cartesian coordinates.  

We can separate variables in the Schr\"odinger equation using the symmetries of (\ref{e:quantum-Hamiltonian-cylindrical}). The potentials $A_\tht$, $A_z$ and $V(r)$ are independent of $z$ and $\tht$ so that $H$ commutes with the momentum $\hat{p}_z = -i \hbar \pdr_z$ and angular momentum $\hat{p}_\tht = -i \hbar \pdr_\tht$. Thus $H, \hat{p}_z$ and $\hat{p}_\tht $ can be chosen to have common eigenstates. Consequently, the $\tht$ and $z$-dependence of the energy eigenfunctions can be taken to be $\exp(i l \tht)$ and $ \exp \left(i p_z z/\hbar \right)$, where $l$ must be an integer on account of the $2 \pi$-periodicity of $\tht$ and $p_z$ a real number. This leads to the separation of variables in the wavefunciton:
\beq
	\psi(r, \tht, z) = \frac{1}{\sqrt{r}} \varrho(r) \exp(i l \tht) \exp \left(\frac{i p_z z}{\hbar} \right).
	\label{e:psi-seperation-of-variables}
	\eeq
Putting $\hat{H} \psi = E \psi$ we get the radial eigenvalue problem
	\beq
	-\frac{\hbar^2 \varrho''(r)}{2 \mu} + U(r) \varrho(r)  = E \varrho(r).
	\label{e:eigenvalue-problem}	
	\eeq
The $1/ \sqrt{r}$ prefactor in (\ref{e:psi-seperation-of-variables}) eliminates the $\varrho'$ arises from the operator $\hat p_r^2$. Here, the effective potential
	\beqs
	U(r) &=& -\frac{\hbar^2}{8 \mu r^2} + \half \frac{\left[ \hbar l - \frac{q A_{\tht}}{c} \right]^2}{ \mu r^2} + \frac{1}{2 \mu} \left[ p_z - \frac{q A_z}{c} \right]^2 + q V(r) \cr
	       &=& \half \left[ \frac{\hbar^2\left[ l^2 - \frac{1}{4}\right]}{\mu r^2} + \frac{q \la k m \hbar l}{\mu c} + \frac{p_z^2}{\mu} + q k^2 m^2  \right. \cr
	     && \left.  + r^2 \left( \frac{ q^2 \la^2 k^2 m^2}{4 \mu c^2} - \frac{q \la k  p_z}{\mu c} + q k^2 \right) + \frac{q^2 \la^2 k^2 r^4}{4 \mu c^2} \right]
	\label{e:effective-potential-radial-part}
	\eeqs
includes centrifugal (inverse-square), quadratic and quartic terms in $r$. The `centrifugal' term is attractive only when $l = 0$. 

This effective potential (in units where $\mu = q = c =1$) differs from that obtained by Rajeev and Ranken (in Eq. 4.8 of \cite{R-R}). More precisely, in the expression for $U(r)$ obtained by Rajeev and Ranken, the quantity $A_\tht$ was wrongly taken as $\la m k r/2$ instead of $-\la m k r^2/2$.  Thus, the corresponding radial equation they obtained in the strong coupling limit and the subsequent analysis to obtain the dispersion relation for quantized screw-type waves needs to be reconsidered. They proposed that the resulting dispersion relation should give a glimpse of the nature of the degrees of freedom of the scalar field theory in the strongly coupled high-energy limit. In addition, they suggested that the strong coupling limit of the scalar field theory could also be interpreted as a `slow-light' post-relativistic regime. However, as we point out in Section \ref{s:Nilpotent-scalar-field-theory-dual-to-the-PCM} (also see \cite{G-V-1}), the ‘slow-light’ limit ($c \to 0$) holding $\la$ fixed is not quite the same as the strong-coupling limit of the scalar field theory. 

To be doubly sure about the formula (\ref{e:effective-potential-radial-part}) for the effective potential in the quantum theory, we re-derive (\ref{e:eigenvalue-problem}) and (\ref{e:effective-potential-radial-part}) through a complementary viewpoint, where the RR model is interpreted as a quartic oscillator. This simple interpretation of the RR model will facilitate our analysis of its quantum theory.

\section[Rajeev-Ranken model as a quartic oscillator]{Rajeev-Ranken model as a quartic oscillator \sectionmark{Rajeev-Ranken model as a quartic oscillator}}
\sectionmark{Rajeev-Ranken model as a quartic oscillator}
\label{s:Mechanical-interpretation-of-the-Rajeev-Ranken-Model}

It is possible to interpret the classical Hamiltonian of the Rajeev-Ranken model expressed in Darboux coordinates (see Eqs. (\ref{e:Hamiltonian-mech-Darboux}, \ref{e:Hamiltonian-mech-R-P}))
	\beqs
	 H &=& \half \left[ \left( k P_1 - \frac{\la m k R_2}{2} \right)^2 + \left( k P_2 + \frac{\la m k R_1}{2} \right)^2 + \left( k P_3 - \frac{\la k}{2}(R_1^2 + R_2^2) \right)^2 \right] \cr
	 && + \frac{k^2}{2}(R_1^2 + R_2^2 + m^2),
	\label{e:Hamiltonian-mech-RP}
	\eeqs
as the Hamiltonian for a particle of mass $\mu = 1$, moving in a cylindrically symmetric quadratic plus quartic potential. To see this, we regard the Darboux coordinates $R_{1, 2, 3}$ and conjugate momenta $k P_{1,2,3}$ as the Cartesian components of the position and conjugate momentum of a particle of mass $\mu$. Using this interpretation we rewrite Eq.~(\ref{e:Hamiltonian-mech-RP}) as a Hamiltonian for a quartic oscillator:
	\beqs
	H &=& \frac{p_x^2 + p_y^2 + p_z^2}{2\mu} + \frac{\la m k (x p_y - y p_x)}{2\mu} + \left( \frac{\la^2 m^2 k^2}{8 \mu} - \frac{\la k p_z}{2\mu} + \frac{k^2}{2} \right) (x^2 + y^2) \cr
	&& + \frac{\la^2 k^2}{8 \mu} (x^2 + y^2)^2 + \frac{k^2 m^2}{2}.
	\label{e:Hamiltonian-quadratic-quartic-Cartesian}
	\eeqs
The second term in $H$ is proportional to the $z$-component of angular momentum ($L_z$). As mentioned in the previous section, the Hamiltonian possesses a translational symmetry along and rotation symmetry about the $z$-axis. Thus, we make a canonical transformation to cylindrical coordinates $(r, \tht, z)$ and their conjugate momenta $(p_r, p_\tht, p_z)$ defined in Section \ref{s:classical-Hamiltonian-in-terms-of-cylindrical-coordinates}, which satisfy the Poisson brackets (\ref{e:canonical-PB-cylindrical}). Upon doing so, the Hamiltonian (\ref{e:Hamiltonian-quadratic-quartic-Cartesian}) becomes
	\beq
	H = \frac{1}{2 \mu} \left[ p_r^2 +  \frac{p_{\tht}^2}{r^2} + p_z^2 \right] + \frac{\la m k}{2 \mu} p_\tht  + \left( \frac{\la^2 m^2 k^2}{8 \mu} - \frac{\la k p_z}{2 \mu} + \frac{k^2}{2} \right) r^2 + \frac{\la^2 k^2}{8 \mu} r^4 + \frac{k^2 m^2}{2}.
	\label{e:Hamiltonian-quadratic-quartic-cylindrical}
	\eeq
Though the terms linear in $p_{\tht}$ and $p_z$ are not conventionally present in an anharmonic oscillator Hamiltonian, the RR model requires them. Notice that, when $k=0$, $H$ reduces to the Hamiltonian of a free particle, while for $\la = 0$ it is a cylindrically symmetric {\it harmonic} oscillator. 

\vspace{.25cm}
\small 

{\bf \fl Remark:} Interestingly, the Hamiltonian of a quartic anharmonic oscillator
	\beq
	H = \half \left( p^2 + \om^2 q^2 \right) + \la q^4
	\eeq
can be re-expressed as a {\it quadratic} Hamiltonian by introducing the new variable $Q = q^2$
	\beq
	H = \half \left( p^2 + \om^2 q^2 \right) + \la Q^2.
	\eeq
However, in contrast to the step-2 nilpotent $q$-$p$ Heisenberg algebra, the new variables satisfy a step-3 nilpotent algebra (see Section \ref{s:Nilpotent-scalar-field-theory-dual-to-the-PCM} for the definition of step $k$ in a nilpotent algebra): 
	\beq
	\{Q, p \} = 2q, \quad \{q, p \} = 1 \quad \text{and} \quad \{q, Q\} = 0.
	\eeq
Similarly, introducing $X = x^2$ and $Y = y^2$, we rewrite (\ref{e:Hamiltonian-quadratic-quartic-Cartesian}) as a quadratic Hamiltonian: 
	\beqs
	H &=& \frac{p_x^2 + p_y^2 + p_z^2}{2\mu} + \frac{\la m k (x p_y - y p_x)}{2\mu} + \left( \frac{\la^2 m^2 k^2}{8 \mu}+ \frac{k^2}{2} \right) (x^2 + y^2) \cr && - \frac{\la k p_z}{2\mu} (X + Y) + \frac{\la^2 k^2}{8 \mu} (X + Y)^2 + \frac{k^2 m^2}{2}.
	\eeqs
The EOM follow from this quadratic form on the step-3 nilpotent algebra:
	\beqs
	\{x, p_x \} &=&1,\quad  \{ X, p_x \} = 2 x, \quad \{ x, X \} = 0, \quad  \{ y, p_y \} = 1, \quad \{ Y, p_y \} = 2y, \cr 
	\{ y, Y \} &=& 0 \quad \text{and} \quad \{z, p_z \} = 1.
	\eeqs
This is similar to the formulation of the RR model in terms of  the variables $L$ and $S$, where the Hamiltonian (\ref{e:Hamiltonian-RR-LS-variables-quantum}) is a quadratic form on a step-3 nilpotent Lie algebra. In this sense, the RR model joins the harmonic and anharmonic oscillators, Maxwell and Yang-Mills theory in their formulation in terms of  quadratic Hamiltonians on nilpotent Lie algebras. As mentioned in \cite{R-R}, this formulation may facilitate finding the spectrum of the Hamiltonian using the representation theory of the underlying nilpotent group \cite{J-K}. 

\normalsize

\vspace{.25cm}

{\fl \bf Dimensional analysis:} Requiring that $H, p_{x,y,z}$ and $x, y, z$ have dimensions of energy, momentum and length, we find that the parameters in (\ref{e:Hamiltonian-quadratic-quartic-Cartesian}) have the following dimensions\footnote{However, this assignment of dimensions differs from that in the RR model \cite{R-R} where $m, R, P$ are dimensionless while 
	\beq
	[k]_{RR} = L^{-1} \quad \text{and} \quad [H]_{RR}= L^{-2}.
	\eeq}:
	\beq
	[\mu] = M, \quad [k] = M^{1/2} T^{-1}, \quad  [m] = L  \quad \text{and} \quad [\la] = M^{1/2} L^{-1}.
	\label{e:dimensions-of-parameters}
	\eeq
In particular in the classical theory, $\tl \la= \la m/ \sqrt{\mu}$ is the only independent dimensionless combination and defines a nondimesional coupling constant. Since $p_z$ and $L_z$ are conserved quantities,  from the structure of (\ref{e:Hamiltonian-quadratic-quartic-Cartesian}), the energy of any classical state can be expressed as 
	\beq
	E = \frac{p_z^2}{2 \mu} + \frac{\la m k L_z}{\mu} + m^2 k^2 f(\tl \la, \tl p_z, \tl L_z), 
	\label{e:classical-energy-dependence}
	\eeq 
for some function $f$ of the three dimensionless variables $\tl \la, \tl p_z = p_z/k m \sqrt{\mu}$ and $\tl L_z = L_z/km^2 \sqrt{\mu}$. Here, $m^2 k^2$ has dimensions of energy.

\section[Quantum Rajeev-Ranken model]{Quantum Rajeev-Ranken model \sectionmark{Quantum Rajeev-Ranken model}}
\sectionmark{Quantum Rajeev-Ranken model}
\label{s:Quantum-RR-model}

In this section, we study the quantum RR model by canonically quantizing the isotropic anharmonic oscillator. Quantum anharmonic oscillators have been studied in various contexts and several results have been obtained in the literature. For instance, the Schr\"odinger eigenvalue problem for the 1D quartic oscillator may be reduced  \cite{D-S-A-C-D} to the triconfluent Heun equation ($[0, 0, 1_6]$ in Ince's classification, see Appendix \ref{E:Singularities-of-second-order-ordinary-differential-equations}). The energy levels of this oscillator display remarkable analytic properties in the complex coupling constant plane \cite{B-W}.  Some exact results are available for the $N$-dimensional isotropic sextic oscillator \cite{D-W}, but they do not extend to the quartic version. Hill determinants have been used to numerically obtain the spectrum of 1D anharmonic oscillators \cite{B-D-S-S-V} as well as 2D isotropic quartic oscillator by truncating a Frobenius series expansion \cite{Taseli}. However, we are not aware of any exact results for the latter system. Here, we examine the analytic properties of the Schr\"odinger eigenvalue problem that follows from treating the RR model as a 3D quartic oscillator and its weak and strong coupling limits.

Even before formally quantizing the RR model, we may infer the possible dependence of energy eigenvalues on parameters from dimensional analysis. From (\ref{e:dimensions-of-parameters}), since $k m^2 \sqrt{\mu}$ has dimensions of action, in the quantum theory $\tl \hbar = \hbar/ k m^2 \sqrt{\mu}$ is a second independent dimensionless combination in addition to the classically present dimensionless coupling constant $\tl \la = \la m/ \sqrt{\mu}$. Thus, generalizing (\ref{e:classical-energy-dependence}), the energy of any quantum state must be of the form
	\beq
	E = \frac{p_z^2}{2 \mu} + \frac{\la m k L_z}{\mu} + m^2 k^2 g(\tl \la, \tl \hbar, \pi_z, l),  
	\label{e:spectrum-quantum-RR}
	\eeq 
for some function $g$ of the four dimensionless combinations $\tl \la, \tl \hbar, \pi_z= m p_z/\hbar$ and $l = L_z/\hbar$.

\vspace{.25cm}

{\fl \bf Canonical quantization:} To quantize the system in Cartesian coordinates we represent the canonical momenta as differential operators:
	\beq
	\hat p_x = -i \hbar \pdr_x, \quad \hat p_y = -i \hbar  \pdr_y \quad \text{and} \quad \hat p_z = -i\hbar  \pdr_z,
	\label{e:canonical-quantization-momenta}
	\eeq
those satisfying the canonical commutation relations $[ x, p_x ] = [ y, p_y ] = [ z, p_z ] = i \hbar$. Thus the Hamiltonian (\ref{e:Hamiltonian-quadratic-quartic-Cartesian}) becomes the operator
	\beqs
	\hat H &=& \half \left[ \frac{\hat p_x^2 + \hat p_y^2 + \hat p_z^2}{\mu} + \frac{\la m k \hat L_z}{\mu} + \left( \frac{\la^2 m^2 k^2}{4 \mu} - \frac{\la k \hat p_z}{\mu} + k^2 \right) (x^2 + y^2)  \right. \cr
	 && \left. + \frac{\la^2 k^2}{4 \mu} (x^2 + y^2)^2 + k^2 m^2 \right].
	\label{e:quantum-Hamiltonian}
	\eeqs
To facilitate separation of variables in the Schr\"odinger equation, we introduce cylindrical coordinates $(r, \tht, z)$ and the corresponding momentum operators \cite{Griffiths, Liboff}
	\beq
	\hat p_r = -i \hbar \frac{1}{\sqrt{r}} \pdr_r \sqrt{r} = -i \hbar \left( \pdr_r +  \ov{2 r}  \right), \quad \hat L_z = \hat p_{\tht} = - i \hbar \pdr_{\tht} \quad \text{and} \quad  \hat p_z = -i \hbar \pdr_z,
	\eeq
which furnish a representation of the canonical commutation relations $[r, p_r] = [\tht, p_\tht] = [z, p_z] = i \hbar$. They are hermitian with respect to the inner product $\bra \phi |  \psi  \ket = \int \phi^* \psi \: r \: dr \: d\tht \: dz$. The Hamiltonian (\ref{e:quantum-Hamiltonian}) then becomes:
	\beq
	\hat{H} = \frac{1}{2 \mu} \left[ \hat p_r^2 +  \frac{ \hat p_{\tht}^2 - \frac{\hbar^2}{4}}{r^2} + \hat p_z^2 \right] + \frac{\la m k}{2 \mu} \hat p_\tht  + \left( \frac{\la^2 m^2 k^2}{8 \mu} - \frac{\la k \hat p_z}{2 \mu} + \frac{k^2}{2} \right) r^2 + \frac{\la^2 k^2}{8 \mu} r^4 + \frac{k^2 m^2}{2}.
	\label{e:quantum-Hamiltonian-cylindrical-quadratic-quartic}
	\eeq
Notably, this Hamiltonian differs from the direct quantization of the classical cylindrical Hamiltonian (\ref{e:Hamiltonian-quadratic-quartic-cylindrical}) by the addition of an attractive quantum `anti-centrifugal' potential energy $-\hbar^2/ 8 \mu r^2$ \cite{C-R-S-S-W} which cancels a similar term in  $\hat p_r^2 = -\hbar^2 \left( \pdr^2_{r} + (1/r) \pdr_r - 1/4r^2 \right)$.

The Hamiltonian (\ref{e:quantum-Hamiltonian-cylindrical-quadratic-quartic}) commutes with $\hat p_z = -i \hbar \pdr_z$ and  $\hat L_z = \hat p_\tht = -i \hbar \pdr_\tht$, so that all three operators can be chosen to have common eigenstates. Hence, the $\tht$- and $z$-dependence of the energy eigenfunctions can be taken to be $\exp(i l \tht)$ and $ \exp \left(i p_z z/\hbar \right)$  leading to the eigenfunction:
	\beq
	\psi = \rho(r) \exp(i l \tht) \exp \left( \frac{i p_z z}{\hbar} \right).
	\label{e:sep-of-var-common-eigenstate}
	\eeq
Here, $p_z$ can be any real number while $p_\tht = l \hbar$, where $l$ must be an integer on account of the $2 \pi$-periodicity of $\tht$. Separating variables in $\hat H \psi = E \psi$, we arrive at a radial eigenvalue problem
	\beq
	-\frac{\hbar^2}{2 \mu} \left( \rho''(r) + \ov{r} \rho'(r) - \frac{l^2}{r^2} \rho(r)  \right) +  U(r) \rho = \left( E -\frac{p_z^2}{2 \mu} - \frac{\hbar l \la m k}{2 \mu} - \frac{k^2 m^2}{2}  \right) \rho,
	\label{e:radial-eigenvalue-problem-for-quad-quartic}
	\eeq
with  the potential\footnote{Instead if we use the wave function (\ref{e:psi-seperation-of-variables}), then this potential agrees (in units $q = c = 1$)  with the effective potential (\ref{e:effective-potential-radial-part}) obtained using the electromagnetic interpretation of Section \ref{s:Electromagnetic-Hamiltonian}. This confirms that the effective potential in \cite{R-R} has an error.} 
	\beq
	U(r) = \al r^2 + \beta r^4 \quad \text{where} \quad \al = \frac{\la^2 m^2 k^2}{8 \mu} - \frac{\la k p_z}{2 \mu} + \frac{k^2}{2} \quad \text{and} \quad \beta = \frac{\la^2 k^2}{8 \mu}.
	\label{e:potential-radial-quad-quartic}
	\eeq
 For the free particle case ($k = 0$), the potential $U(r)$ is absent and (\ref{e:radial-eigenvalue-problem-for-quad-quartic}) reduces to the Bessel equation \cite{B-O}. In this case, $E - p_z^2/2 \mu$ is simply the energy eigenvalue of the free particle in the $x$-$y$ plane (see Eq.~(\ref{e:quantum-Hamiltonian})), so it must be $ \geq 0$ irrespective of the value of $l$.

It is convenient to separate out the free particle motion in the $z$-direction and define the 2D isotropic anharmonic oscillator Hamiltonian
	\beq
	\hat H_1 = \hat H -\frac{p_z^2}{2 \mu} - \frac{\hbar l \la m k}{2 \mu} - \frac{k^2 m^2}{2} = \frac{1}{2 \mu} \left( \hat p_r^2 + \frac{\hat p_\tht^2 - \frac{\hbar^2}{4}}{r^2} \right) + U(r),
	\label{e:Hamiltonian-two-dimensional-anharmonic-oscillator}
	\eeq
with eigenvalue 
	\beq
	E_1 = E - \frac{p_z^2}{2 \mu} - \frac{\hbar l \la m k}{2 \mu} - \frac{k^2 m^2}{2}.
	\label{e:shifted-energy-eigenvalue}
	\eeq
We notice that the coefficient of the quartic term in $U(r)$ (\ref{e:potential-radial-quad-quartic}) is positive ($\beta > 0$) while that of the quadratic term ($\alpha$) can have either sign. Thus the potential is either purely convex or shaped like a Mexican-hat\footnote{However, the minima of the potential are not static solutions because of the $p_{\tht}$ term in the Hamiltonian. See  Appendix \ref{F:Goldstone-mode-of-the-RR-model}.}. In either case, the spectrum of $\hat H_1$ is bounded below and discrete.

\subsection{Quantum RR model in terms of dimensionless variables} 
\label{s:quantum-RR-model-in-terms-of-dimensionless-variables}

Assuming $k, m \neq 0$  ($k = 0$ corresponds to a free particle), we may re-write the Hamiltonian (\ref{e:quantum-Hamiltonian}) in terms of the dimensionless variables:	
	\beqs
	(\tl x, \tl y, \tl z) &=& \frac{1}{m}(x,y,z), \quad \tl p_{x,y,z} = \frac{p_{x,y,z}}{k m \sqrt{\mu}} = -i \tl \hbar \pdr_{\tl x, \tl y, \tl z}, \cr
	\tl \la &=& \la m/ \sqrt{\mu} \quad \text{and} \quad  \tl \hbar = \hbar/ k m^2 \sqrt{\mu}.
	\eeqs
Dividing (\ref{e:quantum-Hamiltonian}) by $k^2 m^2/2$ we get the dimensionless Hamiltonian 
	\beq
	\tl H = \tl p_x^2 + \tl p_y^2 + \tl p_z^2 + \tl \la \tl L_z + \left( \frac{\tl \la^2}{4} - \tl \la \tl p_z + 1 \right)(\tl x^2 + \tl y^2) + \frac{\tl \la^2}{4}(\tl x^2 + \tl y^2)^2 + 1.
	\label{e:Hamiltonian-dimensionless-Cartesian}
	\eeq
Here $\tl L_z  = \tl x \tl p_y - \tl y \tl p_x$. Similarly, the cylindrical coordinate Hamiltonian (\ref{e:quantum-Hamiltonian-cylindrical-quadratic-quartic}) may be written in terms of dimensionless variables ($\tl r = r/m$): 
	\beq
	\tl H = -\tl \hbar^2 \left[  \frac{\pdr^2}{\pdr \tl r^2} + \ov{\tl r} \dd{}{\tl r}  + \frac{1}{\tl r^2} \frac{\pdr^2}{\pdr \tht^2} +  \frac{\pdr^2}{\pdr \tl z^2}\right] - i  \tl \hbar \tl \la \frac{\pdr}{\pdr \tht} + \left(\frac{\tl \la^2}{4} + i \tl \hbar \tl \la \frac{\pdr}{\pdr \tl z} + 1 \right) \tl r^2 + \frac{\tl \la^2 }{4} \tl r^4 + 1.
	\label{e:Hamiltonian-dimensionless-cylindrical-coordinates}
	\eeq
The formulation in terms of dimensionless couplings will facilitate taking strong and weak coupling limits in Section \ref{s:Weak-and-strong-coupling-limits-of-the-Schrodinger-eigenvalue-problem}.

As before, using the symmetries of the Hamiltonian we separate variables in the energy eigenvalue problem $\tl H \psi = \tl E \psi$  (where $\tl E = 2 E/k^2 m^2$) for the wavefunction 
	\beq
	\psi = \rho(\tl r) \exp(i l \tht) \exp\left( \frac{i \tl p_z \tl z}{\tl \hbar} \right).
	\label{e:seperation-of-variables-dimensionless}
	\eeq
Thus the eigenvalue problem becomes
	\beq
	-\tl \hbar^2 \left( \rho''(\tl r) + \ov{\tl r} \rho'(\tl r) - \frac{l^2}{\tl r^2} \rho(\tl r)  \right) +  \tl U( \tl r) \rho = \left( \tl E - \tl p_z^2 - l \tl \hbar \tl \la - 1  \right) \rho,
	\label{e:eigen-value-problem-cylindrical-coordinates-dimensionless}
	\eeq
with  the potential 
	\beq
	\tl U(\tl r) = \tl \al \tl r^2 + \tl \beta \tl r^4 \quad \text{where} \quad \tl \al = \frac{\tl \la^2}{4} - \tl \la \tl p_z + 1 = \frac{2 \al}{k^2} \quad \text{and} \quad \tl \beta = \frac{\tl \la^2}{4} = \frac{2 \beta m^2}{k^2}.
	\eeq
As in (\ref{e:Hamiltonian-two-dimensional-anharmonic-oscillator}), we define a dimensionless Hamiltonian for the 2D anharmoic oscillator by separating out the free particle motion in the $z$-direction:
	\beq
	\tl{H_1} = \tl H - \tl p_z^2 - l \tl \hbar \tl \la - 1 = -\tl \hbar^2 \left[ \frac{\pdr^2}{\pdr \tl r^2} + \ov{\tl r} \dd{}{\tl r}  + \frac{1}{\tl r^2} \frac{\pdr^2}{\pdr \tht^2} \right] + \tl U(\tl r),
	\label{e:Hamiltonian-dimensionless-2-dimension}
	\eeq
with eigenvalue
	\beq
	\tl E_1 = \tl E - \tl p_z^2 - l \tl \hbar \tl \la - 1.
	\label{e:relation-E-E1}
	\eeq
Thus, the radial equation can be rewritten as 
	\beq
	-\tl \hbar^2 \left( \rho''(\tl r) + \ov{\tl r} \rho'(\tl r) - \frac{l^2}{\tl r^2} \rho(\tl r)  \right) + \tl U(\tl r) \rho = \tl E_1 \rho.
	\label{e:radial-equation-dimensionless-coupling}
	\eeq
{\bf \fl Normalizability condition:} From (\ref{e:seperation-of-variables-dimensionless}) and the inner product $\bra \phi |  \psi  \ket = \int \phi^* \psi \; r \: dr \: d\tht \: dz$, we get the normalizability condition for radial bound states: 
	\beq
	\bra \rho |  \rho  \ket = m^2\int \tl r \rho^2(\tl r) \:  d\tl r < \infty.
	\label{e:normalizability-condition-radial-bound-state}
	\eeq
In particular, $\rho(\tl r)$ is normalizable provided it decays faster than $1/\tl r$ as $\tl r \to \infty$ and grows slower than $1/ \tl r$ as $\tl r \to 0$.

\subsection{Weak and strong coupling limits of the Schr\"odinger eigenvalue problem}
\label{s:Weak-and-strong-coupling-limits-of-the-Schrodinger-eigenvalue-problem}

As will be discussed in Section \ref{s:Properties-of-the-radial-Schrodinger-equation}, the radial equations (\ref{e:radial-eigenvalue-problem-for-quad-quartic}) and its dimensionless version (\ref{e:radial-equation-dimensionless-coupling}) are not solvable, in general, in terms of familiar functions. Here, we consider the weak coupling and a suitably defined strong coupling limit of these radial equations. The energy spectrum in the weak coupling limit is explicitly obtained. In the strong coupling limit, we are able to find the dependence of energy levels on the wavenumber $k$. We use these results to deduce dispersion relations for quantized continuous screw-type waves of the scalar field theory in these limits.

\vspace{.25cm}

{\fl \bf Weak coupling limit:} In the weak coupling limit $\tl \la \to 0$, the Hamiltonian (\ref{e:Hamiltonian-dimensionless-Cartesian}) reduces to 
	\beq
	\tl H = \tl p_x^2 + \tl p_y^2 + \tl p_z^2 + \tl x^2 + \tl y^2 + 1.
	\eeq	
By separating the free particle motion in the $z$-direction, we find that in the weak coupling limit the above Hamiltonian reduces to that of a 2D harmonic oscillator with mass 1/2 and angular frequency 2:
	\beq
	\tl H_1 = \tl p_x^2 + \tl p_y^2 + \tl x^2 + \tl y^2. 
	\eeq
Thus we have the spectrum
	\beq
	\tl E_{1_{\tl \la \to 0}} = (n_x + n_y + 1) 2 \tl \hbar, \quad \text{where} \quad  \tl E_{1_{\tl \la \to 0}} = \tl E - \tl p_z^2 - 1.
	\eeq
Here, $n_x$ and $n_y$ are nonnegative integers. Re-expressing this in terms of dimensionful variables we get the energy spectrum in the weak coupling limit:
	\beq
	\lim_{\la \to 0} E =  \frac{k^2 m^2}{2} + (n_x + n_y + 1) \frac{\hbar \, |k|}{\sqrt{\mu}} + \frac{p_z^2}{2 \mu}.
	\label{e:weak-coupling-dispersion-relation}
	\eeq
The spectrum and the corresponding wavefunctions can also be obtained via the radial equation. In the weak coupling limit $\tl \la \to 0$, the radial eigenvalue problem (\ref{e:eigen-value-problem-cylindrical-coordinates-dimensionless}) becomes
	\beq
	-\tl \hbar^2 \left( \rho''(\tl r) + \ov{\tl r} \rho'(\tl r) - \frac{l^2}{\tl r^2} \rho(\tl r)  \right) + \tl r^2 \rho = \tl E_{1_{\tl \la \to 0}} \rho.
	\eeq
This radial equation is a special case of the  confluent hypergeometric equation and can be solved in terms of generalized Laguerre polynomials. The normalizability of the wavefunction implies the spectrum:
	\beq
	\tl E_{1_{\tl \la \to 0}}  = 4 |\tl \hbar| \left( n + \frac{|l| + 1}{2} \right).
	\eeq
In terms of dimensionful parameters, this agrees with (\ref{e:weak-coupling-dispersion-relation}), once we identify $2 n + |l|$ with $n_x + n_y$.

\vspace{.25cm}

{\fl \bf Strong coupling limit:} We now consider a novel strong coupling limit of the radial equation in dimensionless variables (\ref{e:radial-equation-dimensionless-coupling}). We let $\tl \la, \tl \hbar \to \infty$ holding $\tl g = \tl \la/ \tl \hbar$ and $\tl p_z$ finite, so that all terms on the LHS of (\ref{e:radial-equation-dimensionless-coupling}) grow like $\tl \la^2$. To get a nontrivial eigenvalue problem in this limit, we focus on eigenvalues $\tl E_1$ that grow quadratically with $\tl \la$ and consequently define $\tl E_2 = \tl E_1 /\tl \la^2 $. In the anharmonic oscillator, $\tl p_z$ is the dimensionless conserved  $z$-component of momentum and can take any real value. However, in the context of the RR model, 
	\beq
	\tl p_z = \frac{p_z}{k m \sqrt{\mu}} = \frac{k P_3}{k m \sqrt{\mu}} =  \frac{\tl \la}{2 m^2}(2 \mathfrak{c} - m^2) + \frac{1}{ \tl \la \mu}.
	\eeq
To keep $\tl p_z$ finite in this strong coupling limit, the Casimirs $\mathfrak{c}$ and $m$ must be chosen so that $2 \mathfrak{c} - m^2 \to 0$, in such way that $\tl \la (2 \mathfrak{c} - m^2)$ approaches a finite limit. Holding $\tl p_z$ finite in this limit, ensures that the $k$-dependence drops out and the radial equation in the strong coupling limit becomes:
	\beq
	 \rho''(\tl r) + \ov{\tl r} \rho'(\tl r) - \left( \frac{l^2}{\tl r^2} + \frac{\tl g^2}{4}\left(\tl r^2 + \tl r^4 \right)  - \tl g^2 \tl E_2 \right) \rho(\tl r) = 0.
	 \label{e:strong-coupling-radial-equation-dimensionless}
	\eeq
The corresponding Hamiltonian (\ref{e:Hamiltonian-dimensionless-Cartesian}) in this strong coupling limit is given by 
	\beq
	\frac{\tl H}{\tl \hbar^2}  = -\left(\frac{\pdr^2}{\pdr \tl x^2} + \frac{\pdr^2}{\pdr \tl y^2} \right)  - i \tl g \left( \tl x \frac{\pdr}{\pdr \tl y} - \tl y \frac{\pdr}{\pdr \tl x} \right) + \frac{\tl g^2}{4} ( \tl x^2 + \tl y^2 + ( \tl x^2 + \tl y^2 )^2).
	\eeq 
It follows that the finite rescaled energy eigenvalue in this strong coupling limit $\tl E_2 (\tl g , l)$ is independent of $k$. Thus, in this strong coupling limit, the nontrivial dimensionless energy eigenvalues $\tl E \sim \tl \la^2 \tl E_2(\tl g, l) + l \tl \hbar \tl \la  + \tl p_z^2 + 1$ must diverge quadratically in $\tl \la$ with the last two terms being sub-leading. Finally, the original (dimensionful) energy (see Eq.~(\ref{e:eigen-value-problem-cylindrical-coordinates-dimensionless})) $E$ is
	\beqs
	E_{\rm strong} &=& \frac{k^2 m^2}{2} \tl E \sim \frac{k^2 m^2}{2} \tl \hbar^2 \left(  \tl g^2 \tl E_2 (\tl g, l) + \tl g l + \frac{\tl p_z^2 + 1}{\tl \hbar^2}  \right)\cr
	&=& \frac{\hbar^2}{2 \mu m^2} \left( \tl g^2 \tl E_2 (\tl g, l) + \tl g l + \frac{\tl p_z^2 + 1}{\tl \hbar^2} \right).
	\eeqs
Thus, in the strong coupling limit ($\hbar, \la \to \infty$), the energy $E$ is quadratically divergent but has no leading dependence on $k$. Thus, unlike in the weak coupling limit ($\la \to 0$) where we found  $E_{\la \to 0} = m^2 k^2/2 + (n_x + n_y +1) (\hbar |k|/ \sqrt{\mu}) + p_z^2/ 2\mu$ (\ref{e:weak-coupling-dispersion-relation}), at  strong coupling we find $2 \mu m^2(E_{\rm strong}/\hbar^2) \propto k^0$.

Though $k$ does not play the role of a wavenumber in the nonrelativistic quartic oscillator (\ref{e:quantum-Hamiltonian-cylindrical-quadratic-quartic}), it {\it is} a wavenumber in the screw-type wave solutions of the scalar field theory $\phi = e^{Kx} R(t) e^{-Kx} + m K x$ with $K = ik \sig_3/2$. Thus, the above $E$-$k$ relations may be regarded as dispersion relations for quantized  screwons in the weak and strong coupling limits. The term $p_z^2/2 \mu$ in $E_{\la \to 0}$ is a constant addition to the energy of the oscillator. However, in the RR model, it depends on both $k$ and $\la$: 
	\beq
	\frac{p_z^2}{2 \mu} = \frac{k^2}{2 \mu} \left(\frac{ \la}{2} (2 \mathfrak{c} - m^2) + \frac{1}{\la}\right)^2,
	\label{e:z-kinetic-term-in-H}
	\eeq
and is seen to be a divergent constant in the weak coupling limit $\la \to 0$. But  the difference $E - (p_z^2/ 2 \mu)$, has a finite limit:
	\beq
	\lim_{\la \to 0} (E - (p_z^2/ 2 \mu)) = m^2 k^2/2 + (n_x + n_y +1) (\hbar |k|/ \sqrt{\mu}).
	\label{e:weak-coupling-limit-dispersion-relation}
	\eeq
The quadratic $m^2 k^2/2$ term is also a constant (independent of $x$ and $t$) addition to the relativistic energy per unit length of screwons. Indeed, it arises when the screwon ansatz $\phi = e^{Kx} R(t) e^{-Kx} + m K x$ is inserted in the field energy density $(1/2) (\dot \phi^2 + \phi'^2)$. The term linear in $k$  in (\ref{e:weak-coupling-limit-dispersion-relation}) is like the more conventional linear dispersion relation for free relativistic particles.

\vspace{.25cm}

\small

{\bf \fl Remark:} To ensure that $\tl p_z$ (see Eq.~(\ref{e:z-kinetic-term-in-H})) is finite, we need to impose the condition $2 \mathfrak{c} - m^2 \to 0$ among the Casimirs. Thus, taking this strong coupling limit effectively restricts the dynamics of the model to a special submanifold of the phase space on which the coordinates satisfy the relation:
	\beq
	\frac{L_1^2 + L_2^2}{k^2} = -\frac{2S_3}{k \la}.
	\eeq
Given the Casimirs $\mathfrak{c}$ and $m$, the dynamics is confined to a 3D submanifold labelled by $S_{1,2}$ and $r = \sqrt{L_1^2 + L_2^2}/k$.

\normalsize

\vspace{.25cm}

In the strong coupling limit ($\tl \la , \tl \hbar \to \infty$, holding $\tl p_z$ fixed), the terms involving $\tl z$ are sub-leading compared to those involving $\tl x$ and $\tl y$ (it is an anisotropic limit) and we get the Hamiltonian 
	\beq
	\tl H_2 = \tl H/ \tl \hbar^2 = \tl P_x^2 + \tl P_y^2 + \tl g \tl P_{\tht} + \frac{\tl g^2}{4} \left( \tl x^2 + \tl y^2 + (\tl x^2 + \tl y^2)^2 \right).
	\eeq
Here $\tl P_x =  \tl p_x/ \tl \hbar= -i \pdr/ \pdr \tl x$, $\tl P_y = \tl p_y/ \tl \hbar = -i \pdr/ \pdr \tl y$ and $\tl P_{\tht} = \tl x \tl P_y - \tl y \tl P_x$ with the commutation relations:
	\beq	
	[ \tl x, \tl P_x ] = i \quad \text{and} \quad [ \tl y, \tl P_y ] = i.
	\eeq
Thus, the strong coupling limit of the theory is an anisotropic scaling limit resulting in a dimensional reduction to a 2D quartic  anharmonic oscillator with an additional term proportional to the conserved angular momentum. Having arrived at the strong coupling limit, we may further send $\tl g \to 0$ resulting in a free particle moving on a plane. On the other hand, when $\tl g \to \infty$, the potential energy dominates in a manner similar to the strong coupling limit ($\tl \la \to \infty$) of the $\tl \hbar \to 0$ classical theory. It is noteworthy that the classical model has only one coupling $\tl \la$, and its strong coupling limit is defined as the one where $\tl \la \to \infty$. In this strong coupling limit, the potential energy $ (\tl \la^2/4)( \tl x^2 + \tl y^2 + ( \tl x^2 + \tl y^2)^2)$ becomes the dominant term in the dimensionless classical Hamiltonian $\tl H_1 = \tl H - \tl p_z^2 - \tl \la \tl p_z -1$ (see Eq. (\ref{e:Hamiltonian-dimensionless-2-dimension})). Unlike this classical strong coupling limit, our quantum strong coupling limit has a dimensionless free parameter $\tl g$. Moreover, the strong coupling limit of the quantum theory is incompatible with this classical limit ($\tl \hbar \to 0$) since in the former $\tl \hbar \to \infty$.

\subsection{Properties of the radial Schr\"odinger equation}
\label{s:Properties-of-the-radial-Schrodinger-equation}

We now use Ince's classification (see Appendix \ref{E:Singularities-of-second-order-ordinary-differential-equations}) to discuss some properties of the second order radial eigenvalue problem (\ref{e:radial-equation-dimensionless-coupling}). The latter and its strong coupling limit (\ref{e:strong-coupling-radial-equation-dimensionless}) are both of type $[0, 1, 1_6]$. This means they have two singular points: the regular point $\tl r = 0$  and  the irregular point $\tl r = \infty$ which has rank 3 since $K_1 = -1$ and $K_2 = 4$ in Eq.~(\ref{e:rank-of-irregular-singularity}).
The rank 3 irregular singularity at $\infty$ can be thought of as having being formed by the coalescence of four nonelementary regular singular points. Thus, we may regard our radial equations as confluent forms of a differential equation with 10 elementary regular singularities ($[10,0,0]$)  or of what is sometimes called a generalized Lam\'e equation of type $[0, 5, 0]$ \cite{Crowson}.  In particular, our radial equations cannot, in general, be solved in terms of hypergeometric, Heun or Lam\'e functions or their confluent forms. 

By contrast, the weak coupling limit of (\ref{e:radial-equation-dimensionless-coupling}) 	
	\beq
	-\tl \hbar^2 \left( \rho''(\tl r) + \ov{\tl r} \rho'(\tl r) - \frac{l^2}{\tl r^2} \rho(\tl r)  \right) + \tl r^2 \rho = \tl E_{1_{\la \to 0}} \rho
	\label{e:weak-coupling-radial-equation-Ince-type}
	\eeq
is an equation of type $[0,1, 1_4]$. The rank of the irregular singularity at $\tl r = \infty$ in this equation can be reduced from 2 to 1 by the substitution  $\tl r^2 = x$, resulting in an equation of type $[0,1,1_2]$. It is noteworthy that the confluent hypergeometric equation is also of type $[0,1,1_2]$. Not surprisingly, (\ref{e:weak-coupling-radial-equation-Ince-type}) can be solved in terms of generalized Laguerre polynomials which are special cases of the confluent hypergeometric function. 

Returning to the radial equation (\ref{e:radial-equation-dimensionless-coupling}), for large values of $\tl r$, the method of dominant balance \cite{B-O} gives the asymptotic behaviour
	\beq
	\rho(\tl r) \sim \exp\left( -\frac{\sqrt{\tl \beta}}{\tl \hbar} \left( \frac{\tl r^3}{3} + \frac{ \tl \alpha \tl r}{2 \tl \beta} \right) \right) \tl r^{-3/2} a(\tl r), \quad \text{where} \quad a(\tl r) \sim \mathcal{O}(1) \quad \text{as} \quad \tl r \to \infty.
	\eeq 
As noted in (\ref{e:rank-asymptotic-relation-solution}), the rank (three) of the singularity at $\infty$ determines the dominant asymptotic behaviour. The same asymptotic behaviour also arises in the strong coupling limit of Section \ref{s:Weak-and-strong-coupling-limits-of-the-Schrodinger-eigenvalue-problem}, where $\tl \al/ \tl \beta \to 1$ and $\sqrt{\tl \beta}/\tl \hbar  \to \tl g/2$. See Appendix \ref{G:Asymptotic-behaviour-in-the-strong-coupling-limit} for the details of the asymptotic behaviour in the strong coupling limit. 

Now we turn to the behaviour of (\ref{e:radial-equation-dimensionless-coupling}) around the regular singularity $\tl r = 0$.  The Frobenius series $\rho (\tl r) =  \tl r^\eta  \sum_{n = 0}^\infty \rho_n \tl r^n$ leads to the  exponents $\eta_{1,2} = \pm l$ (see (\ref{e:Indicial-equation-general-form})). The condition (\ref{e:normalizability-condition-radial-bound-state}), that the wavefunction be normalizable restricts us to $\eta = \eta_1$. In general $\rho_n$ satisfy a four-term recurrence relation. The formulae are somewhat shorter in the strong coupling limit, where we get (see Appendix \ref{H:Frobenius-method-for-strong-coupling-limit-Local-analysis})
	\beqs
	\rho_1 &=& 0, \quad \rho_2 = \frac{- \tl g^2 \tl E_2}{4 l + 4} \rho_0, \quad \rho_3 = 0, \quad (8 l + 16) \rho_4 = \frac{\tl g^2}{4} \rho_0 - \tl g^2 \tl E_2 \rho_2, \quad \rho_5 =0, \cr
	\text{and} &&(2 n l + n^2)\rho_n + \tl g^2 \tl E_2 \rho_{n-2} - \frac{\tl g^2}{4}(\rho_{n-4} - \rho_{n-6})  = 0,  \quad \text{for} \quad n = 6,8, \ldots,
	\label{e:recursion-relations-around-zero-rho}
	\eeqs
with $\rho_{\rm odd} = 0$. By contrast, one has two- and three-term recurrence relations for the hypergeometric and Heun equations \cite{Ince}. We observe that the number of terms $t$ in these recurrence relations is related to the number of elementary singularities in the parent equations of type $[e, 0, 0]$ via $t = (e - 2)/2$.

\section[Separation of variables and the WKB approximation]{Separation of variables and  the WKB \\approximation \sectionmark{Separation of variables and the WKB approximation}} 
\sectionmark{Separation of variables and the WKB approximation}
\label{s:Separatio-of-variables-and-the-WKB-approximation}

Here we consider the semiclassical approximation of the quantum RR model. We separate variables in the Hamilton-Jacobi equation and obtain the WKB quantization condition in an implicit form. In the weak coupling limit the spectrum from the WKB quantization condition agrees with that obtained previously. It remains to estimate the spectrum for other values of the coupling $\la$.

\subsection{Hamilton-Jacobi equation}
\label{s:Hamilton-Jacobi-equation}

The Hamilton-Jacobi (HJ) equation $S_t + H(q, \pdr S/ \pdr q) = 0$, for the Rajeev-Ranken model is most easily analysed in cylindrical Darboux coordinates. Putting $S = {\cal W} - Et$, we get the time-independent HJ equation from (\ref{e:Hamiltonian-quadratic-quartic-cylindrical}):
	\beqs
	\frac{1}{2 \mu} \left[ (\pdr_r {\cal W})^2 + \ov{r^2} (\pdr_{\tht} {\cal W})^2 + (\pdr_z {\cal W})^2 \right] + \frac{\la m k}{2\mu} \pdr_{\tht}{\cal W} &+& \left( \frac{ \la^2 m^2 k^2}{8 \mu} - \frac{\la k}{2 \mu} \pdr_z {\cal W} + \frac{k^2}{2} \right)r^2 \cr
	&+& \frac{\la^2 k^2 r^4}{8 \mu}  + \frac{k^2 m^2}{2} = E.
	\eeqs
Supposing that ${\cal W}(r, \tht, z) = W(r) + W_{\tht}(\tht) + W_z(z)$, the HJ equation becomes
	\beqs
	 \frac{1}{2 \mu} \left[ W'(r)^2 + \ov{r^2} W_{\tht}'(\tht)^2 + W_z'(z)^2 \right] + \frac{\la m k}{2\mu} W_{\tht}' &+& \left( \frac{ \la^2 m^2 k^2}{8 \mu} - \frac{\la k}{2 \mu} W_z' + \frac{k^2}{2} \right)r^2 \cr
	 &+& \frac{\la^2 k^2 r^4}{8 \mu}  + \frac{k^2 m^2}{2} = E.
	 \eeqs
From (\ref{e:Hamiltonian-quadratic-quartic-cylindrical}), we notice that $\tht$ and $z$ are cyclic coordinates so the momenta $p_{\tht} = \pdr_\tht {\cal W}$ and $p_z = \pdr_z {\cal W}$ are constants of motion. Thus, we must have (using (\ref{e:potential-radial-quad-quartic}))
	\beq
	W_{\tht}'(\tht) = p_{\tht}, \quad 
	W_z'(z) = p_z \quad \text{and} \quad \frac{1}{2 \mu}\left(W'(r)^2 + \frac{p_{\tht}^2}{r^2} \right) + \al r^2 + \beta r^4 = E - \frac{p_z^2}{2\mu} - \frac{\la p_{\tht} m k}{2 \mu} - \frac{k^2 m^2}{2},
	\eeq
for separation constants $p_{\tht}$ and $p_z$, which can have either sign. Changing variables to $s = r^2$, $W(r)$ is expressed as an elliptic integral:
	\beq
	W(r = \sqrt{s}) = \pm \int^{\sqrt{s}} \frac{ds}{2 s} \sqrt{ 2 \mu (E s  - \al s^2 - \beta s^3 ) - (p_z^2 + \la p_{\tht} m k + k^2 m^2 \mu) s - p_{\tht}^2}.
	\label{e:Hamilton-Jacobi-generator-elliptic-integral}
	\eeq
Thus, it is possible to separate variables and obtain a complete solution of the HJ equation involving three separation constants $E, p_{\tht}$ and $p_z$. This is perhaps not surprising since the EOM can be solved in terms of Weierstrass elliptic functions \cite{R-R}. 

\subsection{WKB approximation}
\label{s:WKB-approximation}

Here, we find the WKB quantization condition for the energy spectrum of the RR model (\ref{e:quantum-Hamiltonian-cylindrical-quadratic-quartic}). For this purpose, it is convenient to separate variables in cylindrical coordinates using the factorized wavefunction:
	\beq
	\psi(r, \tht, z) = \frac{\varrho(r)}{\sqrt{r}} \exp\left(\frac{i p_{\tht} \tht}{\hbar} \right) \exp \left(\frac{i p_z z}{\hbar}\right).
	\label{e:common-eigenstate}
	\eeq 
Division by $\sqrt{r}$ ensures that the radial equation formally looks like the Schr\"odinger eigenvalue problem for a particle on the positive half line subject to the potential $U_{\rm eff}$:
	\beq
	\frac{-\hbar^2}{2 \mu} \varrho'' + U_{\rm eff} \varrho = \left(E - \frac{p_z^2}{2 \mu} - \frac{p_{\tht} \la m k}{2 \mu} - \frac{k^2 m^2}{2} \right) \varrho, \quad \text{where} \quad U_{\rm eff} = U(r) + \frac{\gamma}{r^2}
	\label{e:radial-equation}
	\eeq
with $\gamma = (\hbar^2/2\mu)( p_{\tht}^2/\hbar^2 - 1/4)$. Here, $U(r) = \al r^2 + \beta r^4$ is the potential from (\ref{e:potential-radial-quad-quartic}). We look for radial eigenfunctions of the form $\varrho(r) = \exp \left( i W(r)/\hbar \right)$. Substituting in (\ref{e:radial-equation}) gives
	\beq
	i \hbar W''(r) - W'(r)^2 + p(r)^2 = 0, \quad \text{where} \quad
	p(r)^2 = 2 \mu \left(E - \frac{p_z^2}{2 \mu} - \frac{p_{\tht} \la m k}{2 \mu} - \frac{k^2 m^2}{2}  - U_{\rm eff}\right).
	\label{e:radial-equation-WKB-and-classical-momentum}
	\eeq
We now do a semiclassical expansion of $W$ and $E$:
	\beq
	W(r) = W_0 + \hbar W_1+ \cdots \quad \text{and} \quad E = E^{(0)} + \hbar E^{(1)}+ \cdots.
	\eeq 
At $\mathcal{O}(\hbar^0)$, we obtain 
	\beq
	W_0(r) = \pm \int^r  dr \: \left[2 \mu E^{(0)} - p_z^2 - p_{\tht} \la m k - k^2 m^2 \mu - 2 \mu( \al r^2 + \beta r^4)  - \frac{p_{\tht}^2}{r^2} \right]^{1/2}, 
	\eeq
while at $\mathcal{O}(\hbar)$ we get $W_1' = ((iW_0''/2) + \mu E^{(1)})/W_0'$.
$W_0$ agrees with Hamilton's characteristic function (\ref{e:Hamilton-Jacobi-generator-elliptic-integral}) upon changing variables to $s = r^2$. 
Requiring $\varrho(r) \sim \exp \left( iW_0/ \hbar \right)$ to be singlevalued, we obtain the quantization condition
	\beq
	\int_{r_{\rm min}}^{r_{\rm max}} dr \, \left[2 \mu E^{(0)} - p_z^2 - p_{\tht} \la m k - k^2 m^2 \mu - 2 \mu( \al r^2 + \beta r^4)  - \frac{p_{\tht}^2}{r^2} \right]^{1/2} = n \pi \hbar,
	\label{e:quantization-condition}
	\eeq
where the radial quantum number $n$ is a (large) integer. Here, $r_{\rm min} < r_{\rm max}$ are the positive zeros of the quartic polynomial $\lim_{\hbar \to 0} p(r)^2$, enclosing the classically allowed interval (there can only be one such interval). Similarly, for $\psi$ to be singlevalued on the circle we must have $p_{\tht}= l \hbar$ for a large integer $l$. The quantization condition (\ref{e:quantization-condition}) leads to an elliptic integral, but we have not been able to invert it to explicitly obtain the semiclassical spectrum other than in the weak coupling limit.

\vspace{.25 cm}

{\fl \bf Weak coupling limit:}  When $\la \to 0$, $\al = k^2/2, \beta = 0$ and the quantization condition (\ref{e:quantization-condition}) simplifies to a trigonometric integral (here $s = r^2$)
	\beq
	\int_{s_{\rm min}}^{s_{\rm max}} \frac{ds}{2 s} \sqrt{a s^2 + b s + c} = n \pi \hbar,
	\label{e:Integral-for-quatization-condition-isotropic-harmonic-oscillator-redefined}
	\eeq
where 
	\beq
	a = - k^2 \mu, \quad b = 2 \mu E_1 = 2 \mu \left(E^{(0)} - \frac{p_z^2}{2\mu} - \frac{k^2 m^2}{2}\right) \quad \text{and} \quad c = -p_{\tht}^2.
	\eeq 
Notice that  $a < 0$, $b > 0$ ($2 \mu \: \times$ energy of the 2D anharmonic oscillator at weak coupling) and $c < 0$. Using this, the turning points are the roots of $a s^2 + b s + c = 0$:	
	\beq
	s_{\rm min, max} = \ov{k^2} \left[E_1 \mp \frac {\sqrt{\Delta}}{2 \mu}\right] > 0, \quad \text{where} \quad  \Delta = b^2 - 4 a c = 4 \mu^2 \left( E_1^2 - \frac{k^2 p_{\tht}^2 }{\mu} \right). 
	\label{e:classical-turning-points}
	\eeq
The RHS of (\ref{e:classical-turning-points}) is to be interpreted in the classical limit, where commutators and other terms of $\mathcal{O}(\hbar)$ are ignored. To get the allowed energy levels, we evaluate the LHS of (\ref{e:Integral-for-quatization-condition-isotropic-harmonic-oscillator-redefined}) using  Eq. 2.267(1) of \cite{G-R}:
	\beq
	\int_{s_{\rm min}}^{s_{\rm max}} \frac{ds}{2 s} \sqrt{a s^2 + b s + c} = \frac{\pi}{2} \left( \frac{\sqrt{\mu}}{|k|} E_1 - |p_{\tht}| \right).
	\eeq
This leads to the spectrum
	\beq
	E_1 \approx 2 \left( n + \frac{|p_{\tht}|}{2 \hbar} \right) \frac{\hbar |k|}{\sqrt{\mu}} \quad \text{where} \quad p_{\tht} = l \hbar \quad \text{for} \quad l, n \gg 1.
	\eeq
This weak coupling semiclassical result agrees with the previously obtained exact spectrum in the weak coupling limit $E_1 =  (n_x + n_y + 1)\hbar |k|/\sqrt{\mu}$ (\ref{e:weak-coupling-dispersion-relation}), if we identify $n_x + n_y$ with $2 n + |l|$ \cite{Pauli} for $p_{\tht} / \hbar = l$ a large integer.

\section[Unitary representation of nilpotent Lie algebra]{Unitary representation of nilpotent Lie algebra \sectionmark{Unitary representation of nilpotent Lie algebra}}
\sectionmark{Unitary representation of nilpotent Lie algebra}
\label{s:Unitary-representation-of-nilpotent-Lie-algebra}

Here we exploit the canonical quantization of the RR model in Darboux coordinates $(R_a, k P_a)$ (see Section \ref{s:Quantum-RR-model}) to obtain a representation of the Poisson algebra of the $L$-$S$ variables of the model. Classically, the latter satisfy the step-3 nilpotent Poisson brackets\footnote{Interestingly, the classical equations of motion of the RR model can be interpreted as Euler equations for this nilpotent Lie algebra. Notably, the EOM may {\it also} be interpreted as Euler equations for a centrally extended Euclidean algebra, which we obtained by comparing them with the Kirchhoff's equations. (See Appendices \ref{B:Kirchhoff-equations} and \ref{C:RR-equations-as-Euler-equations-for-a-nilpotent-Lie-algebra} for details.)}
	\beq
	\{ L_a, L_b \} = 0, \quad \{S_a, S_b \} = \la \eps_{abc} L_c \quad \text{and}  \quad \{ S_a, L_b \} = -\eps_{abc} K_c.
	\eeq
To relate these to the Darboux coordinates, recall that
	\beq
	L = [ K, R ] + m K \quad \text{and} \quad S = \dot R + \frac{K}{\la}, 
	\eeq
where $K = i k \sig_3/2$. Introducing $kP_{1,2} = \dot R_{1,2} \pm \la m k R_{2,1}/2$ and $kP_3 = \dot R_3 + \la k (R_1^2 + R_2^2)/2$, in component form we have,
	\beqs
	L_1 &=& k R_2, \quad L_2 = -k R_1, \quad L_3 = -m k, \quad K_{1,2} = 0, \quad K_3 = -k \cr 
	S_1 &=& k P_1 - \frac{\la}{2} m k R_2, \quad S_2 = k P_2 + \frac{\la}{2} m k R_1 \quad \text{and} \quad S_3 = k P_3 - \frac{k}{\la} - \frac{\la k}{2}(R_1^2 + R_2^2). \qquad
	\label{e:L-and-S-component-form-Darboux-coordinates}
	\eeqs
In the quantum theory, we wish to represent $L, S$ and $K$ as hermitian operators on a Hilbert space obeying the commutation relations obtained by the replacement $\{ A , B \} \to (1/i \hbar) [A, B]$:
	\beq
	\quad [L_a, L_b] = 0, \quad [S_a, S_b] = i \hbar \la \eps_{abc} L_c,  \quad \text{and} \quad [S_a, L_b ] = - i \hbar \eps_{abc} K_c.
	\label{e:commutation-relations}
	\eeq
More explicitly, the nonzero commutation relations among the generators are:
	\beqs
	[L_1, S_2] &=& -i \hbar K_3, \quad [L_2, S_1] = i \hbar K_3, \quad [S_1, S_2] = i \hbar \la L_3, \cr
	[S_1, S_3] &=& -i \hbar \la L_2 \quad \text{and} \quad [ S_2, S_3 ] = i \hbar \la L_1.
	\label{e:nilpotent-Lie-algebra}
	\eeqs
We now exploit our physical interpretation of the RR model as an anharmonic oscillator to discover a unitary representation of this nilpotent Lie algebra. Indeed, using the canonical Schr\"odinger representation of $R_a$ and $kP_a$ (\ref{e:canonical-quantization-momenta}) and the relations (\ref{e:L-and-S-component-form-Darboux-coordinates}) we are led to the following representation 
	\beqs
	L_1 &=& k y, \quad L_2 = -k x, \quad L_3 = -m k I, \quad K_{1,2} = 0, \quad K_3 = -k I \cr 
	S_1 &=& -i \hbar \dd{}{x} - \frac{\la}{2} m k y, \quad S_2 = - i \hbar \dd{}{y} + \frac{\la}{2} m k x \quad \text{and} \cr
	S_3 &=& -i \hbar \dd{}{z} - \frac{k}{\la} I- \frac{\la k}{2}(x^2 + y^2), \qquad
	\label{e:representation-of-nilpotent-algebra}
	\eeqs
where $I$ is the identity. These hermitian operators on the Hilbert space $L^2(\mathbb{R}^3_{xyz})$ give us an infinite-dimensional unitary representation of the nilpotent algebra (\ref{e:commutation-relations}). 

The dynamics of the quantum RR model is specified by the hermitian and positive Hamiltonian
	\beq
	H  = \frac{S_a^2 + L_a^2}{2} + \frac{k S_3}{\la} + \frac{k^2 }{2 \la^2} = \half \left[ \left(S - \frac{K}{\la} \right)^2 + L^2 \right], 
	\label{e:Hamiltonian-RR-LS-variables-quantum}
	\eeq
which is a quadratic form on this Lie algebra. Using the above commutation relations (\ref{e:commutation-relations}), we find the quadratically nonlinear Heisenberg equations of motion:
	\beq
	\dot S_a = \frac{1}{i\hbar} [S_a, H ] =  \la \eps_{abc} S_b L_c  \quad \text{and} \quad 
	\dot L_a = \frac{1}{i \hbar} [L_a, H ] = \eps_{abc} K_b S_c. 
	\eeq
	
{\fl \bf Reducibility of representation:} As in the classical theory, $L_3 = -m k$ and $\mathfrak{c} k^2 = (L_1^2 + L_2^2 + L_3^2)/2 + k S_3/\la$ are Casimir operators of the nilpotent commutator algebra (\ref{e:commutation-relations}). We may represent them as differential operators on $L^2(\mathbb{R}^3_{xyz})$:
	\beq
	L_3 = -m k I \quad \text{and} \quad \mathfrak{c} k^2 =  \left(\frac{k^2 m^2}{2} - \frac{k^2}{\la^2} \right) I - \frac{i \hbar k}{\la} \dd{}{z}.
	\eeq
Evidently, $L_3$ is a multiple of the identity while $\mathfrak{c} k^2$ is essentially $\pdr_z$. These commute with all the  operators in  (\ref{e:representation-of-nilpotent-algebra}) as the latter do not involve the coordinate $z$. Thus the representation (\ref{e:representation-of-nilpotent-algebra}) is reducible with invariant subspaces given by the simultaneous eigenspaces of $L_3$ and $\mathfrak{c}$. The latter carry sub-representations labelled by the eigenvalues of $L_3$ and $\mathfrak{c} k^2$. The eigenvalue problem for $\mathfrak{c} k^2$ 
	\beq
	\left[- \frac{i \hbar k}{\la} \dd{}{z} + \left(\frac{m^2 k^2}{2} - \frac{k^2}{\la^2} \right) I \right]  \psi(x,y,z) =  \frac{k p_z}{\la} \psi(x,y,z),
	\eeq
leads to the eigenfunctions $\psi(x, y, z) = F(x,y) \exp(i p_z z/\hbar)$ corresponding to the eigenvalue $k p_z/ \la$. Thus, the representation decomposes as a direct sum of sub-representations labelled by the two real numbers $m$ and $p_z$. Since $F(x,y)$ is an arbitrary function, these sub-representations on $L^2(\mathbb{R}^2_{xy})$ are infinite dimensional with the generators represented as:
	\beqs
	L_1 &=& k y, \quad L_2 = -k x, \quad L_3 = -m k I, \quad K_3 = -k I \cr 
	S_1 &=& -i \hbar \dd{}{x} - \frac{\la}{2} m k y, \quad S_2 = - i \hbar \dd{}{y} + \frac{\la}{2} m k x \quad \text{and} \cr
	S_3 &=& \left( p_z - \frac{k}{\la}\right) I - \frac{\la k}{2}(x^2 + y^2),
	\label{e:irreducible-representation-of-nilpotent-Lie-algebra}
	\eeqs 
which continue to satisfy the step-3 nilpotent Lie algebra (\ref{e:nilpotent-Lie-algebra}). Since there are no additional Casimirs, (\ref{e:irreducible-representation-of-nilpotent-Lie-algebra}) now furnishes a unitary irreducible representation of  (\ref{e:nilpotent-Lie-algebra}).

\chapter[Discussion]{Discussion}
\chaptermark{Discussion}
\label{chapter:Discussion}

In this thesis we have discussed the dynamics and integrability of a mechanical system describing a class of nonlinear screw-type wave solutions of a  scalar field theory dual to the 1+1D SU(2) principal chiral model (PCM). Unlike the PCM, this dual scalar field theory has a positive beta function and could serve as a toy model to study strongly coupled field theories with a perturbative  Landau pole. Recently, Rajeev and Ranken found a class of classical nonlinear wave solutions of this field theory. These novel screw-type continuous waves could play a role similar to solitary waves in other field theories.  They defined a consistent reduction of the field theory to this nonlinear wave sector, which is described by a mechanical system with three degrees of freedom \cite{R-R}. We call this mechanical system the Rajeev-Ranken model. 

In Chapter \ref{chapter:Introduction}, we motivated the study of the Rajeev-Ranken model starting from its field theoretic precursors and also summarized the major results of this thesis. In Chapter \ref{chapter:Nilpotent-field-theory-to-the-Rajeev-Ranken-model}, we introduced the Rajeev-Ranken (RR) model as a consistent reduction of the pseudodual scalar field theory. We discussed the Hamiltonian and Lagrangian formulations of the  PCM and its dual field theory. Furthermore, we compared their current algebras which are a semi-direct product of an $\mathfrak{su}(2)$ and an abelian algebra and a nilpotent current algebra. Finally, we obtained a consistent mechanical reduction of the scalar field theory by restricting it to the sector of  nonlinear screw-type wave solutions. In Chapter \ref{chapter:Hamiltonian-formulation-and-Liouville-integrability}, we investigated some integrable features of the classical RR model. The Liouville integrability of the model was discussed using Lax pairs and $r$-matrices leading to a complete set of conserved quantities in involution. Moreover, we found a Poisson pencil associated with the model. In Chapter \ref{chapter:Phase-space-structure-and-action-angle-variables}, we discussed the structure of the phase space, obtaining a foliation by invariant tori of various dimensions.  We classified all possible common level sets of conserved quantities and analyzed the nature of dynamics on them. We also found a set of action-angle variables for the Rajeev-Ranken model. In Chapter \ref{chapter:Quantum-Rajeev-Ranken-model-and-anharmonic-oscillator}, we discussed aspects of the quantum Rajeev-Ranken model by interpreting it as a quartic oscillator. This viewpoint helped us to quantize the model and separate variables in the Schr\"odinger equation. We analyzed the corresponding radial equation in a weak and a novel strong coupling limit to understand the properties of the quantized nonlinear wave. A more detailed summary of the results obtained in this thesis may be found in Section \ref{s:outline-and-summary-of-results}.

While working on the dynamics and integrability of the RR model, we wrote an expository article on the idea of Lax pairs and zero curvature representations in classical mechanical and continuum wave systems \cite{G-V-3, G-V-4, G-V-5}. In this article, we explain the idea of realizing a nonlinear evolution equation as a compatibility condition between a pair of linear equations by considering the examples of the harmonic oscillator, Toda chain \cite{Flaschka, Henon}, Eulerian rigid body \cite{L-L, H-F}, Rajeev-Ranken model \cite{G-V-1, G-V-2, G-V-6}, KdV equation \cite{G-G-K-M, M-G-K, Z-K} and the nonlinear Schr\"odinger equation \cite{F-T, Z-S-1, Z-S-2}. This introductory article can serve as a stepping stone to the vast literature on the theory of integrable systems \cite{A-S, Das, D-J, Dunajski, F-T, M-J-D, N-M-P-Z}.

\vspace{.5cm}

{\fl \bf Comparison with other models and further directions for research:} Comparing and contrasting the RR model and its parent scalar field theory with other (possibly integrable) mechanical systems and field theories is instructive and can help in discovering new features of these models. For instance, we were able to find a new Hamiltonian formulation for the Neumann model, which is an integrable system, by comparing it with the RR model. Moreover, we found a kinship between the EOM of the RR model and the Kirchhoff equations. The latter too is integrable and describes the motion of a rigid body in an ideal fluid. On the other hand, the parent scalar field theory can be viewed as a large level weak coupling limit of the 1+1D WZW model or as a pseudodual of the SU(2) principal chiral model. Comparing these models has and could continue to be instructive.

There are several directions of research which arise from this work. To begin with, there are classical aspects of the model that are yet to be addressed. For instance, we have not yet identified a bi-Hamiltonian formulation of the model. It would also be desirable to find an algebraic-geometric formulation based on the spectral curve and Jacobian  \cite{B-B-T}. This should give an alternate approach for obtaining the $r$-matrix of the model via the associated loop group. In addition, this approach should help in relating our Poisson bracket formulation to the Kostant-Kirillov bracket on the dual of the Lie algebra associated with the loop group. In another direction, bilinearization of the EOM of the scalar field theory in the sense of Hirota could help in discovering other classes of solutions. The study of the stability of various types of solutions of the model also needs further attention. The analysis of the quantum Rajeev-Ranken model is far from complete. A more detailed understanding of the spectrum of excitations is desirable. Going beyond our canonical quantization, we would like to explore quantum R-matrices and path integral approaches to the quantum theory.  The connection between the strong coupling limit and sub-Riemannian geometry (and its quantum counterpart) pointed out in \cite{R-R} is another possible direction for research. Finally, the possible extension of some of our results from the mechanical reduction to the scalar field theory is an interesting but challenging task. 


\appendix

\chapter[Compairson with the Neumann model]{Compairson with the Neumann model} \chaptermark{Compairson with the Neumann model}
\label{A:RR-model-and-Neumann-Model}

The EOM (\ref{e:EOM-LS}) and Lax pair (\ref{e:Lax-pair}) of the RR model have a formal structural similarity with those of the (N = 3) Neumann model. The latter describes the motion of a particle on $S^{N-1}$ subject to harmonic forces with frequencies $a_1, \cdots, a_N$ \cite{B-B-T}. In other words, a particle moves on  $S^{N-1} \subset \mathbb{R}^N$ and is connected by $N$ springs, the other ends of which are free to move on the $N$ coordinate hyperplanes. The EOM of the Neumann model follow from a symplectic reduction of dynamics on a $2N$ dimensional phase space with coordinates $x_1, \cdots, x_N$ and $y_1, \cdots, y_N$. The canonical PBs $\{ x_k, y_l \} = \del_{kl}$ and Hamiltonian
	\beq
	H = \frac{1}{4} \sum_{k \neq l} J_{kl}^2 + \half \sum_{k} a_k x_k^2 
	\label{e:Hamiltonian-Neumann}
	\eeq
lead to Hamilton's equations 
	\beq
	\dot x_k = - J_{kl} x_l \quad \text{and} \quad \dot y_k = - J_{kl} y_l - a_k x_k \quad (\text{no sum over $k$}).
	\eeq
Here, $J_{kl} = x_k y_l - x_l y_k$ is the angular momentum. Introducing the column vectors $X_k = x_k$ and $Y_k = y_k$ and the frequency matrix $\Om = \text{diag} (a_1, \cdots, a_N)$, Hamilton's equations become
	\beq
	\dot X = - J X \quad \text{and} \quad \dot Y = - J Y - \Om X.
	\eeq
It is easily seen that $X^t X$ is a constant of motion. Moreover, the Hamiltonian and PBs are invariant under the `gauge' transformation $(X, Y) \to (X , Y + \eps X)$ for $\eps \in \mathbb{R}$. Imposing the gauge condition $X^t (Y + \eps(t) X) = 0$ along with $X^t X = 1$ allows us to reduce the dynamics to a phase space of dimension $2(N-1)$. If we define the rank 1 projection $P = X X^t$ then $J = X Y^t - Y X^t$ and $P$ are seen to be gauge-invariant and satisfy the evolution equations
	\beq
	\dot J = [\Om, P] \quad \text{and} \quad \dot P = [P, J].
	\label{e:EOM-Neumann-angmom-proj}
	\eeq
The Hamiltonian (\ref{e:Hamiltonian-Neumann}) in terms of $J, P$ and $\Om$ becomes
	\beq
	H_{\rm Neu} =  \tr\left(-\frac{1}{4} J^2 + \half \Om P \right).
	\label{e:Hamiltonian-Neumann-JP} 
	\eeq
The PBs following from the canonical $x$-$y$ PBs 
	\beqs
	\{ J_{kl} , J_{pq} \} &=& \del_{kq} J_{pl} - \del_{pl} J_{kq}  + \del_{ql} J_{kp}  - \del_{kp} J_{ql}, \cr
	\{ P_{kl} , J_{pq} \} &=& \del_{kq} P_{pl} - \del_{pl} P_{kq}  + \del_{ql} P_{kp}  - \del_{kp} P_{ql} \;\; \text{and} \;\; \{ P_{kl} , P_{pq} \} = 0
	\label{e:PB-JP-Neumann}
	\eeqs
and the Hamiltonian (\ref{e:Hamiltonian-Neumann-JP}) imply the EOM (\ref{e:EOM-Neumann-angmom-proj}). This Euclidean Poisson algebra is a semi-direct product of the abelian ideal spanned by the $P$'s and the simple Lie algebra of the $J$'s.

Notice the structural similarity between the equations of the RR model (\ref{e:EOM-LS}) and those of the Neumann model (\ref{e:EOM-Neumann-angmom-proj}). Indeed, under the mapping $(L, S, K, \la) \mapsto (J, P, \Om, 1)$, the EOM (\ref{e:EOM-LS}) go over to (\ref{e:EOM-Neumann-angmom-proj}). The Lax pair for the Neumann model \cite{B-B-T}
	\beq
	L(\zeta) = -\Om + \ov \zeta J + \ov{\zeta^2} P \quad \text{and} \quad M(\zeta) = \ov \zeta P \quad \text{with} \quad \dot L = [M, L]
	\eeq
and that of the RR model $A_{\varepsilon}(\zeta) = -K + L/\zeta + S/ (\la \zeta^2)$ and $B(\zeta) = S/ \zeta$ (\ref{e:Lax-pair}) are similarly related for $\la = 1$. Despite these similarities, there are significant differences. 

(a) While $L$ and $S$ are Lie algebra-valued traceless anti-hermitian matrices, $J$ and $P$ are a real anti-symmetric and a real symmetric rank-one projection matrix. Furthermore, while $K$ is a constant traceless anti-hermitian matrix ($(ik/2) \sig_3$ for $\mathfrak{su}(2)$), the frequency matrix $\Om$ is diagonal with positive entries. 

(b) The Hamiltonian (\ref{e:Hamiltonian-Neumann-JP}) of the Neumann model also differs from that of our model (\ref{e: H-mechanical}) as it does not contain a quadratic term in $P$. However, the addition of $(1/4)\tr P^2$ to (\ref{e:Hamiltonian-Neumann-JP}) would not alter the EOM (\ref{e:EOM-Neumann-angmom-proj}) as $\tr P^2$ is a Casimir of the algebra (\ref{e:PB-JP-Neumann}).

(c) The PBs (\ref{e:PB-JP-Neumann}) of the Neumann model bear some resemblance to the Euclidean PBs (\ref{e:PB-semi-diect-tilde-LS}) of the RR model expressed in terms of the real anti-symmetric matrices $\tl S$ and $\tl L$ of Section \ref{s:Hamiltonian-mechanical}. Under the map $(\tl L, \tl S, \la) \mapsto (J, P, 1)$, the PBs (\ref{e:PB-semi-diect-tilde-LS}) go over to (\ref{e:PB-JP-Neumann}) up to an overall factor of $-1/2$. On the other hand, if we began with the $\{ \tl L_{kl}, \tl S_{pq} \}_{\varepsilon}$ PB implied by (\ref{e:PB-semi-diect-tilde-LS}) and then applied the map, the resulting $\{ J, P \}$ PB would be off by a couple of signs. These sign changes are necessary to ensure that the $J$-$P$ PBs respect the symmetry of $P$ as opposed to the anti-symmetry of $\tl S$. This also reflects the fact that the symmetry $\{ \tl S_{kl}, \tl L_{pq} \} =\{ \tl L_{kl}, \tl S_{pq} \}$ is not present in the Neumann model: $\{ J_{kl}, P_{pq} \} \neq \{ P_{kl}, J_{pq} \}$.

(d) Though both models possess nondynamical $r$-matrices, they are somewhat different as are the forms of the fundamental PBs among Lax matrices. Recall that the FPBs and $r$-matrix (\ref{e:r-matrix-semi-direct}) of the RR model, say, for the Euclidean PBs are (here, $k,l,p,q = 1,2$):
	\beq
	\left\{ A_{\varepsilon}(\zeta) \stackrel{\otimes}{,} A_{\varepsilon}(\zeta') \right\}_{\varepsilon} = \left[ r_{\varepsilon}(\zeta, \zeta') , A_{\varepsilon}(\zeta)\otimes I + I \otimes A_{\varepsilon}(\zeta') \right] \;\; \text{and} \;\; r_{\varepsilon}(\zeta, \zeta')_{k l p q} = -\frac{\la \: \del_{k q} \del_{lp}}{2 (\zeta - \zeta')}.
	\eeq
This $r$-matrix has a single simple pole at $\zeta = \zeta'$. On the other hand, the FPBs of the Neumann model may be expressed as a sum of {\it two} commutators
	\beq
	\{ L(\zeta) \stackrel{\otimes}{,} L(\zeta') \} = [ r_{12}(\zeta, \zeta'), L(\zeta) \otimes I ] - [ r_{21}(\zeta', \zeta) , I \otimes L(\zeta') ].
	\label{e:FPB-Neumann}
	\eeq
The corresponding $r$-matrices have simple poles at $\zeta = \pm \zeta'$ (here, $k,l,p,q = 1, \cdots, N$):
	\beq
	r_{12}(\zeta, \zeta')_{klpq} = -\frac{\del_{kq} \del_{lp}}{\zeta - \zeta'}  - \frac{\del_{kl} \del_{pq}}{\zeta + \zeta'}  
	\quad \text{and} \quad
	r_{21}(\zeta', \zeta)_{klpq} = -\frac{\del_{kq} \del_{lp}}{\zeta' - \zeta}  - \frac{\del_{kl} \del_{pq}}{\zeta' + \zeta} \neq -r_{12}(\zeta, \zeta')_{klpq}.
	\eeq	
Note that the anti-symmetry of (\ref{e:FPB-Neumann}) is guaranteed by the relation $r_{12}(\zeta, \zeta')_{klpq} = r_{21}(\zeta,\zeta')_{lkqp}$.
	
\vspace{.25cm}

{\fl \bf New Hamiltonian formulation for the Neumann model:} An interesting consequence of our analogy is a new Hamiltonian formulation for the Neumann model inspired by the nilpotent RR model PBs (\ref{e:PB-nilpotent-tilde-LS}). Indeed, suppose we take the Hamiltonian for the Neumann model as 
	\beq
	H = H_{\rm Neu} + \ov{4} \tr P^2 = \tr \left( -\ov{4} J^2 + \half \Omega P + \ov{4} P^2 \right)
	\eeq
and postulate the step-3 nilpotent PBs, 
	\beqs
	\{ P_{kl}, J_{pq} \}_{\nu} &=& - \del_{kq} \Omega_{pl} + \del_{pl} \Omega_{kq} - \del_{ql} \Omega_{kp} + \del_{kp} \Omega_{ql}, \cr
	\{ P_{kl} , P_{pq} \}_{\nu} &=& \del_{kq} J_{pl} - \del_{pl} J_{kq} - \del_{ql} J_{kp} + \del_{kp} J_{ql} \quad \text{and} \quad \{ J_{kl} , J_{pq} \}_{\nu} = 0,
	\label{e:nilpotent-PB-Neumann}
	\eeqs
then Hamilton's equations reduce to the EOM (\ref{e:EOM-Neumann-angmom-proj}). These PBs differ from those obtained from (\ref{e:PB-nilpotent-tilde-LS}) via the map $(\tl L, \tl S, \tl K, \la) \mapsto  (J, P, \Omega, 1)$ by a factor of $1/2$ and a couple of signs in the $\{P, P \}_{\nu}$ PB. As before, these sign changes are necessary since $P$ is symmetric while $\tl S$ is anti-symmetric. It is straightforward to verify that the Jacobi identity is satisfied: the only nontrivial case being $\{ \{ P, P \}, P \} + \rm{cyclic} = 0$ where cancellations occur among the cyclically permuted terms. In all other cases the individual PBs such as $\{ \{ P, J \}, J \}$ are identically zero. Though inspired by the $\mathfrak{su}(2)$ case of the RR model, the PBs (\ref{e:nilpotent-PB-Neumann}) are applicable to the Neumann model for all values of $N$.

\chapter[Relation to Kirchhoff's equations and Euler equations]{Relation to Kirchhoff's equations and Euler equations} \chaptermark{Relation to Kirchhoff's equations and Euler equations}
\label{B:Kirchhoff-equations}

Kirchhoff's equations govern the evolution of the momentum $\vec P$ and angular momentum $\vec M$ (in a body-fixed frame) of a rigid body moving in an incompressible, inviscid potential flow \cite{MT}. Here, $\vec P$ and $\vec M$ satisfy the Euclidean $\mathfrak{e}(3)$ algebra:
	\beq
	\{ M_a, M_b \} = \eps_{abc} M_c, \quad 
	\{ P_a, P_b \} = 0 \quad \text{and} \quad 
	\{ M_a, P_b \} = \eps_{abc} P_c. 
	\eeq
The Hamiltonian takes the form of a quadratic expression in $\vec P$ and $\vec M$ \cite{D-K-N}:
	\beq
	2 H = \sum a_i M_i^2 + \sum b_{ij} (P_i M_j + M_i P_j) + \sum c_{ij} P_i P_j.
	\label{e:Kirchhoff's-Hamiltonian}
	\eeq
The resulting equations of motion are 
	\beq
	\dot{\vec P} = \vec P \times \dd{H}{\vec M} \quad \text{and} \quad 
	\dot{\vec M} = \vec P \times \dd{H}{\vec P} + \vec M \times \dd{H}{\vec M}.
	\label{e:Kirchhoff's-equations}
	\eeq
Now taking $a_i =1, b_{ij} =0$ and $c_{ij} = \delta_{ij}$ and using the map $\vec M \mapsto -\vec L$ and $\vec P \mapsto \vec S - \vec K/\la$, we see that the Hamiltonian of the Kirchhoff model reduces to that of the Rajeev-Ranken model (\ref{e: H-mechanical}). However, unlike in the Kirchhoff model, $L$ and $\tl{S} = S - K/\la$ in the Rajeev-Ranken model satisfy a centrally extended $\mathfrak{e}(3)$ algebra following from Eq. (\ref{e:PB-SL-dual}):
	\beq
	\{ L_a, L_b \} = -\la \eps_{abc} L_c, \quad 
	\{ \tl{S}_a, \tl{S}_b \} = 0 \quad \text{and} \quad 
	\{ L_a, \tl{S}_b \} = -\la \eps_{abc}\left( \tl{S}_c + \frac{K_c}{\la} \right).
	\eeq
Thus, the equations of motion of the Rajeev-Ranken model (\ref{e:EOM-LS}) differ from those of the Kirchhoff model (\ref{e:Kirchhoff's-equations}). Nevertheless, this formulation implies that the equations of the Rajeev-Ranken model may be viewed as Euler-like equations for a centrally extended Euclidean algebra with the quadratic Hamiltonian $H = (L^2 + \tl{S}^2)/2$. 

Alternatively, if we use the dictionary $\vec M \mapsto -\vec L$ and $\vec P \mapsto \vec S$, then the Poisson algebras of both models are the same $\mathfrak{e}(3)$ algebra. The differences in their equations of motion may now be attributed to the linear term $\vec K \cdot \vec S/\la$ in the Rajeev-Ranken model Hamiltonian (\ref{e: H-mechanical}), which is absent in  (\ref{e:Kirchhoff's-Hamiltonian}). For more on the Kirchhoff model, its variants and their integrable cases, see for instance \cite{D-K-N, Sokolov, B-M-S}.

\chapter[RR equations as Euler equations for a nilpotent Lie algebra]{RR equations as Euler equations for a nilpotent Lie algebra}
\chaptermark{RR equations as Euler equations for a nilpotent Lie algebra}
\label{C:RR-equations-as-Euler-equations-for-a-nilpotent-Lie-algebra}

The equations of motion of the RR model $\dot L = [K, S]$ and $\dot S = \la [ S, L]$ may be viewed as the Euler equations for a nilpotent Lie algebra. Indeed, they follow from the quadratic Hamiltonian 
	\beq
	H = \frac{(S-K/\la)^2 + L^2}{2},
	\eeq
and the step-3 nilpotent Poisson brackets $\mathfrak{n}_3$:
	\beqs
	\{ L_a, L_b \} &=& 0, \quad \{S_a, S_b \} = \la \eps_{abc} L_c, \quad \{ S_a, L_b \} = -\eps_{abc} K_c \quad \text{and} \cr
	\{ K_a, K_b \} &=& \{ K_a, L_b \} = \{ K_b, S_b \} = 0. \qquad
	\eeqs
This algebra is a central extension by the generators $K_a$ of the step-2 nilpotent algebra
	\beq
	\mathfrak{n}_2: \quad \{ L_a, L_b \} = 0, \quad \{S_a, S_b \} = \la \eps_{abc} L_c, \quad \{ S_a, L_b \} = 0.
	\eeq
The $L_a$ form an abelian ideal of this latter algebra with three-dimensional abelian quotient $\mathfrak{n}_2/\mathfrak{l}$ which is generated by $S_a$.  As before, we take $K_3 = -k, K_{1,2} = 0$ so that $\mathfrak{n}_3$ is seven-dimensional with generators $(L_a, S_a)$ and the identity $\mathbf{I}$. The Hamiltonian is a quadratic form on this Lie algebra. If we use the basis $L_a, \tl S_a = S_a - K_a/ \la$ and $\mathbf{I}$ then the Hamiltonian is 
	\beq
	H = \half (\tl S^2 + L^2)
	\eeq
and corresponds to inverse inertia matrix ${\cal I}_{ij}^{-1} = {\rm Diag }(1,1,1,1,1,1, 0)$. The zero eigenvalue of ${\cal I}_{ij}^{-1}$ in the central direction can be made nonzero by adding a constant term to the Hamiltonian. Thus the RR model can be viewed as an Euler top for the nilpotent Lie algebra $\mathfrak{n}_3$. Similarly, the RR equations can also be viewed as Euler equations for a  centrally extended Euclidean algebra as mentioned in Appendix \ref{B:Kirchhoff-equations} and \cite{G-V-2}.

\chapter[Calculation of $\Tr A^4(\zeta)$ for the Lax matrix]{Calculation of $\Tr A^4(\zeta)$ for the Lax matrix}
\chaptermark{Calculation of $\Tr A^4(\zeta)$ for the Lax matrix}
\label{D:A^4}

In Section \ref{s:conserved-quantities} we found that the conserved quantities $\Tr A^n(\zeta)$ are in involution and obtained four independent conserved quantities ${\mathfrak{c}}, m, s$ and $h$ by taking $n = 2$. Here, we show that the conserved quantities following from $\Tr A^4(\zeta)$ are functions of the latter. We find that 
\scriptsize
	\beqs
	A^4 &=&  \bigg[ \zeta^8 (K_a K_b K_c K_d) 
				  - \zeta^7 (K_a K_b K_c L_d + L_a K_b K_c K_d + K_a L_b K_c K_d + K_a K_b L_c K_d) \cr
			   &+&  \zeta^6 \left(- \frac{K_a K_b K_c S_d}{\la} + L_a K_b K_c L_d + K_a L_b K_c L_d + K_a K_b L_c L_d \right. \cr  && \left.- \frac{S_a K_b K_c K_d}{\la} + L_a L_b K_c K_d - \frac{K_a S_b K_c K_d}{\la} + L_a K_b L_c K_d + K_a L_b L_c K_d - \frac{K_a K_b S_c K_d}{\la} \right) \cr
			   &+&  \zeta^5 \left(\frac{L_a K_b K_c S_d}{\la} + \frac{K_a L_b K_c S_d}{\la} + \frac{K_a K_b L_c S_d}{\la} \right. \cr && + \left. \frac{S_a K_b K_c L_d}{\la} - L_a L_b K_c L_d + \frac{K_a S_b K_c L_d}{\la} - L_a K_b L_c L_d - K_a L_b L_c L_d + \frac{K_a K_b S_c L_d}{\la} \right.\cr && + \left. \frac{S_a L_b K_c K_d}{\la} + \frac{L_a S_b K_c K_d}{\la} + \frac{S_a K_b L_c K_d}{\la} + \frac{K_a S_b L_c K_d}{\la} + \frac{L_a K_b S_c K_d}{\la} + \frac{K_a L_b S_c K_d}{\la} - L_a L_b L_c K_d\right) \cr
	           &+& \zeta^4 \left(\frac{S_a K_b K_c S_d}{\la^2} - \frac{L_a L_b K_c S_d}{\la} + \frac{K_a S_b K_c S_d}{\la^2} - \frac{L_a K_b L_c S_d}{\la} - \frac{K_a L_b L_c S_d}{\la} + \frac{K_a K_b S_c S_d}{\la^2} \right.\cr && - \left. \frac{S_a L_b K_c L_d}{\la} - \frac{L_a S_b K_c L_d}{\la} - \frac{S_a K_b L_c L_d}{\la} - \frac{K_a S_b L_c L_d}{\la} - \frac{L_a K_b S_c L_d}{\la} - \frac{K_a L_b S_c L_d}{\la} + L_a L_b L_c L_d \right. \cr && \left.\frac{S_a S_b K_c K_d}{\la^2} - \frac{S_a L_b L_c K_d}{\la} - \frac{L_a S_b L_c K_d}{\la} + \frac{S_a K_b S_c K_d}{\la^2} - \frac{L_a L_b S_c K_d}{\la} + \frac{K_a S_b S_c K_d}{\la^2}\right)\cr
	           &+& \zeta^3 \left(- \frac{S_a L_b K_c S_d}{\la^2} - \frac{L_a S_b K_c S_d}{\la^2} - \frac{S_a K_b L_c S_d}{\la^2} - \frac{K_a S_b L_c S_d}{\la^2} - \frac{L_a K_b S_c S_d}{\la^2} - \frac{K_a L_b S_c S_d}{\la^2} + \frac{L_a L_b L_c S_d}{\la} \right.\cr && - \left.\frac{S_a S_b K_c L_d}{\la^2} + \frac{S_a L_b L_c L_d}{\la} + \frac{L_a S_b L_c L_d}{\la} - \frac{S_a K_b S_c L_d}{\la^2} + \frac{L_a L_b S_c L_d}{\la} \right.\cr && - \left. \frac{K_a S_b S_c L_d}{\la^2} - \frac{S_a S_b L_c K_d}{\la^2} - \frac{S_a L_b S_c K_d}{\la^2} - \frac{L_a S_b S_c K_d}{\la^2}\right)\cr
	           &+& \zeta^2 \left(- \frac{S_a S_b K_c S_d}{\la^3} + \frac{S_a L_b L_c S_d}{\la^2} + \frac{L_a S_b L_c S_d}{\la^2} - \frac{S_a K_b S_c S_d}{\la^3} + \frac{L_a L_b S_c S_d}{\la^2} - \frac{K_a S_b S_c S_d}{\la^3} \right.\cr && + \left. \frac{S_a S_b L_c L_d}{\la^2} + \frac{S_a L_b S_c L_d}{\la^2} + \frac{L_a S_b S_c L_d}{\la^2} - \frac{S_a S_b S_c K_d}{\la^3}\right) \cr 
	           &+& \zeta \left(\frac{S_a S_b L_c S_d}{\la^3} + \frac{S_a L_b S_c S_d}{\la^3} + \frac{L_a S_b S_c S_d}{\la^3} + \frac{S_a S_b S_c L_d}{\la^3}\right) + \frac{S_a S_b S_c S_d}{\la^4} \bigg] t_a t_b t_c t_d.
	\eeqs 
\normalsize
Evaluating the trace yields the polynomial (\ref{e:trace-A4}) whose coefficients are functions of the conserved quantities ${\mathfrak{c}}, m , s$ and $h$, thus showing that $\Tr A^4$ does not lead to any new conserved quantity.

\chapter[Singularities of second order ordinary differential equations]{Singularities of second order ordinary differential equations}
\chaptermark{Singularities of second order ODEs}
\label{E:Singularities-of-second-order-ordinary-differential-equations}

\section[Singularities of second order ODEs]{Singularities of second order ODEs \sectionmark{Singularities of second order ODEs}}
\sectionmark{Singularities of second order ODEs}
\label{s:Singularities-of-second-order-ODEs}

We notice that the radial equation (\ref{e:radial-equation-dimensionless-coupling}) and its strong coupling limit (\ref{e:strong-coupling-radial-equation-dimensionless}) are second order homogeneous linear ODEs with rational coefficients. To place them in context, we summarize some features of the class of second order ODEs:
	\beq
	y'' + p(z) y' + q(z) y = 0,
	\label{e:second-order-ODE}
	\eeq
for the function $y(z)$. Here $p$ and $q$ are meromorphic functions on the complex plane. If both $p(z)$ and $q(z)$ are regular at a point $z_0$, then $z_0$ is an ordinary point and any other point is a singular point of the equation. A point $z_0 \neq \infty$ is a regular singularity if at least one of $p$ or $q$ has a pole at $z_0$ in such a way that if $p$ has a pole it is a simple pole and if $q$ has a pole it is at most a double pole. On the other hand, $z_0 \neq \infty$ is an irregular singularity if either $p$ has at least a double pole or $q$ has at least a triple pole \cite{A-S-S-W-P-D}.

The nature of the point at infinity $(z_0 = \infty)$ may be determined by writing (\ref{e:second-order-ODE}) in terms of $\zeta = 1/z$:
	\beq
	\frac{d^2 y}{d \zeta^2} + \left[ \frac{2}{\zeta} - \frac{1}{\zeta^2} p \left(\ov{\zeta} \right) \right] \frac{d y}{d \zeta} + \frac{1}{\zeta^4} q \left( \ov {\zeta} \right) y =  0.
	\label{e:transform-linear-ODE}
	\eeq
$z = \infty$ is called an ordinary point/regular/irregular singularity of (\ref{e:second-order-ODE}), if $\zeta = 0$ is a corresponding point of (\ref{e:transform-linear-ODE}). In other words, $z = \infty $ is an ordinary point if the Laurent series of $p$ and $q$ around $z = \infty$ are of the form $p(z) = 2/z + \cdots$ and $q(z) = q_4/z^4 + \cdots$. On the other hand, $z = \infty$ is a regular singularity if the Laurent series of $p$ and $q$ around $z = \infty$ satisfy any one of the following three conditions:
\begin{enumerate}
\item $p(z) = 2/z + \cdots$ and $q(z) = q_2/z^2 + q_3/ z^3 + \cdots$ with $q_2$ and $q_3$ not both zero,

\item $p(z) = p_1/z + \cdots$ with $p_1 \neq 2$ and $q(z) = q_4/z^4 + \cdots$ \text{or}

\item $p(z) = p_1/z + \cdots$ with $p_1 \neq 2$  and $q(z) = q_2/z^2 + q_3/ z^3 + \cdots$ with $q_2$ and $q_3$ not both zero.
\end{enumerate}
Finally, $z = \infty$ is an irregular singularity if it is neither an ordinary nor a regular singular point. Alternatively, it is an irregular singularity if either the Laurent series of $p$ around $z = \infty$ contains at least one nonnegative power ($z^0, z^1 \cdots$) or that of $q$ contains at least one power larger than $-2$ ($1/z, z^0, \cdots$). For example, $y'' + a y' + b y = 0$ with constants $a$ and $b$ not both zero has an irregular singularity at $z = \infty$, while every other point is an ordinary point. Indeed, the solution $y = c_1 e^{r_1 z} + c_2 e^{r_2 z}$ has an essential singularity at $z = \infty$. If $a$ and $b$ are both zero, then $y = c_1 z + c_2$ has a simple pole at $z = \infty$ which is a regular singular point. In general, at an ordinary point, the solution of (\ref{e:second-order-ODE}) is analytic. At a regular singular point, it is either analytic, has a pole of finite order or an algebraic or logarithmic branch point singularity. At an irregular singular point, the solution typically has an essential singularity \cite{A-W, B-O}. 

At a regular singularity $z_0 \neq \infty$ (if $z_0 = \infty$ we work with $\zeta = 1/z$), we may expand the solution in a Frobenius series $y = (z - z_0)^\rho \sum_{0}^{\infty} y_n (z - z_0)^n$ with the possible exponents $\rho = \rho_{1,2}$ determined by the indicial equation
	\beq
	\rho^2 + (A - 1) \rho + B = 0 \quad \text{where} \quad 
	A = \lim_{z \to z_0} (z - z_0) p(z)  \quad \text{and} \quad B = \lim_{z \to z_0} (z - z_0)^2 q(z).
	\label{e:Indicial-equation-general-form}
	\eeq
In fact, $A = B = 0$ iff $z_0$ is an ordinary point while $z_0$ is a regular singularity iff the limits exist with $A$ and $B$ not both zero. Moreover, if $|\rho_1 - \rho_2| = 1/2$, then $z_0$ is called an elementary regular singular point. Otherwise it is nonelementary \cite{Ince}.

\section[Poincar\'e rank and species]{Poincar\'e rank and species \sectionmark{Poincar\'e rank and species}}
\sectionmark{Poincar\'e rank and species}
\label{s:Poincare-rank-and-species}

The Poincar\'e {\it rank} of a singular point is a measure of its irregularity. For definiteness, suppose $z_0 = \infty$ is a singular point, then its rank $g$ is defined as 
	\beq
	g = 1 + {\rm max} \left( K_1, \frac{K_2}{2} \right) \quad \text{where} \quad p(z) = \mathcal{O}(z^{K_1}) \quad \text{and}  \quad q(z) = \mathcal{O}(z^{K_2}) \quad \text{as} \quad z \to \infty.
	\label{e:rank-of-irregular-singularity}
	\eeq
If $z_0 = \infty$ is a regular singularity its rank is either zero or a negative (half) integer while for an irregular singularity it can be $1/2, 1, 3/2, \cdots$. Notably, it is possible to double the rank of an irregular singular point via the quadratic transformation $z = w^2$. Thus it is possible to restrict to integer ranks. For example, the equation $ y'' + (1/z) y' + (1/z) y = 0$ has an irregular singularity of rank 1/2 at $z = \infty$. Upon putting $z = w^2$, it becomes $y''- (1/2w) y' + 4 y = 0$, which has a rank 1 irregular singularity at $w = \infty$. Moving away from $\infty$, a singular point $z_0 \neq \infty$ of (\ref{e:second-order-ODE}) is said to have the rank $g = 1+ {\rm max} (K_1, K_2/2)$ if $p(z)$ and $q(z)$ have poles of order $K_1 + 2$ and $K_2 + 4$ respectively (see Eq.~(\ref{e:transform-linear-ODE})). It is sometimes also useful to define the {\it species} of an irregular singularity as twice its rank \cite{A-S-S-W-P-D}. 

The rank controls the asymptotic behaviour of solutions to (\ref{e:second-order-ODE}) at an irregular singular point. If $z_0= \infty$ is an irregular singular point of integer rank $g$, then we have the exponential asymptotic behaviour
	\beq
	y (z) \sim \exp[A_g z^g + A_{g-1} z^{g-1} + \cdots + A_1 z] Y(z), \quad \text{where} \quad
	Y(z) = z^{-\rho}\sum_{n \geq 0} y_n z^{- n}.
	\label{e:rank-asymptotic-relation-solution}
	\eeq
There is a loose resemblance between the rank of an irregular singularity and the genus of an entire function.

\section[Invariance of rank]{Invariance of rank \sectionmark{Invariance of rank}}
\sectionmark{Invariance of rank}
\label{s:Invariance-of-rank}

Though the quadratic transformation $z = w^2$ doubles the rank of an irregular singularity, there is a class of transformations that preserve it. In fact, under a fractional linear transformation $w = (az + b)/(cz + d)$, the coefficients of (\ref{e:second-order-ODE}) remain meromorphic and the rank of a singularity remains unchanged though its location may be altered. 

On the other hand, under a linear change of dependent variable $y(z) = F(z) a(z)$, (\ref{e:second-order-ODE}) becomes
	\beq
	a''(z) + \left( 2 \frac{F'(z)}{F(z)} + p(z) \right) a'(z) + \left( \frac{F''(z)}{F(z)} + p(z) \frac{F'(z)}{F(z)} + q(z) \right) a(z) = 0.
	\label{e:linear-transformed-ODE}
	\eeq
To ensure that (\ref{e:linear-transformed-ODE}) has meromorphic coefficients, we will restrict to functions of the form $F = z^{\mu} R_1 e^{R_2}$, where $\mu$ is real and $R_{1,2}(z)$ are rational functions. For definiteness, let us suppose that $z = \infty$ is a rank $g$ irregular singular point of (\ref{e:second-order-ODE}). Suppose, further that $R_2(z) \sim z^n$ as $z \to \infty$. Then we find that $z = \infty$ continues to be a rank $g$ irregular singularity of (\ref{e:linear-transformed-ODE}) provided $n \leq g$. In particular, there is no restriction on $\mu$ or $R_1$. This restriction on $n$ is understandable in view of the connection between the rank and the asymptotic behaviour in (\ref{e:rank-asymptotic-relation-solution}).

Interestingly, it is possible to create irregular singular points through the confluence of regular singularities  \cite{Ince}. For instance, the coalescence of two elementary regular singular points produces a nonelementary regular singularity, while the merger of three elementary regular singularities gives an irregular singularity of species 1 (rank 1/2). More generally, an irregular singularity of species $r$ is formed by the coalescence of $r+2$ elementary regular singular points.

\section[Ince's classification]{Ince's classification \sectionmark{Ince's classification}}
\sectionmark{Ince's classification}
\label{s:Ince-classification}

Ince introduced a classification of the ODEs (\ref{e:second-order-ODE}) based on the number and nature of singularities. Such an equation is said to be of type $[a, b, c_i, d_j, \cdots]$, if $a$ is the number of elementary regular singular points, $b$ is the number of nonelementary regular singular points, and $c, d, \cdots$ are the number of irregular singularities of species $i, j, \cdots$. For example, the hypergeometric equation is denoted $[0, 3, 0]$ as it has three nonelementary regular singularities at $z = 0, 1$ and  $\infty$. The confluent hypergeometric equation is denoted $[0, 1, 1_2]$. It has  a regular (nonelementary) singularity at zero and an irregular singularity of rank 1 at $\infty$ formed by the coalescence of regular singularities at $1$ and $\infty$. The Heun equation is denoted $[0, 4, 0]$ as it has four nonelementary regular singular points \cite{A-S-S-W-P-D}. When two of them coalesce we get the confluent Heun equation $([0, 2, 1_2])$ with an irregular singularity of rank one. The biconfluent Heun equation $([0, 1, 1_4])$ has an irregular singularity of rank 2 formed by the merger of three nonelmentary regular singular points. The Lam\'e equation for ellipsoidal harmonics is of type $[3, 1, 0]$, it has three elementary regular singularities and one nonelementary regular singularity at infinity. Thus, it can be viewed as a special case of the Heun equation or as a confluent form of an equation of type $[5,0,0]$.

\chapter[Goldstone mode of the RR model]{Goldstone mode of the RR model}
\chaptermark{Goldstone mode of the RR model}
\label{F:Goldstone-mode-of-the-RR-model}

\vspace{-1cm}

We have seen in Chapter \ref{chapter:Quantum-Rajeev-Ranken-model-and-anharmonic-oscillator} that the Rajeev-Ranken model can be treated as a three dimensional anharmonic oscillator with a  quartic plus quadratic potential in the Darboux coordinates $R_1$ and $R_2$. Using cylindrical symmetry, we arrive at a one dimensional problem for the radial coordinate $r$ with the effective potential $U(r)$ (see Eqn. (\ref{e:potential-radial-quad-quartic})). As mentioned in Section \ref{s:Quantum-RR-model}, when $\alpha < 0$, the potential $U(r) = \alpha r^2 + \beta r^4$ has a nonzero minimum at $r_{*} = \sqrt{ -\alpha/ 2 \beta}$. This corresponds to a family of degenerate minima of the potential along a circle on the $R_1$-$R_2$ plane: $R_1^2 + R_2^2 = r_{*}^2 = 8 p_z/\la k - 2 m^2 - 8 \mu/\la^2  > 0$.
 
One could mistakenly treat these minima of the potential as static solutions of the anharmonic oscillator. This is incorrect because of the additional term proportional to the angular momentum in the Hamiltonian (\ref{e:Hamiltonian-quadratic-quartic-Cartesian}) of the anharmonic oscillator. The true static solutions are given by the solutions of the EOM (here $x,y,z = R_{1,2,3}$ and $p_{x,y,z} = kP_{1,2,3}$):
	\small
	\beqs
	\dot{x} &=& p_x - \frac{\la m k y}{2}, \quad \dot{y} = p_y + \frac{\la m k x}{2}, \quad \dot{z} = p_z - \frac{\la k}{2}(x^2 + y^2), \cr
	\dot{p_x} &=& -\frac{\la m k p_y}{2} - \left( \frac{\la^2 m^2 k^2}{8} -\frac{\la k p_z}{2} + \frac{k^2}{2} \right) 2 x -\frac{\la^2 k^2}{2}(x^2 + y^2) x \cr
	\dot{p_y} &=& \frac{\la m k p_x}{2} - \left( \frac{\la^2 m^2 k^2}{8} -\frac{\la k p_z}{2} + \frac{k^2}{2} \right) 2 y -\frac{\la^2 k^2}{2}(x^2 + y^2) y \quad \text{and} \quad \dot{p_z} = 0,
	\eeqs
	\normalsize
with $\dot R_{1,2,3} = 0$ and $\dot P_{1,2,3} =0$. It is possible to show that there is a one parameter family of static solutions parametrized by arbitrary real values of $R_3(t) = R_3$, while the other variables vanish: $R_{1,2} = P_{1,2,3} = 0$. These static solutions\footnote{In terms of the $L$-$S$ variables, this corresponds to a single point on the static submanifold $\Sigma_2$ (see Section \ref{s:Static-and-trigonometric-solutions}), where $L_3 = -m k$ and $S_3 = -k/\la$. This is because the $L$-$S$ phase space does not capture the $R_3$ degree of freedom.} that lie on the $z$ axis are degenerate in energy which is given by $E= m^2 k^2 /2$. Thus we would expect a zero mode/`Goldstone mode' where $R_3$ varies slowly, while the other variables remain zero. However, we do not expect Goldstone bosons in the field theory since it is two-dimensional.

These degenerate static solutions of the RR model correspond to a family of static solutions of the scalar field theory given by $\phi(x,t) = R_3 t_3 + m K x$, where $t_3 = \sig_3/2i$ and $K = -k t_3$. Thus $\phi(x,t) = (R_3 - m k x) (\sig_3/ 2i)$ is linear in $x$ and points exclusively in the third internal direction. 

\chapter[Asymptotic behaviour of the strong coupling radial equation]{Asymptotic behaviour of the strong coupling radial equation}
\chaptermark{Asymptotic behaviour: strong coupling radial equation}
\label{G:Asymptotic-behaviour-in-the-strong-coupling-limit}

The radial Schr\"odinger equation of the quantum Rajeev-Ranken model in the strong coupling limit (\ref{e:strong-coupling-radial-equation-dimensionless})
	\beq
	 \rho''(\tl r) + \ov{\tl r} \rho'(\tl r) - \left( \frac{l^2}{\tl r^2} + \frac{\tl g^2}{4}\left(\tl r^2 + \tl r^4 \right)  - \tl g^2 \tl E_2 \right) \rho(\tl r) = 0,
	\label{e:strong-coupling-radial-equation-appendix}
	\eeq
has an irregular singular point at $\tl r = \infty$. To find the asymptotic behaviour of the radial wavefunction, we make the substitution $\rho(\tl r) = \exp(S(\tl r))$. This anticipates that the leading asymptotic behaviour is of exponential type. In terms of $S(\tl r)$, the radial equation becomes:
	\beq
	S''(\tl r) + (S'(\tl r))^2 +  \ov{\tl r} S'(\tl r) - \left( \frac{l^2}{\tl r^2} + \frac{\tl g^2}{4}\left(\tl r^2 + \tl r^4 \right)  - \tl g^2 \tl E_2 \right) = 0.
	\label{e:strong-coupling-radial-equation-dimensionless-change-of-variable}
	\eeq
We make the `slowly varying' assumption that $S''(\tl r) \ll (S'(\tl r))^2$ as $\tl r \to \infty$, which will be seen to be a self-consistent assumption. For large $\tl r$, the quartic term in (\ref{e:strong-coupling-radial-equation-dimensionless-change-of-variable}) dominates, so the `asymptotic radial equation' is
	\beq
	 S'(\tl r)^2  \sim  \frac{\tl g^2}{4} \tl r^4.
	 \eeq
This implies 
	\beq
	S'(\tl r) \sim \pm \frac{\tl g}{2} \tl r^2 \quad \text{or } \quad S(\tl r) = \pm \frac{\tl g}{6} \tl r^3 + c(\tl r),
	\eeq
where the constant of integration $c$ of the limiting asymptotic radial equation is allowed depend on $\tl r$, in order to incorporate the subleading behaviour as $\tl r \to \infty$. For consistency, $c(\tl r)$ must satisfy the condition $c(\tl r) \ll \tl g \tl r^3/6$ as $\tl r \to \infty$. For normalizability, the eigenfunction $\rho(\tl r) \to 0$ as $\tl r \to \infty$. Thus, we must choose the negative sign for $S(\tl r) = -\tl g \tl r^3/6 + c(\tl r)$. Substituting this in the radial equation (\ref{e:strong-coupling-radial-equation-dimensionless-change-of-variable}) we get
	\beq
	c''(\tl r) + c'(\tl r)^2 - \left( \tl g \tl r^2 - \ov{\tl r} \right) c'(\tl r) - \frac{3}{2} \tl g \tl r - \frac{l^2}{\tl r^2} - \frac{\tl g^2}{4} \tl r^2 + \tl g^2 \tl E_2 = 0.
	\eeq
As before, in the limit $\tl r \to \infty$ we may use the inequalities $c''(\tl r) \ll \tl g \tl r$ and $c'(\tl r)^2 \ll \tl g \tl r^2 c'(r)$ to obtain an asymptotic equation for $c(\tl r)$:
	\beq
	\tl g \tl r^2 c'(\tl r) \sim -\frac{3}{2} \tl g \tl r - \frac{\tl g^2}{4} \tl r^2.
	\eeq
Thus we have,
	\beq
	c(\tl r) \sim -\frac{3}{2} \ln r - \frac{\tl g}{4} \tl r + {\rm constant}.
	\eeq
This gives the leading asymptotic behaviour of $\rho(\tl r)$:
	\beq
	\rho(\tl r) \sim a (\tl r) \tl r^{-\frac{3}{2}} e^{-\frac{\tl g}{2} \left( \frac{\tl r^3}{3} + \frac{\tl r}{2} \right)} \quad \text{as} \quad \tl r \to \infty.
	\label{e:asymptotic-behaviour-rho}
	\eeq
Here, we once again allowed the constant $a$ to depend on $\tl r$ in order to allow for further subleading behaviour.

{\fl \bf Remark:}	 Note that just like (\ref{e:strong-coupling-radial-equation-appendix}), the radial equation for the variable coefficient $a(\tl r)$ also has an irregular singular point of rank 3 at $\tl r = \infty$. In fact, the radial equation (\ref{e:strong-coupling-radial-equation-appendix}), under the substitution (\ref{e:asymptotic-behaviour-rho}) becomes
	\beq
	a''(\tl r) -\left( \frac{2}{\tl r} + \frac{\tl g}{2} +  \tl g \tl r^2 \right) a'(\tl r) + \left( \frac{1}{\tl r^2} \left( \frac{9}{4} - l^2 \right) + \frac{\tl g}{2 \tl r} + \tl g^2 \left( \ov{16} + \tl E_2 \right) \right) a(\tl r) = 0.
	\label{e:radial-equation-asymptotic}
	\eeq
By the transformation, $\tl r = 1/s$ we obtain the equation
	\beq
	\frac{d^2 a}{ds^2} + \left( \frac{4}{s} + \frac{\tl g}{2 s^2} + \frac{\tl g}{s^4} \right) \frac{da}{ds} + \left( \frac{\left(\frac{9}{4} - l^2 \right)}{s^2} + \frac{\tl g}{2 s^3} + \frac{\tl g^2 \left( \ov{16} + \tl E_2 \right)}{s^4} \right) a(s)= 0,
	\eeq
which by the rules of Appendix \ref{E:Singularities-of-second-order-ordinary-differential-equations} has an irregular singularity of rank 3 at $s = 0$. 

\chapter[Frobenius method for strong coupling limit: Local analysis]{Frobenius method for strong coupling limit: Local analysis}
\chaptermark{Frobenius method for strong coupling limit: Local analysis}
\label{H:Frobenius-method-for-strong-coupling-limit-Local-analysis}

We know that the radial equation (\ref{e:strong-coupling-radial-equation-dimensionless}) has a regular singular point at $\tl r = 0$. We consider a series solution around this point of the form: 
	\beq
	\rho(\tl r)  = \sum_{n = 0}^{\infty} \rho_n \tl r^{\eta + n}.
	\eeq
Substituting this in (\ref{e:strong-coupling-radial-equation-dimensionless}) gives
	\beqs
	\sum_{n=0}^\infty \rho_n (\eta + n)(\eta + n -1) \tl r^{\eta + n -2} &+& \frac{1}{\tl r}  \sum_{n=0}^\infty \rho_n ( \eta + n ) \tl r^{\eta + n - 1} \cr
	&-& \left( \frac{l^2}{\tl r^2} + \frac{\tl g^2}{4} (\tl r^2 + \tl r^4) - \tl g^2 \tl E_2 \right) \sum_{n = 0}^\infty \rho_n \tl r^{\eta + n} = 0.
	\eeqs
We rewrite this equation as 
	\beq
	\sum_{n=0}^\infty ((\eta + n)^2 - l^2)  \rho_n \tl r^{\eta + n -2} - \frac{\tl g^2}{4}\left( \sum_{n =4}^\infty \rho_{n-4} \tl r^{\eta + n -2} + \sum_{n = 6}^\infty \rho_{n-6} \tl r^{\eta + n -2} \right)
		+ \tl g^2 \tl E_2 \sum_{n = 2}^\infty \rho_{n -2} \tl r^{\eta + n - 2}  = 0.
	\eeq
From this, we have the indicial exponents $\eta = \pm l$. Choosing $\eta = l$ in order that the normalizability condition (\ref{e:normalizability-condition-radial-bound-state}) is satisfied, we get the four-term recursion relation (\ref{e:recursion-relations-around-zero-rho}).



\end{document}